# Towards automated symptoms assessment in mental health

Maxim Osipov

Wolfson College

University of Oxford

Supervised by
Prof. Gari Clifford
Dr. David Clifton
Submitted: Michaelmas Term, 2015

July 24, 2016

This thesis is submitted to the Department of Engineering Science, University of Oxford, in partial fulfilment of the requirements for the degree of Doctor of Philosophy

“ – *А я действительно похож на галлюцинацию. Обратите внимание на мой профиль в лунном свете*, – кот полез в лунный столб и хотел еще что-то говорить, но его попросили замолчать, и он, ответив: – *Хорошо, хорошо, готов молчать. Я буду молчаливой галлюцинацией*, – замолчал.”

— Михаил Булгаков, *Мастер и Маргарита*

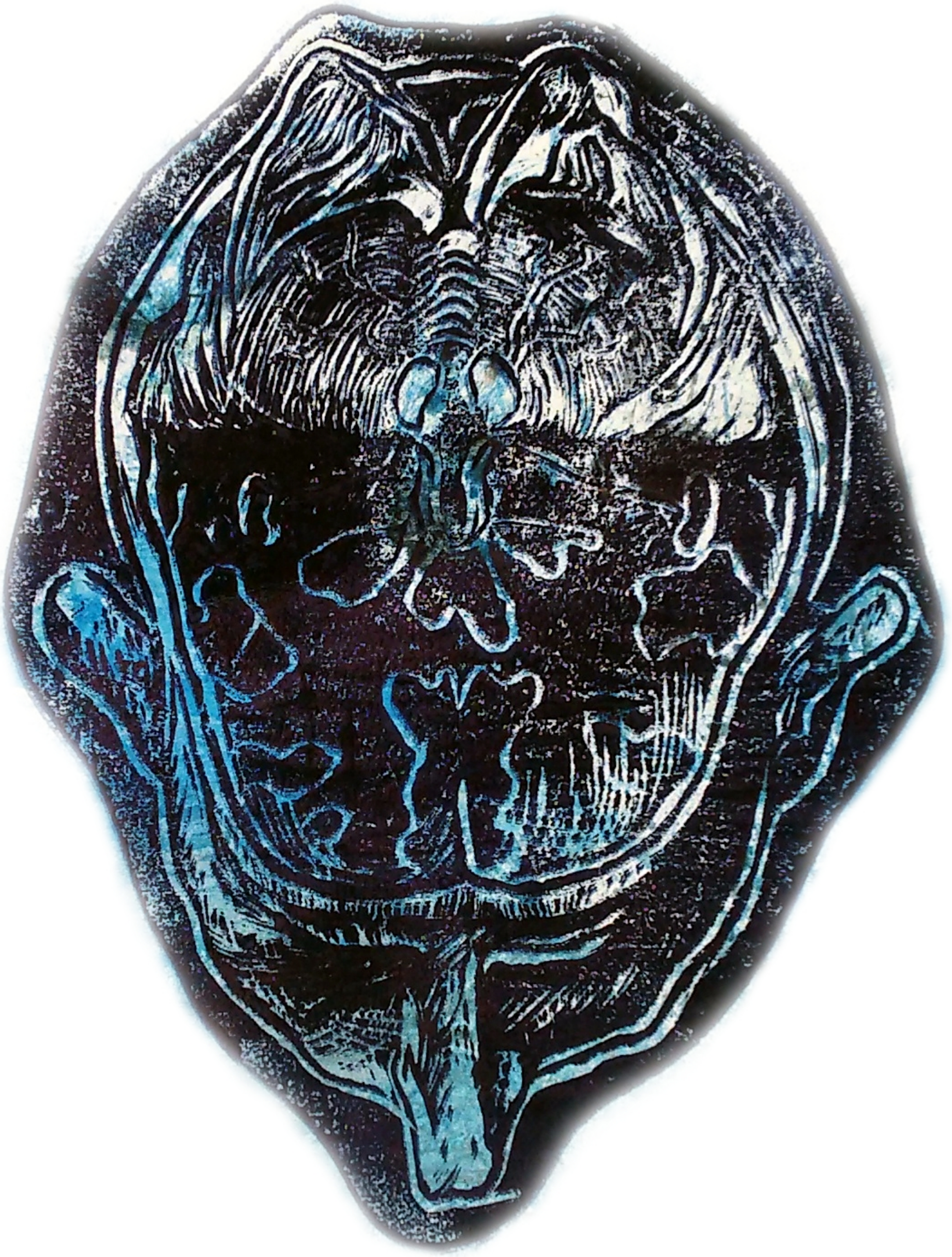

*“The Unknown Mind” by Unknown Artist. Edited.*

# Declaration

I declare that this thesis is entirely my own work, and except where otherwise stated, describes my own research.

Maxim Osipov,

Wolfson College

# Dedication

To my parents and Alina.

**Towards automated symptoms assessment in mental health**

Maxim Osipov
Wolfson College

Doctor of Philosophy
Michaelmas Term, 2015

## Abstract

Mental and behavioural disorders introduce a significant burden on society, estimated to account for 12% of the global burden of disease, with approximately 450 million suffering from them every day, and only a small number of those getting any treatment. The situation will worsen with time, with unipolar depressive disorders predicted by the World Health Organisation to become the leading cause of disabilities by 2030.

Mental disorders affect primarily the mind and the brain, leading to pathological changes in emotions or cognition. Although clinical manifestations of different mental disorders may vary, Liddle *et al.* suggested five principal symptom dimensions, including reality distortion, disorganisation, psychomotor, mood and anxiety dimensions. For assessment of symptoms in clinical practice, the structured clinical interview, alongside standard questionnaires are used, but in many cases are not providing a reliable and objective diagnostic tool due to the complexity of the assessed phenomena.

Activity and motion analysis has the potential to be used as a diagnostic tool for mental disorders. However, to-date, little work has been performed in turning stratification measures of activity into useful symptom markers. The research presented in this thesis has focused on the identification of objective activity and behaviour metrics that could be useful for the analysis of mental health symptoms in the above mentioned dimensions. Particular attention is given to the analysis of objective differences between disorders, as well as identification of clinical episodes of mania and depression in bipolar patients, and deterioration in borderline personality disorder patients.

A principled framework is proposed for mHealth monitoring of psychiatric patients, based on measurable changes in behaviour, represented in physical activity time series, collected via mobile and wearable devices. The framework defines methods for direct computational analysis of symptoms in disorganisation and psychomotor dimensions, as well as measures for indirect assessment of mood, using patterns of physical activity, sleep and circadian rhythms.

An extensive mHealth software tracking system was constructed, and data collected from over 100 individuals. Using the developed framework, the accuracy of differentiation between healthy controls and bipolar disorder was 67%, healthy controls and borderline personality disorder 70%, and bipolar vs. borderline personality disorder 80%. For identification of clinical states of euthymia, mania and depression the accuracy of differentiation of euthymia and mania was 80%, euthymia and depression 85% and mania and depression 90%, when using leave-one-out cross-validation. For personalised mood models, the mean absolute error of symptom scores estimation was in the range of 1.36 to 3.32 points, this corresponds to the ranges reserved in psychiatric questionnaires for a unique identifiable mood state (4-5 points).

Finally, the methods were applied to a new data set (schizophrenia patients and matched controls) and were shown to be 95.3% accurate using leave-one-out cross-validation at classifying the cohort. Both physiological as well as activity features were relevant for classification of this cohort, and so the hypothesis that heart rate added additional predictive power was tested. The combination of HR and locomotor activity features provided almost a 10% increase in classification accuracy above using locomotor features alone, and almost a 17% increase over using heart rate based features alone.

The approach of computational behaviour analysis, proposed in this thesis, has the potential for early identification of clinical deterioration in ambulatory patients, and allows for the specification of distinct and measurable behavioural phenotypes, thus enabling better understanding and treatment of mental disorders.

# Acknowledgements

This thesis would not have been possible without the very valuable input of my supervisors, Prof. Gari Clifford and Dr. David Clifton and that of other collaborators at Oxford, including:

- Prof. John Geddes
- Prof. Guy Goodwin
- Dr. Katharina Wulff
- Dr. Amy Bilderbeck
- Dr. Christopher Hinds
- Dr. Kate Saunders
- and Dr. Yashar Behzadi, Proteus Biomedical, Redwood City, CA, United States

The author would also like to acknowledge the following funding bodies, which provided project or fellowship funding for the research presented in this thesis:

- Hill Foundation Scholarship
- RCUK Digital Economy Programme grant number EP/G036861/1 (Oxford Centre for Doctoral Training in Healthcare Innovation)
- Wellcome Trust Centre Grant No. 098461/Z/12/Z (Sleep, Circadian Rhythms and Neuroscience Institute)
- EPSRC grant EP/K020161/1 (Multiscale markers of circadian rhythm changes for monitoring of mental health)
- Wellcome Trust Strategic Award No. 102616/Z (Collaborative Oxford Network for Bipolar Research to Improve Outcomes)

# Contents

# List of Figures

# List of Tables

# List of Abbreviations

AASM: American Academy of Sleep Medicine.
AL: Activity Level.
ANS: Autonomous Nervous System.
ASRM: Altman Self-Rating Scale for Mania, 5-item scale.
AMoSS: Automatic Monitoring of Mental Symptoms.
BART: Balloon Analogue Risk Task.
BBLOCKS: Bayesian BLOCKS.
BDI-II: Beck Depression Inventory, second edition.
BIS-11: Barratt Impulsiveness Scale, 11-item scale.
BMI: Body Mass Index.
BOCPD: Bayesian Online Change Point Detection.
CBT: Cognitive Behaviour Therapy.
CV: Coefficient of Variation.
DFA: Detrended Fluctuation Analysis.
DMT: Delayed Memory Task.
DNA: Deoxyribonucleic Acid.
DSM: The Diagnostic and Statistical Manual of mental disorders.
DTW: Dynamic Time Warping.
ECG: Electrocardiogram.
EDA: Electrodermal Activity.
EDHMM: Explicit Duration HMM.
EEG: Electroencephalogram.
EMA: Ecological Momentary Assessment.
EN: Euclidean Norm.
ENMO: Euclidean Norm Minus One.
EPSRC: The Engineering and Physical Science Research Council.
FMRI: Functional Magnetic Resonance Imaging.
FN: False Negative.
FP: False Positive.
FP7: Seventh Framework Programme.

GAD-7: Generalised Anxiety Disorder, 7-item scale.

GBI: General Behaviour Inventory.

GP: Gaussian Process.

GPS: Global Positioning System.

HAMD: Hamilton Depression rating scale.

HF: High Frequency of HRV (0.15 to 0.4 Hz).

HFEN: High-pass Filtered Euclidean Norm.

$HFEN_+$: High-pass Filtered Euclidean Norm minus one.

HMM: Hidden Markov Model.

HR: Heart Rate.

HRV: Heart Rate Variability.

IMT: Immediate Memory Task.

IQR: Inter-Quartile Range.

IS: Intra-daily Stability.

IV: Inter-daily Variability.

HTTPS: Hyper Text Transfer Protocol, Secure.

QIDS-SR16: Quick Inventory of Depressive Symptomatology, Self-Report, 16-item scale.

L5: Least active 5 hours of activity.

LAMP: Linux, Apache, MySQL and PHP.

LASSO: Least Absolute Shrinkage and Selection Operator.

LF: Low Frequency of HRV (0.04 to 0.15 Hz).

LOOCV: Leave One Out Cross Validation.

M10: Most active 10 hours.

MAE: Mean Absolute Error.

MAR: Missing At Random.

MB: Megabyte.

MCAR: Missing Completely At Random.

MD5: Message-Digest algorithm, version 5.

MEMS: Micro-Electro-Mechanical System.

MET: Metabolic Equivalent.

MI: Multiple Imputation.

MIP: Mean Immobility Period.

ML: Maximum Likelihood.

MNAR: Missing Not At Random.

mRMR: minimum Redundancy Maximum Relevance.

MRS: Mania Rating Scale.

MSE: Multiscale Entropy.

NHS: The UK National Health Service.

NIH: The US National Institutes of Health.

OLS: Ordinary Least Squares.

PAEE: Physical Activity Energy Expenditure.

PET: Positron Emission Tomography.

PHQ: Patient Health Questionnaire.

PNS: Parasympathetic Nervous System.

PROMIS: Patient-Reported Outcome Measure Information System.

RA: Relative Amplitude.

RBF: Radial Basis Function.

REC: Research Ethic Committee.

REM: Rapid Eye Movement.

RMDM: Recursive Mean Difference Maximisation.

RMSSD: Root Mean Square Successive Difference.

RR: R-to-R peak distance in HR.

RRGEN: McSharry-Clifford 24-hour RR-tachogram generator.

SCID: The Structured Clinical Interview for DSM Disorders.

SD: Standard Deviation.

SE: Squared Exponential.

SI units: International System of units.

SMOTE: Synthetic Minority Over-sampling TEchnique.

SNS: Sympathetic Nervous System.

SSRI: Selective Seratonin Reuptake Inhibitor.

SVM: Support Vector Machines.

TE: Transfer Entropy.

TN: True Negative.

TP: True Positive.

VI: Vulnerability Index.

VLF: Very Low Frequency of HRV (0.0033 to 0.04 Hz).

WHO: World Health Organisation.

UK: The United Kingdom.

ULF: Ultra-Low Frequency of HRV (below 0.0033 Hz).

USM: Under-Sampling Majority.

YMRS: Young Mania Rating Scale.

# Glossary

- Affect - the experience of feeling or emotion, usually indicated by facial, vocal or gestural behaviour.
- Actigraphy - non-invasive method of capturing rest and activity patterns using electronically recorded activity from a wearable unit (actigraph).
- Activity count - unit of activity measurement by an actigraph.
- Accelerometer - a device measuring g-force, including both acceleration and gravity components.
- Bipolar disorder - a mental disorder characterised by periods of elevated and depressed mood.
- Borderline personality disorder - a personality disorder characterised by unstable mood, behaviours, relationships and self-perception.
- Catatonia - a state of motor immobility or stupor, often present in schizophrenia.
- Circadian rhythm - a biological process with oscillations of about 24 hours.
- Chlorpromazine - antipsychotic medication, primarely used in schizophrenia.
- Dynamic programming - a method of solving a complex problem by breaking it down into a set of simpler sub-problems, and re-using solutions of those.
- Dysphoria - a severe state on unease or dissatisfaction.
- Entropy - a measurement of the disorder or randomness of a system.
- Euthymia - a relatively asymptomatic state in mood disorders.
- Impulsivity - a psychological construct, describing tendency to act on the spur of the moment.
- Imputation - substitution of some value for missing data.
- Locomotor activity - a variety of methods animals or humans use to change their location.
- Metabolic equivalent (MET) - measure of the intensity of physical activity, where 1 MET is the amount of energy produced by a person seated at rest.

- Mixed state - a state with simultaneous manifestation of symptoms of depression and mania (usually in bipolar disorder).
- Negative symptom - in schizophrenia, a deficit of normal emotional reactions or other behaviours.
- Sleep apnoea - a sleep disorder, characterised by pauses or severe shallowness in breathing.
- Pieso-electric - an effect of accumulating electrical charge in material in response to mechanical stress.
- Positive symptom - in schizophrenia, a symptom that can be described as behaviour usually not present in normal people (such as hallucinations).
- Psychophysiology - branch of psychology, studying physiological bases of psychological processes.
- Psychomotor - a behaviour, represented in change of physical activity patterns originated from mental processes.
- Schizophrenia - a mental disorder characterised by reality distortion.
- Schizoaffective disorder - a mental disorder characterised by the presence of symptoms of both schizophrenia and mood disorder.
- Stationarity - a property of a system, described by stability of statistical characteristics over time.
- Symptom - a departure from a normal function noticeable by a patient (as opposite to observed by clinician only).
- Syndrome - a combination of symptoms and signs, correlated with each other and a specific disease.
- Transcendental - a phenomena, independent of experience.
- Unipolar depressive disorder - a mental disorder, characterised by a severe and persistent low mood.
- Zeitgeber - an external stimulus that synchronises an organism's biological rhythms to 24 hour cycle.

# Chapter 1

# Introduction

## 1.1 Motivation

Mental and behavioural disorders place a significant burden on society, estimated to account for 12% of the global burden of disease, with one in four people experiencing mental health problems during their lifetime and an estimated 450 million suffering from them every day, although only a small fraction receive treatment, due to the lack of diagnosis (Sayers, 2001). The situation is predicted to get worse with time, and according to the World Health Organisation Global Burden of Disease report unipolar depressive disorders will be the single leading cause of disabilities by 2030, from $3^{rd}$ place in 2004 (Mathers et al., 2008).

In the United Kingdom (UK) mental disorders (including dementia) affect approximately 8.7 million people with direct costs to the UK National Health Service (NHS) of £22.5 billion per year (McCrone, 2008). This number is expected to increase to 9.88 million people, at a cost of £32.6 billion by 2026. Mental illness accounts for nearly half of all ill health for people aged under 65, and the cost of lost employment is estimated at £26.1 billion, with many people not in contact with NHS services and not receiving any treatment (35% of those with depression and 51% with anxiety disorders) (Layard, 2012).

Mental health issues are also associated with physical illnesses, with stress being related to the common cold (Cohen et al., 1991) and delays in wound healing (Kiecolt-Glaser et al., 1998). Mental health issues, in particular depression and bipolar disorder, are also linked with obesity (McElroy et al., 2004), suggesting a range of well-known co-morbidities, including cardiovascular problems, sleep apnoea and increased risk of cancer and diabetes. A common complication of mental health illnesses is suicide. It is estimated that by 2020 1.5 million people a year will commit suicide and up to 30 million will attempt to do so (Bertolote and Fleischmann, 2002).

## 1.2 The problem of diagnosis

The diagnosis and management of mental disorders heavily depends on clinical judgement (Lobbestael et al., 2010) and patients' self-reports, making it a challenging task in situations when patients' cognition is impaired. There is a high degree of similarity between symptoms of different mental disorders (Green, 2006; Cannon et al., 1997) and in some cases it is very difficult, even for a trained clinician, to distinguish between disorders with similar manifestations and thus prescribe the correct treatment (Henry et al., 2001).

Collecting symptoms history is the key part of the diagnostic process, however assessing history during interviews is difficult due to substantial recall bias (Ben-Zeev et al., 2012). Numerous mental health scales have been suggested to bring objectivity into the diagnostic and therapeutic process (Rush et al., 2008). But there is evidence, that reliance on such scales only may also lead to mistaken conclusions (Shedler et al., 1993), so more objective diagnostic criteria may be necessary to create a reliable diagnostic system, and potentially define better disorder boundaries, predictive of treatment response and accurately capturing the mechanisms of mental dysfunction (Insel et al., 2010).

## 1.3 The problem of deterioration

Most mental disorders, such as schizophrenia or bipolar disorder, are life-long conditions often involving relapses and remitting, and cannot be cured completely (Lehman et al., 2004; Geddes and Miklowitz, 2013). Treatment strategy in mental health focuses on avoiding relapses, by starting treatment at the first signs of deterioration. And in many cases, early warning signs can be detected long before the full relapse, enabling early intervention for management of the disorder (Molnar et al., 1988; Klosterkötter et al., 2001).

However, early warning signs of relapse are often missed by patients and clinicians alike, identification of relapse signature in ambulatory patients is not routinely performed in clinical practice, and if performed at all is often based on self-reports and affected by compliance and recall bias issues (Miklowitz et al., 2012). A more systematic and objective approach to management of psychiatric disorders with a focus on early warning signs may therefore be useful in clinical practice.

From the treatment perspective, most psychiatric medications have quite severe side effects. For example the most widely used mood stabiliser, lithium, may lead to tremor and weight gain as side effects, and require careful dose adjustment to avoid poisoning (Schou et al., 1970). Side effects of antipsychotic medication include sedation, weight gain and sexual dysfunctions (Fleischhacker et al., 1994). Reducing the duration of pharmacological treatment by providing it only when first signs of relapse are detected would help to avoid unnecessary inconvenience for patients and improve compliance and outcomes (Tandon et al., 2009; Miklowitz et al., 2012; Geddes and Miklowitz, 2013).

## 1.4 Approach of this work

Automated and objective assessment of mental health may provide more timely interventions and prevent or decrease the severity of significant events. Since the classical work of Lacey in 1959 (Lacey, 1959), psychophysiology has been a heavily debated approach to introduce objectivity into the diagnostic and psychotherapeutic process. Such measures as physical activity, heart rate, electrodermal activity, skin temperature, muscle potentials, as well as clinical indexes based on these measures and features derived from them, may provide valuable information to assess the patients' mental state (Cacioppo, 1991).

While there is some scepticism among practising clinicians about the possibility of automating the diagnostic process (Steiner, 2011), psychophysiological measures are extensively used by researchers to analyse mental conditions, including aggression (Scarpa and Raine, 1997), psychopathy (Rothemund et al., 2012), conduct problems (Lorber, 2004), schizophrenia (Wulff et al., 2012) and others. There are also recent examples of activity-based biomarkers for predicting treatment response (McCall, 2015), thus potentially establishing actigraphy as an important clinical tool in psychiatry.

However, most existing research focuses on small subsets of psychophysiological variables, collected in controlled laboratory settings (Hornero et al., 2006; Minassian et al., 2010; Walther et al., 2014). In order to create a complete picture of the patients' state, longitudinal measurements capturing data from patients in normal living conditions may be required. Such an approach would potentially allow continuous identification of clinical state and deterioration in patients with mental health issues, and thus enable timely intervention (Paradiso et al., 2010; Bardram et al., 2012).

This work proposes a multidimensional framework for continuous ambulatory assessment of symptoms in mental health. The framework is based on evaluation of behaviour patterns, extracted from physical activity recordings, collected using off-the-shelf wearable and mobile devices, so it is potentially suitable for prodrome identification and early detection of deterioration in psychiatric patients.

## 1.5 Contributions

The main contribution of this work is in the introduction of a principled framework for analysis of mental health symptoms, based on objective behavioural features. The framework is based on the five-cluster model of symptoms (Liddle, 2001) and suggests objective measures for symptoms in psychomotor, disorganisation and mood clusters, with a potential to extend to the remaining two clusters (reality distortion and anxiety) by using physiological information, such as heart rate and electrodermal activity.

The framework integrates methods from physical activity research and sleep medicine (including analysis of sleep and circadian rhythm disturbances). The analysis approach is designed to deal with non-stationary human activity data by identifying and treating independently subjects' physiological states, such as sleep and wakefulness, using the Hidden Semi-Markov Models, as well as using non-stationarity and duration of stationary activity segments as features itself.

The multi-dimensional model allows for analysis and differentiation between similar disorders, with bipolar disorder and borderline personality disorder addressed in this work. It also allows for a more objective and complete description of clinical deterioration, applicable across a wide range of disorders.

Novel activity analysis methods are introduced based on definitions of mental health symptoms, including Multiscale Entropy-based activity disorganisation assessment, Detrended Fluctuation Analysis of stationary activity segment durations, obtained using the Bayesian Online Change Point Detection, and assessment of day-to-day activity patterns variability by measuring the Dynamic Time Warping distance between successive 24-hour activity patterns. These analysis methods were shown to be useful in describing clinically relevant changes in behaviour patterns, and in classification between healthy controls, biploar disorder and borderline personality disorder, as well as healthy controls and schizophrenia patients.

## 1.6 List of publications

This thesis was presented in the following publications:

Osipov, M., Behzadi, Y., Kane, J. M., Petrides, G., and Clifford, G. D. (2015). Objective identification and analysis of physiological and behavioral signs of schizophrenia. *Journal of Mental Health*, 24(5):276–282. Appears in Chapters 5 and 9.

Palmius, N., Osipov, M., Goodwin, G. M., Saunders, K., Bilderbeck, A. C., Tsanas, A., and Clifford, G. D. (2014). A multi-sensor monitoring system for objective mental health management in resource constrained environments. In *Appropriate Healthcare Technologies for Low Resource Settings (AHT 2014)*, pages 4–4. Institution of Engineering and Technology. Appears in Chapter 3.

Osipov, M., Wulff, K., Foster, R. G., and Clifford, G. D. (2013). Multiscale entropy of physical activity as an objective measure of activity disorganization in a context of schizophrenia. In *IX Congreso Internacional de Informática en Salud 2013*, pages 1–9. Appears in Chapters 2 and 5.

Roebuck, A., Monasterio, V., Gederi, E., Osipov, M., Behar, J., Malhotra, A., Penzel, T., and Clifford, G. D. (2013). A review of signals used in sleep analysis. *Physiological Measurement*, 35(1):R1–R57. Appears in Chapter 5.

The following publications are in preparation:

Osipov, M., Clifford, G.D. Evaluation of change point detection algorithms for human physical activity analysis. In preparation for *IOP Physiological Measurement.* Based on Chapter 4.

Osipov, M., Saunders, K.E.A., Clifford, G.D. Assessment of behaviour impulsivity using physical activity change point detection. In preparation for *IEEE Journal of Biomedical and Health Informatics.* Based on Chapters 4 and 5.

Osipov, M., Springer, D.B., Clifford, G.D. Automatic bed time segmentation using Explicit Duration Hidden Markov Models. In preparation for *IEEE Transactions on Biomedical Engineering.* Based on Chapter 4.

# Chapter 2

# Automating symptoms assessment in mental health

## 2.1 Brief history of physical activity and physiology in assessment of mental health

### 2.1.1 The roots

Mental disorders affect primarily mind and brain, leading to pathological changes in emotions or cognition. Unfortunately, until very recently there were no instruments to allow us to look into a living brain for a better understanding of pathologies related to mental health problems, to understand the etiology and processes behind mental disorders. And since it is not possible to cut open a living human brain, in order to understand the pathologies related to mental health problems researchers had to rely mostly on post-mortem analysis, studies on animal models and psychopharmacological evidence (Liddle, 2001). Very limited possibilities to observe living human brain activity were provided by EEG, and more recently a better resolution picture has become possible with imaging techniques, such as PET and FMRI, marking a new era in neuroscience (Andreasen, 1988), but to date none of these methods are diagnostically specific.

From a clinical perspective, there is still no way to look into the disorder-related activity of a living brain and mind continuously and for an extended period of time, as may be needed for analysis and assessment of symptoms for clinical intervention.

While direct observation of mind and brain is not really feasible, could we possibly rely on secondary measurements, based on changes in physical activity, behaviours or bodily processes, that are induced by changes in mental state? In other words, can we identify measurable objective signs of a person's mental health? And is there instrumentation that would allow us assess these signs continuously and without imposing a significant burden on the patient?

The first question brings us to a discussion of the relationship between the mind and the body, originating from Greek philosophers - the so called "mind-body problem". Plato believed, that the mind is completely independent from the body and the physical world, so no information measurable from the body could help us with understanding the mind. But already in the times of Plato many philosophers considered that the mind and the body are tightly connected, and affect each other. Such conclusions were often derived from observations of mind disorders. For example the treatise ”On the Sacred Disease“ from the Hippocratic Corpus (Jones, 1923) describes epilepsy as based mostly on physiological causes originating in the brain and manifesting itself with certain physical phenomena, thus eliminating the transcendental component of mental illness:

> *"The hands are paralysed and twisted when the blood is still, and is not distributed as usual. The eyes roll when the minor veins are shut off from the air and pulsate. The foaming at the mouth comes from the lungs; for when the breath fails to enter them they foam and boil as though death were near."*

Hippocrates of Cos in his work "On Regimen in Acute Diseases" (Jones, 1923) takes a step forward, linking lifestyle to both physical and mental health, and declaring that changes in behaviour can induce changes in health and mind.

These ideas were further developed by Diocles of Crystus and Galen, who described in some case reports the physiological and behavioural manifestations of mental illness, such as insomnia and irregular heart rate, probably providing a first documented example of heart rate variability analysis in assessment of mental disorders (Jackson, 1969):

> *"I was called in to see a woman who was stated to be sleepless at night and to lie tossing about from one position into another... After I had diagnosed that there was no bodily trouble, and that the woman was suffering from some mental uneasiness... I applied my hand to her wrist and noticed that her pulse had suddenly become extremely irregular. This kind of pulse indicates that the mind is disturbed."*

Similar ideas were expressed by Aristotle, who described emotions as psycho-physical phenomena, for example anger as *"seething heat in the region of the heart"* (Aristotle, 1984). As with many other fields of science, the understanding of mind, cognition and emotions remained relatively unchanged since antiquity until the $19^{th}$ century, when the concepts of cognition and emotions were developed further based on collected empirical evidences.

The major step forward in the understanding of mind and behaviour was performed by Charles Darwin, who postulated the evolutionary origin of emotional expression as mostly acquired serviceable habit, however sometimes originating purely from the constitution of the nervous system. In his work "The expression of the emotions in man and animals" (Darwin, 1872) Darwin declared three principles of expression:

> 1. *The principle of serviceable associated habits.*
> 2. *The principle of antithesis.*

3. *The principle of actions due to the constitution of the nervous system, independently from the first of the will, and independently to a certain extent of habit.*

These principles did not only declare the tight connection between emotions and their physical manifestations, but also the evolutionary nature of emotions, the tight interconnection of cognitive and affective actions and involuntary (not originating from evolutionary causes or cognitive factors) actions, originating solely from the nervous system.

On the other side of the ocean, probably the earliest modern attempt to associate mental state and physical manifestation was performed by William James (James, 1884), who postulated the theory of emotions, independently also developed by Carl Lange ("James-Lange theory"). The theory suggested that emotions arise in response to specific stimuli, which evoke physiological arousal, such as change of heart rate, electrodermal activity or specific muscle reaction, and this arousal in turn leads to specific emotion.

The James-Lange theory was soon challenged by Walter Bradford Cannon and Philip Bard ("Cannon-Bard theory"), who localised the centre of emotional expression in the brain and verified that physiological changes follow the emotional activation and arousal does not have to occur before emotion in a heartbreaking and disturbing to modern sensibilities experiment on 52 cats (46 survived long enough to prove the theory) (Bard, 1928).

This is where the modern era in understanding the connection between body and mind began. Most further research on interactions between mind and body, in particular in the field of physiological reactions to psychological processes, was developed using experimental methods, and in more narrow areas, such as psychophysiology, psychiatry, and other specific areas of neuroscience and medicine.

### 2.1.2 The modern era

The presence of the conceptual framework for linking bodily processes and mind has led to intensified research in the field. The major driving force behind scientific progress was the development of new medical instrumentation, which would allow analysis of certain aspects of emotional activation with high accuracy and without placing a significant burden on a person, and which made the application of experimental methods possible.

One of the earliest successful attempts to use physiology for the analysis of the mental state led to the creation of the polygraph, invented in 1921 by John Augustus Larson (Larson, 1923). It is probably the most controversial and widely adopted tool for forensic psychiatry. A polygraph measures several physiological signs, such as blood pressure, respiration and skin conductivity and allows the operator to differentiate between genuine and deceptive answers by visualising physiological response to emotional activation.

Such an early and successful application of physiological methods to psychological applications ignited significant interest, but the field of psychophysiology was defined only in 1955 with the creation of "The Psychophysiology Newsletter" by Albert F. Ax and the Society for Psychophysiological Research later in 1959. John I. Lacey, one of the pioneers of psychophysiology in 1959 proposed the use of physiological measures as a clinical tool for mental disorders, specifically in the psychotherapeutic process (Lacey, 1959). However, in the same paper he also expressed concerns about the usefulness of simple physiological measures:

> *"...as far as present evidence goes, individual indicant concordance is so low that single autonomic measures cannot be used to unequivocally rank-order the "arousal-value" of different stimuli for a given individual, or the arousability of different individuals".*

Despite the initial work on the introduction of physiological instrumentation in clinical psychiatry (Gerard, 1949; Altschule, 1953; Lacey, 1959), physiological measurements were not widely adopted in the field. Research suggested a great degree of variability in individual responses and difficulties in assessing the ground truth (mental state). A good critical statement on the subject was provided by John M. Hinton in 1966 (Hinton, 1966), listing the five causes of inconclusive results in physiological assessment of mental phenomena:

> *"**First**, the hypothesised relationship between the physiological and the psychological state may be mistaken. **Secondly**, the physiological phenomenon may be measured by an unreliable method. **Thirdly**, the psychological state under consideration may be ill-defined or insusceptible to accurate measurement. Even if the quality of the mental state could be agreed, few would claim complete reliability for the various rating scales. Nor can it be assumed that different subjects placed in the same experimental situations or given identical stimuli will experience equal psychological changes. **A fourth reason** for inconsistent psycho-physiological relationships is the influence of other variables besides the selected emotion upon the bodily state. **A fifth cause** for the lack of uniformity in results is the idiosyncrasies of the individual's habitual state and his way of reacting."*

All these facts diminished the attention of clinical specialists from the field and left it mostly as a playground for research, where novel data acquisition techniques and analysis methods were developed. With the improvement of medical instrumentation, data analysis methods and introduction of new measurement modalities, some of the problems mentioned by Hinton were solved, making the longitudinal monitoring of activity and physiology feasible, and potentially enabling the creation of personalised models of psychophysiological response.

In particular, the introduction of non-invasive actigraphic units has enabled long-term assessment of sleep, physical activity and rest-activity cycles in research settings since the early 90's (Witting et al., 1990). In 1995 Martin H. Teicher noted that activity and motion analysis has a good potential to be used as a diagnostic tool for mental disorders (Teicher, 1995). However until now the problem has not been addressed properly in clinical practice, due to difficulties introducing instrumentation for continuous physiology assessment and low patient compliance with research-grade instrumentation.

### 2.1.3 The future

Personal activity monitoring devices have now become consumer products. Communication protocols allow continuous upload, visualisation and analysis of data in real time, and algorithms have been developed to identify types of activity patterns from acceleration data. Personal activity monitors now often integrate physiologic monitoring tools, such as heart rate and electrodermal activity sensors, appropriate for estimation of average activity and energy expenditure (El-Amrawy and Nounou, 2015). This creates an opportunity for a more complete and personalised assessment of physiological and behavioural correlates of mental symptoms, with the potential to overcome the difficulties encountered by researchers and psychiatrists in previous years.

One of the enabling factors for the new increasing interest in applications of physiological and behavioural monitoring for diagnostic and therapeutic process in psychiatry was triggered by the large scale adoption of mobile devices and the rapid rise of wearable technologies (Mohr et al., 2014). Most commercial companies on the market, such as Fitbit and Jawbone, were launched less than 10 years ago (Fitbit was founded in 2007 and Jawbone UP lifestyle tracker in 2011). Despite the initial scepticism, fitness tracking and the so-called "quantified self" approach were quickly adopted by a large number of people.

The success of these devices was to a great extent based on the variety of actigraphy data analysis techniques, developed by researchers during previous years, and the evidence base on the importance of activity and sleep for people's health (Ancoli-Israel et al., 2003; Faulkner and Biddle, 2013). However, in the past, research was constrained by technological limitations, such as limited battery life and data storage capacity, with the main results created around very coarse-grained analysis methods, such as minute-by-minute sleep/wake segmentation, sedentary, light, moderate and vigorous physical activity identification, or cosinor analysis of circadian rhythms. These analysis methods were adopted by commercial device manufacturers. For the analysis of mental health, more elaborate data analysis methods may be required (Agelink et al., 2002; Hauge et al., 2011; Indic et al., 2011; Walther et al., 2014), and the rapid rise of commercially available and inexpensive activity tracking solutions enables researchers to conduct large scale studies, which are necessary to validate the effectiveness of these methods in ambulatory settings.

Furthermore, given the gradual introduction of physiology and environment sensors (heart rate, electrodermal activity and ambient light) into the "quantified self" technologies toolkit, as well as capabilities of mobile phone sensors to capture social activity patterns, many of the problems mentioned in Section 2.1.2 can be solved now by using simple and commercially available multi-sensor solutions. There is a growing need for the development and validation of new sensor fusion-based analysis methods, which are capable of using the great amount of diverse data collected from mobile and wearable sensors. In the field of psychiatry, these data need to be translated into clinically useful and useable information, helping both patient and clinical specialist to better understand the problem and improve the management of disorders (Prociow et al., 2012; Torous et al., 2015; Naslund et al., 2015).

Using this information, and benefiting from the wide penetration of activity and physiology monitoring tools, integrated into smart phone applications and wearable devices (Migo et al., 2014), it may be possible to address the problem of continuous activity and physiology monitoring for the early identification of symptoms and clinical states of psychiatric patients.

## 2.2 Diagnostic criteria for mental disorders

Mental health problems are very complex phenomena, often with poorly understood etiology, unclear diagnostic criteria and blurred boundaries between different conditions. Frequently, the very definition of a disorder has a historic origin and it is debatable if such a definition is correct and such a separate diagnostic entity exists (Dahl, 1985; Maj, 1998; Angst et al., 2003). The main reference for the diagnosis of psychiatric disorders is the "Diagnostic and Statistical Manual of Mental Disorders" by the American Psychiatric Association, with the fifth edition released in 2013 (American Psychiatric Association, 2013) (DSM-5). The DSM-5 defines a mental disorder as follows:

> *"A mental disorder is a syndrome characterised by clinically significant disturbance in an individual's cognition, emotion regulation, or behaviour that reflects a dysfunction in the psychological, biological, or developmental processes underlying mental functioning. Mental disorders are usually associated with significant distress or disability in social, occupational, or other important activities. An expectable or culturally approved response to a common stressor or loss, such as the death of a loved one, is not a mental disorder. Socially deviant behaviour (e.g., political, religious, or sexual) and conflicts that are primarily between the individual and society are not mental disorders unless the deviance or conflict results from a dysfunction in the individual, as described above."*

It can be seen that the definition of a mental disorder is based on socially unacceptable disturbances in behaviour, resulting in a disability and not caused by socially approved external factors. The definition includes a large amount of contextual and subjective information, making the diagnostic process a challenging task, especially given that cognition and insight are often impaired in psychiatric patients.

In order to understand the problem in better detail, let us take a look at three typical examples of mental disorders (which are the focus of this thesis), affecting different aspects of emotions, cognition and behaviour, including:

- **Bipolar disorder**, characterised by instability in mood with variations between very depressed and highly exited states with intermediate euthymic periods, usually on a time scale of months or years.
- **Borderline personality disorder**, manifesting itself in impulsivity of behaviour and instability of interpersonal relationships, self-image and affect.
- **Schizophrenia**, characterised by severe disturbance in cognition, such as presence of delusions, hallucinations, and disorganisation in thinking or behaviour.

These three disorders represent important classes of serious mental illnesses in the DSM-5 classification (American Psychiatric Association, 2013), affecting predominantly mood, behaviour and cognition. In order to provide the reader with a better understanding of complexity of behavioural phenomena characterising these mental illnesses, the complete DSM-5 diagnostic criteria for those are provided in the next chapters.

### 2.2.1 Bipolar disorder

Excluding the earliest definitions of bipolar disorder by Avicenna (Gruner, 1970) and Gao Lian (Carpenter, 1990), the first modern attempts to describe the bipolar disorder belong to Jules Baillarger (Baillarger, 1880) and Jean-Pierre Falret (Falret, 1854), followed by Emil Kraepelin, who defined the term "manic depressive psychosis" (Kraepelin, 1987). Carl Gustav Jung has described the bipolar disorder without psychotic episodes and has contributed to our modern understanding, where bipolar type I (with psychotic episodes) and type II (without psychotic episodes) are defined as distinctly separate (Jung, 1921; Angst and Marneros, 2001).

The lifetime prevalence of bipolar disorder ranges from 0.3% to 1.5% depending on country (Weissman, 1996), some studies estimating it as high as up to 5% (Akiskal et al., 2000), with sex ratios almost equal (Weissman, 1996). Rapid cycling (at least four alterations of depression and mania per year) appears in 20% of patients during the course of the disorder and mixed states occur in 40% of bipolar patients (Akiskal et al., 2000). The lifetime risk of suicide attempts in bipolar disorder is 29.2% (Chen and Dilsaver, 1996) or up to 20–30 times greater than that for the general population (Pompili et al., 2013).

The DSM-5 (American Psychiatric Association, 2013) defines the following diagnostic criteria of bipolar disorder. For simplicity, only the bipolar type I is presented here, given that type II diagnostic criteria are essentially the same, but with relaxed requirements for symptoms manifestation and duration of episodes. The DSM-5 approaches the diagnosis of bipolar disorder from the perspective of the presence of manic and depressive episodes, matching certain criteria.

#### 2.2.1.1 Manic episode

A manic episode in the DSM-5 (American Psychiatric Association, 2013) is defined as follows, with at least one manic episode in a lifetime required for the diagnosis of bipolar disorder (type I):

A. *A distinct period of abnormally and persistently elevated, expansive, or irritable mood and abnormally and persistently increased goal-directed activity or energy, lasting at least 1 week and present most of the day, nearly every day (or any duration if hospitalisation is necessary).*

B. *During the period of mood disturbance and increased energy or activity, three (or more) of the following symptoms (four if mood is only irritable) are present to a significant degree and present a noticeable change from usual behaviour:*

1. *Inflated self-esteem or grandiosity.*
2. *Decreased need for sleep (e.g., feels rested after only 3 hours of sleep).*
3. *More talkative than usual or pressure to keep talking.*
4. *Flight of ideas or subjective experience that thoughts are racing.*
5. *Distractibitilty (i.e., attention too easily drawn to unimportant or irrelevant external stimuli), as reported or observed.*
6. *Increase in goal-directed activity (either socially, at work or school, or sexually) or psychomotor agitation (e.e., purposeless non-goal-directed activity).*
7. *Excessive involvement in activities that have a high potential for painful consequences (e.g., engaging in unrestrained buying sprees, sexual indiscretions, or foolish business investments).*

C. *The mood disturbance is sufficiently severe to cause marked impairment in social or occupational functioning or to necessitate hospitalisation to prevent harm to self or others, or there are psychotic features.*

D. *The episode is not attributable to the physiological effects of a substance (e.g., a drug of abuse, a medication, other treatment) or to another medical condition.*

#### 2.2.1.2 Major depressive episode

A major depressive episode is defined in the DSM-5 (American Psychiatric Association, 2013) as follows:

*A. Five (or more) of the following symptoms have been present during the same 2-week period and represent a change from previous functioning; at least one of the symptoms is either (1) depressed mood or (2) loss of interest or pleasure.*

1. *Depressed mood most of the day, nearly every day, as indicated by either subjective report (e.g., feels sad, empty, or hopeless) or observation made by others (e.g., appears tearful).*
2. *Markedly diminished interest or pleasure in all, or almost all, activities most of the day, nearly every day (as indicated by either subjective account or observation).*
3. *Significant weight loss when not dieting or weight gain (e.g., a change of more than 5% of body weight in a month), or decrease or increase in appetite nearly every day.*
4. *Insomnia or hypersomnia nearly every day.*
5. *Psychomotor agitation or retardation nearly every day (observable by others; not merely subjective feelings of restlessness or being slowed down).*
6. *Fatigue or loss of energy nearly every day.*
7. *Feelings of worthlessness or excessive or inappropriate guilt (which may be delusional) nearly every day (not merely self-reproach or guilt about being sick).*
8. *Diminished ability to think or concentrate, or indecisiveness, nearly every day (either by subjective account or as observed by others).*
9. *Recurrent thoughts of death (not just fear of dying), recurrent suicidal ideation without a specific plan, or a suicide attempt or*

*a specific plan for committing suicide.*

*B. The symptoms cause clinically significant distress or impairment in social, occupational, or other important areas of functioning.*

*C. The episode is not attributable to the psychological effects of a substance or another medical condition.*

### 2.2.2 Borderline personality disorder

Borderline personality disorder was initially described by Theophile Bonet in 1684 as well as Charles H. Hughes and Irving C. Rosse and was called "borderline insanity" (Ritschel and Kilpela, 2014). The modern definition of the borderline personality disorder originates from Emil Kraepelin, who called it "excitable personality" (Kraepelin, 1921) with features, closely resembling the modern definition by the DSM-5 (American Psychiatric Association, 2013).

Borderline personality disorder affects about 1-2% of people, but introduces a significant burden on healthcare systems due to patients' behaviour patterns, including self-harm and suicidal tendencies. The suicide rate among patients is very high, with up to 10% of patients committing suicide (Lieb et al., 2004). The borderline personality disorder was considered to be more prevalent in women (70%) (Widiger and Weissman, 1991), however more recent findings indicate no gender bias (Grant et al., 2008; Sansone and Sansone, 2011).

In DSM-5, the borderline personality disorder belongs to a long list of personality disorders, including the paranoid, schizoid, schizotypal, antisocial, borderline, histrionic, narcissistic, avoidant, dependant, obsessive-compulsive and others. The diagnostic criteria include affective, cognitive, behavioural and interpersonal relationship symptoms (Lieb et al., 2004; American Psychiatric Association, 2013):

*A pervasive pattern of instability of interpersonal relationships, self-image, and affects, and marked impulsivity, beginning by early adulthood and present in variety of contexts, as indicated by five (or more) of the following:*

1. *Frantic efforts to avoid real or imagined abandonment.*
2. *A pattern of unstable and intense interpersonal relationships characterised by alternating between idealisation and devaluation.*
3. *Identity disturbance: markedly and persistently unstable self-image or sense of self.*
4. *Impulsivity in at least two areas that are potentially self-damaging (e.g., spending, sex, substance abuse, reckless driving, binge eating).*
5. *Recurrent suicidal behaviour, gestures, or threats, or self-mutilating behaviour.*
6. *Affective instability due to a marked reactivity of mood (e.g., intense episodic dysphoria, irritability, or anxiety usually lasting a few hours and only rarely more than a few days.*
7. *Chronic feelings of emptiness.*
8. *Inappropriate, intense anger or difficulty controlling anger (e.g., frequent displays of temper, constant anger, recurrent physical fights),*
9. *Transient, stress-repeated paranoid ideation or severe dissociative symptoms.*

### 2.2.3 Schizophrenia

Schizophrenia is one of the most controversial and debated psychiatric disorders. The historical record of schizophrenia starts in the 19th century and there is very little evidence of earlier documented schizophrenia cases, or recognition of it as a separate disorder. The definition of schizophrenia was developed by a number of contributors, including Benedict Augustin Morel and Emil Kraepelin, and the term "schizophrenia" was coined by Eugen Bleuler in 1908 (Tsoi et al., 2008).

The risk of developing schizophrenia over one's lifetime is around 0.3-2.0%, with males about 1.4 times more likely to develop schizophrenia then females (Tandon et al., 2008). Schizophrenia is usually diagnosed first based on so called "positive symptoms", involving severely distorted perception of reality (Tandon et al., 2009), so the social functioning of patients with schizophrenia is severely impaired with high costs to society.

The exact definition and diagnostic criteria of schizophrenia are the subjects of heated debates, sometimes going as far as suggesting abandoning the existing concept (Greene, 2007). The diversity of schizophrenia symptoms has led to the definition of subtypes, including catatonic, disorganised, paranoid and others, however this definition is often questioned (Helmes and Landmark, 2003; Regier, 2007). The DSM-5 considers schizophrenia to be a part of a broad schizophrenia spectrum of disorders, defined by such abnormalities as delusions, hallucinations, disorganised thinking, abnormal motor behaviour and negative symptoms (American Psychiatric Association, 2013). The diagnostic criteria for schizophrenia are:

> A. *Two (or more) of the following, each present for a significant portion of time during a 1-month period (or less if successfully treated). At least one of these must be (1), (2), or (3):*
>
> 1. *Delusions.*
> 2. *Hallucinations.*

3. *Disorganised speech (e.g., frequent derailment or incoherence).*
4. *Grossly disorganised or catatonic behaviour.*
5. *Negative symptoms (e.g., diminished emotional expression or avolition).*

B. *For a significant portion of the time since the onset of the disturbance, the level of functioning in one or more major areas, such as work, interpersonal relations, or self-care, is markedly below the level achieved prior to the onset (or when the onset is in childhood or adolescence, there is failure to achieve the expected level of interpersonal, academic, or occupational functioning).*

C. *Continuous signs of the disturbance persist for at least 6 months. This 6-month period must include at least 1 month of symptoms (or less if successfully treated) that meet Criterion A (i.e., active-phase symptoms) and may include periods of prodromal or residual symptoms. During these prodromal or residual periods, the signs of the disturbance may be manifested by only negative symptoms or by two or more symptoms listed in Criterion A present in an attenuated form (e.g. odd beliefs, unusual perceptual experiences).*

D. *Schizoaffective disorder and depressive or bipolar disorder with psychotic features have been ruled out because either 1) no major depressive or manic episodes have occurred concurrently with the active-phase symptoms, or they have been present for a minority of the total duration of the active and residual periods of the illness.*

E. *The disturbance is not attributable to the physiological effects of a substance (e.g., abuse of drugs or medication) or another medical condition.*

F. *If there is a history of autism spectrum disorder or a communication disorder of childhood onset, the additional diagnosis of schizophrenia is made only if prominent delusions or hallucinations, in addition to the other required symptoms of schizophrenia, are also present for at least 1 month (or less if successfully treated).*

### 2.2.4 Summary: dimensional perspective

As we can see from the DSM-5 requirements, the diagnostic criteria for many mental disorders consist of a complex combination of factors, including a person's symptoms, behaviours, cognitive functions, personality traits, physical signs, syndrome combinations, and durations of the above, necessary for the differentiation of abnormal behaviour from normal life patterns.

The narrow definition of diagnostic criteria and the wide variety of possible symptom combinations has led to a large number of diagnostic categories, with many patients demonstrating signs of several mental disorders. The high levels of psychiatric comorbidity (30%-50%) is sometimes argued to be a by-product of the diagnostic system, artificially splitting complex clinical conditions (Maj, 2005). One of the reasons for designing the fifth version of DSM was the introduction of a dimensional approach to diagnosis. The new DSM-5 grouping of disorders is based on eleven indicators, including: shared neural substrates, family traits, genetic risk factors, specific environmental risk factors, biomarkers, temperamental antecedents, abnormalities of emotional or cognitive processing, symptom similarity, course of illness, high comorbidity, and shared treatment response.

Similarly, there are attempts to simplify the analysis of mental symptoms, that are often based on ambiguously defined psychological or behavioural constructs. Symptoms may be organised into several principle dimensions of psychopathology. The work of

**Table 2.1:** *The major clusters of mental symptoms (Liddle, 2001).*

| **Dimension** | **Syndrome** | **Symptom cluster** |
|---|---|---|
| Reality distortion | Reality distortion | Delusions, Hallucinations |
| Disorganisation | Disorganisation | Formal thought disorder, Inappropriate affect, Disorganised or bizarre behaviour |
| Psychomotor | Psychomotor poverty | Flat affect, Poverty of speech, Decreased voluntary motor activity |
| | Psychomotor excitation | Labile affect, Pressure of speech, Motor agitation |
| Mood | Depression | Low mood, Low self-esteem, Hopelessness, Suicidality, Somatic symptoms |
| | Elation | Elevated mood, Elevated self-esteem Decreased need for sleep |
| Anxiety | Anxiety | Feelings of unease, fear or dread, Overactivity of the sympathetic nervous system |

Liddle *et al.* (Liddle, 2001) suggests five major dimensions, as described by Table 2.1, which can be used to define a complete symptoms assessment framework. We can see that certain aspects of behaviour, affect, motor activity, sleep, and activity of the sympathetic nervous system contribute to the broader dimensional picture of symptomatology and in turn are principal diagnostic features for bipolar disorder, borderline personality disorder and schizophrenia as defined above.

While some of the dimensions, such as reality distortion, are mostly subjective and require a clinical interview for proper assessments, and also assume a good knowledge of a patient's history, there are many areas where objective measurement could be of significant benefit for symptom identification and analysis, and improve the clinical picture of disorder. Furthermore, a growing body of evidence suggests the importance of certain symptoms, such as sleep, circadian rhythms (Afonso et al., 2013; Taillard et al., 1993; Wulff et al., 2010) or day time motor activity (Minassian et al., 2010; Burton et al., 2013) in some mental disorders, despite not being included in the DSM-5 diagnostic criteria for particular illnesses.

## 2.3 Diagnostic and assessment tools

### 2.3.1 Clinical interview

The Structured Clinical Interview for DSM Disorders (SCID) is the main diagnostic examination defined by the DSM-IV (American Psychiatric Association, 2000) (the DSM-5 version is under development at the time of writing). There are two main versions, for the Axis I (SCID-I), addressing principal disorders, such as schizophrenia or bipolar disorder, and Axis II (SCID-II), addressing personality disorders, as well as specific versions for research, clinical trials and different subject populations (First et al., 1996; First et al., 1997; First et al., 2002; First et al., 2007). Depending on the individual the interview can take from 15 minutes to several hours and can be administered by a trained mental health professional or a non-clinical research assistant. The SCID includes several modules that could be administered depending on a patient's circumstances. The inter-rater agreement of SCID diagnosis has Cohen's kappa values of 0.61 to 0.83 for the Axis I disorders and 0.77 to 0.94 for the Axis II disorders (Lobbestael et al., 2010).

### 2.3.2 Questionnaires

Alongside clinical interviews, questionnaires remain the main tool for assessment of patient state in research and clinical settings. Questionnaires are used to assess various aspects of well-being and different dimensions of symptoms. The Handbook of Psychiatric Measures (Rush et al., 2008) provides 275 rating methods both for diagnosis and assessment of outcomes. The main benefit offered by questionnaires over the standard clinical interview is the improvement of repeatability and reliability of patient assessment.

However, it was reported that "objective" scales are unable to distinguish between the illusory mental health, demonstrated because of psychological defence mechanisms, and genuine mental health status as identified by clinical judgement (Shedler et al., 1993).

### 2.3.3 Web-based questionnaires

Given the pervasive use and the laborious transcription process involved in the use of paper questionnaires, for most psychiatric scales an attempt is made to create a form for electronic assessment. Electronic delivery also facilitates temporal analysis of reported measures and could be particularly useful for assessment of patient outcomes and self-management. In some cases, implementation of particular tools exists in a standardised form at a country level, such as the Patient-Reported Outcome Measure Information System (PROMIS)[1], developed by the US National Institutes of Health (NIH) and covering the areas of physical, mental, social and global health.

One of the leading examples of integrating the patient-reported outcomes in clinical practice in the UK is the True Colours Self-Management System[2], adopted by the Oxford Health NHS Foundation Trust. The system provides capabilities for online and mobile text-based completion of questionnaires, periodic reminders and patient dashboard, and the visualisation of assessment results in a graphical form. The True Colours system is actively used in clinical practice and research (Miklowitz et al., 2012; Rosa et al., 2014).

Despite the fact that self-reporting is useful for the self-management of mental disorders and for the provision of timely communication of symptoms to clinical specialists, the frequency of assessment and the amount of captured information is not sufficient to allow for the prediction of patient deterioration (Moore et al., 2012). Increasing frequency may lead to diminished compliance, with demonstrated response rates of 88% for weekly and 81% for daily mood prompts (Miklowitz et al., 2012) in research settings.

There is research ongoing to address this compliance issue, for example through the use of virtual agents technology, integrating both assessment and therapy using a realistic animation of a therapist, providing emotional feedback to a patient. The Help4Mood project[3] aimed to facilitate the recovery of patients, suffering from major depression at home using a novel system that monitors patients' health through sensors (Perez-Diaz

[1] http://www.nihpromis.org/
[2] https://oxfordhealth.truecolours.nhs.uk/www/en/
[3] http://help4mood.info/, FP7 project number 248765

de Cerio et al., 2012), questionnaires and diaries, which are integrated with a virtual agent, providing a therapeutic response. The therapeutic agent implementation (Wolters et al., 2013; Bresó et al., 2014) was validated in two small scale studies in different configurations, one involving 5 patients for 7 days and the second involving 4 patients for 15 days. The system usage was on average 94.2% for the first study and 87% for the second study, providing a marginal improvement over the web-based questionnaire approach. Note, that sensors data were not used in these studies.

### 2.3.4 Mobile phone based systems

The benefits of using web-based patient-reported outcome measures could be further increased by collecting such data on mobile devices. However, in many cases psychiatric questionnaires were not designed with small user interfaces in mind, making them difficult to use, and alternative measures need to be introduced, providing patients with short and concise questions, which are accessible on small screens.

The additional benefit created by the usage of mobile phones for psychiatric assessment, compared to desktop computers, is the availability of sensors for the continuous assessment of activity, behaviour (acceleration and location changes) and interpersonal relationship (calls, texts, social networking and voice) patterns.

Although such usage of mobile phone sensors still has not reached clinical practice, a number of research projects have demonstrated the potential benefit of mobile phone sensors for the assessment of different psychological phenomena. For example the EmotionSense mobile platform[4], used for momentary ecological assessment in social psychology research (Rachuri et al., 2010; Lathia et al., 2013), included capabilities for capturing mood using surveys, as well as collecting data from mobile phone sensors and analysing voice for emotional expression.

Other early examples of integrated mobile assessment and intervention using mobile and wearable technologies include the ICT4Depression project[5], developing the Moodbuster mobile application, which facilitates Cognitive Behaviour Therapy (CBT) for treatment of depression. Monitoring of symptoms was performed in a paradigm of Ecological Momentary Assessments (EMA), by providing daily mood rating requests (Santos et al., 2012). A wearable sensor glove for EDA, respiration and ECG was developed as part of the project and medicine intake was also monitored (Silva et al., 2012; Araújo et al., 2012). All information was provided back to the patient via smart phone app and website (van de Ven et al., 2012; Warmerdam et al., 2012). The Moodbuster technology was validated in small scale technology trials in Sweden and the Netherlands (Warmerdam et al., 2012). The trial in Sweden was performed with 24 depressed student participants scoring 5 or more on the PHQ-9 depression scale (Kroenke et al., 2001) and diagnosed with major or minor depression. The Moodbuster treatment was provided for 6 weeks and a significant decrease in PHQ-9 score post treatment was observed (ICT4Depression Consortium, 2013b). The study in the Netherlands was performed with 23 depressed subjects recruited mostly via media advertisement, and the 10-item Kessler's psychological distress scale (Kessler et al., 2002) was used to identify subjects eligible for the study (scoring above 20). The Beck Depression Inventory second edition (BDI-II) (Beck et al., 1988) was used to assess the treatment response and significantly lower post-treatment scores were found (ICT4Depression Consortium, 2013a).

[4]http://emotionsense.org/, funded by the EPSRC Ubhave Project

The MONARCA project[6] focused mostly on behavioural and physiological monitoring, attempting to solve the task of predicting depressive and manic episodes in bipolar disorder patients (Bardram et al., 2012). The monitoring system included a mobile phone (Osmani et al., 2013), wrist-worn activity monitor, sock-integrated physiological sensor for GSR and HR (Gravenhorst et al., 2013), and a stationary EEG measurement system. Physiological information from the sensors was combined with behavioural data from the mobile phone, such as GPS location traces, phone usage patterns, voice and motion anal-ysis results (Gruenerbl et al., 2014). The MONARCA smart phone platform was piloted in small-scale trials in Austria (10 bipolar patients) and Denmark (12 bipolar patients), both collecting data for approximately 3 months. The study in Denmark was focused on the assessment of the system's usability and found that patients demonstrated similar adherence to paper based and mobile data collection (Bardram et al., 2013). The study in Austria focused on the recognition of clinical states, broadly defined as a mood scale with 7 items, from depressive to manic, assessed using the Hamilton Depression Rating Scale (HAMD) (Hamilton, 1960) and Young Mania Rating Scale (YMRS) (Young et al., 1978) and achieved 76.4% recognition accuracy using fusion of location, acceleration and voice features. For state change detection (novelty detection), the accuracy was as high as 97.4% (Grunerbl et al., 2014; Gruenerbl et al., 2014) on testing set with three-fold cross-validation.

[5]http://www.ict4depression.eu/, FP7 project number 248778
[6]http://www.monarca-project.eu/, FP7 project number 248545

### 2.3.5 Wearable devices

While smart phones provide a potentially powerful tool for monitoring changes in behaviour, psychological state (Bardram et al., 2012; Santos et al., 2012; Lathia et al., 2013) and treatment delivery, they lack the capability for assessing physiology and sleep. This is why wearable devices are often considered for the analysis of mental state. Wearable devices can take different forms, starting from conventional wrist-worn bracelets to smart textile clothing solutions. The above-mentioned ICT4Depression and MONARCA projects included design of wearable instrumentation as dedicated work packages, with ICT4Depression focusing on the development of a "smart glove" (measuring EDA and blood volume pulse) (Silva et al., 2012; Araújo et al., 2012) and MONARCA using a "smart sock" (measuring EDA and acceleration) approach (Kappeler-Setz et al., 2011; Gravenhorst et al., 2013). Help4Mood also incorporated eZ430-Chronos smart watch (Texas Instruments, US) as one of the system components (Perez-Diaz de Cerio et al., 2012). However, none of the projects described above reported clinically useful results using the in-house developed instrumentation.

The OPTIMI project[7] used a different approach, focusing mostly on physiological assessment, and aiming to develop behavioural and physiological monitoring tools to detect potential early signs of stress and depression, using EEG, ECG, activity monitoring, voice analysis, cortisol measurements and electronic diaries (Botella et al., 2011). However, the reported results included evaluation of the EEG headset only. It was validated for usability with 10 healthy participants with positive feedback (Zaragozá et al., 2011). An accuracy of 94.2% was reported for detecting depression (BDI-II score of 14 or higher) using $\alpha$-band absolute power, $\beta$-band absolute power, $\theta$-band absolute power, C0-complexity, largest Lyapunov Exponent and Lempel-Ziv complexity features and with three-fold cross-validation on 25 female volunteers (Zhang et al., 2013).

A textile multi-sensor monitoring system for patients with bipolar disorder was developed by the PSYCHE project[8] (Paradiso et al., 2010; Javelot et al., 2014). The developed "sensor shirt" included ECG, respiratory rate and acceleration sensors to perform physiological assessment. The main activities of the project mostly concentrated on the development of tools and methods for the analysis of ECG and identification of mood states (Valenza et al., 2014b; Valenza et al., 2014a), so validation of the technology was performed in controlled conditions with 8 bipolar patients. However, only 3 subjects participated in the evaluation of mood detection during unstructured activity, with a reported accuracy of euthymic state recognition as high as 82.4% (Valenza et al., 2015). In the framework of the same project small scale experiments were performed with EDA analysis (10 bipolar patients for 20 minutes) (Greco et al., 2012; Greco et al., 2014) and voice analysis (11 bipolar patients) (Guidi et al., 2014) and significant differences were identified between mood states.

[7]http://www.optimiproject.eu/, FP7 project number 248544
[8]http://www.psyche-project.org/, FP7 project number 247777

## 2.4 Discussion and conclusions

Psychiatric questionnaires and clinical interviews are the gold standard for the evaluation of clinical states in mental health. However, there are some issues with objectivity and inter-rater agreement of such assessments (Lobbestael et al., 2010). To evaluate the presence of specific symptoms we may also collect information using accepted physiological measurements, which are known to be correlated with these symptoms. Such measurements, important in the context of mental disorders, may include physical, cardiac and electrodermal activity (Cacioppo et al., 2007). Despite the fact, that only limited correlations were reported previously, combining measures from different monitoring modalities and with a focus on assessing specific symptoms and not disorders *per se*, may provide a way to improve diagnosis and management of mental disorders.

**Physical activity:** Actigraphy information can be obtained using mobile phones or body mounted accelerometry measurement devices, located on the wrist, hip or waist, depending on research settings. Changes in physical activity and sleep structure have been shown to correlate with a range of short and long term medical conditions, from cardiac problems to mental health issues (Foster and Wulff, 2005; Wirz-Justice, 2007a; Wulff et al., 2010). Usage of actigraphy for sleep-wake assessment was first proposed in 1982 by Webster *et al.* (Webster et al., 1982) and usage of body-mounted actigraphs for analysis of sleep and circadian rhythms in mental disorders, such as schizophrenia, was pioneered in 1990 by Witting *et al.* (Witting et al., 1990) and has been widely adopted since then. Recent models of actigraphs may include high-precision micro-electro-mechanical sensors (MEMS) for tri-axial acceleration measurement and allow for very precise automated classification of activity (Zhang et al., 2012), opening up the possibility for the development of powerful actigraphy analysis methods.

**Heart rate:** The heart rate is controlled by both the sympathetic nervous system (SNS) and parasympathetic nervous system (PNS) (Appelhans and Luecken, 2006; Lorber, 2004). In the context of mental disorders, the heart rate is especially important for analysing problems related to emotional regulation and was proposed as an objective measure of emotional arousal (Lacey and Lacey, 1974). HR derived features, such as heart rate variability (HRV) (Appelhans and Luecken, 2006; Friedman and Thayer, 1998), are useful in the analysis of such conditions as depression (Carney et al., 2001), aggression (Scarpa and Raine, 1997), anxiety (Lang et al., 1998) and psychopathy (Hare and Quinn, 1971). Heart rate data may be collected using different modalities, including electrocardiogram (ECG), blood pressure monitors, audio acquisition devices (such as electronic stethoscopes) and pulse oximeters. Portable (Holter) ECG monitoring devices are most widely used for cardiac function assessment and HRV analysis, due to their capability in the detection of normal beats for HRV calculation (Task Force of the European Society of Cardiology the North American Society of Pacing Electrophysiology, 1996) and good temporal resolution. A study by Lu *et al.* (Lu et al., 2008) indicates that pulse oximetry can also be used to extract some measures of heart rate variability.

**Electrodermal activity:** Unlike the heart rate, electrodermal activity is influenced only by the SNS (Lorber, 2004; Fowles, 2007). EDA is measured by obtaining skin conductivity and is usually split into the tonic and phasic parts, where tonic is a reference level and phasic corresponds to the reaction to a certain stimulus (Boucsein, 2012). The phasic EDA signal can be further analysed to extract response features, such as frequency, amplitude, rise time, recovery time and other metrics. Although the primary use of EDA is related to emotion regulation research (Birket-Smith et al., 1993; Scarpa and Raine, 1997) and psychopathy (Fung et al., 2005), it is also used in studies of schizophrenia (Öhman, 1981; Dawson et al., 2010).

A number of projects were started in the last decade to evaluate the potential of physiological measurements for use in mental health assessment, described in Table 2.2. These projects mainly benefited from the development of consumer-grade activity monitoring devices, such as mobile phones and commercial fitness trackers, as well as advances in electronic miniaturisation, enabling the development of wearable physiological sensors.

However, the existing studies suffer from a number of shortcomings, including:

**Table 2.2:** *Summary of publications on physiology and activity in analysis of mental disorders.*

| Study description | Study type | Participants | Results |
|---|---|---|---|
| Treatment of depression using virtual agents, Help4Mood (Bresó et al., 2014) | Questionnaires | Major depression, 2 cohorts<br>1) 5 subjects, 7 days<br>2) 4 subjects, 15 days | Only usability evaluated |
| Recognition of emotions using mobile phone sensors, EmotionSense (Rachuri et al., 2010) | Smart phone voice, bluetooth and accelerometer | Healthy subjects<br>12 subjects, 24 hours | 58.7-84.7%% accuracy of emotion recognition (no CV) |
| Evaluation of blood pulse volume sensor, ICT4Depression (Silva et al., 2012) | Blood pulse volume | Healthy subjects<br>10 subjects, 5 minutes | $R^2 > 90\%$ comparing to reference signal |
| Evaluation of localised EEG sensor, ICT4Depression (Araújo et al., 2012) | EEG | Healthy subjects<br>1 subject | Visual EEG evaluation |
| Evaluation of mobile phone based mood intervention, ICT4Depression (ICT4Depression Consortium, 2013a) | Blood pulse volume and ECG | Depression<br>23 subjects, 6 weeks | Only usability evaluation |
| Evaluation of mobile phone based mood intervention, ICT4Depression (ICT4Depression Consortium, 2013b) | Blood pulse volume and ECG | Depression<br>25 subjects, 6 weeks | Only usability evaluation |
| Evaluation of mobile phone for mood monitoring, MONARCA (Osmani et al., 2013) | Smart phone accelerometer | Bipolar disorder<br>9 patients, 3 months | Patient-specific correlation between PA and mood reported |
| Technical feasibility of mobile galvanic skin response sensors, MONARCA (Gravenhorst et al., 2013) | Galvanic skin response | Bipolar disorder<br>11 patients, 7 months<br>(3-4 30 minute sessions) | Only feature differences reported (p-values) |
| Evaluation of smart phone sensors for identification of mood, MONARCA (Grunerbl et al., 2014) | Smart phone GPS, accelerometer and sound | Bipolar disorder<br>12 patients, 12 weeks | 76% mood recognition accuracy (no CV) |
| Evaluation of smart phone questionnaires for identification of mood, MONARCA (Bardram et al., 2013) | Questionnaires | Bipolar disorder<br>12 patients, 12 weeks | 29% increase of compliance reported comparing to paper |
| Technical feasibility of wearable galvanic skin response sensors, MONARCA (Kappeler-Setz et al., 2011) | Galvanic skin response (sock) | Healthy subjects<br>8 subjects, 21 minutes | 88% of reference signal responses detected |
| Technical feasibility of wireless sensor network for mood monitoring, Help4Mood (Perez-Diaz de Cerio et al., 2012) | Smart phone, smart watch, key ring, belt and bed motion sensors | Healthy subjects<br>numbers and durations not reported | Only usability evaluated |
| Evaluation of a system for stress detection, OPTIMI (Zaragozá et al., 2011) | Wearable motion, ECG, stationary EEG and microphone | Healthy subjects<br>12 subjects, 1 day | Only usability evaluated |
| Depression detection using EEG, OPTIMI, (Zhang et al., 2013) | EEG | Healthy and depressed volunteers<br>25 subjects, 1 month | 94.2% accuracy (3-fold CV) |
| Mood states detection using HRV, PSYCHE (Valenza et al., 2014a) | ECG | Bipolar disorder<br>8 subjects, 2 nights | Only feature differences reported (p-values) |
| Mood states detection using EDA, PSYCHE (Greco et al., 2012) | EDA | Bipolar disorder<br>3 subjects, 75 days<br>(2-5 20 minutes sessions) | $> 76\%$ state recognition accuracy (5-fold CV) |
| Mood states detection using EDA, PSYCHE (Greco et al., 2014) | EDA | Bipolar disorder and controls<br>20 subjects, 75 days<br>(1-2 20 minutes sessions) | Mixed results reported for different subjects |
| Mood states detection using speech analysis, PSYCHE (Guidi et al., 2014) | Speech | Bipolar disorder and controls<br>30 subjects, 2 days | Only intra-subject analysis of voice features performed |

- The small number of participants and limited duration of data collection.
- The lack of a principled framework, connecting new results and previous research findings.
- The use of experimental technology without considering long-term usability.
- The results are often mixed and hard to generalise due to the use of custom data collection tools and analysis methods.
- The participants are often from specific research cohorts, e.g. specialist clinic settings.

There is a clear need for development of a more solid evidence base in the objective assessment of mental health. It is necessary to validate if the existing monitoring tools are appropriate for long term monitoring, and can be used in clinical populations. There is also a need for better understanding of the complex picture of individual symptoms and the provision of more personalised care. Large scale and long term behavioural data may allow for the creation of individual phenotypes and the identification of more objective boundaries between different disorders, that address underlying pathologies and are predictive to treatment response.

# Chapter 3

# The Automated Monitoring of Symptoms Severity (AMoSS) study

## 3.1 Motivation and study organisation

### 3.1.1 Objectives

As has been mentioned in Chapter 2, the key limitations of objective mental health research studies produced so far include the small number of subjects, limited duration of data acquisition, and difficulties in translating results into clinical practice due to the usage of research-grade data collection instrumentation, which is not completely suitable for continuous use in ambulatory settings. The most important collaborative research projects, aiming to demonstrate the benefits of mobile technology for mental health, including the ICT4Depression, MONARCA and PSYCHE, were working with cohorts of 10-20 subjects and collected up to 3 months of data (Warmerdam et al., 2012; Bardram et al., 2012; Valenza et al., 2015).

In order to overcome these difficulties and create a more solid evidence base for the introduction of behaviour monitoring in clinical practice, the Automated Monitoring of Symptoms Severity (AMoSS)[1] study was designed, with a protocol written by Athanasios

[1] http://conbrio.psych.ox.ac.uk/the-amoss-study

**Table 3.1:** *Projects using mobile and wearable technologies in mental health applications. Many projects were focused on technology development in a framework of European collaborative research, and included multiple separate validation studies with different numbers of participants.*

| Project | Number and type of participants | Study duration |
|---|---|---|
| AMoSS | 50 bipolar disorder,<br>30 borderline personality disorder,<br>50 healthy controls | 1 year |
| MONARCA | Study 1: 10 bipolar disorder<br>Study 2: 12 bipolar disorder | 3 months |
| PSYCHE | Study 1: 8 bipolar disorder patients<br>Study 2: 10 bipolar disorder patients<br>Study 3: 11 bipolar disorder patients | Up to 1 day |
| OPTIMI | Study 1: 10 healthy participants<br>Study 2: 25 depression | Up to 1 day |
| ICT4Depression | Study 1: 24 depression<br>Study 2: 23 depression | Up to 6 weeks |
| Help4Mood | Study 1: 4 depression<br>Study 2: 5 depression | Up to 7 days |

Tsanas and Kate Saunders (not published) (Clifford et al., 2015) with technical contributions from the author of this thesis. The AMoSS study protocol includes collection of data from bipolar disorder (N=50), borderline personality disorder (N=30) and healthy control subjects (N=50) for a duration of one year, representing a substantial advance in both study duration and number of participants compared to previous research (see Table 3.1). The study was approved by the local NHS research ethics committee (REC reference 13/EE/0288) and all participants provided written informed consent.

The AMoSS study was designed to improve the understanding of bipolar and borderline personality disorders and to explore new ways of ambulatory monitoring and self-management in mental health. The study used off-the-shelf consumer-grade monitoring tools in parallel with research-grade instrumentation to acquire high-quality and long-term physiological and behavioural data from patients and controls. The objectives of the study were (Clifford et al., 2015):

- ***Primary objectives:*** *To measure objective differences in activity and physiological measures, between healthy controls and people with bipolar disorder and borderline personality disorder, and associate these with the symptom severity which is quantified using the patients' self-reported assessments.*
- ***Secondary objectives:*** *Secondary objectives of the research include answering the following questions:*
  - *What are the minimal number of sensors required towards our goals in order to obtain accuracy (sensitivity and specificity) of at least 90%?*
  - *Can we identify, in advance, patterns in the data that are associated with deterioration in mental state? This will be determined by exploratory analysis at the end of the study.*
  - *What is the acceptability of the overall monitoring approach, and regarding specific sensors for each participant? We will also obtain implicitly the acceptability of each sensor by the frequency that participants set specific sensors off.*

The primary study objective was dictated by the need for better clinical understanding of bipolar and borderline personality disorders, where diagnostic criteria may not allow for a simple delineation between two. Identification of reliable objective behaviour features associated with specific symptoms of these disorders, as well as prevalence of these behaviours in clinical populations, allows building personalised models of deterioration, that would account for inter-person variations in euthymic state and symptoms manifestation during clinical episodes.

### 3.1.2 Recruitment of participants

Participants for the study were recruited by research staff at the Department of Psychiatry of the University of Oxford, at the Warneford psychiatric hospital. Healthy controls were recruited through advertisement and from previous study cohorts. Bipolar disorder patients are recruited from the OXTEXT-1 study cohort (Armstrong et al., 2015). Borderline personality disorder patients were recruited both from previous research cohorts and by advertisement; due to the prevalence of the disorder in clinical populations, the borderline personality disorder participants were predominantly female, so in order to balance the cohorts, numbers of participants in bipolar disorder and healthy cohorts were increased by comparison with the borderline, and a higher proportion of female participants were recruited to these groups.

#### 3.1.2.1 Demographics and clinical characteristics

All participants of the study were monitored in an ambulatory environment. Patients received treatment and many were medicated, however all were relatively stable at the time of recruitment. Recruitment and data collection for the study is still ongoing, so the healthy and clinical groups are not yet complete. Due to this fact the AMoSS data set may be unbalanced, for example it can be seen that age, weight and BMI are significantly different between healthy and bipolar disorder cohorts (see Table 3.2).

**Table 3.2:** *Demographic characteristics of the AMoSS study participants. Superscripts indicate that the distributions are different with $p < 0.05$ according to the Wilcoxon rank sum test. BD, BPD and HC superscripts refer to bipolar disorder, borderline personality disorder and healthy controls.*

| | **Healthy controls** | **Bipolar disorder** | **Borderline personality disorder** |
|---|---|---|---|
| Number of participants | 32 | 43 | 22 |
| Gender | 23 females | 28 females | 19 females |
| Age (Mean±SD) | 33.12±11.54 $^{BD}$ | 38.37±11.68 $^{HC}$ | 33.05±10.90 |
| Height (Mean±SD) | 169.78±8.37 | 169.47±9.28 | 167.05±7.57 |
| Weight (Mean±SD) | 69.19±13.09 $^{BD}$ | 79.09±18.15 $^{HC}$ | 77.64±19.47 |
| BMI (Mean±SD) | 23.99±4.04 $^{BD}$ | 27.44±5.47 $^{HC}$ | 27.90±7.18 |

### 3.1.3 Data collection

The AMoSS study is multifactorial by design and includes a phase for preliminary assessment, where key data including occupational and personality characteristics are collected. The phase of objective data collection includes a 12-week period (with the option to continue for one year) of continuous activity data collection using Galaxy S3 or S4 smart phone (Samsung, Korea), Fitbit One commercial activity tracker (Fitbit, US), weekly detailed and daily short mood questionnaires, and a week of high intensity data collection, including high frequency acceleration, ECG, body temperature, blood pressure and a short mood questionnaire on the phone up to 10 times a day (see Table 3.3).

The data collection strategy aimed to capture longitudinal data if possible, but in the same time include all modalities that may be important for identification of symptoms of analysed mental disorders, as based on the prior art discussed in Chapter 2. Therefore for long-term monitoring (from 3 months and up to a year) off-the-shelf devices were used that have a proven history of compliance (Fitbit One and smart phone).

In the same time the quality of data collected with such devices is not validated, and the number of signals is limited. Therefore during a shorter period of time, with duration constrained by characteristics of data collecting devices, research-grade instrumentation were also used, such as GENEActiv accelerometer.

ASRM, QIDS-16SR and GAD-7 questionnaires were delivered with a weekly frequency, as these questionnaires require recollection of last week behaviours to provide mood scoring. Mood Zoom questionnaire was designed to provide a low-overhead insight into momentary subject emotional state suitable for SMS delivery and was delivered on a daily basis.

Some phenomena, such as sleep, require additional information for assessment, such as bed time segmentation. Such segmentation in research settings is usually provided by sleep logs, however Fitbit One includes a simple button press capability for bed time identification, so sleep logs were not used in this study to reduce burden on participants with Fitbit-based bed time reporting compliance described later in this thesis.

**Table 3.3:** *Monitoring modalities of the AMoSS study. During the high intensity week participants were instructed to use a number of physiological sensors, in addition to usual mobile and wearable devices and the Mood Zoom questionnaire was also administered with increased frequency.*

| **Instrument** | **Signals** | **Data collection** |
|---|---|---|
| True Colours questionnaires | QIDS-SR16 (depression)<br>ASRM (mania)<br>GAD-7 (anxiety)<br>EQ-5D (quality of life) | Continuously administered for the period of study, delivered weekly |
| Mobile phone sensors | Tri-axial acceleration<br>Light level<br>Battery level<br>Geolocation<br>Calls and texts | Continuously for the period of study, variable sampling rate |
| Mobile phone questionnaires | Mood Zoom (mood) | Once a day for the period of study, 10 times a day during the high intensity week |
| Fitbit, commercial fitness tracker | Steps<br>Floors<br>Sleep analysis<br>Bed time annotation | Continuously for the period of study, one minute resolution, manual (button press) bed time annotation |
| GENEActiv wrist-worn accelerometer | Tri-axial acceleration<br>Light level<br>Body temperature | High intensity week and optional at other times, 10-50Hz sampling rate for a duration of 10-28 days |
| Shimmer ECG | ECG | High intensity week, up to 48 hours ECG at 125Hz |
| Proteus patch | HR<br>Body temperature<br>Physical activity | High intensity week, HR every 10 minutes, physical activity every 5 minutes |
| Omron home blood pressure monitor | Blood pressure<br>Heart rate | High intensity week, 3 times a day |
| Omron temperature probe (ear or oral) | Body temperature | High intensity week, 3 times a day |

The data acquisition platform was designed to use a custom AMoSS mobile application on an Android smart phone as the main data acquisition instrument, providing a data entry interface for users and transferring the collected data to mobile back-end servers, running Linux, Apache, MySQL and PHP (LAMP) software stack and providing protected file storage for all study data (Palmius et al., 2014).

To ensure data security, study participants and clinical assistants were located outside of the protected internal network, where back-end servers were placed and which only research personnel has access to. The data were uploaded to the internal network as anonymised and only clinical assistants have access to participant's identities, stored at the Warneford psychiatric hospital (Oxford, UK).

To ensure compliance with data collection, participants were offered to keep study-provided smart phones if continued the study for more than one year. The back-end platform provided research personnel with status of data collection for each participant, so that in case of interrupted data transfers the participant could be contacted directly to identify and resolve the issue.

Note, that the study participants did not keep sleep log to reduce the study burden on participants. All data collection modalities selected for the study required minimal user interaction, however together could potentially cover a broad spectrum of possible symptoms, and thus account for inter-subject variability in disorder manifestations.

#### 3.1.3.1 Mobile phone application

The key component of the study is a mobile platform for the acquisition of data from mobile phone sensors and Mood Zoom questionnaire (AMoSS mobile application) (see Figure 3.1). The mobile application collects data from the mobile phone accelerometer (tri-axial), light sensor, battery, coarse location information (obtained by proximity of cell towers and Wi-Fi access points, unless the user has enabled GPS explicitly), as well as phone calls and text messages (addressee, direction and length). All information that may lead to the identification of participants by researchers is eliminated at the stage of mobile phone data acquisition according to the following rules:

- For location data, during the application installation a random seed was generated for latitude (range of $\pm 90°$) and longitude (range of $\pm 180°$), and subsequently subtracted from all collected data, so location has an unknown at the analysis stage offset from the real measured location, defined by the seed.

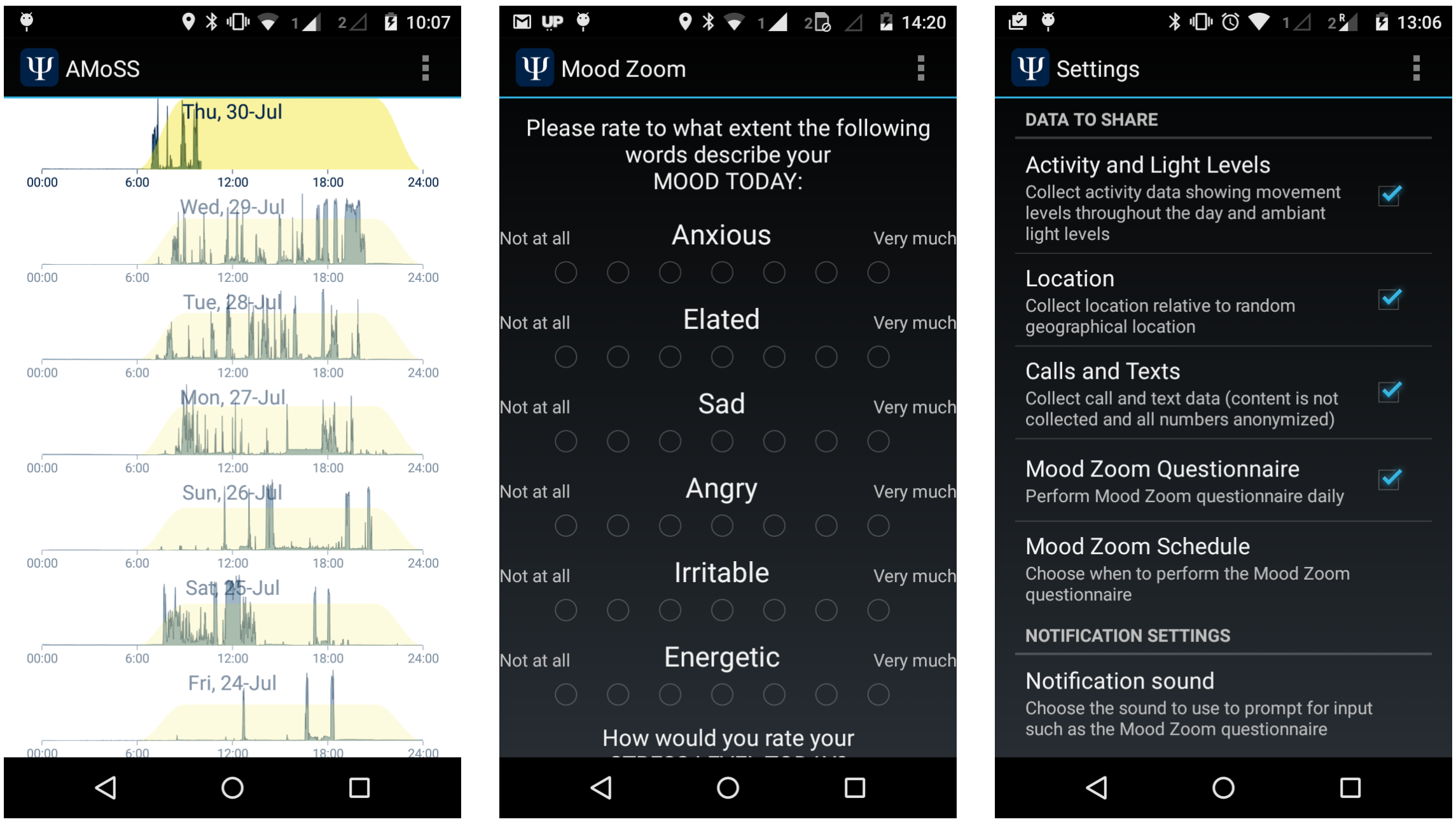

a. b. c.

**Figure 3.1:** *User interface of the AMoSS mobile application. Figure a) shows the main activity visualisation screen, b) shows the Mood Zoom questionnaire prompt and c) presents the application configuration options.*

- For phone calls and texts, the real addressee number is replaced with a MD5 hash, thus eliminating the possibility of identification, but at the same time preserving the addressee's uniqueness, and content of text messages or calls was not collected.

Users agreed with the applications terms of use during installation and required to sign consent form in order to receive credentials to start using the application. The design and functionality of the mobile application were evaluated in 3-months study on healthy volunteers (students) to ensure that the data collection will not undermine usability of smart phones (such as battery life) and ensure data transfer reliability to avoid significant data traffic charges. But no formal comparison were performed between smart phone characteristics with and without AMoSS mobile application.

From the software architecture point of view, the mobile application is organised into a number of asynchronous modules, to facilitate continuous data collection and upload, even if the application itself is not running in the foreground on a user's phone. The following key software components are included in the application:

- **User interface:** Provides visualisation of the last 7 days of collected activity data, an interface for submission of Mood Zoom questionnaire responses and a screen for application configuration (see Figure 3.1).

- **Background sensors data collection service:** The service runs continuously on the mobile phone and collects data from the accelerometer, light sensor, and battery, as well as geolocation data, information about calls and texts and responses from the Mood Zoom questionnaire. All data are saved on the phone's local file system in CSV format for up to 30 days or until uploaded to the AMoSS server.

- **Background data upload service:** The service performs a buffered asynchronous upload of collected data as zip-compressed CSV files and manages the application connectivity. Due to the large amount of data collected (5-10 MB per day per participant) and to protect participants from excessive mobile data payments, server uploads are only performed when a Wi-Fi connection becomes available.

- **Events scheduler:** The scheduler handles the key application events, such as activation of the Mood Zoom questionnaire and repeated data upload attempts.

#### 3.1.3.2 Server side interface

The main design goal of the server side interface was to provide a simple and flexible infrastructure for acquiring and storing data from mobile phone sensors, as well as other types of data collected during the study (Palmius et al., 2014) (see Figure 3.2). The server interface was implemented using the LAMP open-source software stack and included the following components:

- **Web Server:** Apache web server version 2.2.22 with HTTPS support (Rescorla, 2000) and PHP module for implementation of File Server functionality. For improved security, the server is accessible only on an encrypted (HTTPS) connection.

- **Authentication Server**: A basic access authentication facility using user identifier

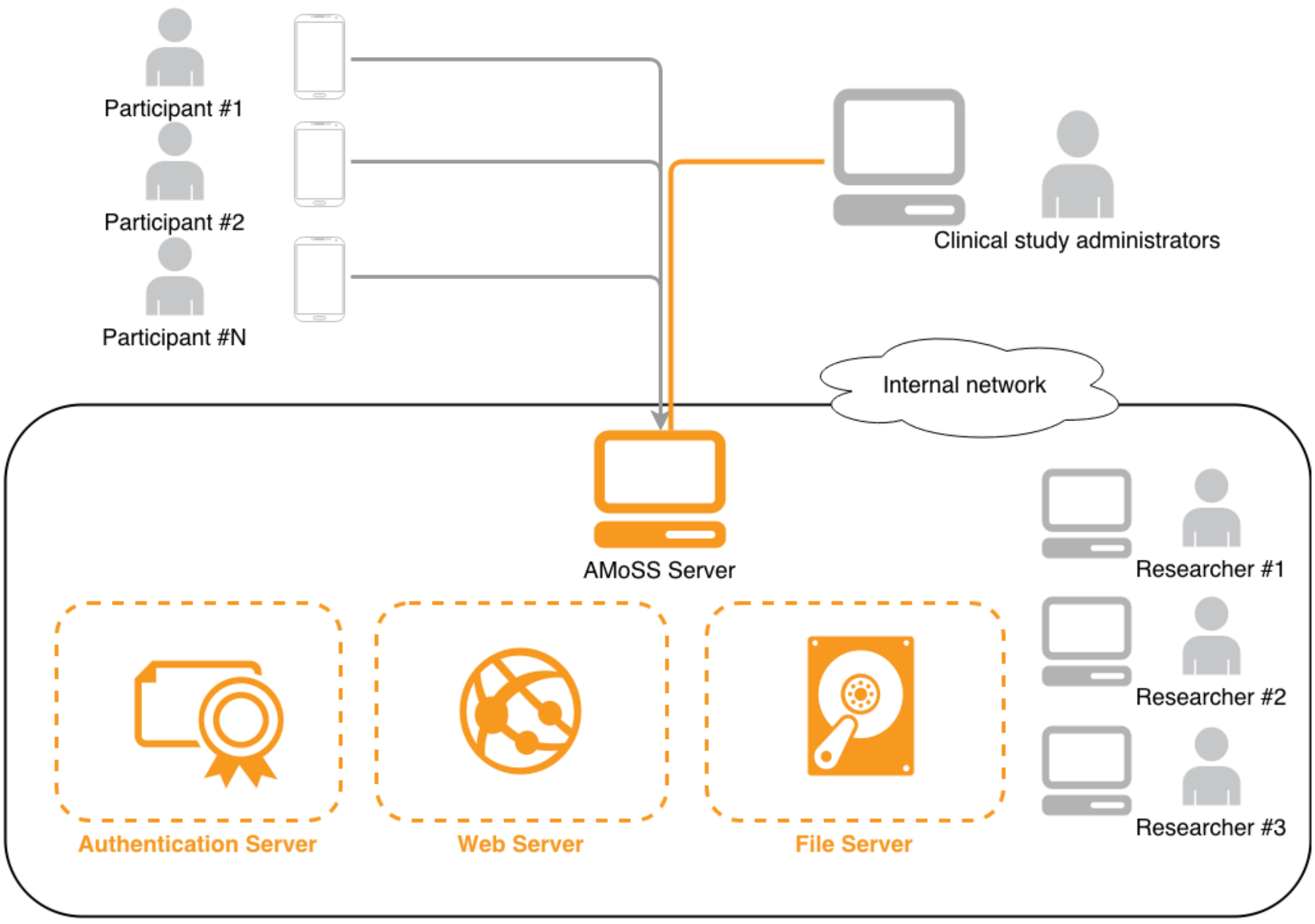

**Figure 3.2:** *Architecture of the AMoSS smart phone data acquisition platform. The password-protected internal computing resources are highlighted in orange. Participants access the internal network using provided smart phones with in-house developed AMoSS application.*

and password (Franks et al., 1999). Each user is assigned a unique study identifier (user name) and a specially generated strong password. The study identifier and password need to be specified within the mobile application to establish a secure server connection.

- **File Server**: A custom PHP scripts infrastructure for unidirectional file transfer (data upload only). Once uploaded, data files are transferred to a secure server area, not accessible by the Apache web server.

The key factor for implementation of server side interface was the limited availability of supporting infrastructure and implementation resources. Therefore a "low-tech" approach to the implementation was taken, where uploaded data were stored in files, thus simplifying server backups and data management by researchers. This approach in the same time allows for a good horizontal scalability as data from different clients may be stored on different file servers and API server can be easily migrated to a facility with automatic load balancing, such as Amazon Elastic Beanstalk. However, for the analysis purpose the data need to be decompressed and converted from CSV to a format with direct data access (Matlab data files used in this thesis), which may need significant computational resources. Source code for all platform components is stored in private Bitbucket Git repositories.

Only users recruited for the study or clinical study administrators are allowed to upload data to the server and only researchers within the internal network can access the uploaded data files. Additionally, a set of support scripts was developed for the creation of new users, generation of strong alphanumeric passwords, conversion and re-sampling of data.

## 3.2 Data analysis subset

Data collected by the AMoSS project are diverse and not all signals listed in Table 3.3 are used in this thesis. Since the main focus of this work is the identification of objective features to continuously assess patients' symptoms, the main emphasis was given to accelerometry, as the least invasive data acquisition tool, and capable of characterising a broad range of symptomatic behaviours (Teicher, 1995).

The selected subset of signals is described in Table 3.4, including the objective measures of physical activity and subjective questionnaire-based ground truth assessment of symptoms. Note, that the ground truth of mental state is represented by subjective assessment of symptoms, however it is the current state-of-art in diagnostics of mental disorders, and shifting towards more objective assessment is the purpose of this thesis.

In the following sections a detailed description is provided for each of the selected data sources.

**Table 3.4:** *Monitoring modalities of the AMoSS study selected for this research, including subjective ground truth (questionnaires) and objective measures of activity and behaviour.*

| **Signals** | **Purpose** |
|---|---|
| BIS-11 (impulsivity) score | Ground truth for identification of trait impulsivity related behaviour features |
| QIDS-SR16 (depression), ASRM (mania) and GAD-7 (anxiety) questionnaires | Ground truth for identification of clinically significant mood episodes in recent two week period |
| Mood Zoom questionnaire | Ground truth for short-term (day-to-day) mood changes |
| Fitbit bed time annotation | Ground truth for development of automated bed/wake segmentation algorithms |
| GENEActiv acceleration and light level | High-quality dataset for identification of features, specific for behaviour of diagnosed patients vs. healthy controls |
| Phone acceleration, light level, battery level, calls and texts | Long-term rich dataset for identification of clinically significant mood episodes |

### 3.2.1 Questionnaires

#### 3.2.1.1 Impulsivity assessment

Impulsivity is widely acknowledged as an important characteristic of many mental disorders and clinical states (Swann, 2009). Impulsivity is generally considered to be a "trait" characteristic, describing the habitual behaviour patterns independently on mental state, however there were very few studies of impulsivity in mania or depression. The International Society for Research on Impulsivity suggests four measurement tools, including the Barratt Impulsiveness Scale (BIS-11) (Barratt and Patton, 1995), Balloon Analogue Risk Task (BART) (Lejuez et al., 2002), Cued Go No-Go (Marczinski and Fillmore, 2003) and Immediate and Delayed Memory Tasks (IMT/DMT) (Dougherty et al., 2002). In the AMoSS study, the impulsivity assessment (BIS-11) is performed once during the enrolment process to identify the following personality traits:

- **Attentional impulsiveness** which assesses task-focus, intrusive thoughts, and

racing thoughts.

- **Motor impulsiveness** which assesses the tendency to act on the spur of the moment and consistency of lifestyle.
- **Non-planning impulsiveness** which assesses careful thinking and planning and enjoyment of challenging mental tasks.

It can be seen that the contributing factors describe human behaviour on multiple scales, from seconds (attentional) to minutes and hours (motor) and days (non-planning), and are strongly linked to physical activity or behaviour. Therefore continuous analysis of physical activity recordings has the potential to characterise a person's impulsivity by analysing both day-time activity patterns and rapid changes in activities (for example frequent changes between sedentary and moderate activities).

The AMoSS study includes healthy, bipolar disorder and borderline personality disorder participants, all having different profiles of impulsivity, as we can see from Figure 3.3. For the bipolar disorder patients increased impulsiveness is expected during manic episodes and for the borderline personality disorder patients it is a persistent distinctive trait characteristic.

#### 3.2.1.2 Weekly mood questionnaires

The presence of clinical episodes of mania and depression in bipolar patients in ambulatory settings is usually assessed with questionnaires. The AMoSS study uses the QIDS-SR16 for the assessment of depression (Rush et al., 2003), ASRM for the assessment of mania (Altman et al., 1997) and GAD-7 for the assessment of anxiety (Spitzer et al., 2006). The questionnaires are delivered on a weekly basis using the True Colours Self-Management System[2], via email or mobile text message.

As we can expect and see from Table 3.5, mood scores vary between the disorder and control groups. However, according to DSM-5 diagnostic criteria symptoms are often

[2] http://truecolours.nhs.uk/

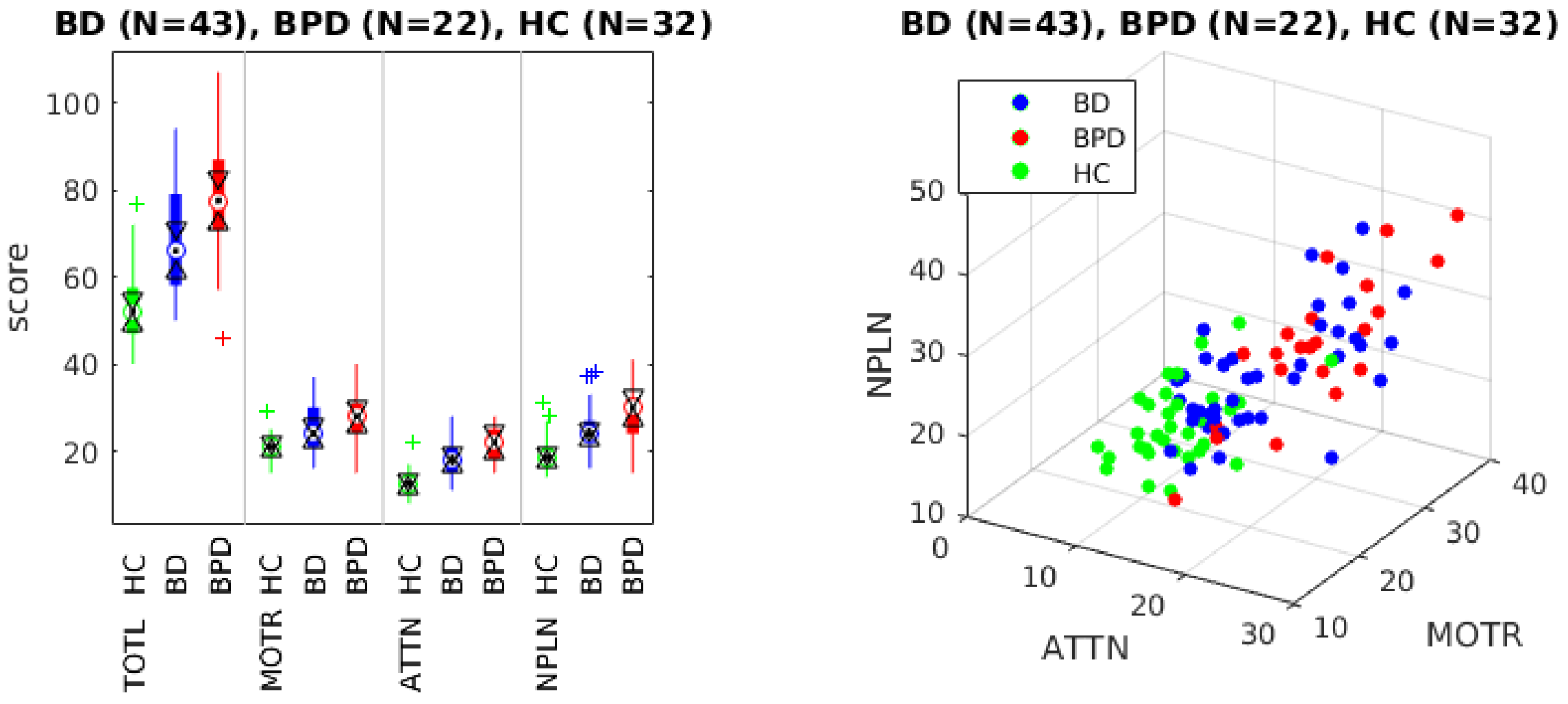

a. b.

**Figure 3.3:** *Boxplots of BIS-11 impulsivity scores of the AMoSS study participants, where TOTL refers to the total score, and MOTR, ATTN, NPLN are the motor, attentional and non-planning* $2^{nd}$ *order factors respectively. BD, BPD and HC refer to biploar disorder, borderline personality disorder and healthy controls.*

required to be present for a certain period of time and to exceed a specific threshold (see Table 3.6), to become a clinically significant episode.

Questionnaire-based weekly mood scores are used as the main ground truth for the identification of bipolar disorder episodes, so compliance with questionnaires is of special importance in this study. As we can see from Table 3.5, compliance among the AMoSS study participants varied considerably, however the mean time between responses is not significantly different between healthy and disorder cohorts at a level of $p < 0.05$ using the Wilcox rank sum test.

**Table 3.5:** *Scores and compliance of AMoSS study participants to QIDS-SR16, ASRM and GAD-7 web-based questionnaires. Superscripts indicate if (and with which) the average value of a given patient distribution is significantly different at a level of* $p < 0.05$ *using the Wilcox rank sum test.*

| | **Healthy controls** | **Bipolar disorder** | **Borderline personality disorder** |
|---|---|---|---|
| QIDS-SR16 (Mean±SD) | $2.24 \pm 2.06^{\text{BD BPD}}$ | $8.01 \pm 3.99^{\text{HC BPD}}$ | $13.85 \pm 4.09^{\text{HC BD}}$ |
| ASRM (Mean±SD) | $1.29 \pm 2.22^{\text{BD BPD}}$ | $2.74 \pm 2.21^{\text{HC}}$ | $2.61 \pm 2.10^{\text{HC}}$ |
| GAD-7 (Mean±SD) | $1.26 \pm 1.43^{\text{BD BPD}}$ | $5.96 \pm 4.40^{\text{HC BPD}}$ | $11.87 \pm 4.94^{\text{HC BD}}$ |
| Response rate, days (Mean±SD) | 7.26±5.03 | 7.60±3.31 | 8.27±3.39 |

**Table 3.6:** *Diagnostic thresholds of weekly mood questionnaires, mood disturbance should persist for a minimum duration of time to be considered a clinically significant episode (see DSM-5 (American Psychiatric Association, 2013)).*

| Questionnaire | Score range | Interpretation |
|---|---|---|
| QIDS-SR16 | 0-5 | No depression |
| | 6-10 | Mild depression |
| | 11-15 | Moderate depression |
| | 16-20 | Severe depression |
| | 21-27 | Very severe depression |
| ASRM | 0-5 | Less likely mania |
| | 6-25 | Higher probability of mania |
| GAD-7 | 0-4 | Minimal anxiety |
| | 5-9 | Mild anxiety |
| | 10-14 | Moderate anxiety |
| | 15-21 | Severe anxiety |

From Figure 3.4 it can be seen that compliance pattern remains stable for each participant for the duration of the study. The relatively high variance in inter-response intervals is explained mostly by participants who stop responding to questionnaires after a period of time, as illustrated by Figure 3.4. It must be also noted, that according to the opinion of the AMoSS clinical team, compliance can be a function of the clinical state, *i.e.* the patient may stop responding during a manic or depressive episode.

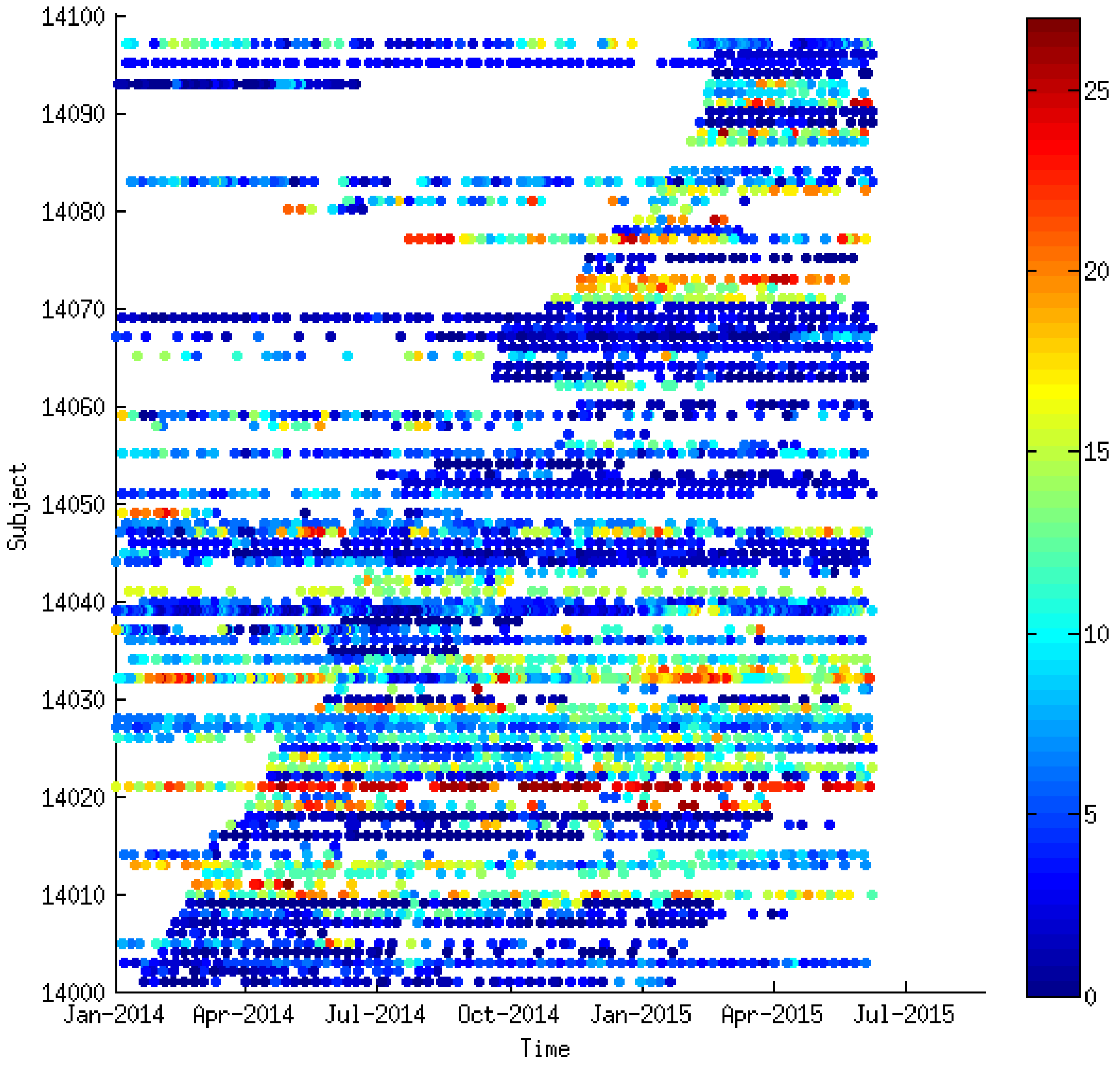

**Figure 3.4:** *Weekly QIDS-SR16 depression scores collected from the AMoSS study participants, demonstrating compliance to email mood prompts. All questionnaires are delivered simultaneously, so similar results are observed also for ASRM and GAD-7 responses. The colour of the plot indicates a score as described by the colour bar on the right side of the plot.*

#### 3.2.1.3 Daily mood questionnaires

Mood questionnaires, delivered on a weekly or bi-monthly basis are reliable and validated clinical instruments, however they may suffer from recall bias in determining the actual objective historic values of measured phenomena (Ben-Zeev et al., 2012). Therefore daily assessment of mood may be useful and included in the AMoSS study protocol. Daily mood assessment was performed using the Mood Zoom questionnaire, implemented as a part of the AMoSS mobile application.

The Mood Zoom questionnaire assesses self-rated mood on six dimensions, namely, anxiety, elation, sadness, anger, irritability and energy, using a 7-item Likert scale, rating each state from "Not at all" applicable to "Very much" applicable. Mood Zoom does not have a diagnostic purpose and is intended to provide a more detailed assessment of mood, irrespective of the underlying disorder, as we can see from Table 3.7.

**Table 3.7:** *Scores and compliance to Mood Zoom questionnaires of AMoSS study participants. Numbers from 1 to 7 correspond to grades of Likert scale with a lower grade representing the lower subjective experience level. Superscripts indicate that the distributions are different with* $p < 0.05$ *according to Wilcoxon rank sum test.*

| | **Healthy controls** | **Bipolar disorder** | **Borderline personality disorder** |
|---|---|---|---|
| Angry (Mean±SD) | 1.39±0.59 $^{BD\ BPD}$ | 1.94±1.20 $^{HC\ BPD}$ | 2.45±0.84 $^{HC\ BD}$ |
| Anxious (Mean±SD) | 1.74±0.82 $^{BD\ BPD}$ | 2.86±1.39 $^{HC\ BPD}$ | 4.07±1.36 $^{HC\ BD}$ |
| Elated (Mean±SD) | 2.71±1.45 | 2.40±1.10 | 2.86±1.10 |
| Energetic (Mean±SD) | 3.50±1.09 $^{BD\ BPD}$ | 2.92±0.94 $^{HC}$ | 2.89±0.87 $^{HC}$ |
| Irritable (Mean±SD) | 1.68±0.69 $^{BD\ BPD}$ | 2.47±1.17 $^{HC\ BPD}$ | 3.13±1.25 $^{HC\ BD}$ |
| Sad (Mean±SD) | 1.56±0.73 $^{BD\ BPD}$ | 2.48±1.27 $^{HC\ BPD}$ | 3.76±1.21 $^{HC\ BD}$ |
| Response rate, days (Median±IQR) | 0.92±0.51 | 1.05±0.64 | 1.01±0.69 |

The Mood Zoom is delivered using the AMoSS smart phone application, and the user is notified about the submission time via the phone alarm, so compliance may be different compared to email-based notifications of weekly questionnaires. From Table 3.7 and Figure 3.5 we can see that the median inter-response interval is not significantly different between groups and the variance can be explained mostly by participants who stopped responding at some point of the study.

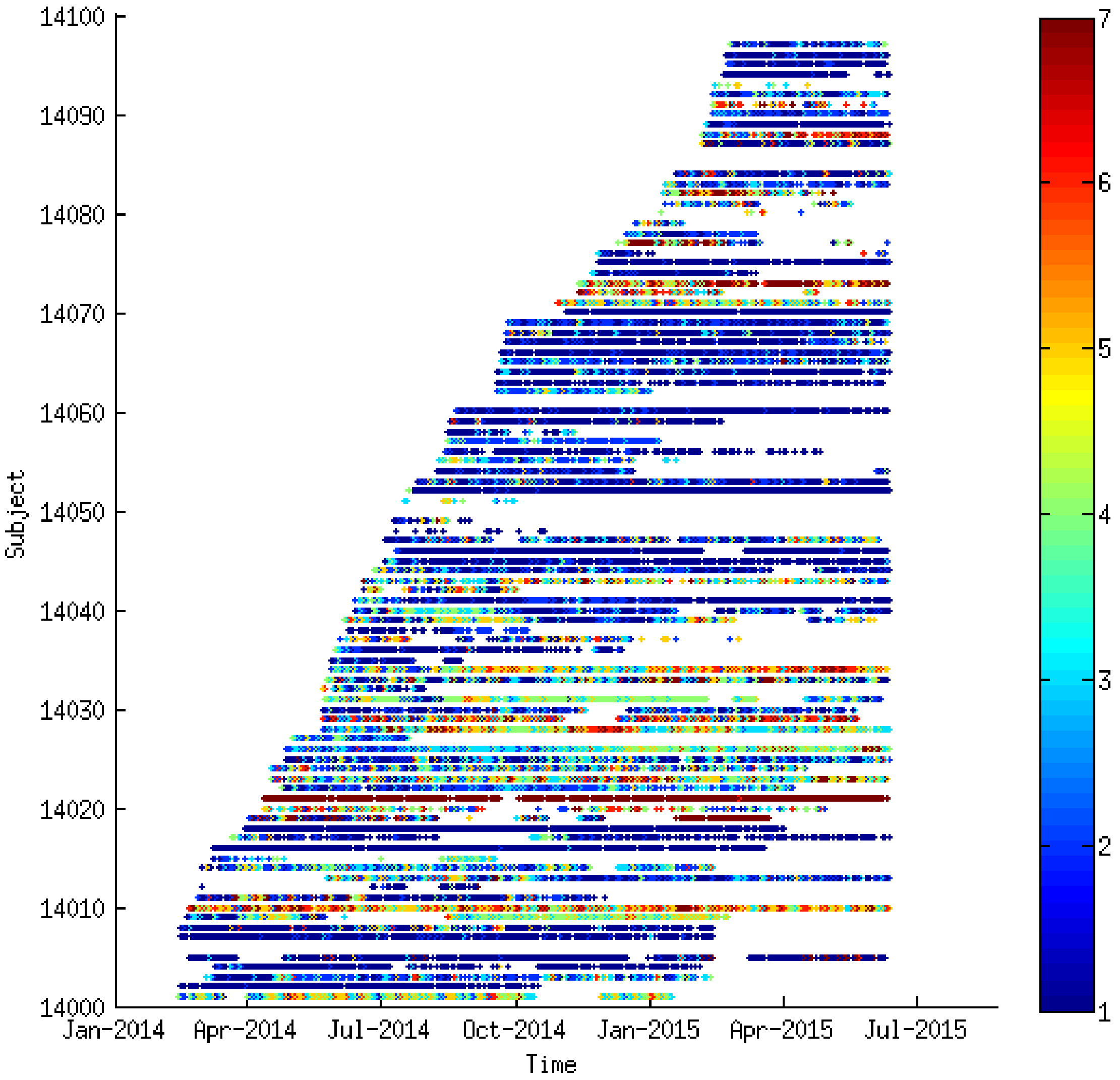

**Figure 3.5:** *Mood Zoom sadness self-rating collected from the AMoSS study participants. According to the application's design, all Mood Zoom questions have to be answered to enable submission of results, so compliance is the same for all mood dimensions. The colour of the plot indicates the score as described by the colour bar on the right side of the plot.*

### 3.2.2 Objective behaviour assessment

Objective assessment of behaviour in this work is based on the analysis of data from accelerometers embedded in mobile phones and GENEActiv wrist-worn devices. Activity measurement using accelerometers (e.g. actigraphy), is an inexpensive and easy-to-use approach, frequently used in sports medicine (Swartz et al., 2000), for sleep (Sadeh, 2011) and circadian research (Ancoli-Israel et al., 2003). Measurement results depend on device location on the human body. For sleep and circadian rhythms analysis (and also in this project), the non- dominant wrist was selected as the preferred location of the accelerometer (Steenari and Aronen, 2002; Berger et al., 2008), but no significant difference in analysis results were reported between dominant and non-dominant wrists, as well as if worn on the waist (Sadeh et al., 1994).

Accelerometers measure changes in speed, typically using piezo-electric sensors. Accelerometer hardware is constantly evolving, with many devices on the market being from an older generation, offering non-directional measurement of acceleration in arbitrary units (counts) rather than SI units ($m/s^2$). Such devices suffer from technological limitations, including a limited amount of data storage, low sampling rates (below 0.1 Hz) and non-linearity of acceleration measurements. Results of activity measurement in such devices are usually collected in epochs of several seconds or minutes (Roebuck et al., 2013). Activity data are often post-processed and available as zero-crossing timestamps (frequency of movement), time-above-threshold (duration of movement) or periodic integration information (intensity of movement) (Hersen, 2006).

Modern micro-electro-mechanical system (MEMS) accelerometers (used by this project), are capable of precise tri-axial acceleration measurement. Such accelerometers only recently started to be used in research (te Lindert and Van Someren, 2013) and allow for a reliable automated classification of activity, with the study of Zhang *et al.* demonstrating 99% classification accuracy of walking, running, household and sedentary activities in a study with 60 healthy participants in laboratory and outdoor environments (Zhang et al., 2012). Modern accelerometers usually integrate a set of sensors, including a tri-axial accelerometer and others (such as light, temperature and pulse rate), however for simplicity the word "actigraph" or "accelerometer" will be used to refer to such devices.

In this work, the main focus is on physical activity data, obtained from GENEActiv (Activinsights Ltd, UK) accelerometers (see Figure 3.6) and Galaxy S3 or S4 mobile phone sensors, that are described in detail in the following sections.

#### 3.2.2.1 Wrist-worn accelerometer

The GENEActiv accelerometer is a waterproof research-grade device, suitable for continuous 24x7 use and includes a tri-axial accelerometer, light sensor and body temperature sensor with operating characteristics presented in Table 3.8. Collected data are stored in internal memory and can be uploaded for analysis using a specialised cradle.

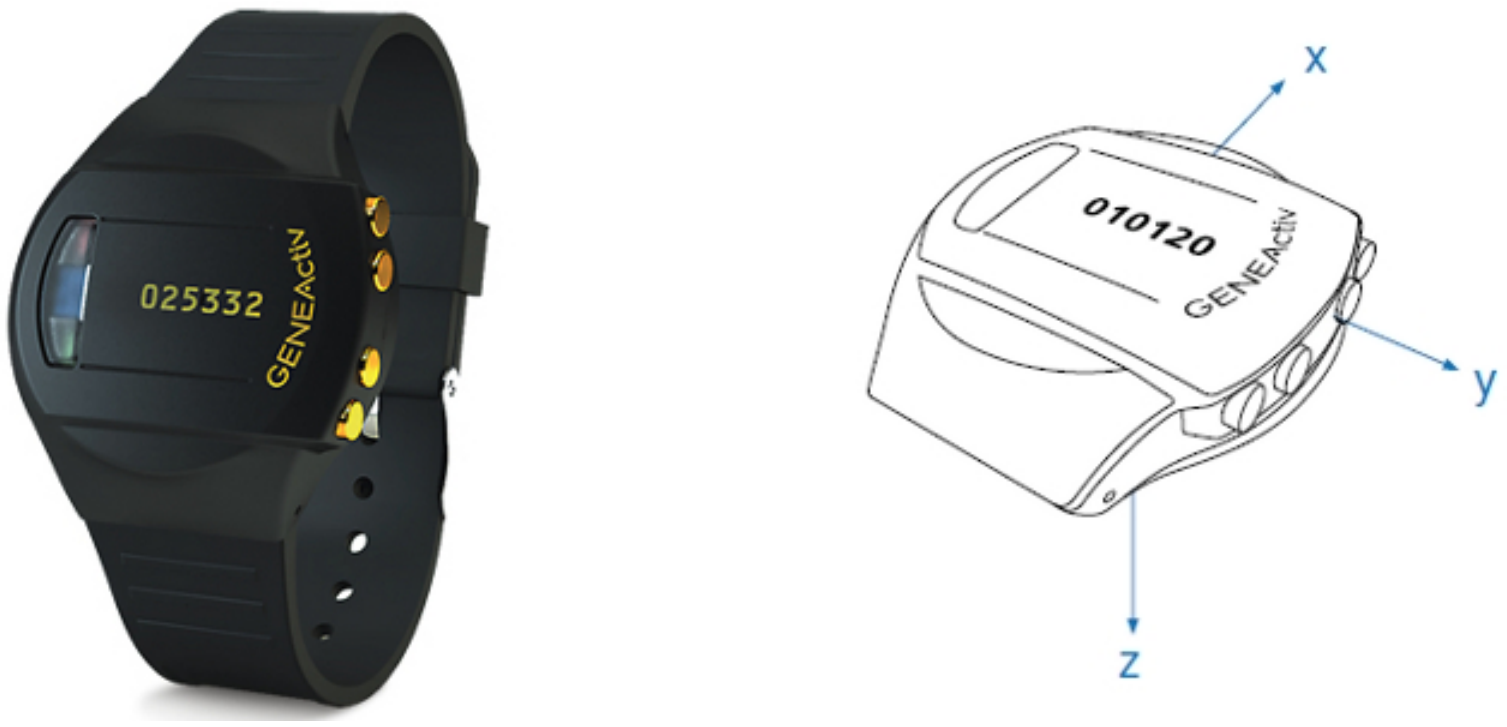

a. b.

**Figure 3.6:** *GENEActiv accelerometer, a) presenting the physical design and b) the orientation of acceleration measurement axes. Adapted from (Activinsights Ltd., 2012).*

The GENEActiv accelerometer provides a very high quality physical activity measurement with configurable sampling frequency (frequencies of 50, 25 and 10 Hz are used in the study). The period of assessment using GENEActiv is limited by the internal memory capacity, and participants need to return the device to study administrators to upload the collected data. Therefore the duration of the GENEActiv assessment is limited to 10-28 days, depending on the configured sampling frequency, with one or more assessment sessions performed per participant during the period of the AMoSS study.

The AMoSS study participants are instructed to use the accelerometer continuously, unless taken off due to a medical procedure or similar. However, not all participants follow the instruction and devices were often taken off at night or not worn during the day time. During the non-wear time the accelerometer measures 1g acceleration (gravity) and it affects (decreases) the cumulative daily activity measurements. Figure 3.7 presents the average daily acceleration for all GENEActiv sessions, with gravity subtracted. We can see that decreased daily acceleration values (dark blue) usually appear closer to the end of the measurement session, as participant compliance decreases towards the end of assessment.

**Table 3.8:** *Operating characteristics of GENEActiv accelerometer (Activinsights Ltd., 2012) and integrated sensors.*

| **Parameter** | **Value** |
|---|---|
| Sampling rate | 10-100 Hz |
| Record duration | 45 days @ 10 Hz, 7 days @ 100 Hz |
| Battery life | 2 months |
| Operating temperature | 5 to 40 deg C |
| | |
| Accelerometer type | MEMS 3-directional |
| Accelerometer range | $+/-8$ g |
| Accelerometer resolution | 3.9 milli g |
| | |
| Light sensor type | Silicon photodiode, 400 to 1100 nm |
| Light sensor range | 0 to 3000 Lux |
| Light sensor resolution | 5 Lux |
| | |
| Temperature sensor type | Linear active thermistor |
| Temperature range | 0 to 60 deg C |
| Temperature resolution | 0.25 deg C |
| | |
| Other features | Event button, Water-resistant to 10 m |

The information, collected with GENEActiv accelerometers, potentially allows identification of objective differences in activity between participant groups, but may not be enough to capture a large number of clinically significant disorder episodes for two main reasons:

- According to the DSM-5 diagnostic criteria, symptoms of depression need to be present for at least two weeks for a clinically significant depressive episode to be detected (or one week for a manic episode).

- In bipolar disorder, disorder episodes are relatively infrequent, and even in rapid cycling bipolar (approximately 10% of patients) episodes may happen up to four times a year.

Patients in the AMoSS study are well-managed and the probability of capturing an episode, while assessing a randomly selected continuous 10-28 day period is quite low. Therefore to capture clinically significant episodes of bipolar disorder in well-managed patients, such as our study cohort, we need to acquire at least one year of data, so that periods of euthymia and deterioration are both present in collected information. Due to usability issues of research-grade instrumentation (Teicher, 1995), such data acquisition is practically possible only with consumer-grade devices.

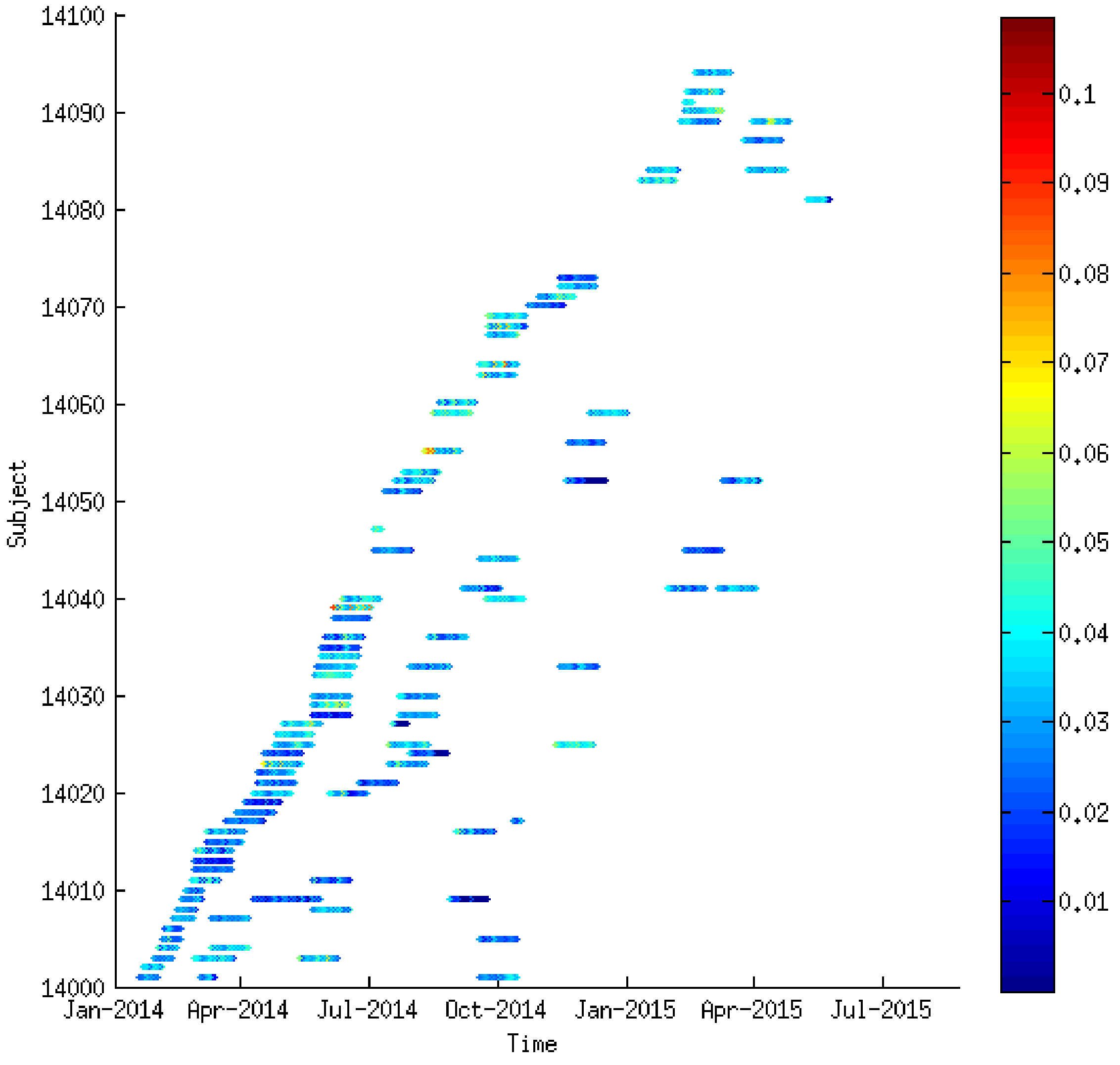

**Figure 3.7:** *GENEActiv activity data collection sessions. Each bar represents a single assessment session. The colour of the plot represents the mean daily acceleration in units of g at sea level, after removing 1 G as the background gravitational field, with deep blue indicating a very low level of activity, such as non-wear or sedentary time.*

#### 3.2.2.2 Mobile phone

Of all consumer devices with the potential to record accelerometry, none is more ubiquitous than the mobile phone, therefore this work focused mostly on the data collected with mobile phone sensors. For continuous long-term physical activity and behaviour assessment in the AMoSS study, Android smart phones were used, making it possible to capture and analyse clinically significant mood episodes.

Participants were given Samsung Galaxy S3 or S4 mobile phones with pre-installed in-house developed AMoSS mobile application, continuously collecting data from phone accelerometer, light sensor, coarse geolocation, as well as time, direction and duration of phone calls and text messages. All participants used the study-provided mobile phone as their primary one.

Key phone sensors characteristics are presented in Table 3.9 and sampling frequencies for all signals are variable and defined by Android platform at run time.

**Table 3.9:** *Operating characteristics of Samsung Galaxy mobile phone sensors.*

| Parameter | Value |
|---|---|
| Sampling rate | variable up to 10 Hz |
| Record duration | unlimited |
| Battery life | depends on usage (1 day typical) |
| | |
| Accelerometer type | LSM330DLC (MEMS 3-directional) |
| Accelerometer range | ±2 g |
| Accelerometer resolution | 0.1 milli g |
| | |
| Light sensor type | CM36651 (CMOS RGB) |
| Light sensor range | 0 to 121240 Lux |
| Light sensor resolution | 1 Lux |

AMoSS mobile application runs in background, providing instrument for truly ambient behaviour monitoring, where non-use of mobile phone is also a useful source of information about participants' behaviour. However, although participants were encouraged to use the study-provided mobile phone as their primary one, not all participants did so. Another issue for mobile phone data collection is data missingness due to sensor failure. Due to the specifics of the Android platform, data collection is performed in a non-deterministic environment and there are several main reasons for missing data from phone sensors:

- Any mobile application can be terminated by the Android platform if phone resources (such as memory) are needed by other applications with a higher priority (for example those running in the foreground and interacting with the user). In such situations data collection may stop for an unspecified period of time until the application is restarted.
- The acceleration sampling rate may decrease or sampling may stop completely in response to external events, for example the sampling rate decreases during periods of phone non-movement to reduce power consumption.
- The user may uninstall the AMoSS application, stop using it, or the internal phone SD card may become full, preventing the phone from saving data collected from the sensors.

Although the AMoSS project clinical team handled issues such as phone damage or accidental uninstallation, it was not possible to eliminate all problems, partially because the developers did not have access to participants' devices. Therefore the phone data collection in the AMoSS project is prone to intermittent interruptions, with frequency and duration influenced partially by phone usage patterns. Figure 3.8 presents all acceleration data collected from the study participants as a percentage of seconds with at least one sample for each day.

We can see that the percentage of missing data varies both between subjects and within the data collection period of each specific subject. Very rarely a single subject would have a data acquisition period with 100% samples collected. This emphasises the need for pre-processing methods, capable of generating an evenly sampled time series by imputing missing values with a reasonable default.

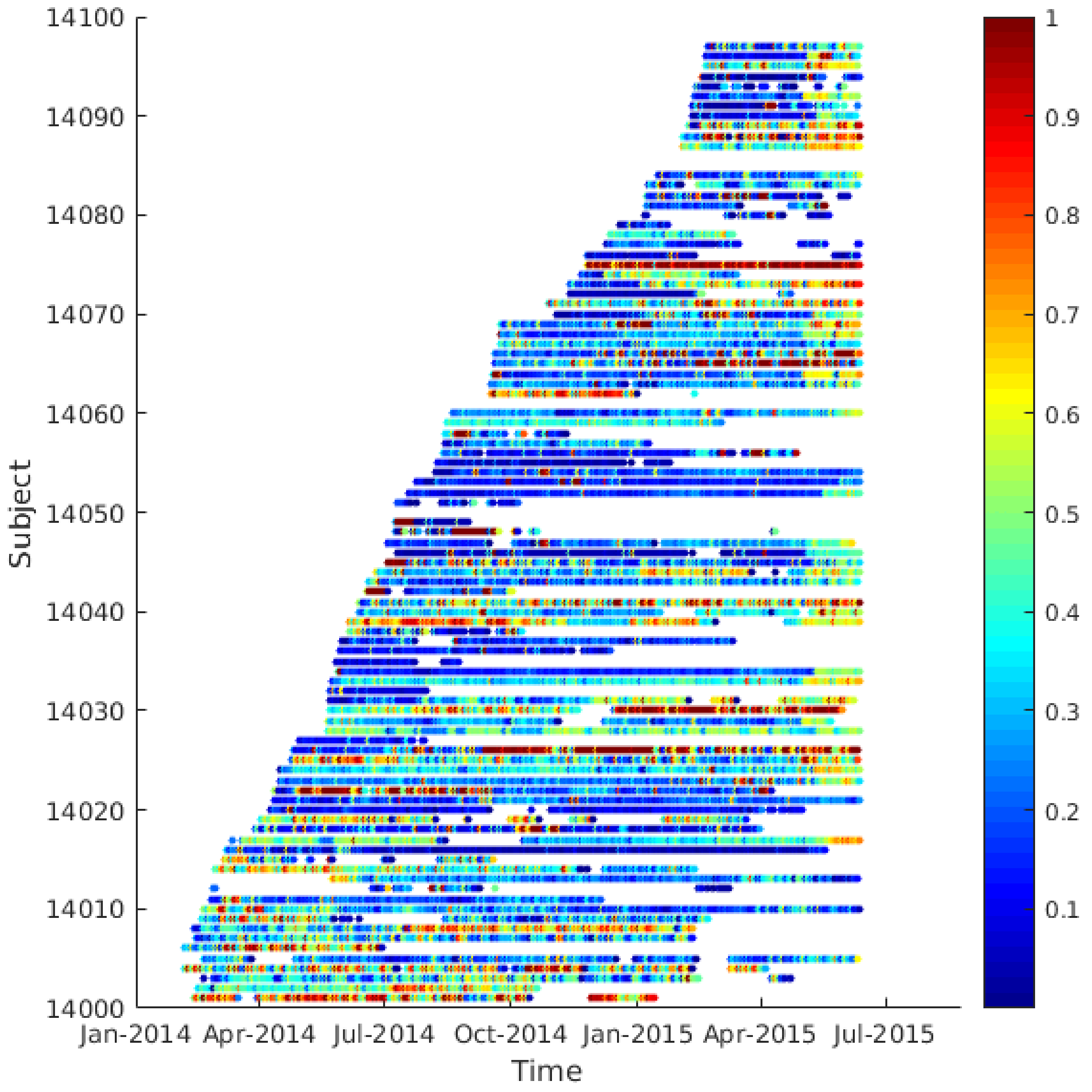

**Figure 3.8:** *Acceleration data collected from a smart phone. Each dot represents a day and the colour of the plot represents the proportion of seconds during a day with at least one sample of acceleration, deep blue indicating a very low sampling rate or missing data.*

## 3.3 Summary

The AMoSS data set includes ground truth information about the subjects' mood, together with physical activity, physiology and behavioural information, which is helpful in identifying clinically significant episodes of disorder. Given the sample size, the number and granularity of collected physiological data, research cohort composition, and the duration of the study, AMoSS stands as one of the most advanced and detailed data sources for longitudinal behavioural assessment of psychiatric disorders.

The composition of the AMoSS data set makes possible the identification of objective differences between cohorts using the captured physical activity patterns, as well as the identification of activity features that would be predictive to clinical symptoms and disorder episodes. The data subset selected for this thesis consists of accelerometer recordings from wearable research-grade devices and the in-house developed smart phone application, together with self-rated mood scores as described by Table 3.4.

The ground truth of emotional state (representing clinical state for our study population) is represented by subjective assessment with ASRM, QIDS-16SR and GAD-7 questionnaires, validated for diagnostic accuracy for assessment of mania, depression and anxiety (Altman et al., 1997; Rush et al., 2003; Spitzer et al., 2006). However, the assessment of physical activity, sleep or other behaviours with questionnaires is known to be inaccurate (Sabia et al., 2014; Busschaert et al., 2015), therefore leading to limited agreement between even perfectly measured subjective and objective data.

GENEActiv accelerometer is widely used for assessment of physical activity and sleep, and is one of the most (independently) validated devices on the market (te Lindert and Van Someren, 2013; Rowlands et al., 2014a; Rowlands et al., 2014b; Schaefer et al., 2014; Hildebrand et al., 2014; Troiano et al., 2014; Chapman et al., 2015). However, the quality and quantity of collected data vary between participants, with one to three GENEActiv recordings per participants, each of 10 to 21 days, and up to 12 months of mobile phone data (depending on recruitment date) with variable sampling rate and missing data periods.

The factors may be limiting for the presented research and were taken into account in the data analysis strategy presented in the next chapters.

# Chapter 4

# Data pre-processing methods

## 4.1 The problem of data quality

The AMoSS data set is diverse and includes different types of information, from self-rated mood questionnaires to continuously collected data from sensors, which are not necessarily evenly sampled. In this work several data analysis methods were applied to these data, extracting different behaviour features. As noted by Enders (Enders, 2006), data quality is not an inherent characteristic of a data set, but is largely defined by analysis methods, each suggesting a different understanding of data quality and potentially different pre-processing strategies. The quality issues specific to our subset are described in Table 4.1, focusing on two main classes of data - questionnaires and sensor measurements.

Data quality issues in the context of questionnaires were discussed by Bowling (Bowling, 2005) with a focus on different modes of questionnaire administration. Many common problems, such as interviewer bias or incomplete responses in the AMoSS study were eliminated by the electronic mode of questionnaire delivery and specific design, which only allowed the submission of complete responses. The remaining issues, including the missing values (non-response) and recall bias, are addressed later in this chapter.

In the case of sensor values, data may suffer from missingness or absence (due to non-wear or sensor failure), noise (inherent characteristic of any sensor) and measurement bias (due to poor factory calibration of a sensor). These issues were analysed extensively in the literature, specifically in the case of accelerometers, and a number of approaches are proposed and described below (Lötters et al., 1998; Catellier et al., 2005; Gietzelt et al., 2008; Curone et al., 2010; van Hees et al., 2013; van Hees et al., 2014).

**Table 4.1:** *The sources of data quality issues in AMoSS data.*

| | Questionnaires | Sensors |
|---|---|---|
| Missing values | Non-response to questionnaires | Non-wear time for wearable sensors, interrupted data collection due to sensor failure |
| Noise | Typos | Noise is often a feature of sensor hardware |
| Bias | Recall bias, subjective bias (often disorder-specific) | Sensor measurement bias may be created by imprecise sensor calibration and hardware heterogeneity |

## 4.2 Noise and bias removal

### 4.2.1 Questionnaires

In the case of AMoSS questionnaires, sources of noise and bias include erroneous responses (typos) and subjective bias, such as recall bias or disorder-specific bias in responses, as noted also by Hinton (Hinton, 1966). For example, the important clinical characteristics of the borderline personality disorder is the identity disturbance, characterised by illogical or incoherent thought, feeling and behaviour, that could affect the objectivity of self-assessment in questionnaire responses (American Psychiatric Association, 2013).

Noise and recall bias could be eliminated by using a second source of observations, which is not always possible in practice. However, the AMoSS study includes collection of objective data from sensors, and one of the analysis goals of this work is to identify behaviour correlates of mood based on these data. Therefore the problem of biased questionnaire responses is partially addressed by this work by using the alternative source of behaviour observations.

### 4.2.2 Accelerometer measurements

While accelerometers are calibrated during manufacturing, there is still the possibility of systematic error in some devices (calibration error). The accelerometer calibration error may include a bias and scaling error for each measurement axis separately, as represented by the offset of accelerometer samples during inactivity periods from a 1 g sphere of a perfectly calibrated device (see Figure 4.1).

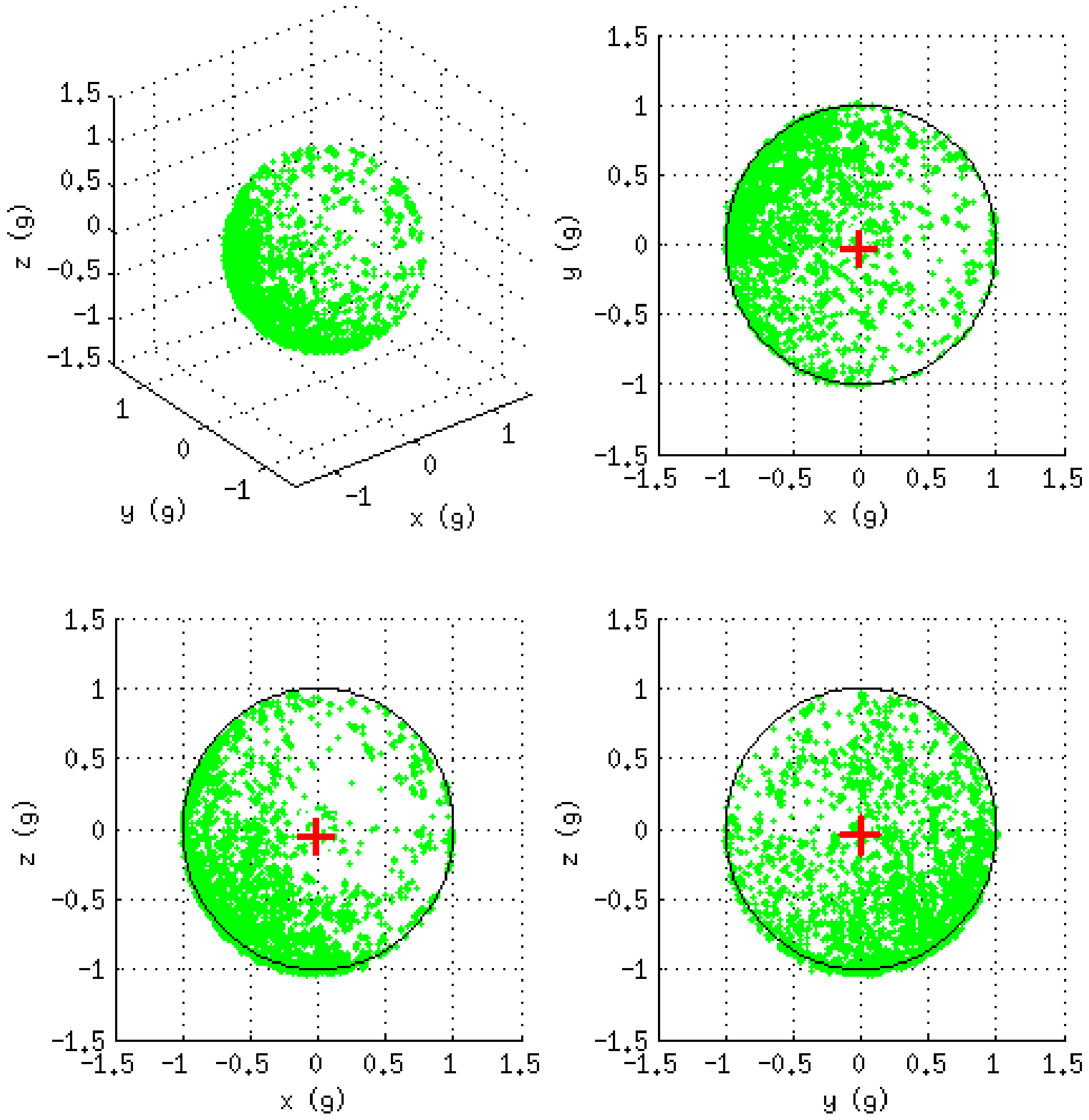

**Figure 4.1:** *GENEActiv accelerometer data plot of a sample subject. Green dots represent accelerometer samples during inactivity periods (identified as described in this section), black circles are projections of a 1 g sphere of perfectly calibrated device measurements and red cross represents the centre of mass. Accelerometer calibration error may include a bias and scaling error for each measurement axis separately.*

The principal framework for addressing this issue is called auto-calibration and includes identification of non-wear episodes, where measured acceleration should be equal to 1 g. Accelerometer measurements during these periods are located on the boundary of a 3-dimensional ellipsoid, with some offset and scaling on each of 3 axes (van Hees et al., 2014). The offset and scaling errors are removed by identifying adjustment factors for every plane, to transform measurements to a perfect sphere with a zero offset (Lötters et al., 1998; Gietzelt et al., 2008; Curone et al., 2010; van Hees et al., 2014). Such a procedure allows the identification of the calibration error directly from captured ambulatory recordings. For GENEActiv accelerometers it was validated by van Hees *et al.* (van Hees et al., 2014) and includes the following steps:

1. For every time window of 10 seconds in each data record the standard deviation and the average acceleration per axis are extracted.

2. Only time windows for which the standard deviation is $<$ 13 milli g (empirically derived) in all three axes are retained.

3. For all retained measurements at times t, axis-specific calibration is defined as $s_i^{'}(t) = d_i + s_i(t)a_i$, where $s_i(t)$ and $s_i^{'}(t)$ are acceleration before and after auto-calibration, $d_i$ is the offset and $a_i$ is the gain factor for axis $i$.

4. Parameters of the model $d_i$ and $a_i$ are optimised by minimising the average calibration error (the difference between 1 g and the euclidean norm for each measurement) using a least squares approach.

The sensor noise is then removed by filtering the transformed accelerometer measurements as described in Section 4.4. Average accuracy improvement after sensor calibration process was $0.0097 \pm 0.0060$ g (Mean±SD).

## 4.3 Dealing with missing data

The choice of strategy for dealing with missing data depends on the underlying assumptions on data missingness. The principal framework for dealing with missing values was developed by Rubin (Rubin, 1976) and describes three missing data scenarios:

- Missing Completely At Random (MCAR): The probability of a missing value is unrelated to both observed and unobserved values.

- Missing At Random (MAR): The probability of a missing value is unrelated to the missing values itself only.
- Missing Not At Random (MNAR): Missingness depends on the missing data.

Traditional strategies for dealing with missing values, including the arithmetic mean imputation, regression imputation, stochastic regression imputation and the last observation carried forward, will provide an unbiased estimate only for MCAR data (Enders, 2006). The approaches that would work under both MCAR and MAR assumptions include the Maximum Likelihood (ML) and Multiple Imputation (MI) (Rubin, 1976; Schafer and Olsen, 1998; Enders, 2006). However, in many (if not all) practical cases the missing data are related to both observed and unobserved values. In such cases the approach for dealing with missing data would depend on the model of data missingness.

In order to understand the nature of data missingness, a separate experiment may need to be performed that require participants to recall the reason for not wearing a certain sensor or not responding to a questionnaire. Due to resource constraints such activity was not performed in the framework of AMoSS study.

### 4.3.1 Questionnaires

For subjective mood questionnaires, extensive analysis of mood time series with a focus on forecasting was performed by Moore (Moore et al., 2012; Moore et al., 2014). Forecasting methods, as well as imputation, use the known time series data to predict unknown values, so forecasting methods may be also used to deal with missing data. Several forecasting methods were investigated by Moore *et al.*, including the unconditional mean, last value carried forward, exponential smoothing and Gaussian Process regression. It was demonstrated that the unconditional mean provides a better forecast in almost half of all analysed mood time series (Moore et al., 2012), so it was selected as the imputation strategy for this study.

### 4.3.2 Accelerometer measurements

For wearable sensors, the main reason for data missingness is the non-wear time, where the data readings are constant as opposed to no data during sensor failure. The identification of missing data periods for GENEActiv and other wearable accelerometers was extensively analysed in the literature (Lötters et al., 1998; Catellier et al., 2005; van Hees et al., 2011; van Hees et al., 2013). Proposed methods of non-wear episodes detection are based on identification of inactivity periods, where the standard deviation of the signal is below a certain threshold. This work adopts the method used by da Silva *et al.* (da Silva et al., 2014) and van Hees *et al.* (van Hees et al., 2013), where for each 30 minute window, the window is classified as non-wear time if for at least two axes the standard deviation of acceleration is below 13 milli g and the signal range is below 50 milli g.

In human activity time series, data imputation is a difficult problem due to the fact that data may be missing during specific activities, such as bathing or excercise (MNAR assumption). Catellier *et al.* demonstrated that data missingness depends on the time of the day and day of the week, therefore a simple imputation strategy is not appropriate (Catellier et al., 2005). Sabia *et al.* (Sabia et al., 2014) and van Hees *et al.* (van Hees et al., 2011) used arithmetic mean imputation, however took for averaging only data from the same time of day, thus incorporating global knowledge into the data model. In this work, following the approach of Sabia *et al.* and van Hees *et al.*, non-wear time was imputed with data from the same time of day, and of the same recording, averaged across the surrounding 5 days, thus preserving local signal characteristics.

## 4.4 Averaging and dimensionality reduction

Modern accelerometers include tri-axial micro-electro-mechanical systems (MEMS) sensors for three-dimensional acceleration measurement with sub-second time resolution. However, for analysis, these measurements are usually converted into a uni-axial representation, measuring cumulative activity for a certain period of time.

### 4.4.1 Euclidean Norm Minus One (ENMO)

Focusing on the day-time activity analysis, van Hees *et al.* (van Hees et al., 2013) performed comparison of different summary activity metrics, derived from tri-axial accelerometers, and analysed the correlation between activity measurement results and direct energy expenditure measurements. The analysed summary metrics included:

- Simple Euclidean norm (EN).
- Euclidean norm minus one, with negative numbers rounded to zero (ENMO).
- Euclidean norm of the high-pass filtered signal (HFEN).
- Euclidean norm of low-pass filtered signal minus one ($HFEN_+$).

The ENMO and $HFEN_+$ were found to explain most variance in daily PAEE, and ENMO was argued to be the preferred metric in both free-living and robot experiments, and was subsequently used in large-scale physical activity studies (da Silva et al., 2014; van Hees et al., 2014). This study used the ENMO as the preferred data representation for day-time activity analysis for compatibility with previous research, and to retain the frequency-dependent information that could be eliminated by filtering.

### 4.4.2 Epoch-based "activity counts"

A large number of actigraphs on the market are older generation devices, usually offering non-directional measurement of acceleration in arbitrary units (counts) rather than gravity units (g). Results of activity measurements in such devices are usually integrated over epochs of several seconds to create so called 'activity counts' (Hersen, 2006).

Many activity analysis methods are developed and validated for such epoch based 'activity counts' data, specifically for sleep analysis, therefore it is important to ensure the compatibility of data formats between modern and traditional devices. For tri-axial GENEActiv accelerometers, te Lindert (te Lindert and Van Someren, 2013) developed a procedure for converting the raw acceleration into activity counts, compatible with Actiwatch (CamNTech, UK) format, and validated it in sleep analysis settings. The procedure includes the following steps:

1. Palmar-dorsal (z) axis of GENEActiv data is selected.
2. Band-pass filter (3-11 Hz, Butterworth order 5) is applied to the data.
3. The resulting signal is rectified and divided into 128 equally spaced bins between 0 and 5 g at the natural sampling frequency (25 or 50 Hz).
4. Within each second, peak value is detected and assumed to be the second count value.

The procedure was validated in a study with 15 healthy adults and demonstrated agreement between the Actiwatch and GENEActiv scored sleep/wake epochs with kappa values of $0.83 \pm 0.07$, and better agreement of analysis results between two simultaneously worn GENEActiv accelerometers, compared with two simultaneously worn Actiwatch accelerometers. This data format, however, may not perfectly represent the subject's physical activity due to the processing of palmar-dorsal axis, therefore in this study "activity counts" were used for sleep analysis only.

## 4.5 Activity segmentation

Non-stationarity is one of the key characteristics of human activity and physiology, driven by either exogenous impulses (such as sleep-wake routine, structured working days, alarms, human interaction, etc.), or intrinsic changes in state variables such as mental activity, blood pressure, heart rate, endogenous activity level and level of consciousness due to sleep and circadian rhythms.

Analysis of non-stationarity is especially important in the context of ambulatory human behaviour, where data are collected from a range of mobile and wearable sensors and non-stationarities describe changes in physiological states, habitual activities or the influence of external factors. Time-related information about transitions between such states is valuable for the assessment of subjects' daily routine, for analysis of sleep and circadian rhythms (Aritake-Okada et al., 2010). Different analysis methods may also be applied to data captured during different physiological states, such as sleep and wakefulness, therefore such states need to be identified (Blackwell et al., 2005; Tracy et al., 2014).

Time series segmentation allows the identification of semi-stationary states within a complete recording. Depending on the states of interests, a number of methods for activity segmentation were proposed in this study, which were helpful for identifying high level states of sleep and wakefulness, as well as transitions between activities within these states, as described in the next sections.

### 4.5.1 Stationary activity segments

One approach for the analysis of non-stationarity of physical activity is splitting the entire recording into stationary segments (each representing a certain behaviour or physiological state) and analysing their duration distribution. Change point detection plays a critical role in the identification of stationary segments.

In order to select the best method of change point detection to be applied to human locomotor activity data, we first evaluated a number of methods on a synthetic data set, created using the physiological heart rate model of McSharry-Clifford (McSharry et al., 2002). The heart rate is known to be correlated with physical activity (Freedson and Miller, 2000), so it is to be expected that segments duration distributions are similar between both. The best performing method was then applied to human data.

#### 4.5.1.1 Recursive Mean Difference Maximisation (RMDM)

This algorithm was proposed by Bernaola-Galvàn (Bernaola-Galván et al., 2001) to study the scaling behaviour of the human heart rate and aims to recursively maximise the difference in mean values between adjacent segments. It was applied to physiological signals, including inter-beat intervals of the human heart (Bernaola-Galván et al., 2001) and DNA analysis (Haiminen et al., 2007; Bernaola-Galván et al., 2012).

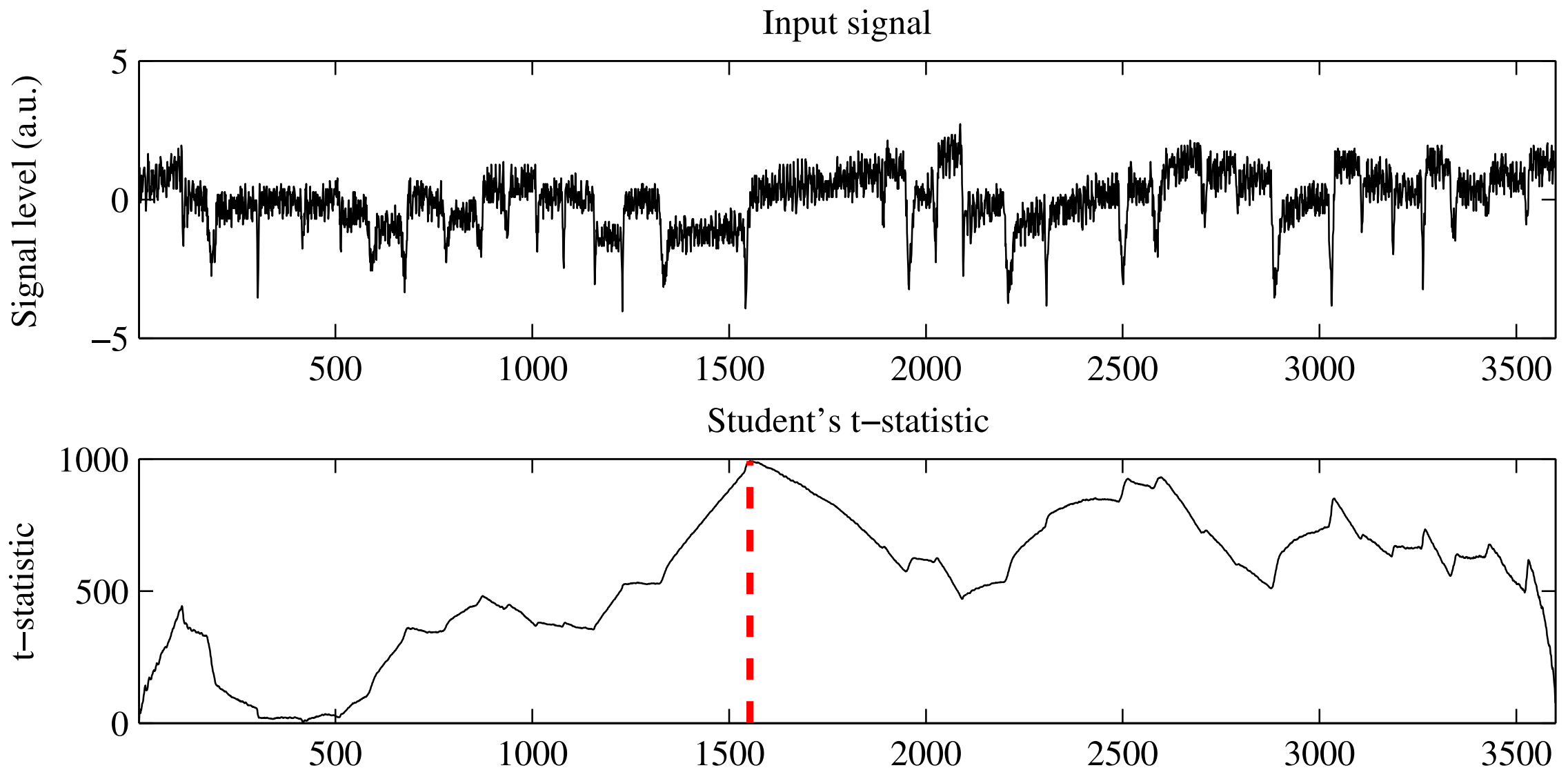

**Figure 4.2:** *The first step of RMDM algorithm. The input signal in normalised units (top plot), together with the Student's t-statistic for the difference of mean between the left and right subsignals (bottom plot) and the candidate change point at $t_{max}$ (red dashed line).*

Given the input signal $S = \{x_1, x_2, ..., x_N\}$ of length $N$, a sliding pointer is moved from left to right, splitting the signal into $S_1 = \{x_1, x_2, ..., x_j\}$ and $S_2 = \{x_{j+1}, x_{j+2}, ..., x_N\}$, where $j$ is the position of the pointer (Figure 4.2). For each point, the means of the left and right signals are calculated as

$$\mu_1 = \frac{1}{N_1} \sum_{x_i \in S_1} x_i, \quad \mu_2 = \frac{1}{N_2} \sum_{x_i \in S_2} x_i. \tag{4.1}$$

The difference between the means is calculated using the Student's $t$-statistics

$$t(S_1, S_2) = \left| \frac{(\mu_1 - \mu_2)}{\sqrt{\sigma_P}} \right|, \tag{4.2}$$

where $\sigma_P$ is the pooled variance, defined as

$$\sigma_P = \frac{N(V(S_1) + V(S_2))}{(N-2)N_1 N_2} \tag{4.3}$$

and $V(S)$ is the sum of squared deviations of the data in the signal $S$:

$$V(S) = \sum_{x_i \in S} (x_i - \mu)^2. \tag{4.4}$$

The Student's $t$-statistic is calculated as a function of the position $j$ in the timeseries and a candidate change point $j_{max}$ is selected, where $t(j)$ reaches the maximum $t_{max}$.

The significance level $\mathcal{P}(\tau)$ of the change point is calculated as $\mathcal{P}(\tau) = \{t_{max} \leq \tau\}$, where $\mathcal{P}(\tau)$ could not be obtained in a closed analytical form and was numerically approximated by Bernaola-Galvàn as

$$\mathcal{P}(\tau) = \left\{1 - I_{\left[\frac{\nu}{\nu+\tau^2}\right]}(\delta\nu, \delta)\right\}^{\gamma}, \tag{4.5}$$

where $\gamma = 4.19lnN - 11.54$, $\delta = 0.40$, $N$ is the length of the signal, $\nu = N - 1$ is the number of degrees of freedom, and $I_x(a, b)$ is the incomplete beta function. Note, that the probability of the change point depends only on the length of the original signal and the selected significance level $\tau$ (usually set to 0.95). The dependency of the length of time series on the selected significance threshold is illustrated by Figure 4.3.

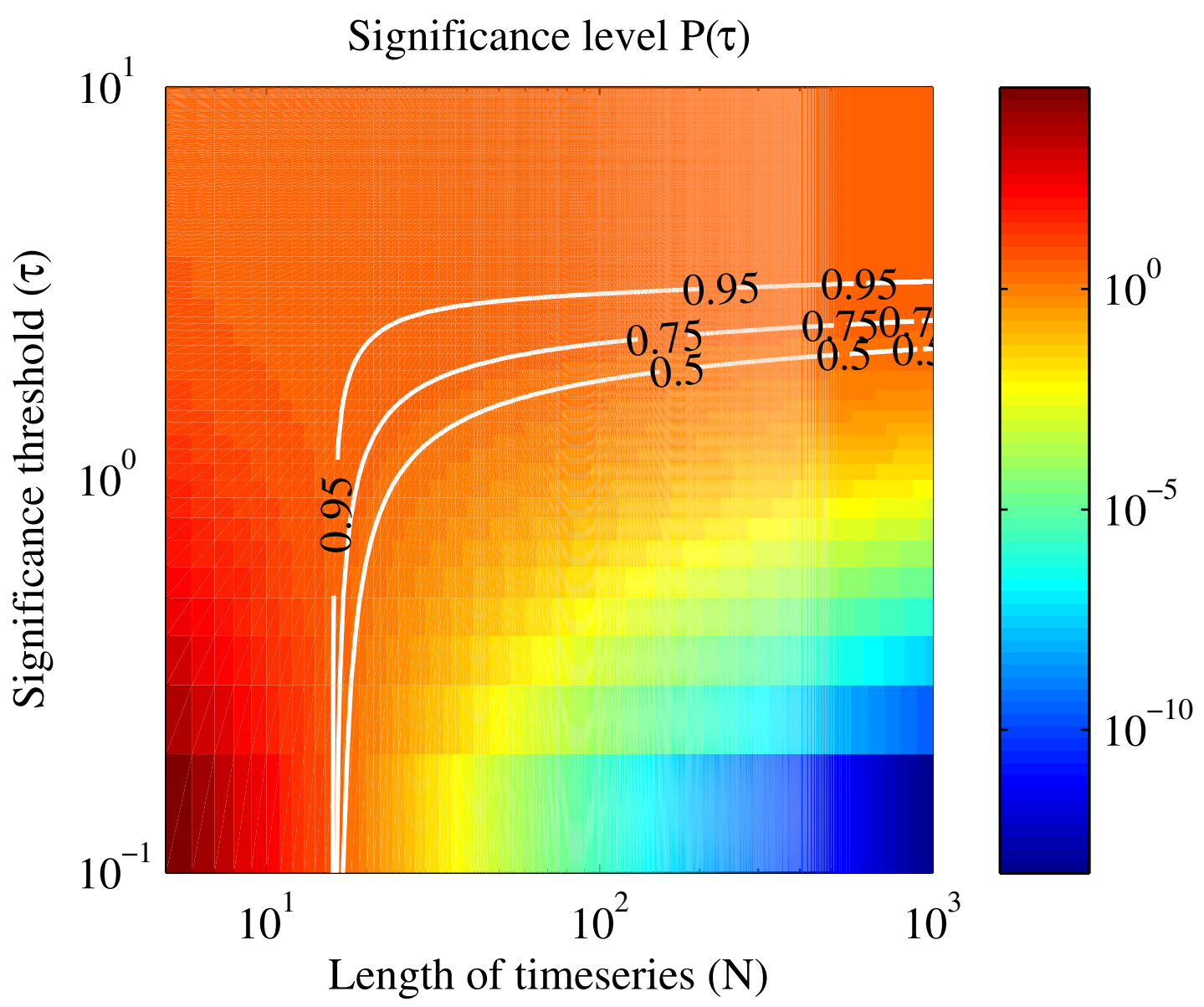

**Figure 4.3:** *The significance $\mathcal{P}(\tau)$ of the change point as a function of the length of timeseries and the selected significance level $\tau$ (50%, 75% and 95% lines are presented).*

If the difference in means is significant, the signal is split into two segments (if the means of the new segments are significantly different from the means of adjacent segments) and the above procedure is repeated for each new segment while significance holds. Note that a weak stationary requires both the mean and variance to be constant and so this technique does not segment non-stationary changes due to shifts in variance.

#### 4.5.1.2 Bayesian Blocks (BBLOCKS)

The Bayesian Blocks algorithm was proposed by Scargle *et al.* (Scargle et al., 2013) and provides a dynamic programming approach for change point detection. The algorithm uses a piecewise constant signal model and performs optimal fitting of model to data. The algorithm requires fitness of segments to be additive, so that

$$F\left[\mathcal{P}(S)\right] = \sum_{k=1}^{N_{segments}} f(S_k), \tag{4.6}$$

where $F\left[\mathcal{P}(S)\right]$ is the total fitness of the segmentation $\mathcal{P}$ of the signal $S$ and $f(S_k)$ is the fitness of the segment $k$, calculated as

$$f(S_k) = \frac{(\sum_N x_i)^2}{4\sum_N \sigma_i^2}, \tag{4.7}$$

where $N$ is the number of data points within a segment, $x_i$ is the individual data point and $\sigma_i$ is the expected error of the data point measurement. Note, that in our case of unknown error of data point measurement and normalised signal, the expected error is taken as $\sigma = 1$.

The optimal segmentation is defined as the one with the maximum fitness, and is calculated using a dynamic programming approach. The algorithm starts with a sub-signal, including only one data point $x_{i=1}$, where only one segmentation is possible. In each step, a new datum $x_{i=i+1}$ is added to the signal and fitness of all new possible segmentations is calculated for each $j = \{1, ..., i\}$ as

$$F_i^j \left[\mathcal{P}(\{x_1, ..., x_i\})\right] = F_{j-1}^{max} + f(\{x_j, ..., x_i\}) - \gamma, \tag{4.8}$$

where the definition of variables is the same as for the Equations 4.6 and 4.7 and $\gamma$ represents prior on the expected number of segments. The change point $j$ is identified at the maximum of $F_i^j \left[\mathcal{P}(\{x_1, ..., x_i\})\right]$ (see Figure 4.4).

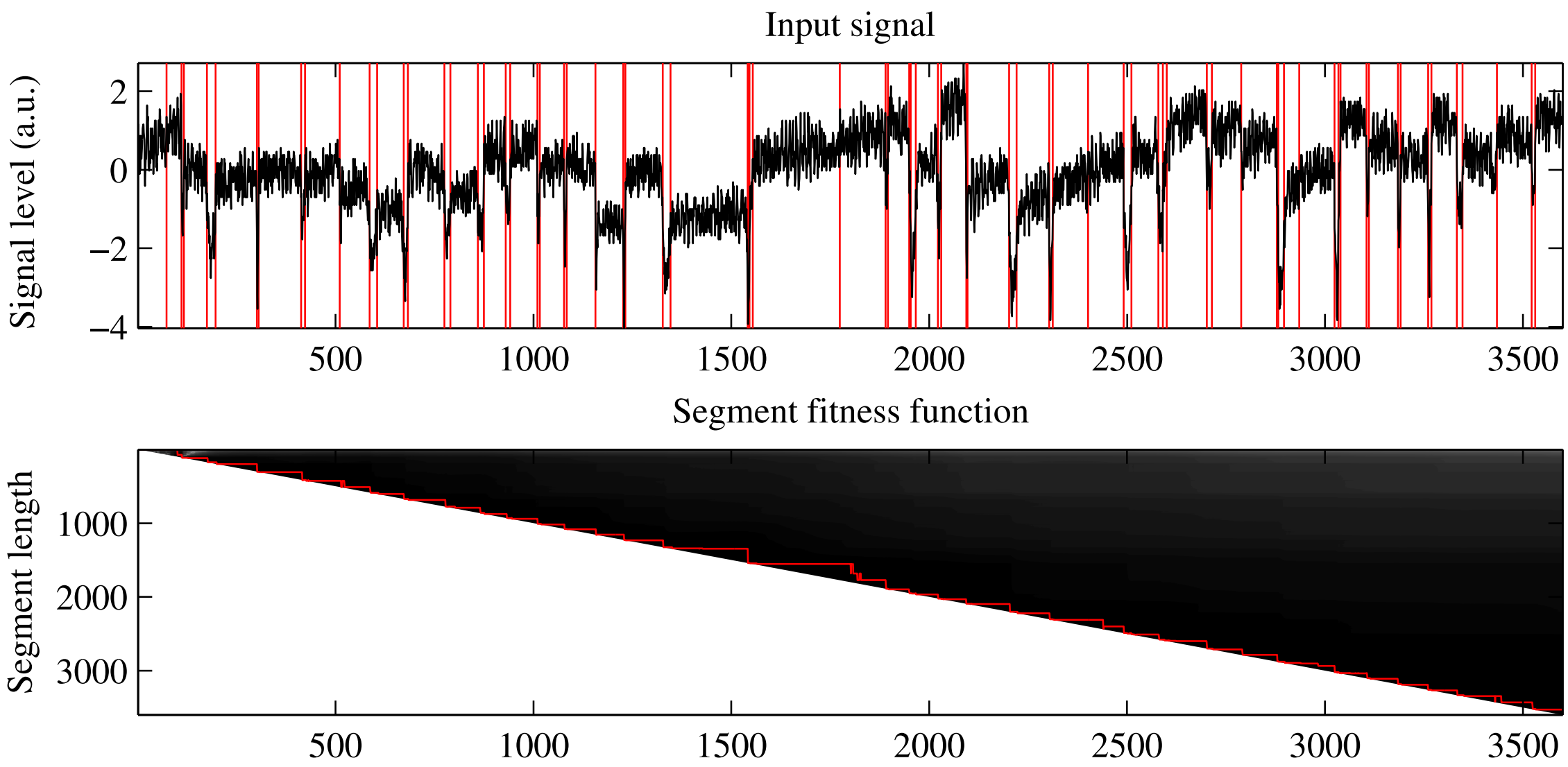

**Figure 4.4:** *The results of BBLOCKS algorithm. The input signal in normalised units (top plot, black) and the resulting segmentation (top plot, red), together with iterative estimation of the segment fitness function (bottom plot, black), where the maximum fitness indicates the last change point (bottom plot, red).*

The process continues for all data points of the signal and the sequence of $j$ values at the $i = N$ defines the detected change points in the signal $S$.

#### 4.5.1.3 Bayesian Online Change Point Detection (BOCPD)

A Bayesian Online Change Point Detection algorithm was proposed by Adams and MacKay (Adams and MacKay, 2007) and further developed by Turner *et al.* (Turner et al., 2009). The method iteratively estimates the probability distribution of segment lengths using the Bayesian inference technique (Figure 4.5).

The method was originally proposed for conjugate-exponential models, where exact inference can be performed. In this work BOCPD was applied to normally distributed data with unknown mean and variance. In this case the conjugate prior distribution takes the form of Normal-inverse-Gamma distribution

$$f(x, \sigma^2|\mu, \nu, \alpha, \beta) = \frac{\sqrt{\nu}}{\sigma\sqrt{2\pi}} \frac{\beta^\alpha}{\Gamma(\alpha)} \left(\frac{1}{\sigma^2}\right)^{\alpha+1} exp\left(-\frac{2\beta + \nu(x-\mu)^2}{2\sigma^2}\right) \tag{4.9}$$

with $\mu$, $\nu$, $\alpha$ and $\beta$ hyperparameters (Bishop, 2006).

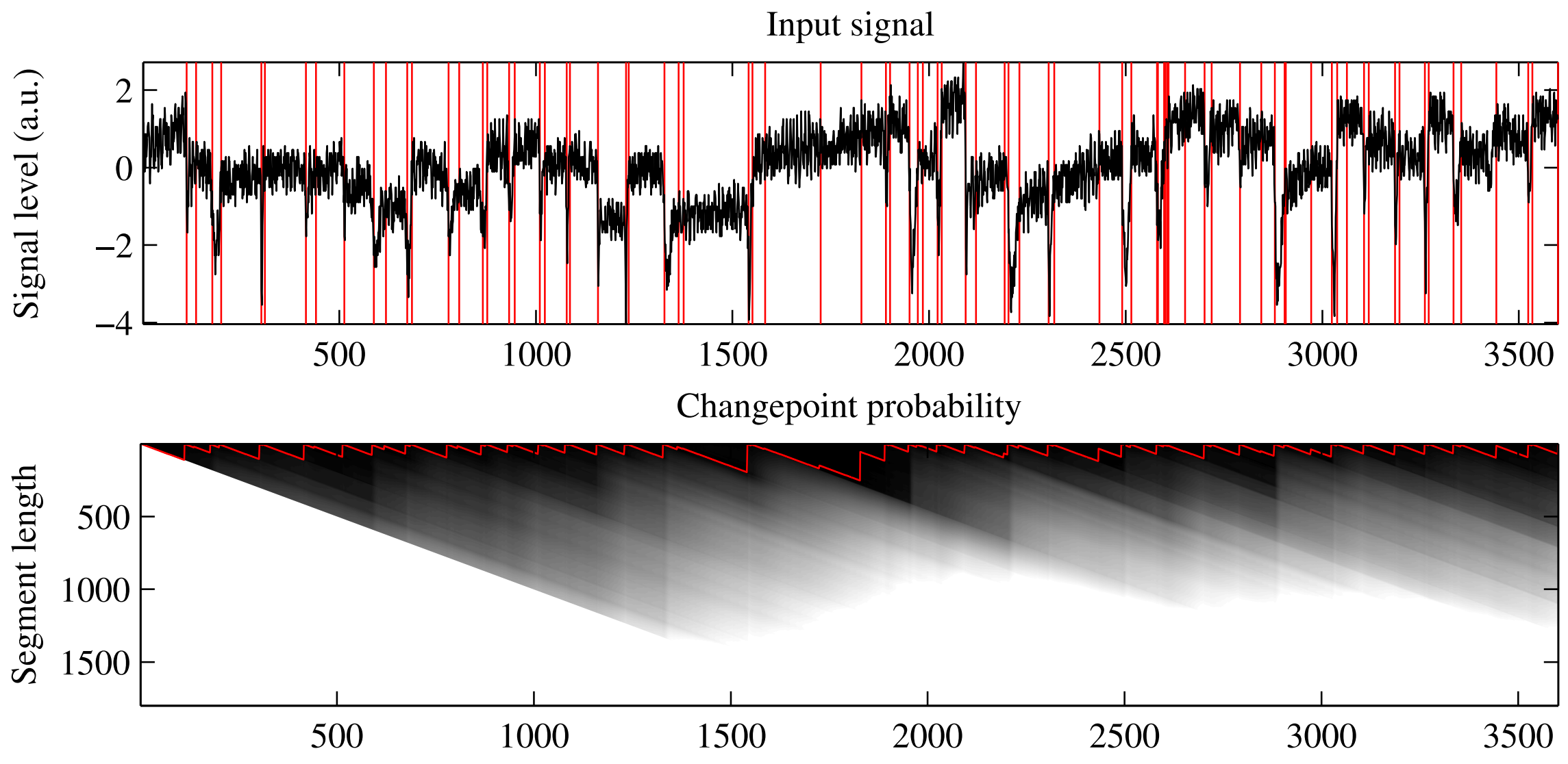

**Figure 4.5:** *The results of BOCPD algorithm. The input signal in normalised units (top plot, black) and the resulting segmentation (top plot, red), together with iterative estimation of segment length probability distribution (bottom plot, black), where the maximum indicates the most likely segment length (bottom plot, red).*

On each step of the algorithm, a new datum $x_{i=i+1}$ is added to the analysed signal. The posterior predictive probability of new datum has the form of a non-standardised Student's t-distribution (Murphy, 2007) with $2\alpha$ degrees of freedom, the centre at $\mu$ and scale parameter equal to $\frac{\beta(\nu+1)}{\alpha\nu}$ is given by

$$\pi_i^{(r)} = t_{2\alpha}(x_i|\mu_i, \frac{\beta_i(\nu_i+1)}{\alpha_i\nu_i}), \tag{4.10}$$

where r is the segment length, $\pi_i^{(r)}$ is the posterior predictive probability of segment length $r$ at data point $i$.

In order to calculate the probability of the change point, we should take into account that with each new datum the segment length may only increase (segment continues) or become zero (change point detected). The growth probabilities are calculated as

$$P(r_i = r_{i-1} + 1, x_{1:i}) = P(r_{i-1}, x_{1:i-1})\pi_t^{(r)}(1 - H(r_{i-1})) \tag{4.11}$$

and the change point probabilities are calculated as

$$P(r_i = 0, x_{1:i}) = \sum_{r_{i-1}} P(r_{i-1}, x_{1:i-1})\pi_i^{(r)} H(r_{i-1}), \tag{4.12}$$

where $r$ is the current segment length and $H(r)$ is a hazard function, defined as $H(r) = 1/\lambda$, where $\lambda$ is a free parameter of the algorithm, representing the expected segment length.

Finally, the hyper-parameters are updated according to

$$\begin{aligned} \mu_{i+1} &= \frac{\nu_i \mu_i + x_i}{\nu_i + 1}, \\ \nu_{i+1} &= \nu_i + 1, \\ \alpha_{i+1} &= \alpha_i + 0.5, \\ \beta_{i+1} &= \beta_i + \frac{\nu_i (x_i - \mu_i)^2}{2(\nu_i + 1)}, \end{aligned} \tag{4.13}$$

as defined in the conjugate Bayesian analysis of the Gaussian distribution (Murphy, 2007) and the process repeats for the remaining data of the analysed signal.

#### 4.5.1.4 Change point detection quality evaluation

Given that the number of change points detected by an algorithm can be less or exceed the number of true change points, and a correctly detected change point may be within the tolerance $\delta$ from a true change point, we used a modified confusion matrix to evaluate the change point detection accuracy, where

- True Positive (TP) is a detected change point within the tolerance $\delta$ from a true change point (each true change point can have only one detected change point),

- False Negative (FN) is a true change point without a detected change point within the tolerance $\delta$,
- False Positive (FP) is a detected change point outside of any true change point tolerance interval $\delta$ and
- True Negative (TN) does not have a meaningful interpretation, so is always 0.

#### 4.5.1.5 Artificial change point data

In order to identify true change points we must use an objective time series where we believe the change points are (almost) perfectly annotated. While there is no available open data set with annotated human activity measured by an accelerometer, there is a physiological model available for the generation of realistic heart rate time series (McSharry et al., 2002). The heart rate is known to be correlated with physical activity (Trost, 2001) and thus the HR model can be used to evaluate the performance of segmentation algorithms.

The McSharry-Clifford (McSharry et al., 2002) algorithm (RRGEN) generates realistic 24-hour RR-tachograms using a model of cardiovascular interactions and transitions between physiological states. It incorporates short range variability due to Mayer waves and respiratory sinus arrhythmia, as well as long range transitions in physiological states by using switching distributions extracted from real data. Ten different 24-hour tachograms were generated from the model together with true change points of model parameters.

All algorithms were evaluated for a range of free parameter values, experimentally selected to cover a wide range of algorithm accuracy. For RDMD the significance level was set to $\mathcal{P}(\tau)$=$\{0.5, 0.6, 0.7, 0.8, 0.85, 0.9, 0.93, 0.95, 0.98, 0.99\}$. For BBLOCKS algorithm the free parameter was set to $\gamma$=$\{0.02, 0.1, 0.2, 0.4, 0.7, 1, 2, 4, 10, 20\}$ and for BOCPD the expected segment length was set to $\lambda$=$\{5, 6, 7, 8, 10, 20, 60, 100, 200, 300\}$, experimentally selected to cover a wide range of algorithm performance. Performance of the algorithms was estimated using the number of correct detections (TP) and the number of incorrect detections (FP).

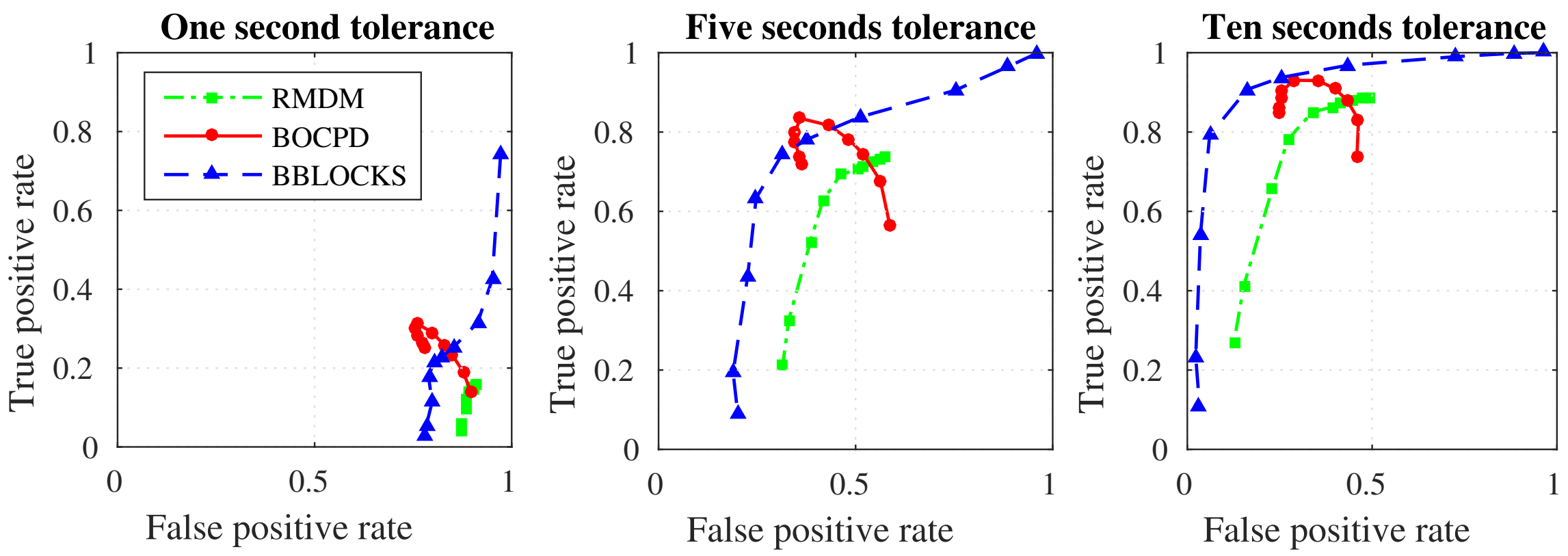

**Figure 4.6:** *Performance of change points detection algorithms on artificial* 24*-hour tachograms. All algorithms were tested with one, five and ten second change point detection tolerance level and for a range of free parameter values.*

#### 4.5.1.6 Evaluation results

The results for a range of free parameter values and different detection thresholds are presented in Figure 4.6 and Table 4.2. All algorithms demonstrated different forms of dependency between free parameter and change point detection accuracy, influenced by the selected change point detection tolerance level. The RMDM algorithm performance was lower compared to Bayesian methods in all cases. The BOCPD method performs better for exact change point detection and BBLOCKS demonstrated performance improvement with increased change point detection tolerance.

Notably, the BOCPD algorithm performance is less sensitive to the selected change point detection tolerance level, as well as free parameter settings. As the main goal of the identification of stationary segments was to detect impulsive behaviours, represented by short interruptions of continuous activities, the results of change point detection with best tolerance (one second) were considered as the most important. BOCPD algorithm also demonstrated arguably the best performance with one second change point detection tolerance and using the expected segment length of 60 seconds (see Table 4.2, selected in magenta colour).

Although the evaluation of the algorithms was performed on HR time series, only having limited correlation with the locomotor activity time series analysed in further research, the BOCPD model was earlier evaluated on other types of data (Adams and MacKay, 2007) and can be applied to any distribution from conjugate-exponential family. Based on this evaluation, the BOCPD algorithm with the expected segment length $\tau = 60$ seconds was selected for the physical activity data segmentation.

**Table 4.2:** *Performance of change points detection algorithms on artificial 24-hour tachograms. All algorithms were tested with one, five and ten second change point detection tolerance level and for a range of free parameter values. Best performance results for each algorithm and tolerance are selected in* ***bold text****.*

| Change point detection accuracy | | | | | | | | |
|---|---|---|---|---|---|---|---|---|
| RMDM | | | BBLOCKS | | | BOCPD | | |
| $\mathcal{P}(\tau)$ | TPR (%) | FPR (%) | $\gamma$ | TPR (%) | FPR (%) | $\lambda$ | TPR (%) | FPR (%) |
| One second tolerance | | | | | | | | |
| 0.50 | 0.16 | 0.91 | 0.02 | 0.74 | 0.97 | 5.00 | 0.14 | 0.90 |
| 0.60 | 0.16 | 0.91 | 0.10 | 0.42 | 0.95 | 6.00 | 0.19 | 0.88 |
| 0.70 | 0.15 | 0.91 | 0.20 | 0.32 | 0.91 | 7.00 | 0.23 | 0.85 |
| **0.80** | **0.15** | **0.90** | 0.40 | 0.25 | 0.85 | 8.00 | 0.26 | 0.83 |
| 0.85 | 0.14 | 0.90 | **0.70** | **0.23** | **0.82** | 10.00 | 0.29 | 0.80 |
| 0.90 | 0.14 | 0.89 | 1.00 | 0.21 | 0.80 | 20.00 | 0.31 | 0.76 |
| 0.93 | 0.12 | 0.89 | 2.00 | 0.18 | 0.79 | **60.00** | **0.30** | **0.75** |
| 0.95 | 0.10 | 0.88 | 4.00 | 0.11 | 0.80 | 100.00 | 0.28 | 0.76 |
| 0.98 | 0.06 | 0.88 | 10.00 | 0.05 | 0.78 | 200.00 | 0.26 | 0.77 |
| 0.99 | 0.04 | 0.87 | 20.00 | 0.03 | 0.78 | 300.00 | 0.25 | 0.78 |
| Five seconds tolerance | | | | | | | | |
| 0.50 | 0.74 | 0.58 | 0.02 | 1.00 | 0.96 | 5.00 | 0.56 | 0.59 |
| 0.60 | 0.73 | 0.56 | 0.10 | 0.96 | 0.89 | 6.00 | 0.68 | 0.56 |
| 0.70 | 0.73 | 0.54 | 0.20 | 0.90 | 0.75 | 7.00 | 0.74 | 0.52 |
| 0.80 | 0.71 | 0.52 | 0.40 | 0.84 | 0.51 | 8.00 | 0.78 | 0.48 |
| 0.85 | 0.71 | 0.50 | 0.70 | 0.78 | 0.38 | 10.00 | 0.82 | 0.43 |
| **0.90** | **0.69** | **0.46** | **1.00** | **0.74** | **0.31** | **20.00** | **0.83** | **0.36** |
| 0.93 | 0.63 | 0.42 | 2.00 | 0.63 | 0.25 | 60.00 | 0.80 | 0.35 |
| 0.95 | 0.52 | 0.39 | 4.00 | 0.44 | 0.23 | 100.00 | 0.77 | 0.35 |
| 0.98 | 0.33 | 0.33 | 10.00 | 0.19 | 0.19 | 200.00 | 0.74 | 0.36 |
| 0.99 | 0.21 | 0.31 | 20.00 | 0.09 | 0.20 | 300.00 | 0.72 | 0.37 |
| Ten seconds tolerance | | | | | | | | |
| 0.50 | 0.89 | 0.49 | 0.02 | 1.00 | 0.96 | 5.00 | 0.74 | 0.46 |
| 0.60 | 0.88 | 0.47 | 0.10 | 1.00 | 0.88 | 6.00 | 0.83 | 0.46 |
| 0.70 | 0.88 | 0.45 | 0.20 | 0.99 | 0.73 | 7.00 | 0.88 | 0.43 |
| 0.80 | 0.87 | 0.42 | 0.40 | 0.97 | 0.43 | 8.00 | 0.91 | 0.40 |
| 0.85 | 0.86 | 0.39 | 0.70 | 0.94 | 0.25 | 10.00 | 0.93 | 0.35 |
| **0.90** | **0.85** | **0.34** | **1.00** | **0.91** | **0.16** | 20.00 | 0.93 | 0.29 |
| 0.93 | 0.78 | 0.27 | 2.00 | 0.79 | 0.06 | **60.00** | **0.90** | **0.26** |
| 0.95 | 0.66 | 0.23 | 4.00 | 0.54 | 0.04 | 100.00 | 0.88 | 0.25 |
| 0.98 | 0.41 | 0.15 | 10.00 | 0.23 | 0.02 | 200.00 | 0.86 | 0.25 |
| 0.99 | 0.27 | 0.13 | 20.00 | 0.11 | 0.03 | 300.00 | 0.85 | 0.25 |

### 4.5.2 Sleep and wakefulness time (L5/M10)

The basic approach for sleep and wakefulness segmentation, widely used in actigraphic studies (Witting et al., 1990; Wirz-Justice et al., 2010; Ortiz-Tudela et al., 2010; Wulff et al., 2012), is the identification of the least active 5 hours ($H_5$) and most active 10 hours ($H_{10}$) of activity for each day, where $H_{10}$ activity data approximately represents the period when the subject is awake and active, and $H_5$ correspond to the (assumed) sleep period (see Figure 4.7).

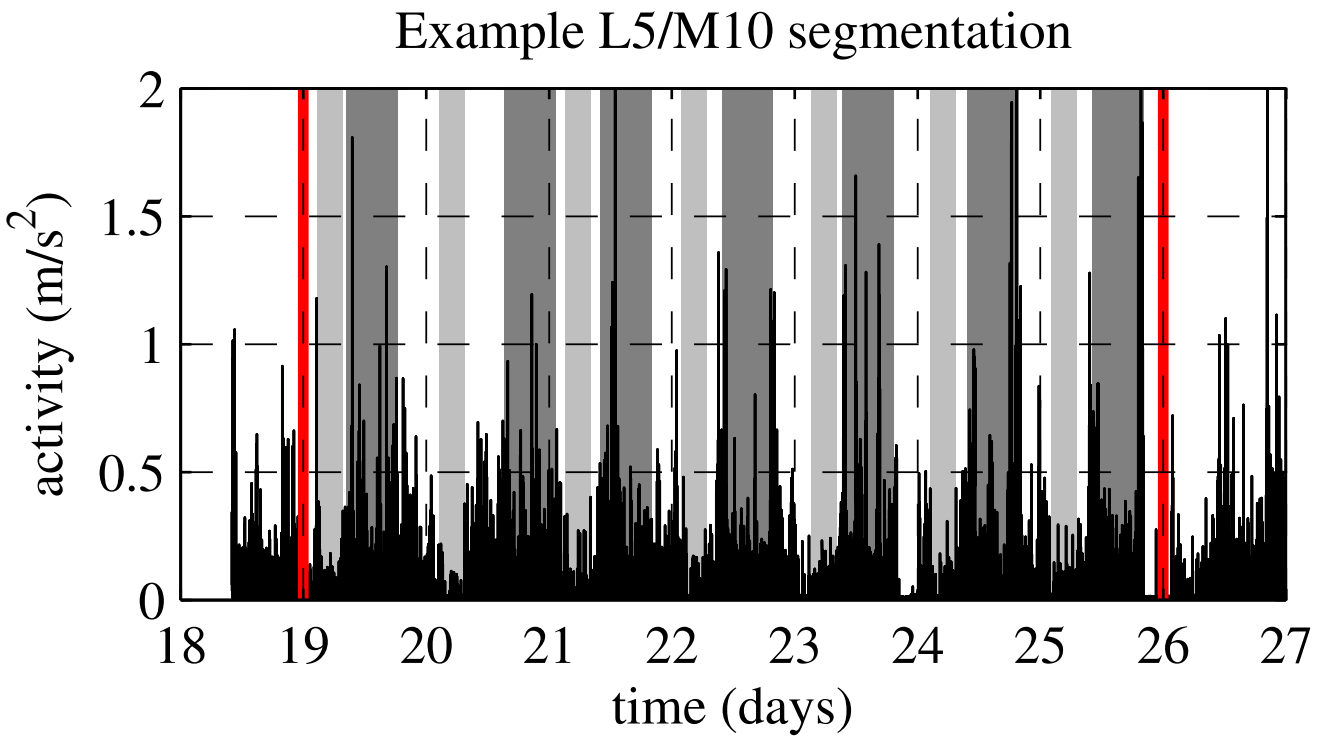

**Figure 4.7:** *Example segmentation of L5 and M10 periods. Activity time series in black, dark grey regions represent the most active 10 hours and light grey the least active 5 hours of activity. Red lines shows the beginning and end of the analysis period.*

The identification was performed by analysing a window of 24 hours and alternating search for $H_5$ and $H_{10}$ intervals. After the $H_5$ or $H_{10}$ interval was identified, a new window was analysed starting at the beginning of the just identified interval and the searched for $H_{10}$ or $H_5$ respectively, according to the following equations

$$H_{10}(i) = \underset{i}{\operatorname{argmax}} \sum_{i}^{i+|H_{10}|} x_i,$$
$$H_5(i) = \underset{i}{\operatorname{argmin}} \sum_{i}^{i+|H_5|} x_i, \tag{4.14}$$

where $x_i$ is the data point within $H_{10}$ or $H_5$ interval and $|H_{10}|$ and $|H_5|$ are the numbers of data points within these intervals (depend on sampling frequency).

### 4.5.3 Bed/wake time segmentation

Analysis of sleep is an important area of physical activity research, and particularly relevant for mental disorders (Littner et al., 2003; Van de Water et al., 2011). Analysis of sleep offers methods for calculation of such important behaviour measures as sleep onset latency or total sleep time (among others), but requires precise segmentation of bed and wake times not offered by simple L5/M10 segmentation.

In research settings, to identify the times of going to bed and waking up, diaries and manual bed time segmentation are traditionally used (Wulff et al., 2006). But in long-term studies, such as the AMoSS project, the use of diaries is difficult due to compliance issues. In the AMoSS study the bed time annotation was performed by manually pressing the Fitbit button when going to bed and after waking up. If the button was pressed, the operating mode of Fitbit changed and sleep analysis was performed by identifying sleep and wake episodes using a proprietary algorithm. The output of the proprietary algorithm was not used because there has been no published scientific validation of the algorithm, and the manufacturer may change the algorithm from device to device without informing the user. This invalidates the notion of scientific repeatability.

Furthermore, overall compliance to Fitbit data collection was at a level of 60%, and for about 60% of all nights when Fitbit was used there was bed time annotation, equating to about 36% of nights with sleep analysis, with no significant difference between study groups using Wilcoxon rank-sum test, as presented in Table 4.3, and with the following

**Table 4.3:** *Compliance of AMoSS study participants to Fitbit activity data collection (% days worn) and manual (button press) bed time annotation (% nights annotated of days worn). Demographic characteristics are presented in Table 3.2. No significant difference between study groups identified using Wilcoxon rank-sum test.*

| | **Healthy controls** | **Bipolar disorder** | **Borderline personality disorder** |
|---|---|---|---|
| % days worn (Mean±SD) | 0.60±0.47 | 0.59±0.67 | 0.74±0.48 |
| % nights annotated (Mean±SD) | 0.72±0.39 | 0.61±0.58 | 0.61±0.27 |

main issues:

- The participants forgot to press the button when going to bed or waking up. In such cases sleep analysis for the night was not performed by the Fitbit algorithm.
- The button was accidentally pressed during the day, leading to sporadic erroneous sleep annotation.
- The button was pressed too early before going to bed or too late after waking up, leading to overestimation of sleep time.

This created a problem in analysis of sleep from GENEActiv activity recordings and using Fitbit bed time annotation, as in many cases no Fitbit bed time annotation were present during the period of GENEActiv data collection. Author couldn't find any published algorithm for automatic identification of bed time from activity data. Since no published algorithm was available for sleep/wake segmentation using GENEActiv data, a novel approach was designed in this thesis based on Hidden Markov Models and validated using existing Fitbit bed time annotation, where available. Note, that Fitbit annotation suffers from compliance and bias issues, therefore only annotations that were physiologically plausible, with bed time more then 4 hours but not exceeding 16 hours were selected.

#### 4.5.3.1 Hidden Markov Models

Hidden Markov Models (HMM) provide a fundamental framework for automatic state identification in real-world processes, as well as estimation of process parameters based on captured data. An HMM is a statistical model, with a system switching between states which are not directly known to an external observer, and available only through indirect observations that are generated with a state-defined probability. The complete HMM model is defined by (Rabiner, 1989):

1. $N$, the number of states in the model, with individual states denoted as $S = \{S_1, S_2, ..., S_N\}$ and the state at time $t$ as $q_t$.

2. $M$, the number of distinct observation symbols per state, with individual symbols denoted as $V = \{v_1, v_2, ..., v_M\}$.

3. $A$, the state transition probability distribution $A = \{a_{ij}\}$ where

$$a_{ij} = P[q_{t+1} = S_j | q_t = S_i], \quad 1 \leq i, j \leq N. \tag{4.15}$$

4. $B$, the observation symbol probability distribution $B = \{b_j(k)\}$ in state $j$ where

$$b_j(k) = P[v_k \ at \ t | q_t = S_j], \quad 1 \leq j \leq N, \quad 1 \leq k \leq M. \tag{4.16}$$

5. $\pi$, the initial state distribution $\pi = \{\pi_i\}$ where

$$\pi_i = P[q_1 = S_i], \quad 1 \leq i \leq N. \tag{4.17}$$

Given the HMM model $\lambda = (A, B, \pi)$ specified by $N, M, A, B$ and $\pi$, the observations $O = O_1 O_2 ... O_T$ are generated at times $t = 1, 2, ..., T$. In the case of continuous observations Equation 4.16 takes the form

$$b_j(O) = \sum_{m=1}^{M} c_{jm}\mathcal{N}[O, \mu_{jm}, U_{jm}], \quad 1 \leq j \leq N \tag{4.18}$$

where $O$ is the continuous vector of observations, $c_{jm}$ is the mixture coefficient for $m^{th}$ mixture in state $j$ and $\mathcal{N}$ is the probability density function (usually Gaussian) with mean vector $\mu_{jm}$ and covariance matrix $U_{jm}$.

The computational power of HMM is based on the capability of solving the so-called "Three Basic Problems" (Rabiner, 1989):

**Problem 1:** Given the observation sequence $O$ and a model $\lambda$, how do we compute $P(O|\lambda)$, the probability of the observation sequence, given the model?

**Problem 2:** Given the observation sequence $O$ and a model $\lambda$, how do we choose the corresponding state sequence $Q$ which best explains observations?

**Problem 3:** How do we adjust the model parameters $\lambda$ to maximise $P(O|\lambda)$?

The solution for Problem 1 is provided by the dynamic programming algorithm known as the "Forward-Backward Procedure", based on inductively calculating the forward variable $\alpha$, defined as

$$\alpha_t(j) = P(O_1 O_2 ... O_t, q_t = S_j|\lambda) \tag{4.19}$$

and includes the following steps:

1. Initialisation:

$$\alpha_1(i) = \pi_i b_i(O_1), \quad 1 \leq i \leq N. \tag{4.20}$$

2. Induction:

$$\alpha_{t+1}(j) = \left[\sum_{j=1}^{N} \alpha_t(j) a_{ij}\right] b_j(O_{t+1}), \quad 1 \leq t \leq T-1, \quad 1 \leq j \leq N. \tag{4.21}$$

3. Termination:

$$P(O|\lambda) = \sum_{j=1}^{N} \alpha_T(j) \tag{4.22}$$

To solve Problem 2 it is required to specify the "optimal" state sequence criterion, usually defined as maximising the $P(Q|O, \lambda)$, i.e. finding the single best state sequence for the given model and observations. The problem is solved using the Viterbi algorithm (Rabiner, 1989), using the best score variable $\delta$, the highest probability along the single path at time $t$ which accounts for the first $t$ observations and ends in state $S_i$

$$\delta_t(j) = \max_{q_1,q_2,\dots,q_{t-1}} P[q_1, q_2, ..., q_t = i, O_1 O_2 ... O_t|\lambda], \tag{4.23}$$

$$\delta_{t+1}(j) = [\max_i \gamma_t(j) a_{ij}] b_j(O_{t+1}). \tag{4.24}$$

Using the best score variable $\delta$ and tracking array $\psi_t(j)$ the procedure for finding the best state sequence includes the following steps:

1. Initialisation:

$$\delta_1(i) = \pi_i b_i(O_1), \quad 1 \leq i \leq N \tag{4.25a}$$

$$\psi_1(i) = 0. \tag{4.25b}$$

2. Recursion:

$$\delta_t(j) = \max_{1\leq i\leq N} [\delta_{t-1}(i) a_{ij}] b_j(O_t), \quad 2 \leq t \leq T, \quad 1 \leq j \leq N \tag{4.26a}$$

$$\psi_t(i) = \operatorname*{argmax}_{1\leq i\leq N} [\delta_{t-1}(i) a_{ij}], \quad 2 \leq t \leq T, \quad 1 \leq j \leq N. \tag{4.26b}$$

3. Termination:

$$p^* = \max_{1\leq i\leq N} [\delta_T(i)] \tag{4.27a}$$

$$q_T^* = \operatorname*{argmax}_{1\leq i\leq N} [\delta_T(i)]. \tag{4.27b}$$

4. Backtracking:

$$q_t^* = \psi_{t+1}(q_{t+1}^*), \quad t = T-1, T-2, ..., 1. \tag{4.28}$$

Problem 3 was not relevant to this work, so the solution for it is not discussed here.

#### 4.5.3.2 Explicit Duration Hidden Semi-Markov Models

The main limitation of the classical Hidden Markov Model described above is the absence of explicit modelling of time spent in each state. It could be introduced by non-zero self-transition probability, however in such cases state duration distributions are limited to an exponential form. The extension of HMM to allow independently (from state transitions) modelling duration of each state is called the Explicit Duration Hidden (Semi-)Markov Model (EDHMM). In the EDHMM, the model $\lambda$ is extended with (usually Gaussian) per-state duration distribution

$$p_i(d) = \mathcal{N}(d, \mu_i, \sigma_i^2) \tag{4.29}$$

with mean $\mu_i$ and standard deviation $\sigma_i^2$ for state $i$. For computational reasons, the maximum state duration is usually limited by $D$, the maximum duration within any state, and self-transitions are not allowed.

Solutions to the problems described in the previous section keep the same form for the EDHMM, but for each state transition $i \to j$ we essentially consider a number of possible previous states, each defined by a pair $\{S_i, d_k\}$, where $S_i$ is the previous state and $d_k$ is the time spent in this state ($1 < d_k < D$).

#### 4.5.3.3 Model for bed/wake segmentation

The model for bed/wake segmentation was defined with two states ("in bed" and "not in bed"), continuous observations, and explicitly specified state duration probability distributions, specified based on bed and wakefulness times.

The activity features, selected as hidden state observations, included the mean level of log-activity at time $t$ (in units of activity counts described in Section 4.4.2) and duration of the stationary activity segment to which the time $t$ belongs. The time granularity was selected as 2 minutes due to computational limitations. The stationary activity segments were identified using the Bayesian Online Change Point Detection method, as described in Section 4.5.1.

Fitbit bed time annotation was used to estimate parameters of HMM, including the state duration distributions $p_i(d)$ and observation probability distributions $b_j(\boldsymbol{O})$. The resulting observation distributions for "in bed" and "not in bed"' states are presented in Figure 4.8.

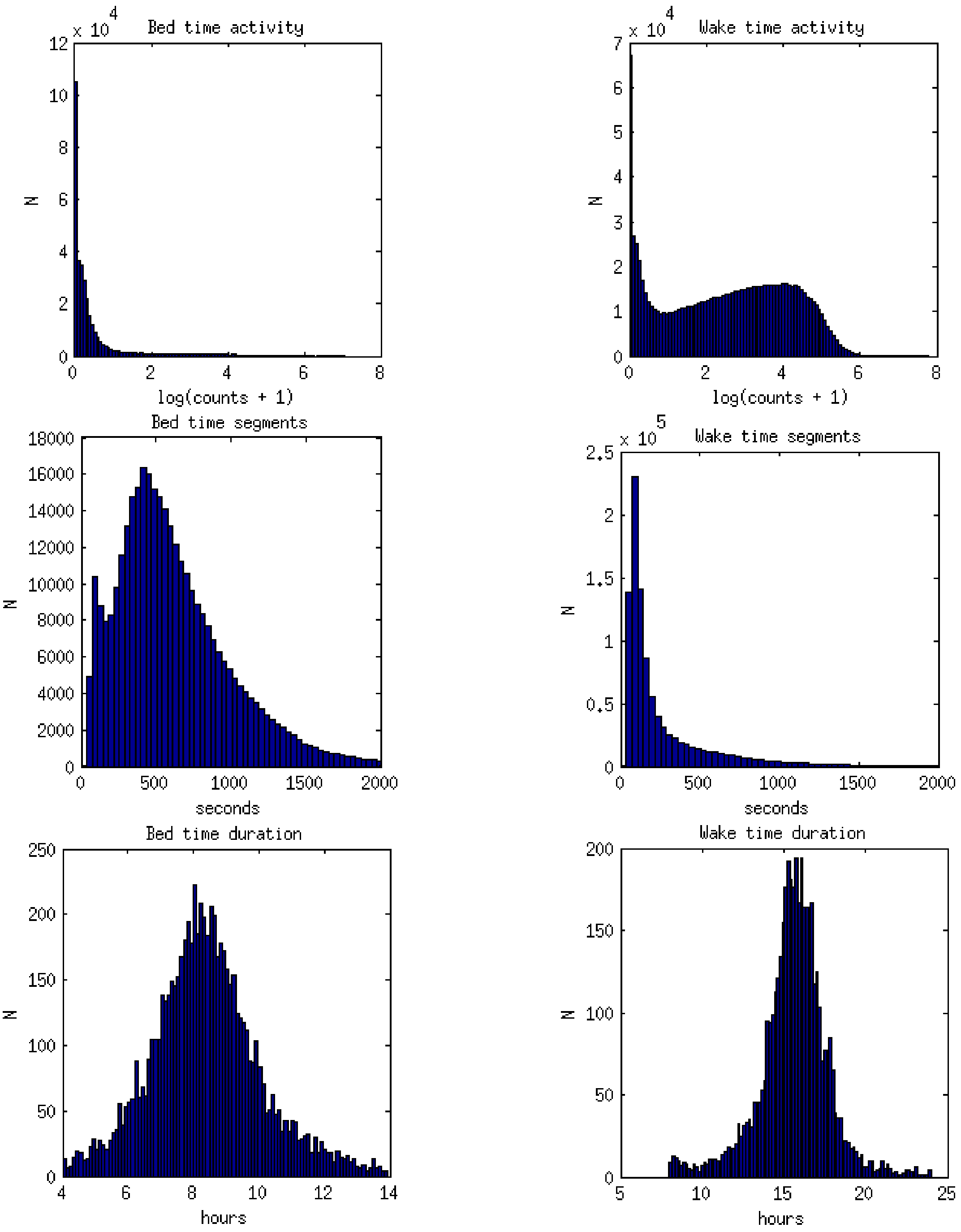

**Figure 4.8:** *Probability distributions of bed- and wake- time activity observations and state durations, identified based on Fitbit bed/wake segmentation.*

#### 4.5.3.4 Segmentation results

Bed/wake segmentation was performed using the proposed Hidden Markov Model and the accuracy of segmentation evaluated using manual Fitbit bed/wake annotation ground truth. Although the accuracy of Fitbit manual bed time segmentation is questionable, manual bed time logging is the state-of-art method in sleep research (Lauderdale et al., 2008). HMM segmentation quality is affected by the missingness of activity data, therefore the annotations with the amount of missing data exceeding 2 hours (the empirically identified threshold) were removed. The resulting accuracy of bed and wake time detection is presented in Figures 4.9 and 4.10.

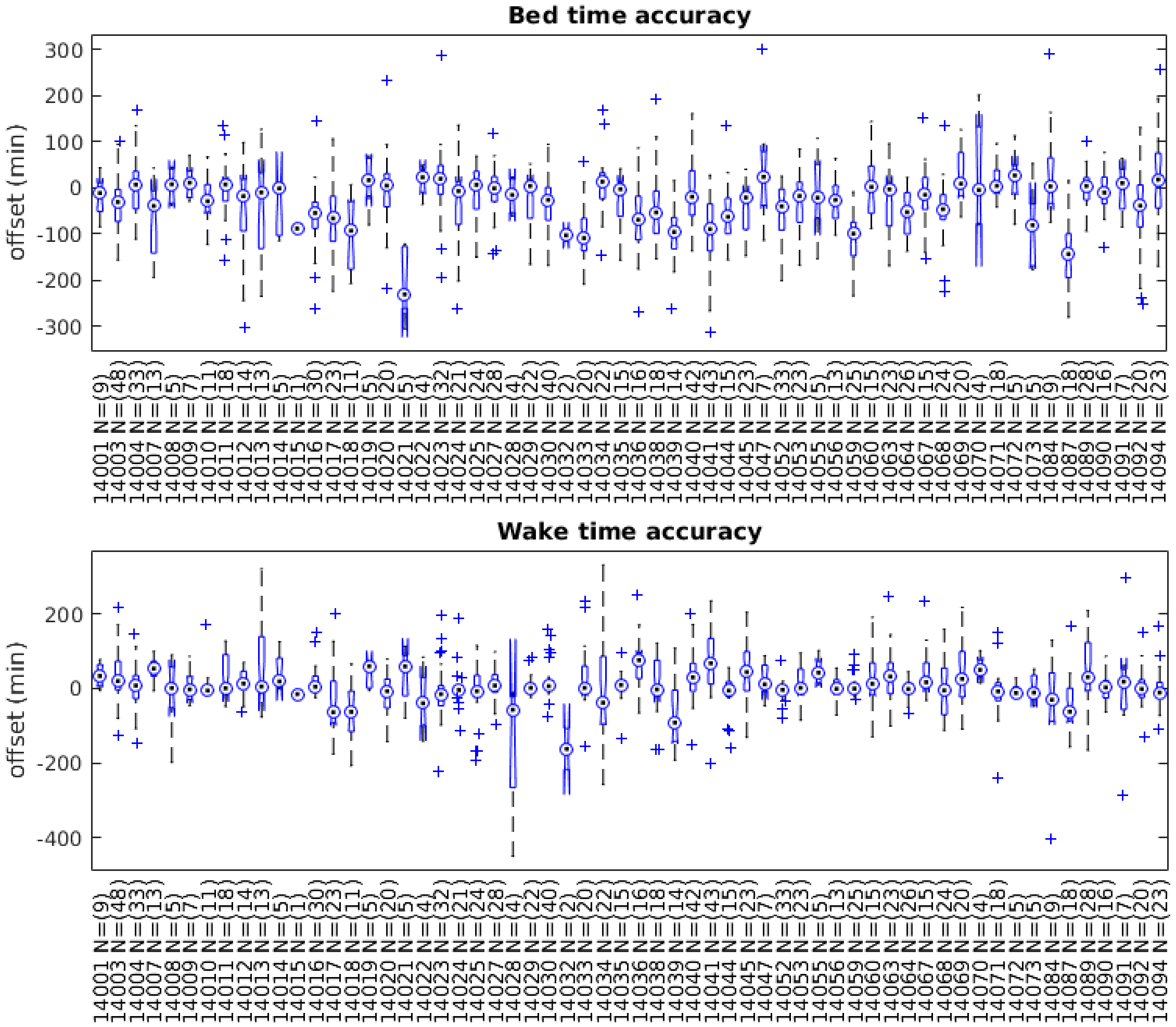

**Figure 4.9:** *Bed and wake time estimation error for each subject, representing offset of HMM-estimated state change comparing to Fitbit reference. On each box, the central mark is the median, the edges of the box are the 25th and 75th percentiles, the whiskers extend to the most extreme data points not considered outliers, and outliers are plotted individually.*

The bed times were estimated with an error of $-34.77 \pm 80.30$ (Mean ± SD) and wake times with an error of $12.36 \pm 76.84$ (Mean ± SD) minutes, compared to the reference Fitbit segmentation, reflecting the issue of negative bias for manual bed time annotation and positive for wake time annotation, mentioned in Section 4.5.3. The reported bed time accuracy identification using sleep reports is 0.80 hours (Lauderdale et al., 2008), therefore the achieved results are within the range acceptable for sleep studies.

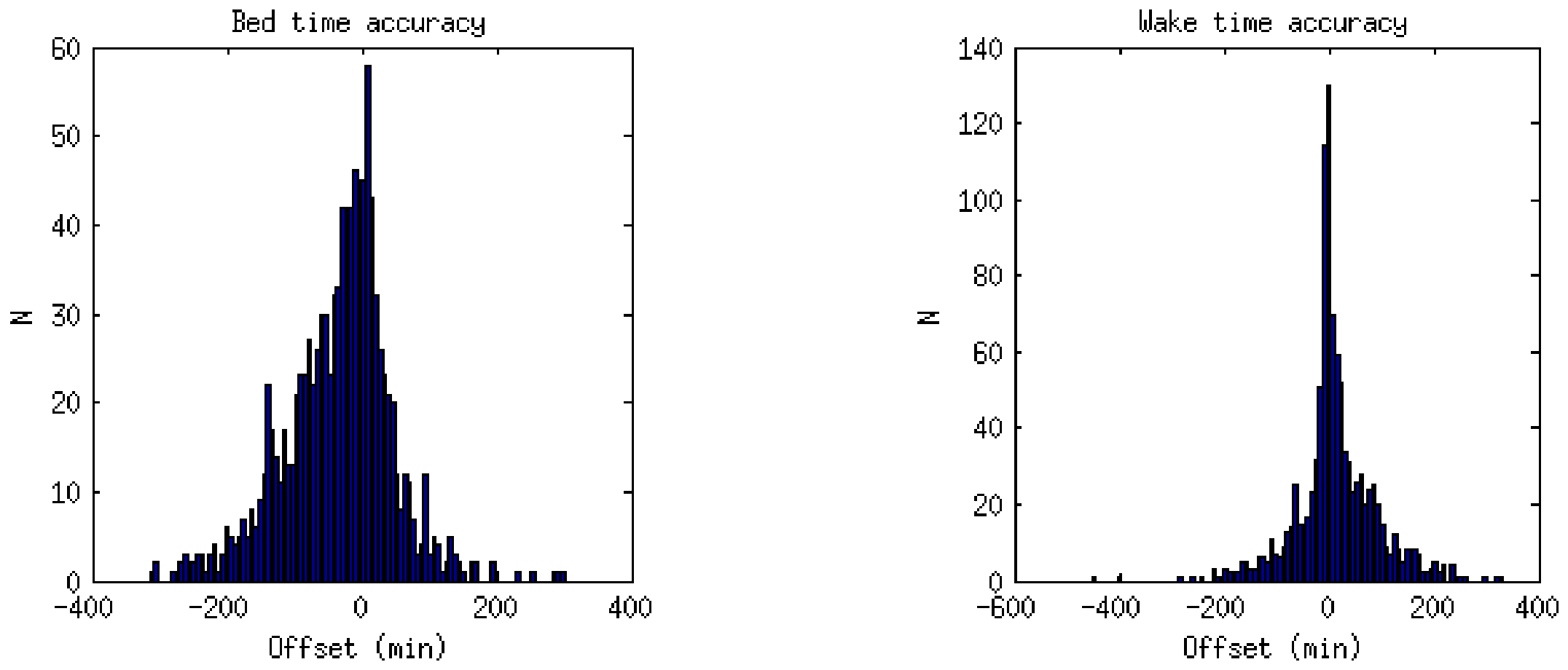

**Figure 4.10:** *Bed and wake time estimation error, representing offset of HMM-estimated state change comparing to Fitbit reference.*

## 4.6 Summary

The AMoSS data set is diverse, often unevenly sampled and suffers from data missingness due to levels of participant compliance and technical difficulties of implementing data collection instrumentation. This chapter proposed data pre-processing approaches, based on previous research findings in the field of ambulatory activity monitoring, to prepare the AMoSS data set for further analysis, keeping in mind the types of applied analysis methods.

Accelerometer-based ambulatory observation of physical activity also captures a very complex non-stationary signal, where non-stationarities themselves contain potentially useful information about subjects' behaviour. In order to perform a meaningful analysis on large volumes of such data, automated methods were proposed in this chapter, which would be helpful in extracting physiological states and identifying segments of data with similar statistical characteristics, as well as non-movement periods or continuous exercises sessions.

# Chapter 5

# Symptoms analysis framework

## 5.1 Objectivity of symptoms

Accelerometer-based data collection allows the characterisation of long-term physical activity and behaviour, including the two main human physiological states, wakefulness and sleep, as well as transitions between these states (circadian rhythm). For the purpose of psychiatric symptoms assessment a number of objective measures can be extracted from actigraphic recordings that allow the description of behaviours associated with specific symptoms of mental health. Some symptoms can be directly mapped to actigraphic measures, based on the DSM-5 diagnostic criteria for certain disorders (for example psychomotor poverty or decreased need for sleep in mania), but in many cases there are no obvious symptom measures based on objective data.

The framework proposed by Liddle and described in Table 5.1 is helpful in formulating a comprehensive set of clinically important objective measures of mental health, which can be potentially useful for a broad range of mental illnesses and are based on recordings of long term physical activity or physiology. Measures could be suggested for clusters of symptoms, and it can be seen that some clusters can be assessed using just physical activity data, while others may require physiological information, and only the dimension of reality distortion is difficult to assess by non-invasive sensors.

**Table 5.1:** *The major clusters of mental symptoms (Liddle, 2001). Clusters where sensor-based objective measurements of symptoms are possible using actigraphy are selected in green, and clusters where other sensors can be used are in magenta.*

| **Dimension** | **Syndrome** | **Symptom cluster** |
|---|---|---|
| Reality distortion | Reality distortion | Delusions, Hallucinations |
| Disorganisation | Disorganisation | Formal thought disorder, Inappropriate affect, Disorganised or bizarre behaviour |
| Psychomotor | Psychomotor poverty | Flat affect, Poverty of speech, Decreased voluntary motor activity |
| | Psychomotor excitation | Labile affect, Pressure of speech, Motor agitation |
| Mood | Depression | Low mood, Low self-esteem, Hopelessness, Suicidality, Somatic symptoms |
| | Elation | Elevated mood, Elevated self-esteem Decreased need for sleep |
| Anxiety | Anxiety | Feelings of unease, fear or dread, Overactivity of the sympathetic nervous system |

Three key dimensions specifically relevant to bipolar and borderline personality disorders are disorganisation, psychomotor and mood, and include symptoms potentially measurable with continuous actigraphic recordings. The main interest of this research is therefore in defining the objective ambulatory symptom measures for these dimensions, to complement or substitute questionnaires and interview observations.

In the next sections features are proposed which are potentially relevant to these symptoms, based on existing physical activity, sleep and circadian rhythm research. Novel methods of analysis are also described that could be useful specifically in the area of mental health. The proposed features are logically organised into two categories, including day-time behaviour and sleep characteristics and three dimensions, including the psychomotor, disorganisation and mood.

Note, that the features are defined based on hypothesis of what the expected behaviours look like, and data mining approaches not explored in this work. Such approaches as neural networks, although very powerful, may require large amounts of train-ing data, and often produce results difficult to interpret by clinical specialists.

## 5.2 Experimental foundations

Several studies have attempted to identify and measure clinically relevant aspects of behaviour in mental health, with the ultimate goal of creating behavioural markers of specific disorders and clinical states. Most existing research has used physical activity analysis techniques, where several established usage scenarios exist in exercise and sleep medicine, so analysis methods are already available for use in mental health research.

These studies covered a wide range of disorders and focused on different analysis methods, mostly addressing only subsets of activity measures in psychomotor and disorganisation symptom dimensions, or focusing on sleep and circadian rhythms. However, a small number of studies also attempted to create a principled multi-dimensional framework for the assessment of mental health, similar to the approach proposed in this work.

### 5.2.1 Psychomotor symptoms

Psychomotor symptoms are prominent features of many disorders, including the bipolar disorder, schizophrenia and clinical depression, and may be easily measured with levels of day-time physical activity. A number of studies have looked at activity levels in mental health, using different analysis methods and in different clinical conditions.

Walther *et al.* (Walther et al., 2009) performed activity analysis of paranoid (N=35; age 39.6±9.64 years), catatonic (N=12; age 45.5±12.19 years) and disorganised (N=13; age 36.23±10.26 years) schizophrenia subtypes. Activity was monitored for 24 hours with 2 seconds resolution and data collected during sleep removed from the main analysis. Three activity parameters were extracted, including:

1. Mean number of activity counts per hour (Activity Level, AL).

2. Percentage of epochs with non-zero activity (Mobility Index, MI).

3. Mean duration of periods with zero activity (Mean Immobility Period, MIP).

General linear models were created for AL, MI and MIP as dependent variables and schizophrenia subtype, age, gender, type of medication, chlorpromazine equivalents, number of episodes and duration of illness as predictors. Only schizophrenia subtype and type of anti-psychotic were found to have significant association with the dependent variables.

Minassian *et al.* (Minassian et al., 2010) performed activity analysis of manic bipolar disorder patients (N=28; age 34.1±13.3 years), schizophrenia patients (N=17; age 36.7±12.4 years) and healthy controls (N=21; age 30.3±9.5 years). The activity measurement experiment was performed on subjects, placed in a novel environment for 15 minutes. Manic bipolar disorder patients were found to have significantly higher levels of motor activity compared to both schizophrenia and healthy controls.

Hauge *et al.* (Hauge et al., 2011) analysed differences in activity between schizophrenia, major depression and healthy controls in two settings, including a) one minute epochs with 5 hours recorded length and b) one hour epochs with two weeks of recording. The study included schizophrenic patients (N=24; age 47.4±11.1 years), depressed patients (N=25; age 42.9±10.7 years) and control group (N=32; age 38.2±13.0 years). The data were analysed and several characteristics were calculated and compared between groups, including the mean activity level, standard deviation (SD), root mean square successive difference (RMSSD), RMSSD/SD, Sample Entropy ($m = 2, r = 0.2$) and Fourier analysis.

For 5 hours recordings, both schizophrenia and depression mean activity levels were significantly lower compared to healthy controls, standard deviation was significantly higher in depression, RMSSD/SD ratio significantly higher in both depression and schizophrenia and Sample Entropy significantly lower in depression. For two week recordings, mean activity levels in both schizophrenia and depression were significantly lower compared to healthy controls, SD of activity was higher in depression compared to both controls and schizophrenia, RMSSD was higher in depression compared to schizophrenia and the RMSSD/SD ratio was significantly lower in schizophrenia compared to healthy controls. In another study on the same data (Berle et al., 2010), it was also identified that schizophrenia patients have significantly lower intra-daily variability and higher inter-daily stability of activity patterns.

In a systematic review of physical activity correlates of schizophrenia, Vancampfort *et al.* (Vancampfort et al., 2012) identified that lower physical activity is associated with the presence of negative symptoms. A systematic review of activity monitoring in patients with depression by Burton *et al.* (Burton et al., 2013) identified decreased day time activity in depression (increased post-treatment), and no difference in night time activity level (but decreased activity level and increased sleep efficacy post-treatment).

### 5.2.2 Disorganisation symptoms

Behaviour disorganisation is one of the observable features of many mental disorders, and initially came up as a separate syndrome in factor-analytic studies of schizophrenia (Liddle, 1987; Ventura et al., 2000), but no objective behaviour measures of disorganisation are used in clinical practice. Among other aspects, disorganisation syndrome is characterised by impulsive behaviour (Kiehl et al., 2000; Ouzir, 2013), where structural organisation of activity is impaired on different temporal scales. Impulsive behaviours were extensively analysed in the context of bipolar disorder, borderline personality disorder, as well as schizophrenia and other mental illnesses (Henry et al., 2001; Lieb et al., 2004; Swann, 2009; Ouzir, 2013).

In actigraphic studies, disorganisation symptoms are often analysed together with psychomotor symptoms, for example in the previously described study of Hauge *et al.* (Hauge et al., 2011) a number of structural activity measures are included together with simple activity level characteristics.

Walther *et al.* (Walther et al., 2014) performed analysis of disorganisation symptoms using actigraphy on schizophrenia spectrum disorder patients (N=100; age 39.2±11.1 years) with data sampled every 2 seconds for 60 minutes. A partial autocorrelation function was computed and for each set of data the longest lag was extracted, where the partial autocorrelation function was significant at $p < 0.05$. Reduced lags were found to be associated with positive symptoms, disorganisation and excitement, and a negative syndrome was associated with reduced mean activity level.

To analyse abnormalities of structural activity organisation in bipolar disorder, Indic *et al.* (Indic et al., 2011) proposed the actigraphic Vulnerability Index (VI) to distinguish between a) high and low risk of mood disorders, b) bipolar disorder patients and healthy controls and c) clinical states in bipolar disorder patients. The vulnerability index was designed to assess multi-scale behaviour of physical activity amplitude and calculated using the following steps:

1. Continuous wavelet analysis of 7 days of activity was performed on a range of time scales from 0.2 to 26 hours.

2. Gamma distribution was fitted into the amplitude of wavelet transform in scales from 0.2 to 2 hours.

3. The Vulnerability Index was defined as $\mathrm{VI} = \int_{0.2}^{2} \gamma(s)ds$.

The index was validated on actigraphy data from three studies. **Study 1** included healthy young adults with the potential risk of bipolar disorder, assessed using the General Behaviour Inventory (GBI) (Depue et al., 1989), and divided into high and low deciles with low (N=35; age 20.9±2.3 years) and high risk (N=35; age 22.3±3.0 years). **Study 2** included type-I bipolar disorder diagnosed subjects (N=15; age 46.8±12.4 years) and healthy controls (N=15; age 46.7±14.1 years). **Study 3** included type-I diagnosed subjects (N=51; age 43.8±10.5 years) monitored for 3 days during maniac, major depression and recovery states. The VI was shown to be associated with increasing risk of bipolar disorder and mood states of patients. Mean VI increased linearly across the groups, ranking major depression < minor depression < euthymia < minor mixed-states < hypomania < major mixed-states < mania (goodness of fit was not reported).

Other types of activity (non-locomotor) were also shown to be useful in the analysis of mental health. In non-actigraphic studies, an eye movement analysis model in schizophrenia was utilised on 88 healthy controls and 88 schizophrenia patients and was found to be capable of achieving accuracy of 87.8% in separation between schizophrenia subjects and controls on 34 9-month retest subjects, 36 novel schizophrenia cases and 52 control subjects (Benson et al., 2012). Analysis of activity disorganisation in schizophrenia was also performed using user-generated time series (Hornero et al., 2006), where 20 schizophrenia patients and 20 control subjects were instructed to press a button as randomly as possible. It demonstrated classification with 80% sensitivity and 90% specificity using Limpel-Ziv complexity measure, 90% sensitivity with 60% specificity with Approximate Entropy and 70% sensitivity with 70% specificity using the Central Tendency Measure using 2-fold cross-validation. A significant difference between schizophrenia and normal control groups ($p < 0.0001$) was also found using chirological (hand) testing with 100 hospitalised schizophrenia patients and 50 healthy control subjects (Pomrock and Ginath, 1982).

### 5.2.3 Mood symptoms (sleep and circadian rhythms)

The recent review paper of Kahn, Shappes and Sadeh (Kahn et al., 2013) described the bi-directional links between sleep and emotions, and concludes that almost all characteristics of sleep, including duration, quality and REM sleep characteristics (Benca, 1992; Benca et al., 1997) may influence emotional state, and emotional state may in turn affect sleep, however this connection is complex and varies between different subject groups. For example, sleep deprivation is known to negatively affect positive mood and increase negative mood intensity (Franzen et al., 2008). But at the same time reduced sleep duration is characteristic of manic states and one of the diagnostic criteria of mania in bipolar disorder (American Psychiatric Association, 2013), and total sleep deprivation shows clinical effect in depressed bipolar patients (Smeraldi et al., 1999). Similarly, the effect of stress and anxiety on sleep duration can be bi-directional, mediated by the individuals' coping style producing a "turn on" response leading to decreased sleep time and increased sleep latency or a "shut off" response to stress leading to withdrawal and increased sleep duration (Sadeh et al., 2004). For low mood, excessive sleepiness is an important predictor of depression (Mayers and Baldwin, 2015; American Psychiatric Association, 2013).

While the influence of sleep on emotions is relatively well understood from the phenomenological perspective, in recent years evidence has emerged of tight connections between disrupted circadian rhythms and mood disorders (Wulff et al., 2010; McClung, 2013; Benedetti and Terman, 2013). Evidence exists of a direct correlation between mood and the severity of rhythm disruption (Wirz-Justice, 2006), chronobiology-inspired successful treatment strategies (Bromundt et al., 2013) and a solid theoretical framework connecting the circadian system and mood (McClung, 2013). The most recent publications suggest that sleep- and circadian rhythm-associated pathways can be promising therapeutic targets for bipolar disorder (Bellivier et al., 2015) and may be used as biomarkers to predict responses to SSRIs treatment in depression (McCall, 2015).

Jones *et al.* (Jones et al., 2005) assessed both circadian activity and sleep patterns in remitted bipolar disorder. The study included bipolar disorder patients ($N = 19$; age $44.37 \pm 13.1$ years) and age- and gender-matched controls ($N = 19$; age $46.89 \pm 14.82$ years). The data were collected for 7 day with 15 seconds resolution and significant differences in circadian characteristics were found (IS and IV), but there was no difference in sleep indices.

Sleep in remitted bipolar disorder was studied by Geoffroy *et al.* (Geoffroy et al., 2014) on remitted bipolar patients ($N = 26$; age $53.5 \pm 11.49$ years) and matching healthy controls ($N = 29$; age $54.1 \pm 9.11$ years). The data were collected for 21 days with 1 minute resolution. Patients demonstrated significantly longer sleep latency, sleep duration, poorer sleep efficiency, a higher fragmentation index, poorer inter-daily stability and more variability in sleep characteristics.

Wulff *et al.* (Wulff et al., 2010) analysed activity, light exposure and melatonin of schizophrenia outpatients ($N = 20$; age $38.8 \pm 8.6$ years) vs. unemployed controls ($N = 21$; age $37.5 \pm 9.6$ years). The data were collected for 6 weeks with a 2 minutes resolution. Analysis of data included sleep characteristics and non-parametric circadian rhythm analysis. All schizophrenia patients had significant sleep and circadian rhythm abnormalities, including phase shifts, non-24 hour periods or highly irregular and fragmented sleep.

### 5.2.4 Multi-dimensional approach

While the above studies attempted to discover objective markers of certain clinical states, many studies analysed only a few selected activity features, and did not define or follow a principled approach for identifying a comprehensive set of objective diagnostic measures of mental disorders. The principled framework for analysis of mental health symptoms would need to be based on the accepted in clinical practice diagnostic markers (American Psychiatric Association, 2013), as well as take into account the possible measurable symptom correlates arising from the known interactions between emotional states and behaviours or physiology.

In attempting to identify behaviour disorder markers from first principles, i.e. observed symptoms, Prociow *et al.* (Prociow et al., 2012) followed the approach similar to the one adopted by this thesis, suggesting new measures and sensors based on clinical syndromes of bipolar disorder, however he only described a system and did not validate it via a study involving patients.

A principled framework was also proposed in the framework of the PSYCHE project (Paradiso et al., 2010), but results of the project indicated a greater focus on sensor technology for the acquisition of physiological signals, rather than identifying behaviour features that are correlated to known symptoms of mental health and capable of improving diagnosis (Valenza et al., 2013).

The study by Grünerbl *et al.* (Grunerbl et al., 2014), used a similar principled approach in the MONARCA project, analysing mood in bipolar disorder using objective features, including phone call features, speech and voice characteristics, as well as acceleration and location changes captured from mobile phone sensors. The study demonstrated an accuracy of 76% in clinical state change identification for bipolar patients (10 subjects and 17 state changes) with the model trained on 66% of samples and tested on the remaining 33%.

### 5.2.5 Discussion

Day-time activity patterns are available for direct observation during clinical interview and important aspects of day-time behaviour are mostly defined by two factors of the 5-factor model of psychopathology, including activity disorganisation or psychomotor poverty and excitation. The interview, however, covers only a short period of time that does not necessarily represent a true clinical picture of patients' states, thus suggesting the need for a longitudinal objective assessment.

Sleep characteristics are recognised diagnostic criteria for mania and depression, and have been demonstrated to be important in other psychiatric conditions, contributing to the mood dimension of symptoms. However, objective sleep assessment is not usually performed as a part of clinical diagnostic assessment of mood disorders, and sleep is not objectively assessed on a regular basis, while known as contributing to relapse prodromes (Scott, 2011).

All together, the activity-based symptoms, including day-time activity patterns and sleep, represent a substantial proportion of DSM-5 diagnostic criteria for many disorders, and contribute to three out of five dimensions of psychopathology (Liddle, 2001), thus forming an important part of the clinical picture. Despite that, there are no attempts at objective assessment of physical activity in the current clinical practice of mental health.

## 5.3 Psychomotor symptoms cluster

Psychomotor symptoms cluster represents levels of activity of a person during periods of wakefulness, and, therefore from the analytical perspective are relatively easy to evaluate. For longitudal continuous data collection however, the question amounts to selecting the characteristics of activity that are important from the clinical perspective.

The most extensive analysis of the influence of physical activity on health was per-formed within the framework of public health research (Troiano et al., 2008; Tudor-Locke et al., 2010; van Hees et al., 2011), therefore the activity measures accepted in public health studies are also used in this work.

### 5.3.1 Distribution-based measures of activity level

A traditional approach for the analysis of activity levels is to use distribution-based measures, for example the average level of ENMO, as was performed in the study of a Brazilian cohort with 8974 subjects (da Silva et al., 2014) and in the Whitehall II study with 3975 participants (Sabia et al., 2014).

Physical activity of longer duration or a higher intensity may equally contribute to a higher value of the average level of ENMO activity. In order to distinguish between these scenarios, a distribution could be fitted into the activity data and parameters of this distribution used as activity level measures. Following the approach of Indic *et al.* (Indic et al., 2011), a Gamma distribution was used in this work, and shape $k$ and scale $\theta$ parameters were selected as activity measures.

### 5.3.2 Metabolic equivalent based measures of activity intensity

The metabolic equivalent (MET) is a unit of energy consumption, where 1 MET = mean measured resting energy expenditure during lying down, calculated for adults as oxygen consumption $\dot{V}O_2 = 3.5mL\ O_2 \cdot kg^{-1} \cdot min^{-1}$ by Hildebrand *et al.* (Hildebrand et al., 2014). In epidemiological studies, activities with values of

- lower than 3 MET are considered as mild physical activity,
- between 3 and 6 MET are considered to be moderate and
- 6 MET or above are vigorous activities.

These cut-off thresholds were originally defined for calculating physical activity energy expenditure based on activity logs, where each type of physical activity was assigned a category. In the case of tri-axial accelerometers, similar to the one used in the AMoSS study, the regression for estimating oxygen consumption from wrist accelerometry for adults was developed by Hildebrand *et al.* (Hildebrand et al., 2014) as

$$\dot{V}O_2 = 0.0323 \cdot ENMO + 7.49 \quad (5.1)$$

where ENMO represents average ENMO in milli g and cut-off thresholds for MET were estimated as 3 METs = 93.2 milli g and 6 METs = 418.3 milli g. The estimations were validated in a study on 30 adults completing 8 activities. Following the approach of da Silva *et al.* (da Silva et al., 2014) the average time per day in moderate and vigorous activities were included in this study in addition to distribution-based activity level measures.

### 5.3.3 Summary of psychomotor features

Based on the above-defined analysis methods, the features described in Table 5.2 were selected as potentially contributing to the psychomotor clusters of symptoms.

**Table 5.2:** *Psychomotor activity features, selected for further analysis.*

| Feature | Description |
|---|---|
| LEVEL_AVG | Average level of ENMO activity |
| LEVEL_GAMMA_K | Shape parameter of Gamma distribution fit into ENMO activity |
| LEVEL_GAMMA_THETA | Scale parameter of Gamma distribution fit into ENMO activity |
| TIME_MODERATE | Time spent in moderate-intensity activities (3-6 MET) |
| TIME_VIGOROUS | Time spent in vigorous activities (>6 MET) |

## 5.4 Disorganisation symptoms cluster

Behavioural disorganisation is one of the most interesting features of mental health, and it includes many aspects, impulsive behaviour being one of them. As can be seen from the factor structure of BIS-11 (Barratt and Patton, 1995), impulsivity includes components of behaviour instability on different temporal scales, from seconds and minutes (attention) to days (non-planning). The following sections propose criteria for assessing the structural organisation of activity on each of these scales, using biomedical community accepted analysis methods.

### 5.4.1 Multiscale entropy analysis

One of the most popular methods for measuring signal disorganisation is entropy (Richman and Moorman, 2000; Sung et al., 2005; Iverson et al., 2005). In the case of actigraphic data, entropy is capable of characterising the stability of motor activity, or the highest-resolution temporal scale of impulsivity. Most of the research described previously used simple sample-based measures of entropy, and demonstrated that it could be useful in the context of mental health (Hornero et al., 2006; Hauge et al., 2011). However, due to the fact that simple entropy measures are dependent on sampling frequency and disorganisation may exist on different time scales, the Multiscale Entropy Analysis (MSE) provides a more comprehensive assessment method.

MSE (Costa et al., 2005) applies the Sample Entropy ($\mathcal{H}_{SE}$) (Richman and Moorman, 2000) calculation to a range of scales, each representing the down-sampled original signal. Sample Entropy ($\mathcal{H}_{SE}$) provides an efficient way to estimate disorganisation from time series data. $\mathcal{H}_{SE}$ is a signal complexity estimation, derived from the negative logarithm of the conditional probability of the appearance of longer patterns in a signal, considering the presence of a shorter pattern. It is estimated by statistic

$$\mathcal{H}_{SE}(m, r, N) = -ln\left[\frac{A^m(r)}{B^m(r)}\right], \tag{5.2}$$

where $m$ is the template length, $r$ is the similarity threshold (or quantisation level), $A^m(r)$ is a probability of matching $(m+1)$-length template, $B^m(r)$ is a probability of matching $m$-length template and $N$ is the length of the record. Two patterns of length $m$ are considered as similar, if each point of a pattern in one part of the signal is within a distance $r$ from the respective point in the other part of the signal.

Sample Entropy calculation for a specific template and similarity threshold is illustrated by Figure 5.1.

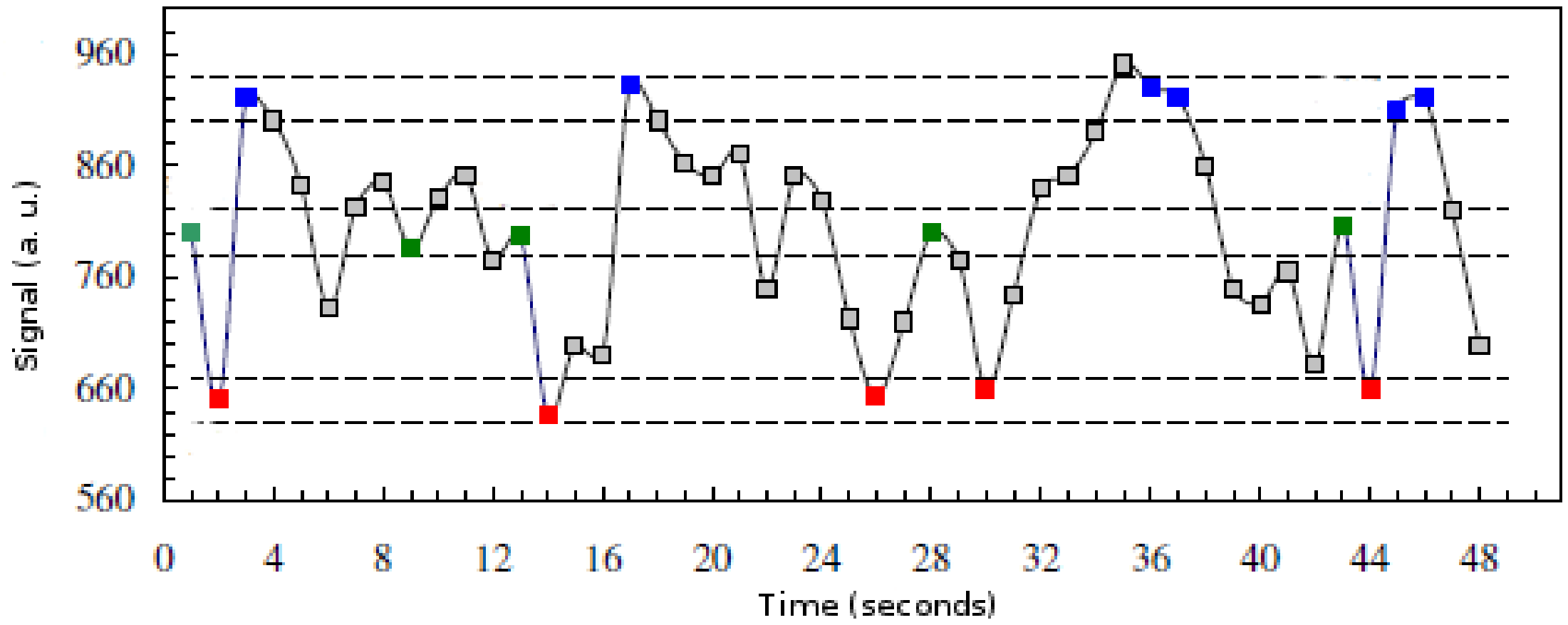

**Figure 5.1:** *Example of Sample Entropy calculation for a specific template. Red, green and blue squares represent measurements of a specific template within a similarity threshold r (shown with dashed lines). The probability of matching the template of length 3, given the template of length 2, equals to the number of green-red-blue sequences divided by the number of green-red sequences. Adapted from (Costa et al., 2005).*

For a Multiscale Entropy calculation the value of each scale $n$ of MSE is the Sample Entropy, calculated over the down-sampled original signal (by taking the mean of $n$ samples).

### 5.4.2 Activity persistence

To assess the attentional impulsiveness, reflecting an individual's task-focus, analysis of continuity of activity was performed, where continuity was measured as the time spent performing a certain physical activity without interruptions. In order to estimate such continuity, a change point detection method (BOCPD) was applied to the activity time series, to find statistically coherent activity segments (see Figure 5.2).

Based on this segmentation, an activity duration distribution was created, which consists of durations of stationary periods of activity (or time spent in individual continuous activities) as data points. Using the same approach as previously for psychomotor cluster analysis, the average duration of activity segment, as well as shape $k$ and scale $\theta$ parameters of Gamma distribution were calculated as activity persistence measures for each day of recording.

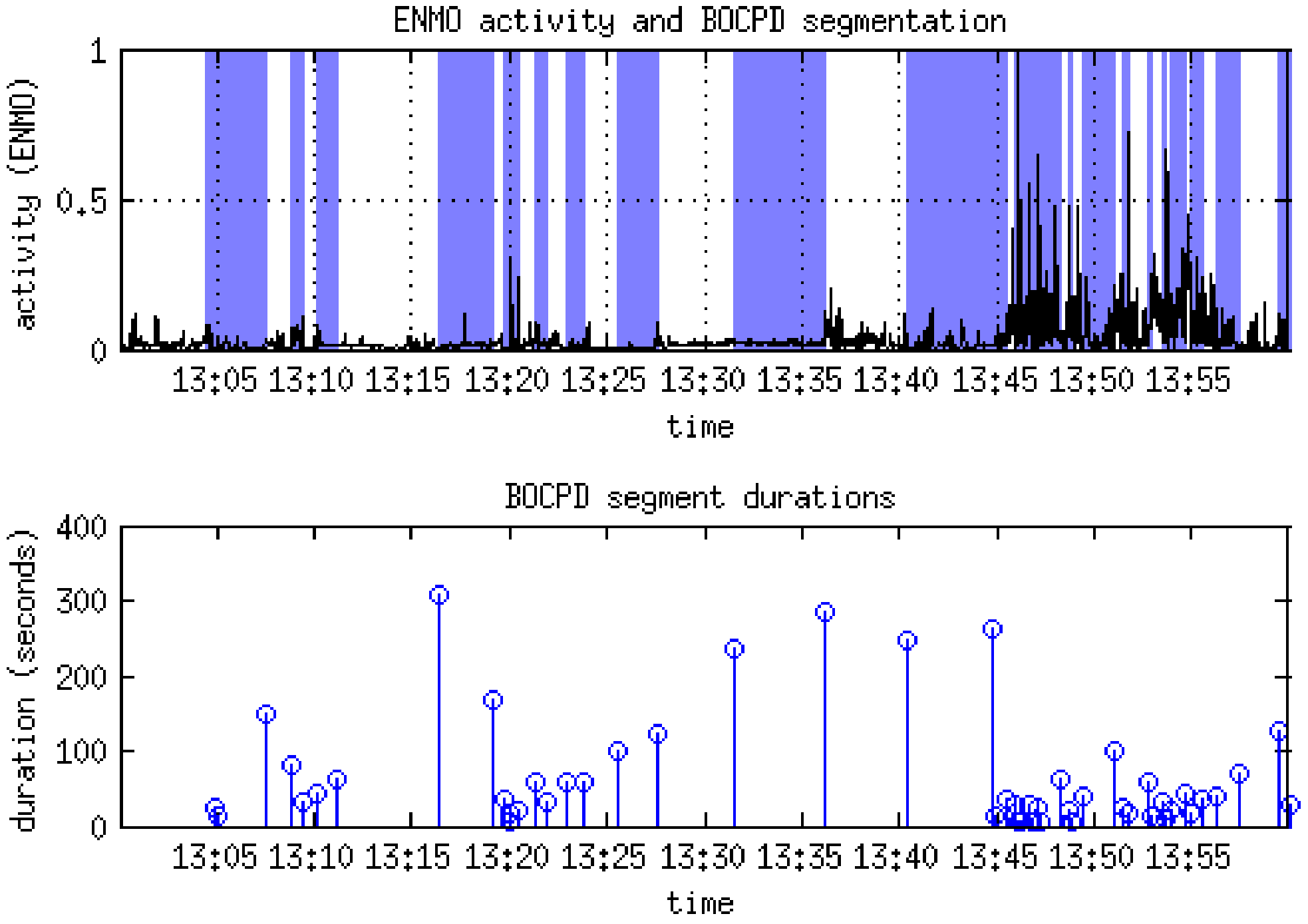

**Figure 5.2:** *Example of Bayesian segmentation of one hour of locomotor activity data. On the top plot one hour of activity is presented in black, and the identified stationary segments as white and purple overlays. On the bottom plot the durations of identified segments are presented.*

#### 5.4.2.1 Scaling of activity (Detrended fluctuation analysis)

Long-range correlations in biological signals were reported to be an important factor for physical and mental health (Bernaola-Galván et al., 2001; Hu et al., 2009; Henry et al., 2010; Indic et al., 2011; Valenza et al., 2013; Valenza et al., 2014a). For example, changes in heart rate regimes demonstrate the same functional form with different parameters between healthy individuals and heart failure patients (Bernaola-Galván et al., 2001). Scaling behaviour of human locomotor activity was also shown to distinguish between risk and clinical states of bipolar disorder (Indic et al., 2011). Detrended Fluctuation Analysis (DFA) provides a method for identification of long range correlations in time series and includes the following steps (Peng et al., 1994):

1. Divide the input signal of length $N$ into $N/l$ non-overlapping segments and define the "local trend" as the ordinate of linear least-squares fit in this segment.

2. Define the detrended signal as the difference between the original signal and the "local trend". Calculate the average of variances of detrended signal over all the boxes $F_d^2(l)$.

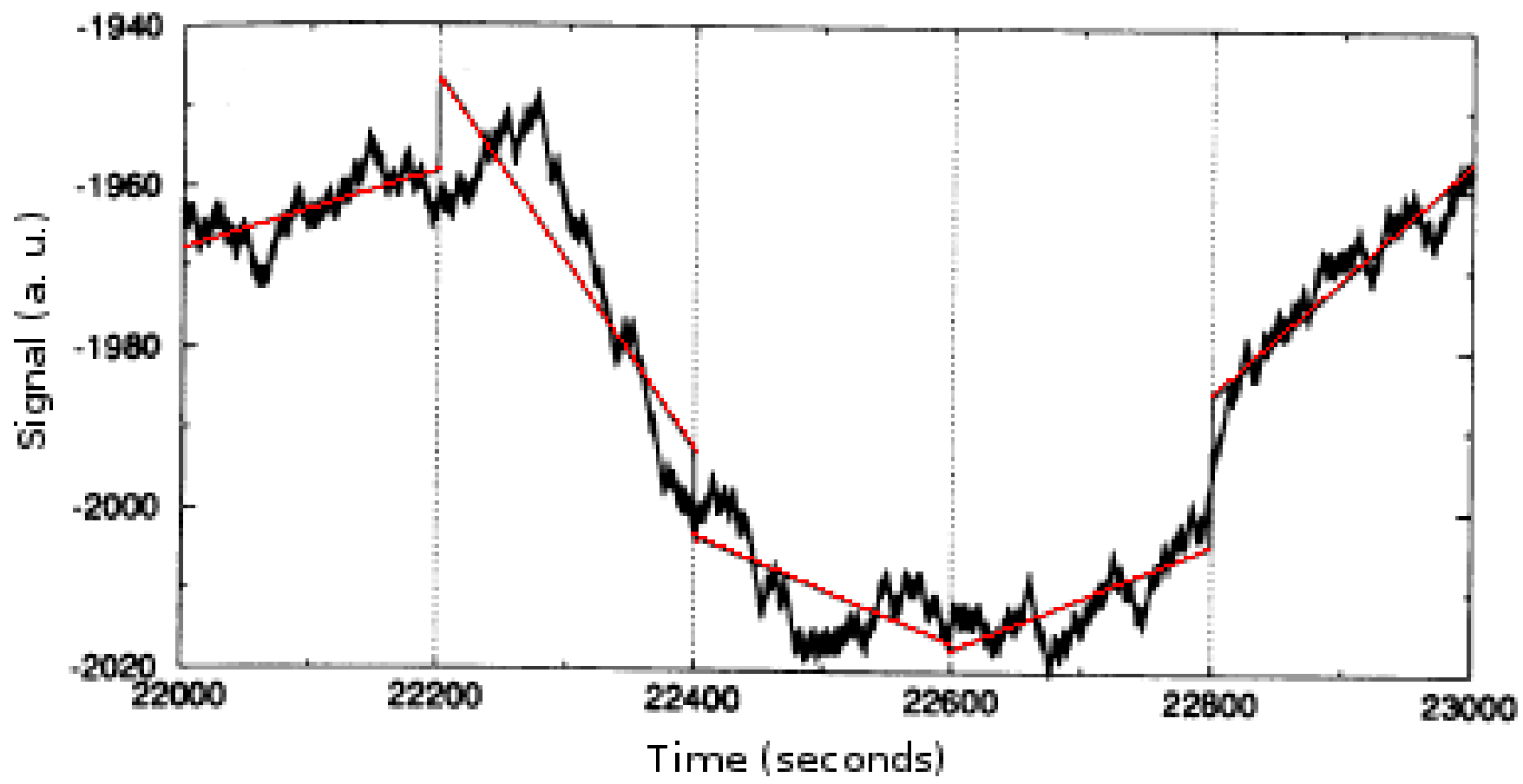

**Figure 5.3:** *Example of Detrended Fluctuation Analysis calculation for a specific segment length. In each segment, the local trend is presented as a red line. Adapted from (Peng et al., 1994).*

If the resulting signal $F_d^2(l)$ is proportional to the $l^\alpha$, the $\alpha$ is called the "scaling exponent" and defines the presence of long-range correlation in the original signal, with the value giving information about the signal self-correlations, where

- $\alpha = 1/2$ for white noise,
- $\alpha = 1$ for pink noise
- and $\alpha = 3/2$ for Brownian noise.

The variance across segments is expected to increase with the increased segment length, and $\alpha$ defines the rate of such increase, which is different for various types of signals, as described above. Note, that $F_d^2(l)$ is a signal statistic dependant on segment length, thus a number of segments of equal length necessary to calculate it.

### 5.4.3 Day-to-day activity patterns variation

Another key factor, contributing to a person's impulsivity and activity disorganisation, is motor impulsiveness, defined as a tendency to act on the spur of the moment and consistency of lifestyle. This behavioural characteristic can potentially be represented by the stability of day-to-day activity patterns. This study proposes to measure the behavioural stability directly, by performing day-to-day comparison of 24-hour activity patterns. Given that activities are not likely to align perfectly between days, the Dynamic Time Warping (DTW) algorithm (Senin, 2008) was used to measure the distance between daily activity patterns.

DTW is a dynamic programming algorithm, that calculates optimal matching between two signals by measuring pairwise distance between all points of these signals and selecting the optimal matching with a minimal cumulative distance (see Figure 5.4).

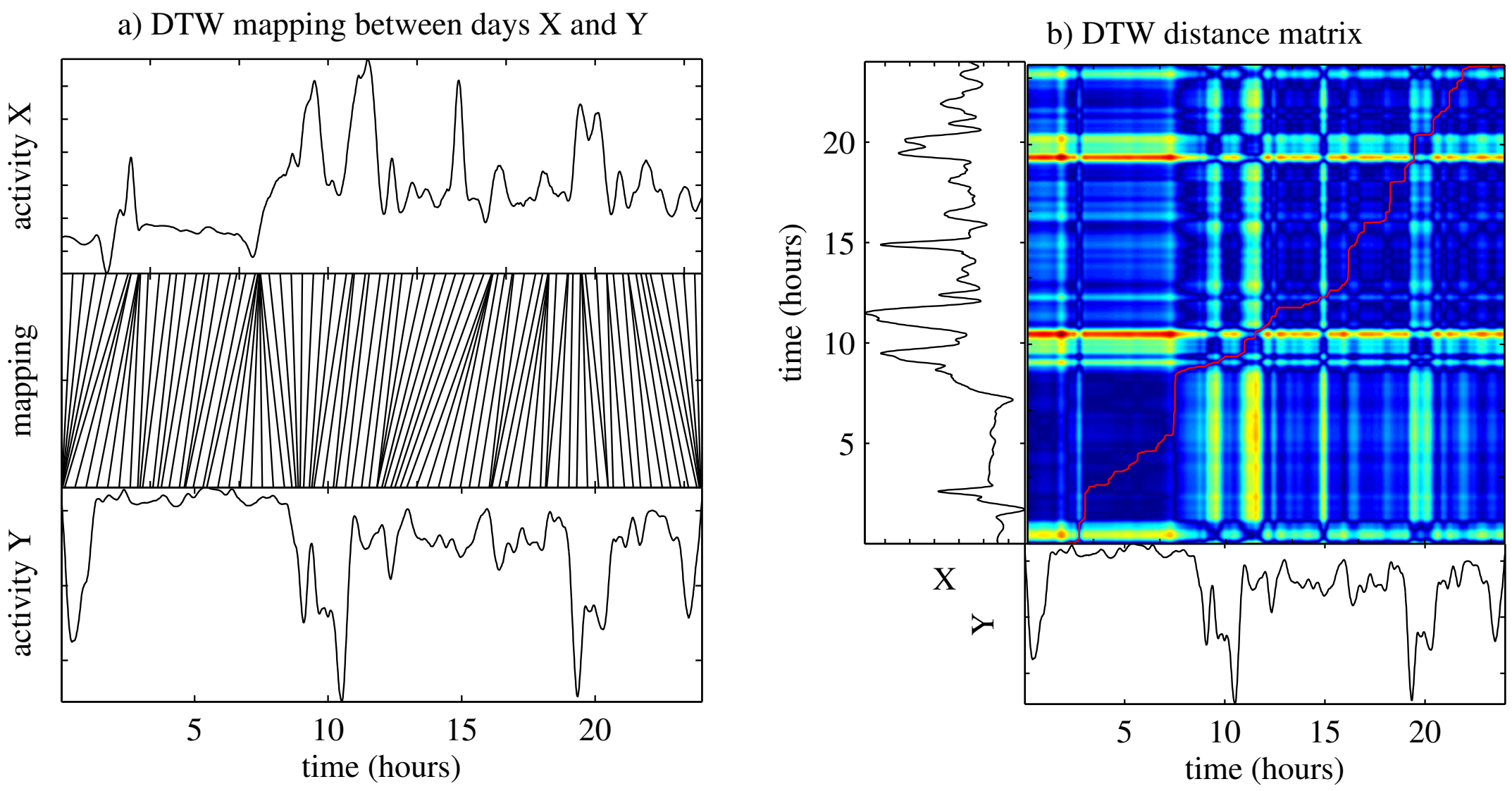

**Figure 5.4:** *Example of matching activity patterns of days X and Y using the Dynamic Time Warping. Picture a) represents mapping between the activity patterns and picture b) shows the distance matrix between activity levels on days X and Y (colour) together with the optimal matching (red line).*

### 5.4.4 Summary of disorganisation features

Based on the above-defined analysis methods, the features described in Table 5.3 were selected as potentially contributing to the disorganisation cluster of symptoms.

**Table 5.3:** *Activity disorganisation features, selected for further analysis.*

| Feature | Description |
| --- | --- |
| LEVEL_MSE_{N} | $N^{th}$ scale of Multiscale Entropy ($N = 1..5$) |
| DUR_AVG | Average duration of BOCPD segments |
| DUR_GAMMA_K | Shape parameter of Gamma distribution fit into BOCPD length |
| DUR_GAMMA_THETA | Scale parameter of Gamma distribution fit into BOCPD length |
| DUR_DFA | DFA $\alpha$ (scaling exponent) of BOCPD length |
| LEVEL_DTW | Average day-to-day DTW distance of activity level patterns |

## 5.5 Mood cluster

The indirect objective measure of mood, used in this study is the quality of sleep. Despite being an important feature of depression and mania, it is not objectively assessed in the context of psychiatric examination. At the same time, the field of sleep analysis is very well developed and actively uses actigraphic sleep assessment techniques (Ancoli-Israel et al., 2003; Sadeh, 2011). A number of sleep analysis methods were created for different types of actigraphs and successfully applied in studies of mental health, including affective disorders (Wirz-Justice et al., 2001; Wulff et al., 2006; Wirz-Justice, 2007a; Wirz-Justice et al., 2010; Scott, 2011).

Key sleep characteristics include such measures as sleep onset latency and sleep duration, which require knowledge of bed and wake times. Given that in an ambulatory environment compliance to manual bed time annotation is low (60% for Fitbit-based annotation in the AMoSS study), a novel method of automatic bed time annotation was developed and described in the previous chapter, and estimated bed times used to calculate sleep characteristics described below.

### 5.5.1 Sleep-wake segmentation

Analysis of sleep is based on identification of wakefulness episodes during bed time, in the case of actigraphic sleep assessment characterised by short motion periods. In order to identify wake-related motion, sleep/wake segmentation of bed-time actigraphic recording is performed by an expert or by using an automatic algorithm (Blackwell et al., 2005).

There are several published algorithms to segment sleep/wake states in actigraphy data. For example, Cole *et al.* (Cole et al., 1992) proposed an algorithm, based on a weighted sum of previous and subsequent epochs, and claimed an accuracy of 88% for distinguishing wakefulness from sleep on a sample of 41 subjects monitored during 1 laboratory night. Sadeh *et al.* (Sadeh et al., 1994) developed a metric to segment sleep/wake states, based on four activity features with overall reported agreement with polysomnographic analysis of 91% − 93% on a sample of 36 subjects monitored during 1 laboratory night. However, due to differences in hardware implementations, most actigraphs implement their own sleep detection algorithms.

In this study the sleep/wake scoring algorithm proposed by Oakley (Oakley, 1997) for Actiwatch (CamNTech, UK) activity monitoring platform was used. The algorithm estimates each 15 seconds activity epoch as wake or sleep based on previous and future values:

$$A_0 = 0.04E_{-(8-5)} + 0.2E_{-(4-1)} + 4E_0 + 0.2E_{+(1-4)} + 0.04E_{+(5-8)} \quad (5.3)$$

where $A_0$ is the score of activity of the current epoch, $E_0$ is the activity counts of the current epoch and $E_n$ is the activity counts of the epochs before and after the current one. The epoch is scored as sleep if $A_0$ exceeds the threshold of 40 counts.

The scoring algorithm relies on conversion of raw tri-axial activity data into the Actiwatch-compatible counts, using a procedure developed by te Lindert *et al.* (te Lindert and Van Someren, 2013) and described in Section 4.4.2.

### 5.5.2 Sleep analysis

Sleep analysis uses bed time derived from HMM segmentation and sleep / wake segmentation from Oakley algorithm to calculate a number of sleep characteristics. Clinical practice parameters for analysis of sleep were originally described by Rechtschaffen and Kales (Rechtschaffen and Kales, 1968), and currently maintained by the American Academy of Sleep Medicine (AASM). The parameters include such characteristics as:

- **To bed time**: The time the subject gets into bed, as identified by bed/wake segmentation.
- **Out of bed time**: The time the subject rose out of bed, as identified by bed/wake segmentation.
- **Time in bed (min)**: Time between "To bed time" and "Out of bed time".
- **Sleep onset latency (min)**: Time it took the subject to fall asleep.
- **Wake after sleep onset (min)**: Time scored as wake between "Sleep onset time" and "Final wake time".
- **Actual sleep time (min)**: Time scored as sleep between "Sleep onset time" and "Final wake time".
- **Sleep efficiency (%)**: "Actual sleep time" divided by 'Time in bed'.
- **Number of wake bouts**: Number of continuous blocks scored as wake.
- **Mean wake bout time (min)**: Mean duration of wake block.
- **Number of sleep bouts**: Number of continuous blocks scored as sleep.
- **Mean sleep bout time (min)**: Mean duration of sleep block.
- **Mobile time (min)**: Total duration of epochs with activity (>0).
- **Immobile time (min)**: Total duration of epochs with activity (=0).

### 5.5.3 Circadian rhythm

The circadian rhythm in humans is characterised by two main variables, $\tau$ describing the period of circadian clock under zeitgeber-free conditions and $\psi$ characterising phase shift between the rhythm and a zeitgeber, such as dusk and dawn. Additionally, the change in day length with season influences the ratio of activity and rest ($\alpha : \rho$) (Wirz-Justice, 2007b).

There are several methods described in the literature for numerical analysis of circadian rhythms, with the Cosinor method and Lomb-Scargle periodogram being the most accepted and used in research, where the Lomb-Scargle method is used for unevenly sampled data (Wirz-Justice et al., 2001; Wulff et al., 2006; Refinetti et al., 2007). The Cosinor method for identification of $\tau$ was used in this research, although it should be noted that in studies of humans it is very difficult to reliably identify $\tau$ due to the influence of social and environmental zeitgebers (Wirz-Justice, 2007b).

For the identification of $\psi$ the most reliable actigraphic measure is the sleep midpoint (Wirz-Justice, 2007b). This, as well as the activity/rest ratio ($\alpha : \rho$) is measured from the HMM based bed/wake identification described in Section 4.5.3.

### 5.5.4 Circadian amplitude

The activity levels during the least active 5 hours, most active 10 hours and relative amplitude between those give insight into the circadian amplitude, or distribution of activity between sleep and wakefulness states, and can be defined as

$$L5 = \frac{\sum_{i=1}^{|H_5|} x_i}{|H_5|}, \tag{5.4}$$

$$M10 = \frac{\sum_{i=1}^{|H_{10}|} x_i}{|H_{10}|}, \tag{5.5}$$

$$RA = \frac{M10 - L5}{M10 + L5} \tag{5.6}$$

where $x_i$ is the activity measurement, $|H_5|$ and $|H_{10}|$ are the numbers of measurements in the least active 5 and most active 10 hours respectively (Witting et al., 1990; Van Someren, 1999), thus L5 and M10 values effectively represent the mean activity levels during these periods.

### 5.5.5 Non-parametric rest-activity characteristics

Inter-daily Stability (IS) and Intra-daily Variability (IV) are so-called non-parametric rest-activity characteristics, and were originally proposed to detect day-to-day variations of activity patterns (IS) and fragmentation of rest-activity rhythms (IV) (Sokolove and Bushell, 1978; Witting et al., 1990) and calculated using the following equations

$$IS = \frac{n \sum_{i=1}^{p} (x_i - \overline{x})^2}{p \sum_{i=1}^{n} (x_i - \overline{x})^2} \tag{5.7}$$

$$IV = \frac{n \sum_{i=2}^{n} (x_i - x_{i-1})^2}{(n-1) \sum_{i=1}^{n} (x_i - \overline{x})^2} \tag{5.8}$$

where $n$ is the total number of data, $p$ is the number of data samples per day, $x_i$ are individual measurements, $\bar{x}$ is the mean of all data and $i$ is the index of a given data point (Van Someren, 1999).

### 5.5.6 Summary of mood features

Based on the above-defined analysis methods, the features described in Table 5.4 were selected as potentially contributing to the mood cluster of symptoms.

**Table 5.4:** *Activity-based sleep and mood features, selected for further analysis.*

| **Feature** | **Description** |
|---|---|
| IN_BED_TIME | The time the subjects gets into bed |
| OUT_OF_BED_TIME | The time the subject got out of bed |
| TIME_IN_BED | Time in bed |
| SLEEP_ONSET_LATENCY | Time it took the subject to fall asleep |
| WAKE_AFTER_SLEEP_ONSET | Wake after sleep onset |
| ACTUAL_SLEEP_TIME | Actual sleep time |
| SLEEP_EFFICIENCY | 'Actual sleep time' divided by 'Time in bed' multiplied by 100 |
| NUMBER_OF_WAKE_BOUTS | Number of continuous blocks scored as WAKE |
| MEAN_WAKE_BOUT_TIME | Mean wake bout time |
| MEAN_SLEEP_BOUT_TIME | Mean sleep bout time |
| MOBILE_TIME | Total duration of epochs with activity (>0) during sleep |
| IMMOBILE_TIME | Total duration of epochs with no activity during sleep |
| TAU | Cosinor-derived $\tau$ |
| PHI | Sleep midpoint as offset to midnight in minutes |
| L5 | Level of activity during the least active 5 hours |
| M10 | Level of activity during the most active 10 hours |
| RA | Relative amplitude between L5 and M10 |
| IS | Inter-daily Stability |
| IV | Intra-daily Variability |

## 5.6 Summary

In this chapter, based on the five-cluster model of mental symptoms (Liddle, 2001), a number of activity features were proposed that have a potential to characterise three dimensions of the model, including psychomotor, disorganisation and mood cluster symptoms. The features were introduced in a principled way, based on clinical practice parameters accepted in sleep medicine and physical activity research, as well as drawing on the body of research on analysis of physical activity and behaviour in the fields of psychiatry and clinical psychology.

# Chapter 6

# Differentiation between disorders

## 6.1 The problem of diagnosis

One of the main problems in treatment of mental health disorders is the reliability of diagnosis, when a disorder is not correctly identified in those who have it, or found in subjects who don't have it. This problem in the current diagnostic framework of DSM-5 was studied by analysing the test-retest reliability of diagnosis, and for most diagnoses moderate-to-good inter-rater agreement with inter-class Kappa of 0.40-0.79 was found (Regier et al., 2013), which is in the range of results achieved in medicine in general (Kraemer et al., 2012). However, between different mental health institutions the inter-rater agreement results were substantially different, with Mayo Clinic demonstrating a Kappa value of 0.73 for diagnosis of bipolar I disorder and the University of Texas at San Antonio showing a Kappa value of 0.27 for the same disorder. A similar picture was observed for borderline personality disorder and major depressive disorder (Regier et al., 2013), and indicates a potentially high subjectivity of diagnosis by some clinical specialists.

More objective criteria are needed for diagnosis of mental disorders to eliminate the diagnostic inaccuracy due to varying levels of expertise and qualification of clinical specialists. The recently proposed DSM-5 cross-cutting symptom assessment tools, based on patient-reported measures, aim to partially address this problem and facilitate the shift towards a more dimensional approach in the diagnosis of mental disorders. In the adult population, the DSM-5 dimensional cross-cutting symptom assessment questionnaire (23 questions covering 13 domains) demonstrated good-to-excellent test-retest reliability for all domains (Narrow et al., 2013). But the questionnaire-only assessment is still not mature enough to provide a reliable tool for diagnostics, and may be prone to subjective bias due to impaired patient self-perception (Stanley and Wilson, 2006).

Objective assessment of behaviours with mobile and wearable sensors could potentially facilitate diagnostics by providing accurate and unbiased measures of clinically important behaviour characteristics. In this chapter the behaviour measures, described previously in Chapter 5, are evaluated for the capability of distinguishing between the bipolar disorder, borderline personality disorder and healthy control subjects.

## 6.2 Data subset selection

For the analysis of diagnostic accuracy of objective activity measures, GENEActiv accelerometer data were analysed together with categorical assignment of subjects to their diagnostic classes (bipolar disorder, borderline personality disorder and healthy controls) as ground truth. Given the complex nature of mental disorders, where symptoms may wax and wane during the course of disease, a relatively asymptomatic short duration period, where both GENEActiv data and mood self-ratings were present, was selected from each participant, following the approach of Huynh *et al.* (Huynh et al., 2015). Contrary to Huynh *et al.* who used a nine day period to capture 2 weekends, this study used a one week period due to limited data availability, to capture a stable emotional state and to avoid potential bias due to different ratios of work and free days between different participants.

By doing so, every group was taken in a relative “normality” state, so that the objective activity signatures of the analysis period were characteristic for the disorder itself, and not a transient clinical state (such as mania or depression in bipolar disorder or deterioration in borderline personality disorder). Such signatures could be potentially used as "behavioural markers" of specific mental health conditions for clinical diagnostic purposes.

### 6.2.1 Euthymic period identification

In order to identify the objective difference between clinical groups, one week of activity was selected from each participant, where symptoms were least presented, as identified by the lowest scores on ASRM and QIDS-SR16, and the least non-wear time was detected by the non-wear detection algorithm described in Section 4.3.2. The selection of a one week interval was based on the granularity of mood scoring (weekly), the amount of data available (up to 3 weeks for most participants), and to avoid the bias due to weekday/weekend variability in activity characteristics. The selected data subset is described in Table 6.1.

**Table 6.1:** *Demographics and data set characteristics of the AMoSS euthymic data subset. Superscripts indicate that the distributions are different with $p < 0.05$ according to Wilcoxon rank sum test. BD, BPD and HC superscripts refer to bipolar disorder, borderline personality disorder and healthy controls.*

| | **Healthy controls** | **Bipolar disorder** | **Borderline personality disorder** |
|---|---|---|---|
| | **Demographic characteristics** | | |
| Number of participants | 23 | 24 | 15 |
| Gender | 18 females | 16 females | 13 females |
| Age (Mean±SD) | 30.61±9.31 $^{BD}$ | 40.04±11.61 $^{HC\ BPD}$ | 32.13±9.58 $^{BD}$ |
| Height (Mean±SD) | 168.87±7.84 | 170.83±10.48 | 168.20±8.27 |
| Weight (Mean±SD) | 65.97±9.27 $^{BD\ BPD}$ | 80.40±15.20 $^{HC}$ | 79.13±18.06 $^{HC}$ |
| BMI (Mean±SD) | 23.15±2.92 $^{BD\ BPD}$ | 27.51±4.53 $^{HC}$ | 27.99±6.11 $^{HC}$ |
| | **Data set characteristics** | | |
| ASRM (Mean±SD) | 1.43±2.23 | 2.17±3.05 | 2.13±2.20 |
| QIDS-SR16 (Mean±SD) | 2.00±2.17 $^{BD\ BPD}$ | 5.71±5.40 $^{HC\ BPD}$ | 10.13±6.15 $^{HC\ BD}$ |
| GAD-7 (Mean±SD) | 1.35±1.85 $^{BD\ BPD}$ | 5.75±4.90 $^{HC\ BPD}$ | 10.20±6.72 $^{HC\ BD}$ |
| BIS-11 (Mean±SD) | 52.30±7.86 $^{BD\ BPD}$ | 64.96±9.78 $^{HC\ BPD}$ | 78.00±12.85 $^{HC\ BD}$ |
| Non-wear time, days (Mean±SD) | 0.14±0.11 | 0.15±0.08 | 0.11±0.09 |

### 6.2.2 Dealing with imbalanced data set

Recruitment for the AMoSS project was still ongoing at the time of writing, so the AMoSS data set was not balanced in terms of numbers of participants per clinical group, as well as age, gender and BMI, as can be seen from Table 6.1. Differences in age, gender and BMI may influence actigraphic measurements (Tudor-Locke et al., 2010; Tucker et al., 2011), and number of subjects in group (class) is important for the definition of unbiased classifiers. In order to compensate for the imbalance in the AMoSS data set two strategies of re-sampling were employed and described below:

- **Under-sampling the majority class (USM):** Removing samples from the majority class until classes are balanced in number of samples and clinical characteristics.
- **Synthetic minority over-sampling technique (SMOTE):** Adding synthetic samples to the minority class until classes are balanced.

#### 6.2.2.1 Under-sampling the majority class

Under-sampling the majority class provides an opportunity for eliminating samples from majority classes that are most dissimilar from samples of the minority class, and thus improving the balance between classes in both demographic characteristics and numbers of samples. The number of samples in the least populated class (borderline personality disorder) was selected as the target under-sampling number, and both healthy controls and bipolar disorder classes were under-sampled by eliminating subjects with lowest (healthy controls) or highest (bipolar disorder) BMI, until the target number of samples was achieved. The resulting BMI limits were set as $> 22.5$ for healthy controls and $< 29$ for bipolar disorder classes, resulting in the data subset described in Table 6.2.

It can be seen that after USM re-samping the imbalance in group demographic characteristics was eliminated and the difference only remains in self-rated mood symptoms, that are specific to a particular disorder. It can also be seen that a QIDS-SR16 rating of borderline personality disorder group exceeds the clinical threshold for moderate depression (> 10) at least for some participants. However, this self-reported low mood does not represent the actual clinical condition of depression, and characteristic of the euthymia state in borderline personality disorder as described by Stanley and Wilson (Stanley and Wilson, 2006).

**Table 6.2:** *Demographics and data set characteristics of the AMoSS euthymic data subset after under-sampling the majority class to match cohorts. Superscripts indicate that the distributions are different with $p < 0.05$ according to the Wilcoxon rank sum test. BD, BPD and HC superscripts refer to bipolar disorder, borderline personality disorder and healthy controls.*

| | **Healthy controls** | **Bipolar disorder** | **Borderline personality disorder** |
|---|---|---|---|
| | | **Demographic characteristics** | |
| Number of participants | 15 | 15 | 15 |
| Gender | 13 females | 10 females | 13 females |
| Age (Mean±SD) | 30.20±6.67 | 37.40±11.95 | 32.13±9.58 |
| Height (Mean±SD) | 166.80±8.03 | 171.40±11.17 | 168.20±8.27 |
| Weight (Mean±SD) | 69.31±9.33 | 72.49±11.75 | 79.13±18.06 |
| BMI (Mean±SD) | 24.83±1.93 | 24.57±2.34 | 27.99±6.11 |
| | | **Data set characteristics** | |
| ASRM (Mean±SD) | 1.67±2.58 | 1.80±3.36 | 2.13±2.20 |
| QIDS-SR16 (Mean±SD) | 1.87±2.64 $^{\text{BD BPD}}$ | 4.40±4.64 $^{\text{HC BPD}}$ | 10.13±6.15 $^{\text{HC BD}}$ |
| GAD-7 (Mean±SD) | 1.33±1.91 $^{\text{BD BPD}}$ | 3.87±2.83 $^{\text{HC BPD}}$ | 10.20±6.72 $^{\text{HC BD}}$ |
| BIS-11 (Mean±SD) | 52.07±9.29 $^{\text{BD BPD}}$ | 64.29±10.34 $^{\text{HC BPD}}$ | 78.00±12.85 $^{\text{HC BD}}$ |
| Non-wear time, days (Mean±SD) | 0.13±0.12 | 0.16±0.09 | 0.11±0.09 |

#### 6.2.2.2 Synthetic Minority Over-sampling Technique

Given the cost of acquiring new samples for the AMoSS data set, it is also beneficial to estimate the accuracy of classification using a complete set of data, and account for the imbalance in numbers of participants in each class by adding synthetic samples to minority classes. The Synthetic Minority Over-sampling Technique (Chawla et al., 2002) is a widely accepted technique for dealing with imbalanced data sets by generating synthetic samples and includes the following steps:

1. Unsupervised classification of complete data set is performed using $k$-nearest neighbours algorithm (five nearest neighbours proposed in the original algorithm and used in this work).

2. Synthetic samples are introduced at a random position along the line segments joining a minority sample with randomly selected nearest neighbours until the target number of samples is reached.

For the AMoSS euthymic data subset, the under-represented class of borderline personality disorder (15 subjects) was over-sampled to match the number of subjects in the bipolar disorder class (24 subjects). The healthy controls class was not over-sampled as the number of subjects in this class (23) was close to the number of bipolar disorder subjects. The transformation was performed in feature space, so demographic characteristics of the data set did not change.

## 6.3 Data analysis approach

The data analysis approach of this chapter included the steps of feature selection and classification, performed in Leave One Out Cross Validation (LOOCV) folds to avoid the curse of dimensionality issues related to over-estimation of classification accuracy when the number of features is comparable to the number of samples. On each step of cross validation the best features were selected, a classifier trained on the best features of the training set, and classification performance estimated on the testing set. For the subset of features, where best classification accuracy was achieved, the prevalence of features in classification folds was also reported. All features were normalised before the analysis by subtracting the mean and dividing by standard deviation.

### 6.3.1 Dimensionality reduction methods

In order to find the best features for separation between disorders, a number of dimensionality reduction techniques were used. There are two types of dimensionality reduction approaches in existence, including *feature extraction* and *feature selection*.

Feature extraction attempts to find the best features by transforming existing features into a lower dimensional space, and feature selection refers to methods of finding optimal subsets of existing features without transformation. Feature extraction methods may be agnostic to the type of problem, for example Linear Discriminant Analysis, or use domain specific knowledge to derive new meaningful features from existing ones.

Feature selection methods attempt to find the best subset of existing features without transformation, and can be broadly divided into filters and wrappers (Guyon and Elisseeff, 2003), where filters optimise some measure of feature usefulness and wrappers use the predictive model itself as a measure of feature selection quality.

In the following analysis a number of state-of-the-art dimensionality reduction methods were evaluated, from simple feature ranking technique to more complex wrapper methods. Based on features identified by these techniques, classifiers were built and their predictive power was estimated in terms of classification accuracy, specificity and sensitivity.

#### 6.3.1.1 Naïve feature ranking method

Naïve feature ranking method sorts all available features according to their $p$ values for differentiating between target classes, and sequentially adds top rated features to the "optimal" subset. This filter method takes into account only the relevance of the features to a particular classification target, but ignores potential cross-correlation between features.

In this study the best features to distinguish between target classes were selected by ordering features according to the absolute value of a standardised u-statistic of two-sample unpaired Wilcoxon rank-sum test.

#### 6.3.1.2 Minimum Redundancy Maximum Relevance

Multivariate minimum-Redundancy-Maximum-Relevance (mRMR) criterion takes into account both individual relevance of features to target class, as well as the complimentarity of selected subsets of features (Peng et al., 2005). This filter method allows the selection of features that are predictive for the target class (maximum relevance), but at the same time orthogonal to each other (minimum redundancy), according to Equations 6.3, 6.1 and 6.2. The relevance criterion is given by

$$maxD(S,c), \quad D = \frac{1}{S}\sum_{x_i \in S} I(x_i;c), \tag{6.1}$$

where $S$ is a feature set with features $x_i$ and $c$ is a target class (with $c = 0$ indicating normal, and $c = 1$ otherwise). The redundancy criterion is given by

$$minR(S), \quad R = \frac{1}{|S|^2}\sum_{x_i,x_j \in S} I(x_i;x_j) \tag{6.2}$$

where $S$ is a feature set with features $x_i$ and $x_j$, and mutual information is defined as

$$I(x;y) = \iint p(x,y)log\frac{p(x,y)}{p(x)p(y)}dxdy, \tag{6.3}$$

where $I(x;y)$ is the mutual information between variables $x$ and $y$. $p(x)$, $p(y)$ and $p(x,y)$ are probability densities of these variables.

The relevance and redundancy criteria were combined using Mutual Information Difference scheme and incremental search was then used to find features which satisfy the above criteria. For the application of mRMR, features were discretized into five states between values of Mean±SD, where SD is one of -1, -0.5, 0.5, 1 as suggested by Peng *et al.* (Peng et al., 2005).

#### 6.3.1.3 Least Absolute Shrinkage and Selection Operator

The Least Absolute Shrinkage and Selection Operator (LASSO) was proposed by Robert Tibshirani as a new method for regression estimation in linear models (Tibshirani, 1996). LASSO minimises L2 norm (residual sum of squares) subject to L1 constraint on the model as

$$\min_{\beta_0,\beta}\left(\frac{1}{2N}\sum_{i=1}^{N}(y_i-\beta_0-x_i^T\beta)^2+\lambda\sum_{j=1}^{p}|\beta_j|\right), \tag{6.4}$$

where $N$ is the number of observations, $y_i$ is the response at observation $i$, $x_i$ is the data vector of observation $i$, $p$ is the length of data vector, $\lambda$ is the regularisation parameter, and $\beta_0$ and $\beta$ are scalar and $p$-length vectors of regression coefficients.

As a result, it may produce regression model coefficients that are exactly 0, or irrelevant for the model. For feature selection LASSO was used as a wrapper method, maximising cross-validation accuracy of classification with selected features.

### 6.3.2 Classification techniques

To analyse the capability of selected features to differentiate between subject classes (and disorders), a number of classification techniques were used and described below. All feature selection and classification experiments were performed for pair-wise separation between study groups and using LOOCV to estimate unbiased classification performance.

#### 6.3.2.1 Logistic regression

Binary logistic regression can be used to predict responses based on independent variables, with the additional benefit of providing a probabilistic result. The logistic regression method is based on fitting the logistic function into data to achieve best separation between target classes. The logistic function is specified as

$$F(x)=\frac{1}{1+e^{-\left(\beta_0+\sum_{i=1}^{m}\beta_i x_i\right)}} \tag{6.5}$$

where $xi$ are the explanatory variables, $\beta_i$ are the coefficients of regression model and $F(x)$ is interpreted as the probability of dependant variable belonging to the "success" class.

#### 6.3.2.2 Support Vector Machines

The Support Vector Machines (SVM) classifier attempts to create a hyperplane with the largest distance to the nearest points in a feature space to separate target classes (Cortes and Vapnik, 1995). This is achieved by maximising the expression

$$\left[\frac{1}{n}\sum_{i=1}^{n} max(0, 1 - y_i(w \cdot x_i + b))\right] + \lambda ||w||^2, \tag{6.6}$$

where $x_i$ are data points of feature vector of length $n$, $y_i$ are $+1$ or $-1$ indicating to which class $x_i$ belongs, $w$ and $b$ are model parameters to optimise and $\lambda$ is the parameter defining the misclassification tradeoff. The example of SVM classification is presented in Figure 6.1.

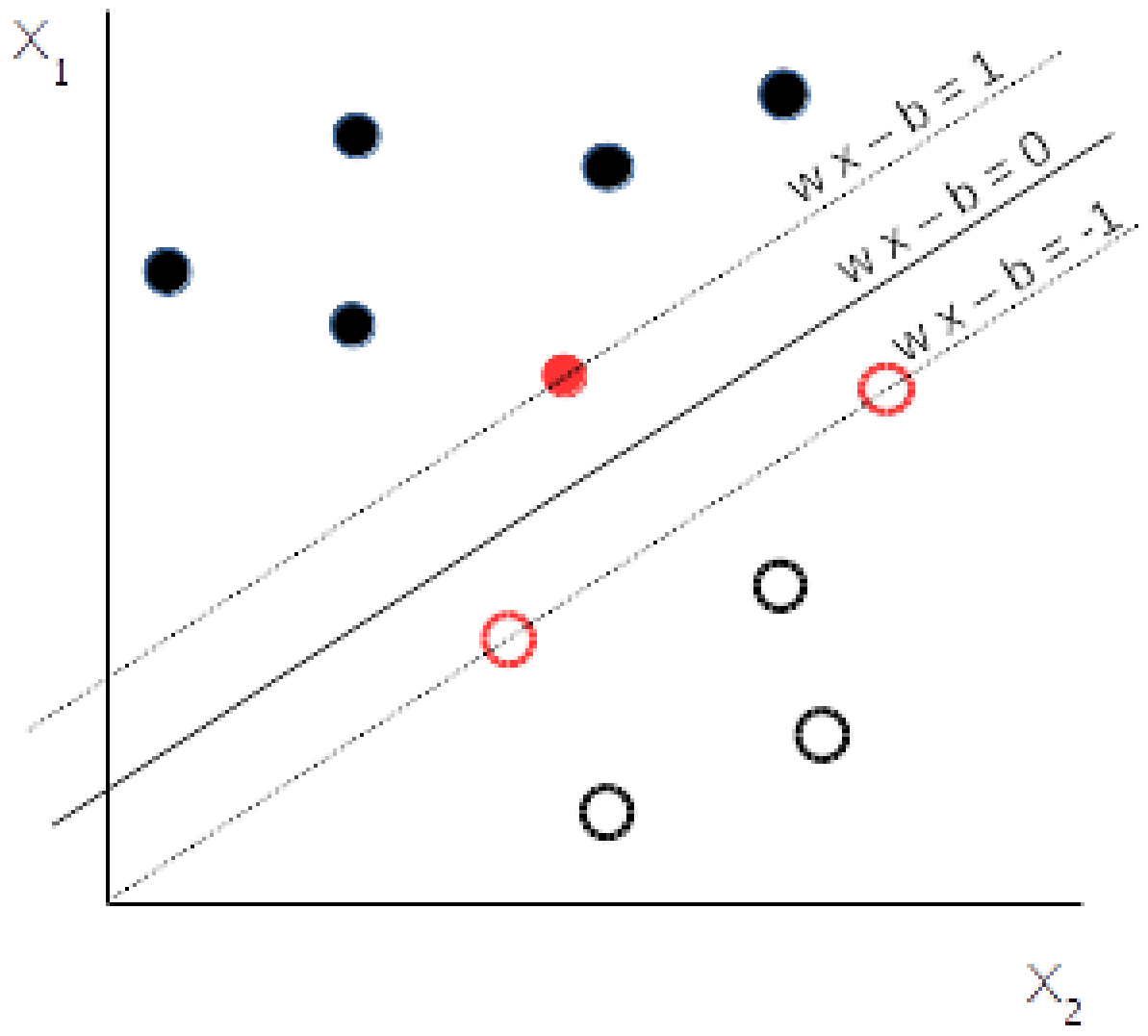

**Figure 6.1:** *Decision boundary of SVM classifier. Filled and empty dots represent two different classes with support vectors selected in red. The solid line represents the maximum-margin hyperplane and dashed lines represent margins.*

The SVM classifier allows for more flexibility in definition of class boundaries compared to logistic regression by using a so-called "kernel trick". With the kernel trick, if linear separation in the original feature space is not possible, features can be mapped into a higher dimensional space using the kernel function. The separation is then performed in a higher dimensional space.

The Gaussian Radial Basis Function (RBF) kernel used by this study, and in order to optimise the RBF kernel parameters $\alpha$ (scaling factor) and $C$ (box constraint), on the first fold of cross-validation an additional LOOCV sub-fold was performed where $\alpha$ and $C$ were identified using grid search.

## 6.4 Analysis results

Presented in this chapter exploratory analysis of differences between clinical groups aims to identify the most important features for separation between the diagnostic categories, using the proposed dimensional framework. Differences between groups are described first and subsequently the proposed features are assessed for their classification accuracy.

### 6.4.1 Comparison between cohorts

Comparison between the proposed behavioural features of mental disorders, corresponding to psychomotor, disorganisation and mood symptom dimensions is presented in Table 6.3. The comparison was performed on a USM data subset, thus the influence of imbalance in age and BMI between study groups was minimised.

The bipolar disorder participants were characterised by slightly elevated levels of physical activity as described by the time in moderate and vigorous activities, slightly lower entropy of activity and longer segments of continuous activity. Time to bed of bipolar disorder participants was slightly earlier, but all sleep characteristics comparable to healthy controls and patterns of activity were better aligned between days as shown by lower DTW distance between daily activity patterns.

Interestingly, no statistically significant difference in objective activity features was identified between healthy controls and bipolar disorder patients. It is consistent with the definition of bipolar disorder, where symptoms are present only during episodes of mania and depression, but contradicts with previous research findings, where differences in behaviour were reported also during euthymic state (Indic et al., 2011; Rock et al., 2014). The lack of statistically significant differences may be due to combined effect of the small number of subjects, short (one week) period of analysis and successful disorder management strategy of bipolar patients.

**Table 6.3:** *Group differences between objective features of mental health as described by dimensional framework. To minimise the influence of cohorts imbalance, the data are shown after under-sampling the majority classes (15 subjects in each group with no significant differences in demographic characteristics). Superscripts indicate that the distributions are different with $p < 0.05$ according to the Wilcoxon rank sum test. BD, BPD and HC superscripts refer to bipolar disorder, borderline personality disorder and healthy controls.*

| | **Healthy controls** | **Bipolar disorder** | **Borderline personality disorder** |
|---|---|---|---|
| | | **Psychomotor dimension** | |
| LEVEL_AVG | 0.03±0.01 | 0.03±0.01 | 0.03±0.01 |
| LEVEL_GAMMA_K | 0.17±0.04 | 0.19±0.06 | 0.15±0.04 |
| LEVEL_GAMMA_THETA | 0.19±0.09 | 0.22±0.19 | 0.18±0.07 |
| TIME_MODERATE | 111.99±26.26 | 124.41±31.32 | 105.86±26.12 |
| TIME_VIGOROUS | 7.72±5.53 | 11.95±18.62 | 7.70±10.22 |
| | | **Disorganisation dimension** | |
| LEVEL_MSE_1 | 0.09±0.04 $^{BPD}$ | 0.08±0.05 $^{BPD}$ | 0.06±0.04 $^{HC\ BD}$ |
| LEVEL_MSE_2 | 0.12±0.03 | 0.11±0.04 | 0.10±0.02 |
| LEVEL_MSE_3 | 0.13±0.03 | 0.12±0.04 | 0.11±0.02 |
| LEVEL_MSE_4 | 0.14±0.03 | 0.13±0.04 | 0.12±0.03 |
| LEVEL_MSE_5 | 0.14±0.03 | 0.14±0.05 | 0.13±0.03 |
| DUR_AVG | 1.02±0.11 | 1.06±0.20 | 1.04±0.13 |
| DUR_GAMMA_K | 0.73±0.04 | 0.72±0.07 | 0.71±0.05 |
| DUR_GAMMA_THETA | 1.42±0.22 | 1.52±0.47 | 1.50±0.30 |
| DUR_DFA | 0.62±0.05 | 0.63±0.04 | 0.66±0.03 |
| LEVEL_DTW | 170.29±15.67 | 167.53±20.63 | 180.49±16.94 |
| | | **Mood dimension (sleep)** | |
| IN_BED_TIME | -58.83±85.25 | -91.49±50.99 $^{BPD}$ | 0.78±100.43 $^{BD}$ |
| OUT_OF_BED_TIME | 477.39±59.19 $^{BPD}$ | 438.11±42.47 $^{BPD}$ | 540.58±97.87 $^{HC\ BD}$ |
| TIME_IN_BED | 536.47±49.59 | 529.85±46.48 | 540.04±64.13 |
| SLEEP_ONSET_LATENCY | 31.64±20.37 | 26.82±16.26 | 29.38±15.15 |
| WAKE_AFTER_SLEEP_ONSET | 48.51±15.84 | 47.63±14.35 $^{BPD}$ | 62.63±19.05 $^{BD}$ |
| ACTUAL_SLEEP_TIME | 456.07±46.44 | 455.16±49.52 | 447.78±55.06 |
| SLEEP_EFFICIENCY | 85.54±3.30 | 86.23±3.49 $^{BPD}$ | 83.20±3.34 $^{BD}$ |
| NUMBER_OF_WAKE_BOUTS | 52.92±12.68 | 48.36±11.64 $^{BPD}$ | 62.80±21.43 $^{BD}$ |
| MEAN_WAKE_BOUT_TIME | 0.91±0.28 | 0.98±0.22 | 1.01±0.20 |
| MEAN_SLEEP_BOUT_TIME | 9.58±2.29 | 10.39±2.59 | 8.57±2.65 |
| MOBILE_TIME | 214.00±36.82 | 199.73±35.25 $^{BPD}$ | 242.46±64.36 $^{BD}$ |
| IMMOBILE_TIME | 290.83±38.11 | 303.31±52.17 | 268.21±48.98 |
| TAU | 1.01±0.01 | 1.01±0.01 | 1.00±0.01 |
| PHI | 209.28±69.07 | 173.31±40.76 $^{BPD}$ | 270.68±93.83 $^{BD}$ |
| L5 | 2.79±2.18 $^{BPD}$ | 2.82±1.37 $^{BPD}$ | 3.88±1.54 $^{HC\ BD}$ |
| M10 | 116.14±27.72 | 134.15±46.86 | 113.21±30.26 |
| RA | 0.95±0.03 $^{BPD}$ | 0.95±0.02 $^{BPD}$ | 0.93±0.04 $^{HC\ BD}$ |
| IS | 0.02±0.01 | 0.01±0.01 | 0.02±0.01 |
| IV | 0.75±0.14 | 0.71±0.17 | 0.75±0.08 |

The borderline disorder participants were characterised by slightly lower levels of daytime activity, significantly lower entropy of activity, later times of going to bed and longer bed times with lower sleep efficiency and higher levels of activity during the least active 5 hours. Also, the variability of daily activity patterns was slightly higher for the borderline personality disorder patients.

It can be seen that most differences between disorders were observed in mood dimension features, related to patterns of circadian rhythms, night time activity and sleep. Significant differences between borderline personality disorder group and both bipolar disorder and healthy controls were reported in wake up times, levels of activity during the least active 5 hours and relative amplitude of day-night activity.

### 6.4.2 Importance of objective features

The importance of objective features for separation between diagnostic categories was assessed using three feature selection algorithms, including the simple ranking of features according to their relevance to target using the p-value of Wilcoxon rank-sum (RANK), mRMR criteria and LASSO, with results presented in Table 6.4.

If using simple relevance criteria (RANK), most of the top features for separation between clinical groups are concentrated in the dimension of mood, while the criteria that take into account complimentarity of features (mRMR) or select features for best separation between groups using cross-validation (LASSO) propose top features belonging to different dimensions of symptoms, suggesting the validity of the dimensional approach.

**Table 6.4:** *Ranking of features for separation between clinical groups using Wilcoxon rank-sum test criteria, mRMR and LASSO, with top five features in the ranking selected in* ***bold font****.*

| | **Healthy controls vs. bipolar disorder** | | | **Healthy controls vs. borderline personality disorder** | | | **Bipolar disorder vs. borderline personality disorder** | | |
|---|---|---|---|---|---|---|---|---|---|
| | RANK | mRMR | LASSO | RANK | mRMR | LASSO | RANK | mRMR | LASSO |
| | | | | | **Psychomotor dimension** | | | | |
| LEVEL_AVG | 11 | 19 | 18 | 23 | 28 | 32 | 16 | 24 | 7 |
| LEVEL_GAMMA_K | 13 | 24 | 7 | 25 | 20 | 23 | 11 | 22 | 13 |
| LEVEL_GAMMA_THETA | 21 | **3** | 11 | 33 | 8 | 18 | 32 | **4** | 18 |
| TIME_MODERATE | 8 | 6 | **2** | 26 | 23 | 24 | 17 | 7 | **5** |
| TIME_VIGOROUS | 29 | 7 | 24 | 21 | **2** | 11 | 25 | 13 | 19 |
| | | | | | **Disorganisation dimension** | | | | |
| LEVEL_MSE_1 | 33 | 22 | **4** | **2** | 6 | **1** | 6 | **3** | 20 |
| LEVEL_MSE_2 | 31 | 29 | 29 | 13 | 17 | 8 | 20 | 16 | 26 |
| LEVEL_MSE_3 | 32 | 33 | 30 | 16 | 30 | 25 | 21 | 28 | 31 |
| LEVEL_MSE_4 | 30 | 17 | 32 | 15 | 32 | 28 | 26 | 33 | 32 |
| LEVEL_MSE_5 | 25 | 34 | 16 | 18 | 34 | 26 | 27 | 34 | 11 |
| DUR_AVG | 26 | 11 | 13 | 28 | 24 | 29 | 34 | 31 | 14 |
| DUR_GAMMA_K | 15 | 27 | 12 | 22 | 26 | 19 | 30 | 30 | 21 |
| DUR_GAMMA_THETA | 22 | 31 | 31 | 24 | 33 | 14 | 33 | 26 | 15 |
| DUR_DFA | 17 | 9 | 19 | 6 | **4** | **3** | 14 | 8 | **4** |
| LEVEL_DTW | 27 | **4** | 22 | 7 | 11 | 20 | 15 | 20 | 9 |
| | | | | | **Mood dimension (sleep)** | | | | |
| IN_BED_TIME | 12 | 14 | 33 | 11 | 13 | 33 | **5** | 10 | 22 |
| OUT_OF_BED_TIME | **1** | 16 | **1** | **4** | 19 | **4** | **1** | **1** | **1** |
| TIME_IN_BED | 23 | 25 | 34 | 29 | 25 | 34 | 23 | 23 | 33 |
| SLEEP_ONSET_LATENCY | 14 | 13 | 6 | 34 | 31 | 21 | 24 | 27 | 23 |
| WAKE_AFTER_SLEEP_ONSET | 34 | 8 | 14 | **5** | 21 | **5** | 8 | 21 | 27 |
| ACTUAL_SLEEP_TIME | 28 | 32 | 27 | 27 | 22 | 30 | 28 | 32 | 34 |
| SLEEP_EFFICIENCY | 16 | 30 | 21 | 12 | 18 | 6 | **3** | 18 | 6 |
| NUMBER_OF_WAKE_BOUTS | 10 | 12 | 28 | 14 | 29 | 9 | 7 | 15 | 8 |
| MEAN_WAKE_BOUT_TIME | **2** | **5** | 17 | 9 | 16 | 31 | 29 | 11 | 24 |
| MEAN_SLEEP_BOUT_TIME | **3** | **1** | 15 | 20 | 15 | 12 | 12 | 25 | 12 |
| MOBILE_TIME | **4** | 23 | **3** | 19 | 7 | 15 | 9 | **5** | 28 |
| IMMOBILE_TIME | 18 | 26 | **5** | 17 | **1** | 10 | 18 | **2** | 16 |
| TAU | 19 | 10 | 9 | 10 | 14 | 7 | 13 | 17 | 10 |
| PHI | **5** | **2** | 25 | 8 | **3** | 13 | **2** | 6 | **2** |
| L5 | 9 | 20 | 26 | **1** | **5** | 27 | 10 | 14 | 29 |
| M10 | 6 | 28 | 23 | 31 | 10 | 16 | 19 | 19 | 25 |
| RA | 20 | 15 | 20 | **3** | 9 | **2** | **4** | 9 | **3** |
| IS | 7 | 18 | 8 | 32 | 27 | 22 | 22 | 29 | 17 |
| IV | 24 | 21 | 10 | 30 | 12 | 17 | 31 | 12 | 30 |

### 6.4.3 Accuracy of separation between study groups

The accuracy of separation between the diagnostic categories is presented in Table 6.5. If a certain feature selection and classification strategy demonstrated superior performance compared to others, it was selected in bold. For USM data balancing strategy simple RANK feature selection demonstrated the best results, indicating that very few highly correlated features were the most relevant for classification.

As expected from the cohort differences described in Section 6.4.1, the best classification accuracy of 80% was achieved for separation between bipolar and borderline personality disorder groups, with 80% sensitivity and specificity when a mRMR feature selection was used. The features used for classification were:

- OUT_OF_BED_TIME
- IMMOBILE_TIME
- LEVEL_MSE_1
- LEVEL_GAMMA_THETA
- MOBILE_TIME

Notably, this result persists for both SVM and logistic regression classifiers, indicating linear separation between classes. Classification accuracy drops only marginally when using SMOTE balancing strategy.

A classification accuracy of 70% with 67% specificity and 73% sensitivity was achieved for separation between healthy controls and borderline personality disorder patients using RANK feature selection strategy. The features used for classification were:

- L5
- LEVEL_MSE_1
- RA

**Table 6.5:** *Summary of classification accuracy for separation between AMoSS clinical groups using different data set balancing, feature selection and classification strategies, with best results selected in* ***bold font****.*

| | Data set balancing strategy | | | | | |
|---|---|---|---|---|---|---|
| | USM | | | SMOTE | | |
| | Feature selection | | | | | |
| | RANK | mRMR | LASSO | RANK | mRMR | LASSO |
| | Healthy controls vs. bipolar disorder | | | | | |
| Logistic regression | | | | | | |
| Accuracy | 0.60 | 0.57 | 0.50 | 0.62 | 0.45 | 0.60 |
| Specificity | 0.53 | 0.53 | 0.47 | 0.54 | 0.46 | 0.58 |
| Sensitivity | 0.67 | 0.60 | 0.53 | 0.70 | 0.43 | 0.61 |
| SVM | | | | | | |
| Accuracy | **0.67** | 0.53 | 0.53 | 0.57 | 0.45 | **0.64** |
| Specificity | **0.67** | 0.53 | 0.60 | 0.42 | 0.42 | **0.54** |
| Sensitivity | **0.67** | 0.53 | 0.47 | 0.74 | 0.48 | **0.74** |
| | Healthy controls vs. borderline personality disorder | | | | | |
| Logistic regression | | | | | | |
| Accuracy | 0.67 | 0.57 | 0.67 | **0.72** | 0.64 | 0.72 |
| Specificity | 0.67 | 0.47 | 0.67 | **0.75** | 0.67 | 0.75 |
| Sensitivity | 0.67 | 0.67 | 0.67 | **0.70** | 0.61 | 0.70 |
| SVM | | | | | | |
| Accuracy | **0.70** | 0.60 | 0.67 | **0.72** | 0.66 | **0.72** |
| Specificity | **0.67** | 0.33 | 0.67 | **0.71** | 0.62 | **0.71** |
| Sensitivity | **0.73** | 0.87 | 0.67 | **0.74** | 0.70 | **0.74** |
| | Bipolar disorder vs. borderline personality disorder | | | | | |
| Logistic regression | | | | | | |
| Accuracy | 0.77 | **0.80** | 0.73 | 0.77 | 0.77 | 0.77 |
| Specificity | 0.73 | **0.80** | 0.67 | 0.79 | 0.79 | 0.79 |
| Sensitivity | 0.80 | **0.80** | 0.80 | 0.75 | 0.75 | 0.75 |
| SVM | | | | | | |
| Accuracy | **0.80** | **0.80** | 0.77 | **0.79** | **0.79** | **0.79** |
| Specificity | **0.67** | **0.80** | 0.60 | **0.75** | **0.75** | **0.75** |
| Sensitivity | **0.93** | **0.80** | 0.93 | **0.83** | **0.83** | **0.83** |

- OUT_OF_BED_TIME
- WAKE_AFTER_SLEEP_ONSET

With synthetic samples the accuracy of classification marginally improved to 72% when using RANK and LASSO feature selection. Again, the logistic regression classifier demonstrated similar accuracy as a SVM classifier.

For separation between healthy controls and bipolar disorder participants the accuracy of classification was 67% with 67% specificity and sensitivity when using RANK feature selection strategy. The features used for classification were:

- OUT_OF_BED_TIME
- MEAN_WAKE_BOUT_TIME
- MEAN_SLEEP_BOUT_TIME
- MOBILE_TIME
- PHI

The accuracy of classification dropped to 64% when using SMOTE oversampling.

## 6.5 Discussion and conclusions

The analysis in this chapter covered two main topics. First, differences between objective behaviour characteristics of clinical cohorts were identified by analysing the behaviour characteristics of euthymic periods of activity. Second, the capability of activity features to differentiate between disorders was evaluated by analysing the cross-validated classification accuracy of diagnostic groups with different data set balancing, feature selection and classification strategies.

For both patient groups, bipolar disorder and borderline personality disorder, the mood dimension, represented by sleep characteristics, appears to be the most important factor separating between groups. For borderline personality disorder patients many of the sleep characteristics are significantly different compared to both healthy controls and bipolar disorder patients (see Table 6.3). While there are no individually significant features of difference between healthy controls and bipolar disorder patients, after feature selection the top features appear to belong to mood dimension.

Comparing performance of different classifiers and feature selection strategies, it can be observed that accuracy of simple logistic regression is similar to a nonlinear approach like an SVM classification, indicating homogenity of diagnostic groups, with linear separation possible between classes.

Finally, an interesting observation to make is that bipolar disorder patients are more different to borderline personality disorder than healthy controls in terms of the activity metrics defined in this chapter. This could be explained by a well-executed disorder management strategy, where bipolar disorder patients actively adhere to a healthy and regular life routine, as demonstrated by low inter-daily variability and differences between successive daily activity patterns.

# Chapter 7

# Differentiation between clinical states

## 7.1 The problem of deterioration

In many psychiatric disorders symptoms may wax and wane over time, and relapse is a continual threat (Sierra et al., 2009; Dawson et al., 2010). For example, in the case of bipolar disorder patients they may slide into a manic or depressive state, in borderline personality disorder there may be attempts at suicide or self-harm, or psychotic relapse in schizophrenia. Clinical deterioration poses a significant problem and may happen undetected due to impaired patient insight, and can lead to hospitalisation (Lieb et al., 2004; Green, 2006; Pompili et al., 2013).

It was demonstrated that recognition of the early warning signs of upcoming clinical episodes are helpful for timely treatment, and helps to achieve better patient outcomes (Birchwood, 2000; Morriss, 2004; Blum et al., 2008). However, the treatment strategies that allow the incorporation of recognition of early warning signs into the management of disorder require close collaboration between patient, family, primary and specialist care (Falloon et al., 1996; Mueser et al., 2002; Mueser et al., 2003). Organising such an interaction is a complex and costly task, and not always possible in a particular patient's circumstances.

The current practice of early warning signs identification includes patients' self-observations facilitated by family, and sometimes also involves clinical specialists using web-based mood self-reporting tools, such as True Colours Self-Management platform (Miklowitz et al., 2012). Self-reports are known to suffer from compliance issues, especially if insight or motivation are impaired, as often happens in mental disorders. In such situations ambient symptoms monitoring using using a mobile and wearable sensors may be of great benefit.

In this chapter the difference in objective activity features between clinically relevant mood states will be analysed to assess the feasibility of objective mood state identification. The task of identification of clinically relevant mood states was approached here as a classification of weekly objective behaviour indicators.

## 7.2 Data subset selection

To identify the objective mood features, the high quality data collected with GENEActiv wearable sensors were used. Such data cover both day and night activity, but are available only for a short period of time from each participant, usually from one to three weeks of continuous recording with one or two data acquisition sessions per participant. Given the DSM-5 diagnostic criteria for mania and depression, where for a clinical episode to be detected it has to be present for one (mania) or two (depression) weeks, and the fact that the frequency of episodes in bipolar disorder is low (four episodes per year in a rapid cycling variant, present in approximately 10% of patients), the probability of capturing a mood episode together with euthymia from a single participant with GENEActiv data acquisition is low. The GENEActiv data can be used therefore only to compare cohorts and not to build personalised regression models, due to the limited number of weekly observations per subject.

### 7.2.1 Distinct mood state identification

In order to perform the analysis, one week periods were selected from data, where mood self-assessment was available and objective GENEActiv activity data present. Using standard diagnostic thresholds of QIDS-16SR and ASRM, the four mood states for classification were defined as

- euthymia, where the ASRM score is below 6 and QIDS-16SR is below 11,
- depression, where QIDS-16SR is above 10,
- mania, where ASRM is above 5, and
- mixed state, where both QIDS-16SR is above 10 and ASRM above 5.

From the analysis in Chapter 6 it can be seen that for the borderline personality disorder patients the connection between the objectively-assessed behaviours and self-reported emotional state is not reliable. Therefore for the present classification only the groups of healthy controls and bipolar disorder patients were used.

Following the approach of Section 6.2, from each participant single weeks of activity were selected, where no more then 10% of GENEActiv data were missing and subjective mood score was reported on both QIDS-16SR and ASRM scales.

A total of 109 episodes were identified with statistical characteristics presented in Table 7.1. There were no mixed state episodes, so the mixed state was excluded from further analysis and binary classification was performed to separate euthymia, depression and mania. It is necessary to mention that the selected weeks do not necessarily represent clinically significant episodes of disorder, as the individual symptoms manifestation may not match the DSM-5 diagnostic criteria.

### 7.2.2 Dealing with imbalanced data set

It can be seen that the mood states data set was unbalanced towards euthymia periods, with the number of weeks five to ten times in excess of the number of weeks with mania and depression. Additionally, the data on depressive state represented a high percentage of recordings from female participants. Therefore the re-sampling strategies described in detail in Section 6.2.2 were used.

**Table 7.1:** *Demographics and data set characteristics of the AMoSS mood classification data subset. Superscripts indicate that the distributions are different with* $p < 0.05$ *according to the Wilcoxon rank sum test. EU, DEP, MAN and MIX superscripts refer to euthymia, depression, mania and mixed states.*

| | **Euthymia** | **Mania** | **Depression** | **Mixed state** |
|---|---|---|---|---|
| Number of weeks, total | 109 | 10 | 16 | 0 |
| healthy controls | 54 | 4 | 0 | 0 |
| bipolar disorder | 55 | 6 | 16 | 0 |
| | | **Demographic characteristics** | | |
| Number of unique subjects | 37 | 6 | 7 | 0 |
| Gender | 27 females | 3 females | 5 females | 0 females |
| Age (Mean±SD) | 34.84±10.59 | 28.00±3.29 | 30.43±5.83 | N/A |
| Height (Mean±SD) | 170.46±7.14 | 171.50±2.74 | 172.29±2.29 | N/A |
| Weight (Mean±SD) | 73.62±12.24 | 66.00±12.05 | 71.29±16.13 | N/A |
| BMI (Mean±SD) | 25.49±4.90 | 22.56±4.82 | 24.08±5.68 | N/A |
| | | **Data set characteristics** | | |
| ASRM (Mean±SD) | 1.31±1.57 $^{\text{MAN}}$ | 10.20±3.79 $^{\text{EU DEP}}$ | 0.81±1.28 $^{\text{MAN}}$ | N/A |
| QIDS-16SR (Mean±SD) | 3.17±2.47 $^{\text{DEP}}$ | 3.10±3.11 $^{\text{DEP}}$ | 17.50±5.82 $^{\text{EU MAN}}$ | N/A |
| GAD-7 (Mean±SD) | 2.89±3.43 $^{\text{DEP}}$ | 4.90±4.38 $^{\text{DEP}}$ | 12.94±6.76 $^{\text{EU MAN}}$ | N/A |
| BIS-11 (Mean±SD) | 58.15±10.82 $^{\text{DEP}}$ | 64.80±16.37 | 67.06±9.33 $^{\text{EU}}$ | N/A |
| Non-wear time, days (Mean±SD) | 0.17±0.13 | 0.20±0.15 | 0.19±0.06 | N/A |

#### 7.2.2.1 Under-sampling the majority class

The strategy for USM re-sampling was selected to minimise the imbalance in gender and age, therefore samples of male participants at an older age ($> 40$) were discarded, keeping at the same time at least one week of data from each unique subject to avoid the data subset being dominated by a few participants with relatively large amount of acquired data, and to analyse all possible subject-specific predictors of mood. The resulting subset is described in Table 7.2.

#### 7.2.2.2 Synthetic Minority Over-sampling Technique

For SMOTE re-sampling, the minority classes of mania and depression were oversampled to match the number of participants in the euthymia class, with no change in demographic

**Table 7.2:** *Demographics and data set characteristics of the AMoSS GENEActiv mood data subset after USM. Superscripts indicate that the distributions are different with $p < 0.05$ according to Wilcoxon rank sum test. EU, DEP, MAN and MX superscripts refer to euthymia, depression, mania and mixed states.*

| | Euthymia | Mania | Depression | Mixed state |
|---|---|---|---|---|
| Number of weeks, total | 10 | 10 | 10 | 0 |
| healthy controls | 7 | 4 | 0 | 0 |
| bipolar disorder | 3 | 6 | 10 | 0 |
| | | **Demographic characteristics** | | |
| Number of unique subjects | 10 | 6 | 7 | 0 |
| Gender | 5 females | 3 females | 5 females | 0 females |
| Age (Mean±SD) | 31.70±10.06 | 30.00±9.55 | 30.71±9.36 | N/A |
| Height (Mean±SD) | 170.40±6.31 | 170.00±7.07 | 171.71±4.82 | N/A |
| Weight (Mean±SD) | 77.60±18.72 | 72.33±19.14 | 78.29±18.79 | N/A |
| BMI (Mean±SD) | 27.04±7.71 | 25.34±7.90 | 26.84±7.67 | N/A |
| | | **Data set characteristics** | | |
| ASRM (Mean±SD) | 1.60±1.43 $^{MAN}$ | 10.20±3.79 $^{EU\ DEP}$ | 0.90±1.45 $^{MAN}$ | N/A |
| QIDS-16SR (Mean±SD) | 2.50±1.78 $^{DEP}$ | 3.10±3.11 $^{DEP}$ | 15.60±5.08 $^{EU\ MAN}$ | N/A |
| GAD-7 (Mean±SD) | 1.60±1.96 $^{DEP}$ | 4.90±4.38 | 11.60±6.69 $^{EU}$ | N/A |
| BIS-11 (Mean±SD) | 58.40±8.76 | 64.80±16.37 | 67.50±10.09 | N/A |
| Non-wear time, days (Mean±SD) | 0.20±0.15 | 0.20±0.15 | 0.17±0.06 | N/A |

or data set characteristics after the re-sampling process.

## 7.3 Data analysis approach

The data analysis approach followed the general structure of cohorts differentiation analysis described in Section 6.3, including the following steps:

1. Data set re-sampling (using USM and SMOTE strategies) described in detail in Sections 6.2.2 and 7.2.2.

2. Feature selection using three alternative methods, including RANK (see Section 6.3.1.1), mRMR (see Section 6.3.1.2) and LASSO (see Section 6.3.1.3).

3. Classification using LR (see Section 6.3.2.1) and SVM (see Section 6.3.2.2) classifiers. For classification LOOCV hold out strategy was used.

## 7.4 Analysis results

The analysis results of this chapter explore the differences in objective behaviour features of mental (emotional) states. The differences in objective state features are presented first, followed by the assessment of their classification accuracy.

### 7.4.1 Comparison between states

Objectively collected features were analysed to compare their distributions in three previously described clinically important mood states, including depression, euthymia and mania, with results presented in Table 7.3. The comparison was performed on a USM data subset.

The manic state was characterised by a significantly different distribution of activity levels as characterised by the parameters of fitted Gamma distribution. Also, almost all disorganisation cluster characteristics were significantly different for mania compared to euthymia state, including higher levels of MSE, decreased duration of continuous activity segments, and decreased DTW distance between daily activity patterns. Also, consistent with theoretical expectations, the manic state was characterised by decreased time in bed, sleep time and immobile time.

The depressed state was characterised (compared to euthymia state) by significantly decreased average levels of daily activity, decreased levels of MSE (only for the first two scales), increased time-to-wake after sleep onset, increased wake bout time and differences in non-parametric circadian rhythm characteristics, including the M10, RA, IS and IV.

As expected, the difference between the manic and depressed states was more pronounced compared to the difference between each of these states and euthymia.

Interestingly, the disorganisation cluster characteristics were more pronounced in differentiating between the euthymia and manic state, while sleep and circadian rhythm characteristics were more often significant between the euthymia and depressed states. Psychomotor characteristics were approximately equally relevant for both mania and de-pression.

**Table 7.3:** *State differences between objective features of mental health. Superscripts indicate that the distributions are different with $p < 0.05$ according to Wilcoxon rank sum test. EU, DEP and MAN superscripts refer to euthymia, depression and mania.*

| | **Euthymia** | **Mania** | **Depression** |
|---|---|---|---|
| | | **Psychomotor dimension** | |
| LEVEL_AVG | 0.03±0.01 DEP | 0.03±0.01 DEP | 0.02±0.01 EU MAN |
| LEVEL_GAMMA_K | 0.18±0.06 MAN | 0.23±0.06 EU DEP | 0.17±0.04 MAN |
| LEVEL_GAMMA_THETA | 0.18±0.11 MAN DEP | 0.12±0.02 EU | 0.12±0.03 EU |
| TIME_MODERATE | 116.05±33.85 DEP | 127.36±24.64 DEP | 92.21±38.94 EU MAN |
| TIME_VIGOROUS | 8.17±10.55 MAN DEP | 3.20±1.69 EU | 2.43±2.48 EU |
| | | **Disorganisation dimension** | |
| LEVEL_MSE_1 | 0.08±0.04 MAN DEP | 0.11±0.01 EU DEP | 0.04±0.06 EU MAN |
| LEVEL_MSE_2 | 0.11±0.03 MAN DEP | 0.13±0.02 EU DEP | 0.09±0.03 EU MAN |
| LEVEL_MSE_3 | 0.12±0.03 MAN | 0.15±0.02 EU DEP | 0.10±0.03 MAN |
| LEVEL_MSE_4 | 0.13±0.03 MAN | 0.16±0.02 EU DEP | 0.11±0.03 MAN |
| LEVEL_MSE_5 | 0.13±0.03 MAN | 0.17±0.03 EU DEP | 0.12±0.04 MAN |
| DUR_AVG | 1.06±0.15 MAN | 0.93±0.10 EU | 1.10±0.22 |
| DUR_GAMMA_K | 0.72±0.06 MAN | 0.77±0.04 EU DEP | 0.69±0.06 MAN |
| DUR_GAMMA_THETA | 1.52±0.35 MAN | 1.23±0.19 EU DEP | 1.66±0.48 MAN |
| DUR_DFA | 0.62±0.04 | 0.62±0.04 | 0.63±0.04 |
| LEVEL_DTW | 171.59±16.28 MAN DEP | 158.39±16.83 EU DEP | 182.03±15.25 EU MAN |
| | | **Mood dimension (sleep)** | |
| IN_BED_TIME | -69.65±78.76 | -29.75±61.19 | -76.15±63.02 |
| OUT_OF_BED_TIME | 478.20±63.19 | 461.58±77.35 | 473.25±129.62 |
| TIME_IN_BED | 548.10±55.11 MAN | 491.57±52.04 EU DEP | 549.65±95.73 MAN |
| SLEEP_ONSET_LATENCY | 26.19±17.18 | 21.32±11.13 | 29.35±25.38 |
| WAKE_AFTER_SLEEP_ONSET | 54.37±19.21 DEP | 52.00±14.83 DEP | 72.87±24.62 EU MAN |
| ACTUAL_SLEEP_TIME | 467.29±51.08 MAN | 418.00±43.43 EU | 447.17±102.79 |
| SLEEP_EFFICIENCY | 85.57±4.54 DEP | 85.36±3.50 | 80.77±8.08 EU |
| NUMBER_OF_WAKE_BOUTS | 54.70±13.04 | 54.74±13.55 | 56.86±16.75 |
| MEAN_WAKE_BOUT_TIME | 0.99±0.28 DEP | 0.95±0.24 DEP | 1.28±0.25 EU MAN |
| MEAN_SLEEP_BOUT_TIME | 9.42±2.35 | 8.57±2.69 | 8.96±3.13 |
| MOBILE_TIME | 221.14±41.24 | 213.89±45.95 | 230.55±50.68 |
| IMMOBILE_TIME | 300.78±43.50 MAN | 256.36±33.50 EU DEP | 289.75±86.68 MAN |
| TAU | 1.00±0.02 | 1.50±1.58 | 1.00±0.01 |
| PHI | 204.28±65.87 | 215.91±64.70 | 198.55±89.98 |
| L5 | 3.46±2.14 | 3.11±1.07 | 3.67±1.41 |
| M10 | 118.95±34.43 DEP | 122.58±24.53 DEP | 90.81±32.21 EU MAN |
| RA | 0.94±0.03 DEP | 0.95±0.02 DEP | 0.91±0.04 EU MAN |
| IS | 0.01±0.01 DEP | 0.01±0.01 DEP | 0.02±0.01 EU MAN |
| IV | 0.74±0.15 DEP | 0.77±0.13 | 0.84±0.08 EU |

### 7.4.2 Importance of objective features

The importance of objective features for identification of mood states was assessed using three feature selection algorithms, including the simple ranking of features according to their relevance to target using the p-value of Wilcoxon rank-sum (RANK), mRMR criteria and LASSO, with results presented in Table 7.4.

Similarly to the difference between disorders, the top rated features for separation between disorders were observed mostly in the mood (sleep) cluster. Features selection methods that take into account complementarity of features tend to select subsets of features belonging to different clusters, including psychomotor, disorganisation and mood.

### 7.4.3 Accuracy of separation between mood states

The accuracy of separation between the mood states is presented in Table 7.5. If certain feature selection and classification strategies demonstrated superior performance compared to others, they were selected in bold.

As expected from the cohort differences presented in Table 7.3 and theoretical considerations, the best results in separation between mood states (using the USM re-sampling strategy) were achieved for separation between mania and depression (90% accuracy with 90% specificity and 90% sensitivity). The LASSO feature selection method demonstrated the best results (with SVM classification), however all other methods have shown higher performance compared to other classification experiments.

Interestingly, for SMOTE oversampling, all feature selection algorithms and classification methods allowed the achievment of perfect separation between mania and depression.

The best classification accuracy of 85% with 100% specificity and 70% sensitivity for separation between euthymia and depression was achieved using LASSO feature selection and SVM classification (USM re-sampling).

**Table 7.4:** *Ranking of features for separation between mood states using the Wilcoxon rank-sum test criteria, mRMR and LASSO, with the top five features in the ranking selected in **bold font**.*

| | Euthymia vs. mania | | | Euthymia vs. depression | | | Mania vs. depression | | |
|---|---|---|---|---|---|---|---|---|---|
| | RANK | mRMR | LASSO | RANK | mRMR | LASSO | RANK | mRMR | LASSO |
| | | | | **Psychomotor dimension** | | | | | |
| LEVEL_AVG | 24 | **4** | 23 | 32 | 10 | 14 | 16 | 13 | 17 |
| LEVEL_GAMMA_K | 25 | 27 | 15 | 15 | 18 | 18 | 25 | 17 | 6 |
| LEVEL_GAMMA_THETA | 31 | 8 | 16 | 17 | 7 | 19 | 29 | **5** | 14 |
| TIME_MODERATE | 18 | 26 | 18 | 33 | 32 | 23 | 21 | 30 | 18 |
| TIME_VIGOROUS | 32 | **5** | 24 | 13 | **5** | 24 | 24 | **4** | 26 |
| | | | | **Disorganisation dimension** | | | | | |
| LEVEL_MSE_1 | 29 | 31 | 19 | 22 | 20 | 7 | 8 | 7 | 27 |
| LEVEL_MSE_2 | 23 | 33 | 9 | 20 | 6 | 25 | 11 | 32 | 7 |
| LEVEL_MSE_3 | 21 | 16 | 25 | 27 | 14 | 26 | 12 | 19 | 28 |
| LEVEL_MSE_4 | 16 | 22 | 26 | 31 | 12 | 27 | 13 | 24 | 29 |
| LEVEL_MSE_5 | 14 | 30 | 10 | 34 | 9 | 20 | 17 | 33 | 30 |
| DUR_AVG | 5 | 21 | 12 | 24 | 34 | 28 | 23 | 16 | 8 |
| DUR_GAMMA_K | 15 | 29 | 17 | 10 | 28 | **4** | 30 | 34 | 15 |
| DUR_GAMMA_THETA | 8 | 7 | 27 | 19 | 17 | 29 | 34 | 23 | 31 |
| DUR_DFA | 17 | 25 | 20 | 11 | 21 | **5** | 9 | 20 | 9 |
| LEVEL_DTW | 10 | 15 | 13 | 12 | 8 | 10 | 19 | 31 | 11 |
| | | | | **Mood dimension (sleep)** | | | | | |
| IN_BED_TIME | 7 | 11 | 28 | **3** | **3** | **2** | 6 | **3** | **1** |
| OUT_OF_BED_TIME | **2** | **1** | **1** | 7 | 29 | 30 | 14 | 25 | 21 |
| TIME_IN_BED | 13 | 13 | 8 | 25 | 25 | 11 | 31 | 21 | 22 |
| SLEEP_ONSET_LATENCY | 33 | 20 | 14 | 28 | 23 | 6 | 20 | 11 | 12 |
| WAKE_AFTER_SLEEP_ONSET | 27 | 28 | 21 | **2** | 16 | 21 | **3** | 8 | 32 |
| ACTUAL_SLEEP_TIME | 9 | 10 | **4** | 21 | 26 | 31 | 26 | 28 | 33 |
| SLEEP_EFFICIENCY | 22 | **3** | 22 | 8 | **4** | 8 | 27 | 14 | 19 |
| NUMBER_OF_WAKE_BOUTS | 11 | 9 | 29 | 18 | 30 | 15 | 18 | 6 | 23 |
| MEAN_WAKE_BOUT_TIME | **1** | 6 | **2** | **1** | **1** | **1** | **1** | **1** | **2** |
| MEAN_SLEEP_BOUT_TIME | 19 | 32 | 30 | 29 | 27 | 32 | 15 | 22 | 24 |
| MOBILE_TIME | 12 | 19 | **5** | 23 | 19 | 33 | 28 | 12 | **5** |
| IMMOBILE_TIME | 30 | 14 | 31 | 30 | 22 | 12 | **5** | 10 | 13 |
| TAU | 6 | **2** | 6 | 9 | **2** | **3** | 22 | **2** | **3** |
| PHI | **4** | 18 | 32 | 6 | 24 | 34 | **4** | 9 | 34 |
| L5 | 34 | 34 | 33 | 16 | 11 | 16 | 10 | 27 | 20 |
| M10 | **3** | 12 | **3** | 26 | 31 | 22 | 32 | 26 | 10 |
| RA | 26 | 23 | 34 | 14 | 33 | 13 | 33 | 18 | 25 |
| IS | 28 | 24 | 7 | **4** | 13 | 17 | **2** | 15 | **4** |
| IV | 20 | 17 | 11 | **5** | 15 | 9 | 7 | 29 | 16 |

**Table 7.5:** *Summary of LOOCV classification accuracy for separation between AMoSS participants' mood states using different data set balancing, feature selection and classification strategies, with best results selected in* ***bold font***.

| | Data set balancing strategy | | | | | |
|---|---|---|---|---|---|---|
| | USM | | | SMOTE | | |
| | Feature selection | | | | | |
| | RANK | mRMR | LASSO | RANK | mRMR | LASSO |
| **Euthymia vs. mania** | | | | | | |
| Logistic regression | | | | | | |
| Accuracy | **0.80** | 0.50 | 0.75 | 0.82 | 0.75 | 0.83 |
| Specificity | **0.80** | 0.60 | 0.80 | 0.73 | 0.76 | 0.74 |
| Sensitivity | **0.80** | 0.40 | 0.70 | 0.91 | 0.73 | 0.93 |
| SVM | | | | | | |
| Accuracy | 0.70 | 0.60 | 0.75 | 0.82 | **0.87** | 0.83 |
| Specificity | 0.90 | 0.20 | 0.80 | 0.68 | **0.73** | 0.72 |
| Sensitivity | 0.50 | 1.00 | 0.70 | 0.95 | **0.98** | 0.94 |
| **Euthymia vs. depression** | | | | | | |
| Logistic regression | | | | | | |
| Accuracy | 0.80 | 0.70 | 0.80 | 0.82 | 0.85 | 0.82 |
| Specificity | 0.80 | 0.70 | 0.90 | 0.85 | 0.89 | 0.84 |
| Sensitivity | 0.80 | 0.70 | 0.70 | 0.78 | 0.82 | 0.80 |
| SVM | | | | | | |
| Accuracy | 0.75 | 0.60 | **0.85** | 0.84 | **0.90** | 0.86 |
| Specificity | 0.70 | 0.40 | **1.00** | 0.87 | **0.97** | 0.94 |
| Sensitivity | 0.80 | 0.80 | **0.70** | 0.79 | **0.81** | 0.75 |
| **Mania vs. depression** | | | | | | |
| Logistic regression | | | | | | |
| Accuracy | 0.85 | 0.80 | 0.85 | **1.00** | **1.00** | **1.00** |
| Specificity | 0.80 | 0.90 | 0.80 | **1.00** | **1.00** | **1.00** |
| Sensitivity | 0.90 | 0.70 | 0.90 | **1.00** | **1.00** | **1.00** |
| SVM | | | | | | |
| Accuracy | 0.85 | 0.80 | **0.90** | **1.00** | **1.00** | **1.00** |
| Specificity | 0.80 | 0.80 | **0.90** | **1.00** | **1.00** | **1.00** |
| Sensitivity | 0.90 | 0.80 | **0.90** | **1.00** | **1.00** | **1.00** |

The best classification accuracy of 80% with 80% specificity and sensitivity for separation between euthymia and mania was achieved using simple RANK feature selection and logistic regression classification.

The results demonstrated for SMOTE re-sampling consistently superseded the results for USM.

## 7.5 Discussion and conclusions

The analysis of this chapter investigated the differences between objective characteristics of activity in different clinically relevant mood states and the possibility to identify these mood states using only data collected from passive activity sensors using classification techniques.

The first important result is that the classification accuracy demonstrated in identification of clinical states substantially (by approximately 10%) exceeds the classification accuracy achieved in separation between different disorders, presented in Chapter 6. This agrees with the expectation of increased symptom manifestation during clinical episodes, while indicating that the clinical states of patients should be carefully considered when differentiating between mental disorders using short-term recordings. Secondly, the theoretical expectation of observing higher differences between mania and depression rather than any of these states and euthymia was also confirmed in practice.

It is interesting to note that while mood (sleep) characteristics are important features for classification in all cases, the manic state appears to demonstrate elevated characteristics from the disorganisation cluster of symptoms and the depression state has more pronounced changes in sleep characteristics, while psychomotor cluster symptoms are equally relevant in both cases.

These results justify the selected dimensional approach to characterising mental health and indicate the need and importance of further research in this area. Using appropriate change point detection techniques on collected objective behaviour data could potentially allow for early identification of behaviour abnormalities related to changes in clinical state.

The main limitation of presented research is in limited number of clinical episodes captured with GENEActiv recordings. This, together with the fact that the number of subjects were comparable with the number of captured episodes, led the choice of the analysis using binary classification as opposed to regression between the mood scores and values of behaviour features. Such regression however, could be attempted once more clinical mood episodes are captured.

# Chapter 8

# Personalised mood modelling

## 8.1 Diversity of individual symptoms

As was demonstrated in Chapter 7, different mood states can be distinguished from each other based on behavioural characteristics, calculated from physical activity time series. But a universal description of a mental state in terms of these behaviour characteristics is hardly possible due to the diversity of possible symptom combinations and behaviour impairments. For example, the borderline personality disorder patients are known to rate their mood and behaviour as very depressed, however no sign of depression may be obvious to an external observer (Westen et al., 1992). This makes the task of creating a universally valid association between behaviour and mood almost impossible.

While the symptoms that characterise mental illnesses are well defined, the exact patterns of symptom combinations may vary from person to person, as described by the DSM-5 diagnostic criteria (American Psychiatric Association, 2013). More important is the fact that prodrome and early warning signs of mental disorder episodes may also vary quite significantly between different patients (Molnar et al., 1988; Keitner et al., 1996; Perry et al., 1999; Birchwood, 2000; Klosterkötter et al., 2001; Jackson et al., 2003; Berk et al., 2007; Goossens et al., 2010; Sierra et al., 2009).

Early research by Molnar *et al.* (Molnar et al., 1988) on 20 bipolar patients con-

cluded that the duration and symptoms of manic and depressive prodrome vary a lot between individuals, but are consistent for the same individual and the same episode type. A more extensive study by Keitner *et al.* (Keitner et al., 1996) on 74 bipolar patients identified that 78% of patients reported prodromal depressive symptoms and 87% reported prodromal manic symptoms (while according to family reports these numbers were even higher). Cognitive, mood, behavioural, neurovegatative and social symptoms were reported in different combinations.

The systematic review of manic and depressive prodromes by Jackson *et al.* (Jackson et al., 2003) identified that at least 80% of individuals with mood disorder can detect symptoms of prodrome. The most important early symptom of mania was sleep disturbance and early symptoms of depression were inconsistent. Importantly, the randomised control trial on efficacy of teaching patients to recognise early symptoms of relapse performed by Perry *et al.* (Perry et al., 1999) identified that such teaching was associated with important clinical improvements, in particular at the time of first manic relapse.

All these facts indicate the importance of creating personalised models of mood deterioration for each individual specifically, that would account for inter-individual differences in prodrome symptomatology. This problem can be solved by identifying the relationship between objective behaviours and mood for each subject individually and building personalised regression models to reconstruct the respective scoring system (such as QIDS-16SR, ASRM or GAD-7) using collected objective activity data.

In this chapter regression models have been constructed to identify the relationship between objective activity features and reported subjective mood state. If accurate, such models may allow the meaningful tracking of patients and identification of early warning signs of deterioration.

## 8.2 Data subset selection

### 8.2.1 Selection of participants

Unfortunately the GENEActiv data collected in the AMoSS study cannot be used to build a personalised regression model due to the limited number of weekly observations per subject. Despite the fact that correlation between GENEActiv and mobile phone collected data is limited, mostly because activity during sleep cannot be captured with a mobile phone, longitudinal changes in day time behaviours can be inferred from mobile phone data and personalised regression models can be built using data from participants' smart phones, where data acquisition was continuous for a period of up to one year.

From all collected data the individuals were selected who captured at least 12 mood points (or three months), where data including phone-derived physical activity and self-reported mood was captured simultaneously. Additionally, the only subjects who were retained in the data set were those who had at least two distinct mood states during the recording, according to thresholds defined earlier in Section 7.2.1. The resulting data subset is described in Table 8.1 and distributions of self-reported mood scores presented in Figure 8.1.

**Table 8.1:** *Demographics and data set characteristics of the AMoSS personalised mood regression data subset. For each mood questionnaire the range is represented as the difference between maximum and minimum scale values for each participant.*

| | Healthy controls | Bipolar disorder | Borderline personality disorder | Total |
|---|---|---|---|---|
| | | **Demographic characteristics** | | |
| Number of participants | 7 | 24 | 12 | 43 |
| Gender | 5 females | 15 females | 11 females | 31 females |
| Phone (Trousers, Jacket, Handbag) | 3,0,4 | 13,3,8 | 4,1,7 | 20,4,19 |
| Age (Mean±SD) | 25.57±2.88 | 36.46±9.71 | 32.00±9.26 | 33.44±9.56 |
| Height (Mean±SD) | 171.14±6.04 | 171.04±9.64 | 166.67±8.28 | 169.84±8.83 |
| Weight (Mean±SD) | 62.00±7.16 | 80.81±16.39 | 80.75±18.26 | 77.73±17.06 |
| BMI (Mean±SD) | 21.24±2.87 | 27.50±4.61 | 29.20±6.94 | 26.95±5.70 |
| | | **Data set characteristics** | | |
| Mood points (Mean±SD) | 49.14±10.32 | 51.08±47.75 | 37.25±19.71 | 46.91±37.46 |
| ASRM (Mean±SD) | 8.57±3.82 | 9.67±5.06 | 8.92±4.27 | 9.28±4.59 |
| QIDS-16SR (Mean±SD) | 10.57±4.12 | 14.33±4.67 | 14.25±4.00 | 13.70±4.53 |
| GAD-7 (Mean±SD) | 10.43±6.08 | 13.42±3.86 | 11.58±2.71 | 12.42±4.10 |

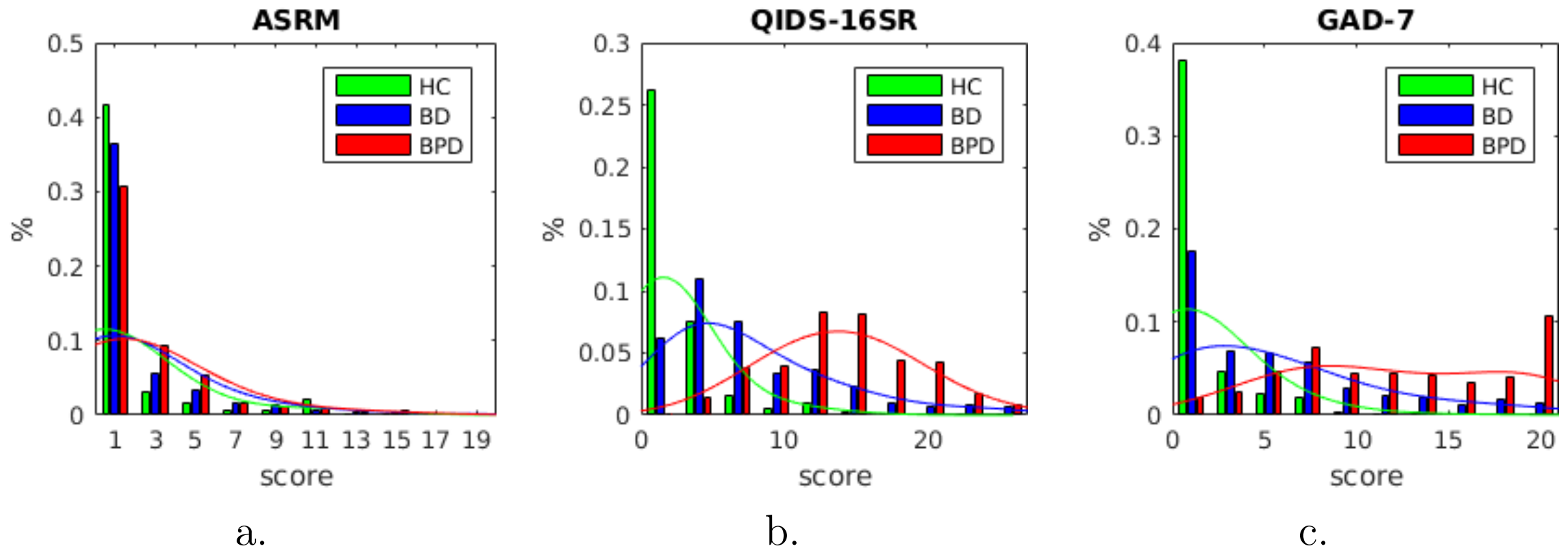

**Figure 8.1:** *Distribution of self-reported mood scores between participant groups. Figure a) shows the ASRM scores, b) shows the QIDS-16SR scores and c) shows the GAD-7 scores.*

For each collected mood score the preceding week of phone activity data was analysed to identify objective behaviour features most predictive to subjective mood scores and regression models were built to estimate severity of mood symptoms (represented by QIDS-16SR, ASRM and GAD-7 scores) from phone behaviour data.

Note, that due to the personalised nature of regression models, it was possible to analyse all clinical groups, including borderline personality disorder participants. As regression models were built for each subject individually, differences between cohorts were not relevant for the analysis.

### 8.2.2 Selection of features

Phone data are limited to day-time only collection, capturing activity during the time when the phone was located on the subject's body. These data do not allow a perfectly accurate reconstruction of activity levels, because phones are not constantly attached to the human body, and may be left at home or located in different pockets, personal carry bags, or left on tables periodically.

Because of these limitations it was not possible to calculate all activity features described in Sections 5.3.3, 5.4.4 and 5.5.6. Psychomotor parameters were excluded as there are no validated thresholds for moderate and vigorous activity derived for phone-based activity data. Sleep-related parameters were excluded as sleep analysis was not possible to perform using phone data, where the phone was not typically adjacent to a subjects' body during the sleep period. The excluded features are presented in Table 8.2 and the features used in further analysis are presented in Table 8.3.

**Table 8.2:** *Phone accelerometer-derived activity features, excluded from the analysis due to limitations of phone data acquisition.*

| **Feature** | **Description** |
|---|---|
| | **Psychomotor features** |
| TIME_MODERATE | Time spent in moderate-intensity activities (3-6 MET) |
| TIME_VIGOROUS | Time spent in vigorous activities (>6 MET) |
| | **Sleep (mood) features** |
| IN_BED_TIME | The time the subjects gets into bed |
| OUT_OF_BED_TIME | The time the subject got out of bed |
| TIME_IN_BED | Time in bed |
| SLEEP_ONSET_LATENCY | Time it took the subject to fall asleep |
| WAKE_AFTER_SLEEP_ONSET | Wake after sleep onset |
| ACTUAL_SLEEP_TIME | Actual sleep time |
| SLEEP_EFFICIENCY | 'Actual sleep time' divided by 'Time in bed' multiplied by 100 |
| NUMBER_OF_WAKE_BOUTS | Number of continuous blocks scored as WAKE |
| MEAN_WAKE_BOUT_TIME | Mean wake bout time |
| MEAN_SLEEP_BOUT_TIME | Mean sleep bout time |
| MOBILE_TIME | Total duration of epochs with activity (>0) during sleep |
| IMMOBILE_TIME | Total duration of epochs with no activity during sleep |
| TAU | Cosinor-derived $\tau$ |
| PHI | Sleep midpoint as offset to midnight in minutes |

## 8.3 Data analysis approach

Firstly, daily-calculated objective activity features were pre-processed using smoothing techniques to produce weekly summaries corresponding to mood self-report.

Secondly, regression models for identification of mood were built for each participant separately. Objective weekly activity measures (described in Table 8.3) were used as predictors and self-reported mood scores (total scores of ASRM, QIDS-16SR and GAD-7) as regression targets.

The quality of regression was evaluated using LOOCV strategy, where all but one data points were used to build regression model and the accuracy of regression was estimated

**Table 8.3:** *Phone accelerometer-derived activity features retained for further analysis of personalised mood-activity models.*

| Feature | Description |
|---|---|
| | **Psychomotor features** |
| LEVEL_AVG | Average level of ENMO activity |
| LEVEL_GAMMA_K | Shape parameter of Gamma distribution fit into ENMO activity |
| LEVEL_GAMMA_THETA | Scale parameter of Gamma distribution fit into ENMO activity |
| | **Disorganisation features** |
| LEVEL_MSE_{N} | $N^{th}$ scale of Multiscale Entropy ($N = 1..5$) |
| DUR_AVG | Average duration of BOCPD segments |
| DUR_GAMMA_K | Shape parameter of Gamma distribution fit into BOCPD length |
| DUR_GAMMA_THETA | Scale parameter of Gamma distribution fit into BOCPD length |
| DUR_DFA | DFA $\alpha$ (scaling exponent) of BOCPD length |
| LEVEL_DTW | Average day-to-day DTW distance of activity level patterns |
| | **Sleep (mood) features** |
| L5 | Level of activity during the least active 5 hours |
| M10 | Level of activity during the most active 10 hours |
| RA | Relative amplitude between L5 and M10 |

using the remaining one (repeating the process for each data point, for each subject).

### 8.3.1 Smoothing techniques

Data from phone sensors are the subject of quality concerns and behavioural artifacts as described in Chapter 4, making calculation of daily activity features potentially unreliable. To minimise the influence of errors of daily-calculated features on mood regression models a number of smoothing techniques were applied to the time series of activity features with the purpose of reducing the variance in feature values due to feature calculation error.

The objective activity features were computed for every day, subject to data availability. However, mood self-reports were collected once a week and described the subject's perception of his / her mental state and behaviour during during the previous week. Therefore the objective features for regression should aggregate information from seven days prior to questionnaire delivery and the smoothing techniques described in the follow-ing sections were parametrised by using a one week analysis window in order to provide a meaningful comparison with self-reported data.

As a baseline for comparison of smoothing techniques, the feature values for the last 24 hours before mood assessment were also used.

#### 8.3.1.1 Median filtering

A simple, but robust to outliers, method of smoothing was proposed by Tukey (Tukey, 1977) and uses a running median to calculate the filtered signal. In our case that of behaviour data with ground truth describing the preceding week, and possible weekly behaviour regularities, a fixed seven day window length for the filter was selected, with a smoothing equation defined as

$$x'_i = Median_{j=i-7}^{i}\{x_j\}, \tag{8.1}$$

where $x_i$ and $x'_i$ are the original and smoothed feature values respectively. The median filter, along with a number of improvements to enhance robustness and performance in specific applications is extensively used in image processing (Brownrigg, 1984) and also adopted in clinical time series data analysis (Davies et al., 2004).

#### 8.3.1.2 Exponential filtering

A well-known phenomena in subjectively recollected reports is the so-called recall bias (Firth and Torous, 2015), when more recent events are perceived as more important compared to older ones. This phenomena may affect self-reported mood scores and result in objectively identified activity features not offering the best explanation for biased self-reports, despite correctly representing the actual behaviour in the mathematical sense.

The class of filtering techniques that may help address this problem is called exponential smoothing (Hunter, 1986), where more recent samples are weighted higher compared to older samples, thus incorporating the potential recall bias into the model. Exponential smoothing is defined as

$$x'_{i+1} = \lambda x_i + (1 - \lambda) x'_i, \tag{8.2}$$

where $x_i$ and $x'_i$ are the original and smoothed measurements respectively and $\lambda$ is a smoothing factor, describing the relative contribution of current and past measurements and set to $2/(N+1)$ with $N$ (or filter length) equal to seven days, as for the median filter. Exponential smoothing is equivalent to an Autoregressive Integrated Moving Average (ARIMA) (0,1,1) model with no constant term.

Exponential smoothing has proven to be useful in physical activity time series analysis (Løvendahl and Chagunda, 2010). In addition to basic smoothing described above, versions to account for trend (double exponential smoothing) and seasonality (triple exponential smoothing) were proposed (Winters, 1960), but not used in this work due to increased complexity of the model and necessity for (possibly subject specific) prior information to infer model parameters.

#### 8.3.1.3 Gaussian Process filtering

The above filtering techniques assume signal stationarity, however it is well-known that human activity is highly non-stationary and may be self-correlated on different scales. For example, reduced night sleep may result in increased sleep pressure on the following day and earlier sleep onset with longer sleep duration on the following night (Birchler-Pedross et al., 2009). Similarly the external zeitgebers, such as weekly work-rest routine, are likely to influence daily physical activity (Fuster-Garcia et al., 2015).

A good filtering algorithm should incorporate these non-stationarities. A popular framework for signal modelling that allows dealing with various kinds of non-stationary signals in a principled way is offered by Gaussian Processes (GP). A GP is a collection of random variables, any finite number of which have a joint Gaussian distribution (Rasmussen and Williams, 2006). GP defines a family of functions fully specified by mean function $m(\boldsymbol{x})$ and covariance function $k(\boldsymbol{x}, \boldsymbol{x}')$ as

$$f(\boldsymbol{x}) = \mathcal{GP}(m(\boldsymbol{x}), k(\boldsymbol{x}, \boldsymbol{x}')),$$
$$m(\boldsymbol{x}) = \mathbb{E}[f(\boldsymbol{x})], \quad (8.3)$$
$$k(\boldsymbol{x}, \boldsymbol{x}') = \mathbb{E}[(f(\boldsymbol{x}) - m(\boldsymbol{x}))(f(\boldsymbol{x}') - m(\boldsymbol{x}'))].$$

The definition of mean and covariance functions allows the modelling of different kinds of non-stationarities, with the covariance function being the most interesting topic. In our case, the signal feature of interest was the dependency of current activity on activity on previous days. Similarly to the family of exponential smoothing techniques, many covariance functions were proposed to model different interactions with Gaussian Processes. However, if interactions are simple or not well defined, the squared exponential (SE) is most commonly used (Roberts et al., 2013). The SE covariance function is defined as

$$k_{SE}(x_i, x_j) = h^2 exp\left[ - \left( \frac{x_i - x_j}{\lambda} \right)^2 \right], \quad (8.4)$$

where $x_i$ and $x_j$ represent the time points of data collection, $h$ is an output-scale amplitude and $\lambda$ is an input scale (of time in our case).

### 8.3.2 Personalised mood regression

The smoothing techniques described in Section 8.3.1, applied to calculated daily activity features, allow the reduction of potential errors introduced by quality issues of phone-based activity data.

These activity features are hypothesised to be associated with mood state, however not all variables may add additional information to the model or even decrease the model's accuracy due to the curse of dimensionality. Different features may be important as mood predictors for different subjects.

Therefore it is essential to select the best features for regression, to characterise the individual subject. Regression techniques that allow selecting only subsets of variables that best describe relationships between predictors and dependent variables are known as regularised regression methods.

#### 8.3.2.1 Regularised regression

The Least Absolute Shrinkage and Selection Operator (LASSO) of Tibshirani (Tibshirani, 1996) extends the linear regression with the Ordinary Least Squares (OLS) method, proposed originally by Legendre (Legendre, 1805) and Gauss (Gauss, 1963), by adding L1 penalty by the number of non-zero regression coefficients.

In multiple linear regression the dependent variable is expressed as a linear combination of $p$ independent variables:

$$y_i = \beta_0 + \beta_1 x_{i1} + \beta_2 x_{i2} + ... + \beta_p x_{ip} + \epsilon_i, \tag{8.5}$$

where $x_{ij}$ are observations of $p$ independent variables, $y_i$ are dependent variables, $\beta_i$ are regression coefficients, $\epsilon_i$ is the observation error and $i$ indexes the particular observation.

The method of LASSO obtains parameter estimates $\beta$ by minimising the square of difference between the dependent variable and estimate, subject to L1 constraint

$$\boldsymbol{\beta} = \underset{\beta}{\operatorname{argmin}} \left( \frac{1}{2N} \sum_{i=1}^{N} (y'_i - y_i)^2 + \lambda \sum_{j=1}^{p} |\beta_j| \right), \tag{8.6}$$

where $\boldsymbol{\beta}$ is the $p$-length vector of regression parameter estimates, $y'_i$ and $y_i$ are the observed estimated dependent variable values respectively, and $N$ is the number of observations.

LASSO regression favours zero $\beta$ coefficients, thus essentially performing feature selection and eliminating non-relevant features from the model. In order to select the best parametrisation for regression in terms of $\lambda$ (and therefore the best features), LASSO uses a cross-validation approach, with 5-fold CV used in this work. Based on cross-validation results, the value of $\lambda$ is selected with the minimum MSE (see Figure 8.2). This allows selecting a parametrisation with the number of features providing the best accuracy.

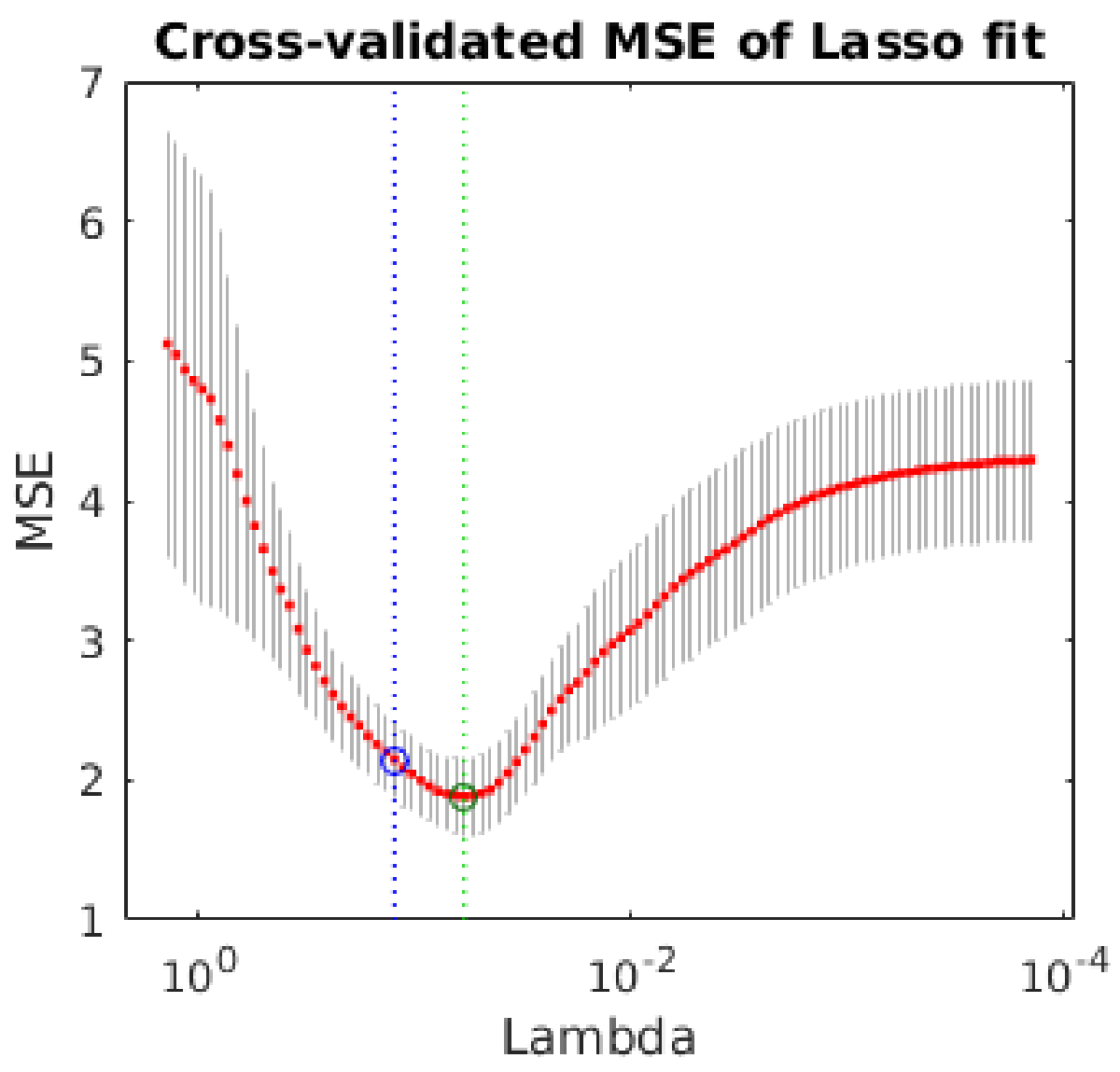

**Figure 8.2:** *Example of LASSO lambda parameter selection using the MSE of cross-validated regression. The green dotted line represents the parametrisation with minimal MSE and blue dotted line represents the parametrisation with the least number of features and MSE within one SD from the minimum.*

The regression procedure was performed with ASRM, QIDS-16SR and GAD-7 scores as target variables and objective activity features as predictors. The result of this procedure including the set of $\beta$ coefficients for predictors constitute the personalised mood prediction model.

#### 8.3.2.2 Regression quality estimation

LOOCV cross-validation procedure was applied to estimate the generalisation error of regression models, where for each subject the accuracy of regression was calculated by building a number of models on data sets, each excluding one data point, and estimating the model performance on this data point (for each data point).

The model performance was estimated using the Mean Squared Error (MSE)

$$MSE = \frac{1}{N}\sum_{i=1}^{N}(y'_i - y_i)^2, \tag{8.7}$$

and the Mean Absolute Error (MAE)

$$MAE = \frac{1}{N}\sum_{i=1}^{N}|y_i' - y_i|, \tag{8.8}$$

where $y_i'$ and $y_i$ are the observed estimated dependent variable values respectively, and $N$ is the number of observations.

The MAE estimate was chosen as it provides the estimate error in units of the original mood scale and is easily interpretable by clinical experts, for example the study by Tsanas *et al.* used MAE to estimate the quality of the model for the prediction of Parkinson's disorder symptom scoring based on voice analysis (Tsanas et al., 2010).

## 8.4 Analysis results

### 8.4.1 Smoothing techniques

To minimise the features calculation error, time series of daily objective activity behaviour features for all subjects were processed by three different methods of data smoothing. The example plot of smoothing results for a single activity feature of a single subject is presented in Figure 8.3.

To estimate the quality of smoothing, LASSO regression was performed using data produced by all three methods of smoothing and also the original data from last 24 hours before mood score was acquired. The MAE of regression was calculated using LOOCV approach for each subject, where N-1 data points were used to build regression model and the error was estimated on the remaining data point, with results presented in Table 8.4.

**Table 8.4:** *Comparison of time series smoothing techniques using the MAE of time series prediction across all participants. The MAE for each subject was calculated using LOOCV approach. No statistically significant differences were found between smoothing methods.*

| | **Last 24 hours** | **Median smoothing** | **Exponential smoothing** | **Gaussian Process smoothing** |
|---|---|---|---|---|
| ASRM (Mean±SD) | 2.13±1.63 | 2.14±1.74 | 2.05±1.45 | 2.23±1.98 |
| QIDS-16SR (Mean±SD) | 2.65±1.47 | 2.68±1.44 | 2.80±1.74 | 2.67±1.36 |
| GAD-7 (Mean±SD) | 2.51±1.31 | 2.53±1.39 | 2.61±1.46 | 2.55±1.42 |

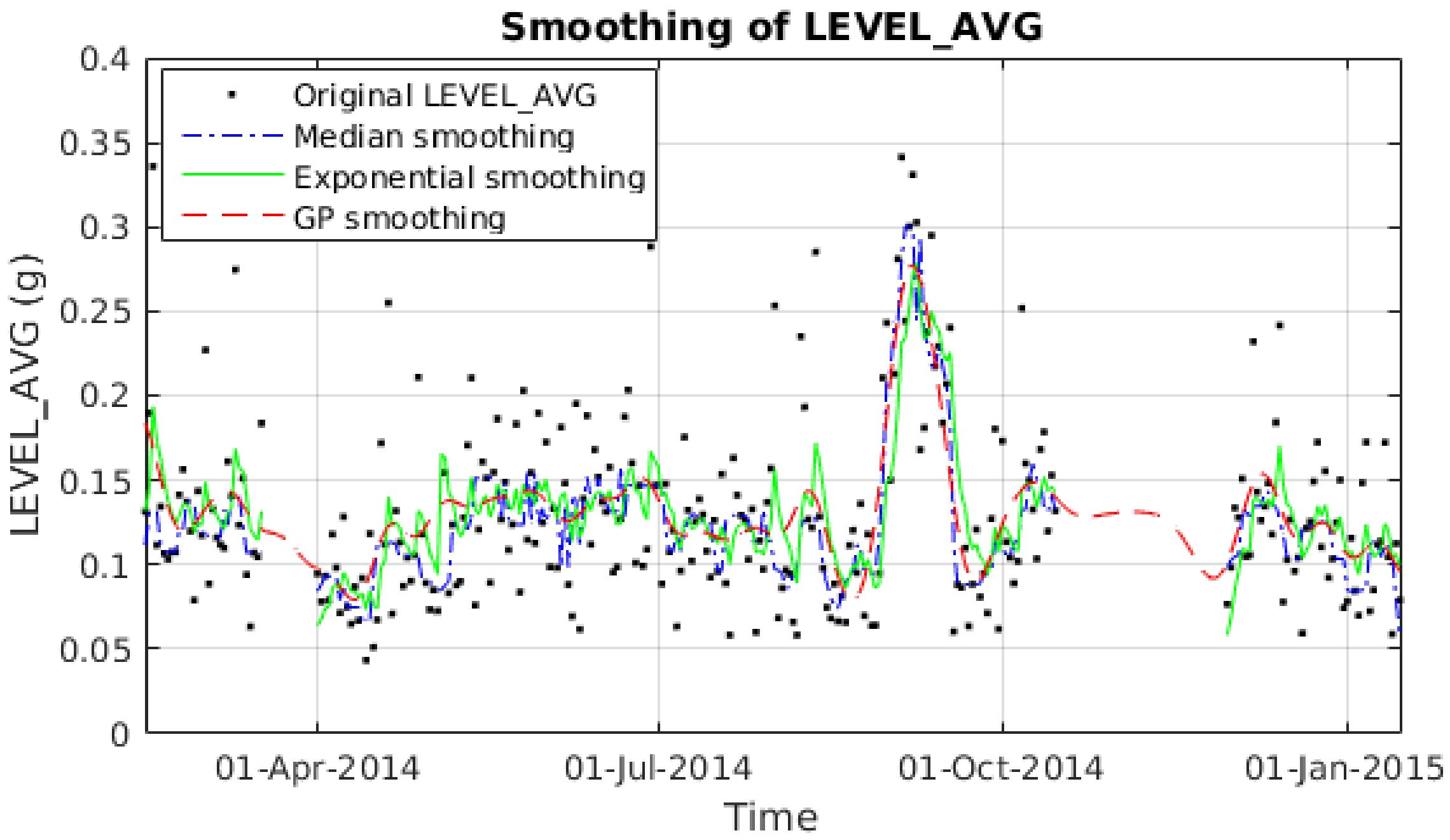

**Figure 8.3:** *Comparison of time series smoothing techniques for a single activity feature (LEVEL_AVG, defined in Section 5.3.1) of a single subject. Note, that in the case of LEVEL_AVG smoothing of daily average activity is equivalent to smoothing of the original activity signal.*

It can be seen that no single method is superior comparing to all others, with the simple last 24 hours value providing slightly better regression results for QIDS-16SR and GAD-7 models, and exponential smoothing providing better results for ASRM models. The GP smoothing provides a similar fit to the other methods, but also has the attractive property of being able to extrapolate the measurements and fill in missing data. Due to simplicity of the model and similarity of results, the last 24 hours activity method was selected for reporting the regression results.

### 8.4.2 Personalised regression models

For each subject and each questionnaire (ASRM, QIDS-16SR and GAD-7) regression models were identified as described in Section 8.3.2.1 and using the objective activity features acquired in the last 24 hours before mood score was collected (see example results at Figure 8.4).

Across all subjects and using the features of the last 24 hours the regression MAE was 2.13±1.63 for ASRM, 2.65±1.47 for QIDS-16SR and 2.51±1.31 for GAD-7 questionnaires as described in Table 8.4.

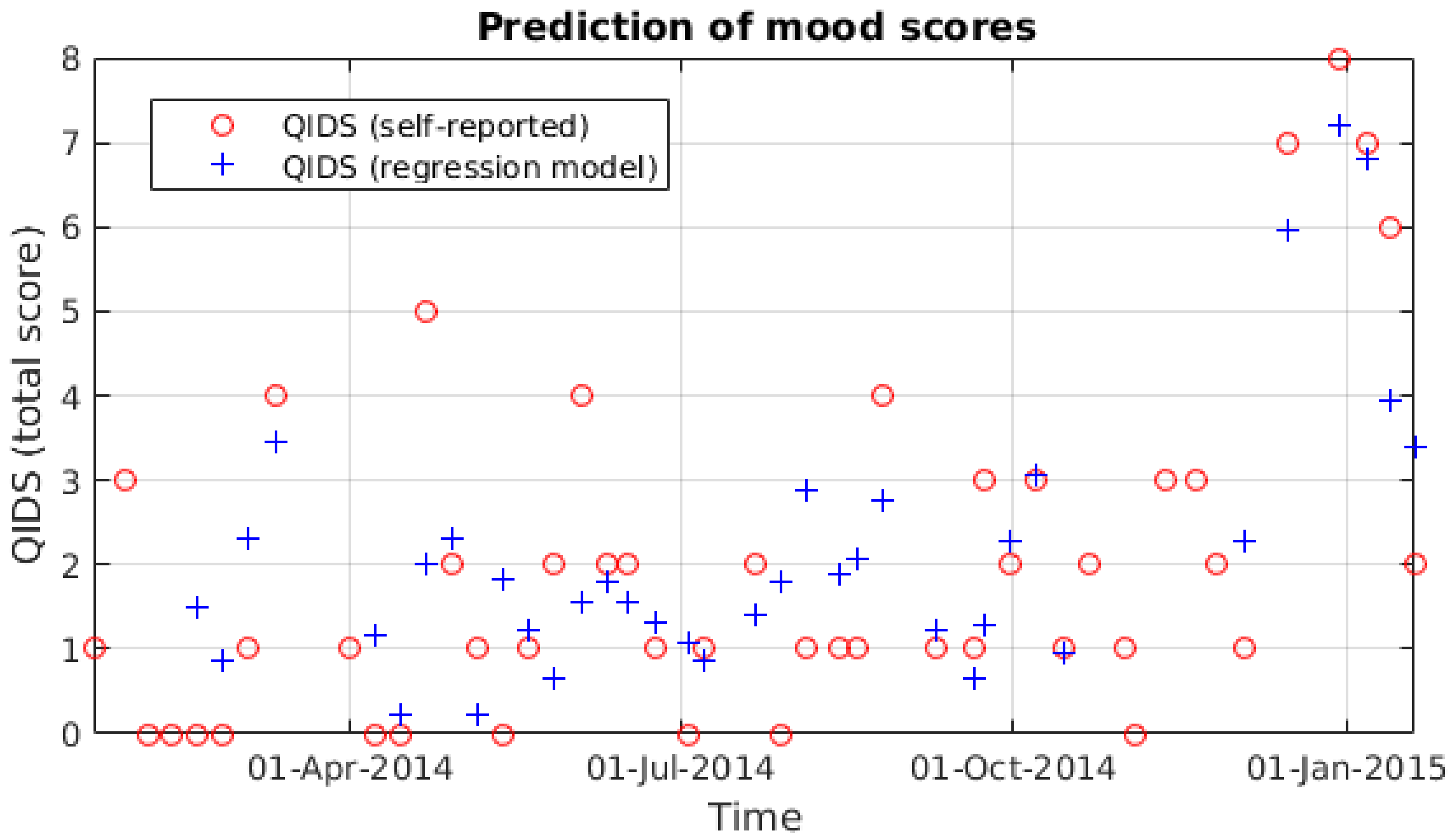

**Figure 8.4:** *Example of results of mood regression model for a single subject. Regression was performed for all self-reported mood scores and smoothed activity features of a single subject.*

Between different study groups the accuracy of prediction models was similar, with healthy controls demonstrating the lowest MAE for all mood characteristics as presented in Table 8.5, and significantly lower MAE comparing to bipolar disorder and borderline personality disorder groups for QIDS-16SR scores.

Pearson's correlation between the self-reported and estimated mood scores was on average between 0.27 and 0.54 for all questionnaires and participants groups with the highest values of 0.54±0.35 for QIDS-16SR of bipolar disorder participants and the lowest values of 0.27±0.33 for ASRM of healthy controls, with no significant difference between groups.

The number of features providing the best LASSO regression performance for personalised mood models was also not significantly different between the groups using the Wilcoxon rank sum test. The average number of features was between 1.6 and 2.97 for all questionnaires and participants groups, with the highest values of 2.97±3.24 for GAD-7 of healthy controls and the lowest values of 1.60±2.12 for ASRM of healthy controls.

**Table 8.5:** *Results of mood score prediction across all participant groups and mood questionnaires. MAE represents the prediction error, r is a Pearson's correlation coefficient between the target and predicted time series and N is the number of features used for the model. Superscripts indicate that the distributions are different with $p < 0.05$ according to the Wilcoxon rank sum test. BD, BPD and HC superscripts refer to bipolar disorder, borderline personality disorder and healthy controls, so a BD superscript in the healthy control column indicates that the parameter was significantly different between healthy controls and bipolar patients only.*

| | **Healthy controls** | **Bipolar disorder** | **Borderline personality disorder** |
|---|---|---|---|
| ASRM | | | |
| MAE (Mean±SD) | 1.38±0.85 | 2.16±1.87 | 2.49±1.40 |
| MSE (Mean±SD) | 3.72±3.41 | 10.71±19.79 | 11.63±11.97 |
| r (Mean±SD) | 0.27±0.33 | 0.39±0.29 | 0.35±0.29 |
| N (Mean±SD) | 1.60±2.12 | 2.27±2.15 | 2.37±2.63 |
| QIDS-16SR | | | |
| MAE (Mean±SD) | 1.36±0.38 $^{BD\ BPD}$ | 2.68±1.44 $^{HC}$ | 3.32±1.50 $^{HC}$ |
| MSE (Mean±SD) | 2.82±1.22 $^{BD\ BPD}$ | 13.62±14.01 $^{HC}$ | 19.92±16.42 $^{HC}$ |
| r (Mean±SD) | 0.36±0.29 | 0.54±0.35 | 0.40±0.39 |
| N (Mean±SD) | 2.87±2.79 | 2.75±2.76 | 2.47±2.37 |
| GAD-7 | | | |
| MAE (Mean±SD) | 1.77±1.43 | 2.72±1.39 | 2.52±0.97 |
| MSE (Mean±SD) | 9.14±16.54 $^{BPD}$ | 13.46±13.51 | 10.54±7.06 $^{HC}$ |
| r (Mean±SD) | 0.39±0.29 | 0.39±0.29 | 0.31±0.34 |
| N (Mean±SD) | 2.97±3.24 | 2.74±2.48 | 1.94±1.94 |

The objective activity features most often selected as predictors in personalised regression models are presented in Table 8.6. For ASRM models the feature most often selected for the models was the difference between activity of successive days measured using the Dynamic Time Warping (LEVEL_DTW). For QIDS-16SR and GAD-7 models the shape parameter of activity duration distribution (DUR_GAMMA_K) was the most prevalent feature.

To estimate the possible influence of phone usage styles on the regression results, MAE of regression was compared between users keeping phone in trousers pocket, jacket pocket and handbag, regardless of grouping into disorder categories. The only significant difference in MAE was found for GAD-7 questionnaire between trousers pocket and handbag (see Table 8.7).

**Table 8.6:** *Prevalence of features selected by LASSO regression as the percentage of personalised regression models where each feature was used. The most prevalent features are selected in* ***bold text****.*

| | ASRM (%) | QIDS-16SR (%) | GAD-7 (%) |
|---|---|---|---|
| LEVEL_AVG | 13.95 | 18.60 | 6.98 |
| LEVEL_GAMMA_K | 23.26 | 23.26 | 16.28 |
| LEVEL_GAMMA_THETA | 20.93 | 23.26 | 16.28 |
| LEVEL_MSE_1 | 20.93 | 23.26 | 9.30 |
| LEVEL_MSE_2 | 6.98 | 11.63 | 11.63 |
| LEVEL_MSE_3 | 11.63 | 16.28 | 4.65 |
| LEVEL_MSE_4 | 9.30 | 18.60 | 4.65 |
| LEVEL_MSE_5 | 11.63 | 16.28 | 13.95 |
| DUR_AVG | 9.30 | 9.30 | 6.98 |
| DUR_GAMMA_K | 18.60 | **37.21** | **27.91** |
| DUR_GAMMA_THETA | 6.98 | 20.93 | 2.33 |
| DUR_DFA | 16.28 | 32.56 | 16.28 |
| LEVEL_DTW | **27.91** | 30.23 | 16.28 |
| L5 | 25.58 | 20.93 | 20.93 |
| M10 | 9.30 | 9.30 | 6.98 |
| RA | 6.98 | 20.93 | 6.98 |

**Table 8.7:** *Comparison of mood score prediction accuracy depending on the phone usage pattern. Superscripts indicate that the distributions are different with* $p < 0.05$ *according to the Wilcoxon rank sum test. T, J and H superscripts refer to trousers pocket, jacket pocket and handbag respectively.*

| | Trousers pocket | Jacket pocket | Handbag |
|---|---|---|---|
| ASRM (Mean±SD) | 2.25±1.53 | 1.76±1.09 | 2.07±1.86 |
| QIDS-16SR (Mean±SD) | 3.02±1.64 | 1.79±0.48 | 2.44±1.34 |
| GAD-7 (Mean±SD) | 2.92±1.34 $^{H}$ | 3.07±1.69 | 1.97±1.04 $^{T}$ |

## 8.5 Discussion and conclusions

As was discussed in Section 6.1, the reliability of diagnosis in psychiatry can be quite low and for most diagnoses moderate-to-good inter-rater agreement between different experts was found (Regier et al., 2013). The correlation between different psychiatric scales depends on the type of scale, for ASRM and MRS r=0.72 (Altman et al., 1997) and for QIDS-16SR and HAM-D24 r=0.84.

Therefore, despite the average correlation between predicted values of psychiatric questionnaires and self-reported scores was in the range of 0.27 and 0.54, it presents an important and promising result, indicating the possibility for identification of more powerful features, given that the most predictive features of mood, such as sleep characteristics (as identified in Chapter 7), were not included in mood prediction models of this chapter. The MAE of scores estimation was in the range of 1.36 to 3.32 points, this corresponds to the ranges reserved in psychiatric questionnaires for a unique identifiable mood state (4-5 points) (as described in Table 3.6), and indicating that clinically relevant results are possible to achieve.

Given that for the analysis of this chapter many of important features, in particular related to sleep, were excluded due to the limitations of data acquisition using mobile phones, improvements in results may be produced by improvements in data acquisition techniques, such as using wearable technology, once the compliance issues described in Section 4.5.3 are resolved.

An interesting observation from the comparison of smoothing techniques is that using the value of objective behaviour features in the last 24 hours directly preceding the collection of subjective behaviour evaluation provides slightly more accurate (although not statistically significant) regression models. This phenomena was expected because weekly self-reports may tend to over-emphasise more recent feelings from the last 24 hours or less (Baranowski, 1988; Stone and Shiffman, 2002; Bowling, 2005; Saunders et al., 2015), and indicates the potential inaccuracies of evaluating symptoms using subjectively assessed mood, which can be prone to artifacts related to mood sampling timing.

Under such circumstances the objective measures of behaviour may provide a more precise (comparing to questionnaires) tool for assessment of symptomatic behaviours and early detection of clinical episodes of disorder, as well as more accurate estimation for the episode duration, important for both diagnostic and treatment purposes.

# Chapter 9

# Fusing activity and physiological data to improve classification

## 9.1 Introduction

As was shown in Section 2.4, physiological signals may be helpful in assessing mental health symptoms that are associated with autonomous nervous system activation, specifically in the anxiety dimension but potentially also including reality distortion, mood and other domains, where symptoms include strong subjective experiences. Electrodermal activity (EDA) and heart rate (HR) were found to be useful in studies of mental disorders (Carney et al., 2001; McCraty et al., 2001; Agelink et al., 2002; Cohen et al., 2003), since HRV has been shown to be a good proxy for the state of the autonomous nervous system (Bär et al., 2008) and EDA is known to be linked with parasympathetic nervous system activation (Boucsein, 2012).

The HR signal has attracted increasing attention from the psychophysiological research community due to the ability to characterise both branches of ANS and the availability of standardised analysis techniques, where variability analysis plays the most significant role (Rechlin et al., 1994; Friedman and Thayer, 1998; Agelink et al., 2002; Cohen et al., 2003; Valkonen-Korhonen et al., 2003; Valenza et al., 2014b). Heart rate variability is also known to be affected by medication (O'Regan et al., 2015), potentially providing an indirect measure of treatment compliance. This thesis therefore presents a preliminary analysis of the potential for including heart rate based metrics as an objective marker of psychiatric health.

### 9.1.1 Experimental foundations

Studies using HR and EDA signals in the area of mental health cover a wide range of conditions, including panic disorder, depression, schizophrenia and bipolar disorder, with most of the studies performed on a small number of subjects. A number of projects that are most relevant to this research are described below.

McCraty *et al.* (McCraty et al., 2001) analysed the HRV in patients with panic disorder. 24-hour ECG records from panic disorder patients ($N = 38$, age $38 \pm 7.65$ years) and normal controls ($N = 38$, age $38 \pm 7.86$ years) were analysed and both time (mean, RMSSD) and frequency (HF, LF, VLF, ULF) HRV characteristics were collected. It was found that the VLF and LF power were significantly lower in panic disorder patients compared to normal controls.

Carney *et al.* (Carney et al., 2001) studied clinical depression in post myocardial infarction patients. Patients with ($N = 307$, age $57.1 \pm 12.3$ years) and without ($N = 366$, $60.9 \pm 11.1$ years) depression had a 24-hour ambulatory ECG before their hospital discharge. Four HRV indices were calculated, including the power in ULF, VLF, LF and HF bands. All HRV indices were significantly lower in patients with depression compared to patients without depression.

Agelink *et al.* (Agelink et al., 2002) analysed differences in heart rate variability between moderate ($N = 16$, age $44.3 \pm 12.6$ years) and severe ($N = 16$, age $53.5 \pm 15.8$ years) major depression patients, as well as normal controls ($N = 64$, age $46.6 \pm 11.9$ years). For HRV analysis, the coefficient of variation (CV), RMSSD and the power in LF and HF bands were calculated. Patients with severe major depression showed significantly higher heart rate and lower CV, RMSSD and HF, compared to controls, while in moderately depressed patients the HRV characteristics were lower, but not significantly.

Heart rate variability in bipolar disorder was studied by Cohen *et al.* (Cohen et al., 2003). Euthymic bipolar patients ($N = 39$, age $38.4 \pm 12.8$ years), and matched by sex and age, normal controls ($N = 39$, age $39.5 \pm 12.8$ years) were recruited for the study. ECG recordings were taken at resting state and HRV indices were calculated, including mean and standard deviation of beat-to-beat intervals, power in VLF, LF, HF bands and total power (TP, $0.01 - 0.40$ Hz). It was found that bipolar patients are characterised by markedly low HRV, independent of specific drug treatments.

Valenza *et al.* (Valenza et al., 2013) analysed the heart rate in bipolar patients ($N = 3$, ages 38, 55 and 37 years) in the PSYCHE project. Up to 6 data acquisition sessions were performed for each subject in different emotional states and classification accuracy of severe clinical mood states from 68.31% to 97.96% (depending on a subject) was achieved using 5-fold cross validation for intra-subject analysis using heart rate features.

Lanata *et al.* (Lanata et al., 2015) performed analysis of RR-intervals of bipolar patients ($N = 10$, age not reported) using Sample Entropy and reported the decreased entropy in depression and mania states compared to euthymia (significance tests not performed).

Valenza *et al.* (Valenza et al., 2014a) performed RR-intervals analysis of bipolar patients ($N = 8$, age not reported) using Multiscale Entropy and Detrended Fluctuation Analysis and identified statistically significant differences between features in depression, hypomania and euthymic states.

However, it is important to point out that many of the above studies failed to do two crucial things. First, they did not attempt to classify patients and therefore provide any indication of the potential accuracy of these features to diagnose or monitor patients. Second, they performed very limited out of sample analysis in order to provide a notion of generalisation of the approaches in diagnosing or monitoring patients outside the very small cohorts analysed.

The question therefore remains if these signals are complimentary to behaviour anal-ysis described in this thesis, and what would be the predictive accuracy of HR for diagnosis. In order to address this point, in this chapter an analysis of HR and actigraphy is presented for a new patient population, specifically for schizophrenia, where changes in locomotor activity and HR have been shown to be indicative of the disease. This validates the methods on an external population and provides an evaluation of the potential for combining physiology with actigraphy for monitoring mental health.

### 9.1.2 The case of schizophrenia

Schizophrenia is a chronic illness, affecting about one percent of the population. It is characterised by delusions, hallucinations, disorganisation of speech and behaviour. Onset of schizophrenia usually occurs in early adult years and during the course of the disorder symptoms may wax and wane, or remain relatively stable. Common complications include depression and suicide (about 50% of patients attempt and approximately 10% succeed), usually early in the course of the illness (American Psychiatric Association, 2013). The broad range and diverse combinations of symptoms make differential diagnosis of schizophrenia a complex task that requires a good knowledge of a patient's history. In addition, the early detection of psychotic relapse/exacerbation often depends on patient self-report which is a particular challenge in an illness in which insight is frequently impaired. Psychotic relapse is a continual threat, largely because of high rates of non-adherence in medication taking. Objective measures which could alert clinicians and caregivers to early signs of relapse would have enormous public health significance.

Walther *et al.* (Walther et al., 2009) performed a quantitative activity analysis in patients with different schizophrenia subtypes, including paranoid (N=18, age 39.60±9.64 years), catatonic (N=6, age 45.50±12.19 years) and disorganised (N=9, age 36.23±10.26 years). Activity was monitored for 24 hours every two seconds, but only data during wakefulness were included in the main analysis. Three activity parameters were extracted: the mean number of activity counts per hour (AL), the percentage of epochs with non-zero activity (mobility index, MI), and the mean duration of immobility periods (MIP).

Only schizophrenia subtype and type of anti-psychotic were found to have significant effect on AL, MI and MIP.

Hauge *et al.* (Hauge et al., 2011) analysed differences in locomotor activity of schizophrenia patients (N=24, age 47.4±11.1 years), depressed patients (N=25, age 42.9±10.7 years) and healthy subjects (N=32, age 38.2±13.0 years) in two settings: a) one minute epochs recorded over five hours, and b) one hour epochs recorded over two weeks. Sample Entropy and Fourier analysis were applied and have shown significantly different profiles of activity between all three groups.

Wulff *et al.* (Wulff et al., 2012) analysed sleep and circadian rhythm disruption in schizophrenia outpatients (N=20, age 38.8±8.6 years) in comparison to healthy unemployed individuals (N=21, age 37.5±9.6 years). Activity and light exposure were recorded every two minutes over six weeks. Activity data were analysed using rest-activity characteristics, cosinor and periodogram analysis. Significant disruption in the sleep and rest-activity cycle was detected in all schizophrenia patients, independently of mood, mental state and anti-psychotic treatment.

Bär *et al.* (Bär et al., 2008) studied the HRV in schizophrenia patients. ECG data were collected for 2 hours from unmedicated patients with schizophrenia ($N = 40$, age $34.7 \pm 7.1$ years) and matched healthy controls ($N = 58$, age $33.8 \pm 8.4$ years). Heart rate as well as the time domain (RMSSD) and frequency domain (VLF, LF, HF power) indices of HRV were calculated. The heart rate was significantly higher and the HF power was lower for the schizophrenia patients group.

Rachow *et al.* (Rachow et al., 2011) analysed the interrelation of skin conductivity levels and cardiac autonomic dysfunction with 18 unmedicated schizophrenia patients and 18 matched controls. They found significantly increased HR (consistent with results of Bär), decreased Root Mean Square of the Successive Differences (RMSSD) in heart beats and decreased complexity (measured with compression entropy) of HR of schizophrenia patients in comparison to controls.

Studies of Walther *et al.*, Hauge *et al.* and Wulff *et al.* demonstrated the usefulness

of actigraphy (Walther et al., 2009; Hauge et al., 2011; Wulff et al., 2012), and studies by Bär *et al.* and Rachow *et al.* have shown that the heart rate is also a useful signal in the analysis of schizophrenia (Bär et al., 2008; Rachow et al., 2011). This chapter presents the analysis of physical activity (Osipov et al., 2013), as well as combined analysis of physical activity and HR (Osipov et al., 2015) for differentiation between schizophrenia and healthy controls using two different data sets described below.

## 9.2 Description of the data sets

### 9.2.1 The Nuffield data set

The first data set was provided by the Nuffield Department of Clinical Neurosciences, University of Oxford and included only actigraphy information from schizophrenia patients and matched healthy controls. Motor activity and light exposure of schizophrenia patients (N=20, age 38.8±8.6 years) and unemployed controls (N=21, age 37.5±9.6 years), were monitored at home for a maximum of 6 weeks (see Table 9.1). For data collection, Actiwatch-L actigraph by CamNTech (Cambridge, UK) was used. The device was continuously worn on the non-dominant wrist with exception of short watch-off intervals due to shower or medical procedures. Actiwatch-L was configured for data collection with one or two minutes epochs.

**Table 9.1:** *Demographics and data set characteristics of the Nuffield schizophrenia data set.*

| | **Healthy controls** | **Schizophrenia** |
|---|---|---|
| Number of participants | 21 | 20 |
| Gender | 8 female | 5 female |
| Age (Mean±SD) | 37.5±9.6 | 38.8±8.6 |
| Medication | N/A | All medicated |

### 9.2.2 Proteus data set

The second data set was provided by Proteus Digital Health (Redwood City, CA) and included both physical activity and heart rate data, collected from schizophrenia patients and healthy controls. Outpatients with diagnosis of schizophrenia ($N = 16$, age $45.1 \pm 12.3$ years), medicated and in relative symptomatic remission and healthy control subjects ($N = 19$, age $51.7 \pm 8.8$ years) without a history of mental disorders were included in the analysis (see Table 9.2). Accelerometry and the heart rate of study participants were monitored for a maximum of four weeks using an adhesive patch manufactured by Proteus Digital Health (Redwood City, CA) (see Figure 9.1). The device was continuously worn on the body and configured for the acquisition of ECG-derived heart rate (every 10 minutes as a mean HR at 15 second intervals) and locomotor activity (every 5 minutes as a mean acceleration at 15 second intervals).

**Table 9.2:** *Demographics and data set characteristics of the Proteus schizophrenia data set.*

| | **Healthy controls** | **Schizophrenia** |
|---|---|---|
| Number of participants | 19 | 16 |
| Gender | 25% female | 42% female |
| Age (Mean±SD) | 51.7±8.8 | 45.1±12.3 |
| Medication | N/A | All medicated |
| Length of disease (years) | N/A | 6.4±3.4 |
| Symptoms | No history of mental disorders | All in relative symptomatic remission |

Schizophrenia patients were prescribed anti-psychotic medication, including Olanzapine, Risperidone, Aripiprazole, Perphenazine, Fluphenazine, Ziprasidone, Haloperidol and Quetiapine.

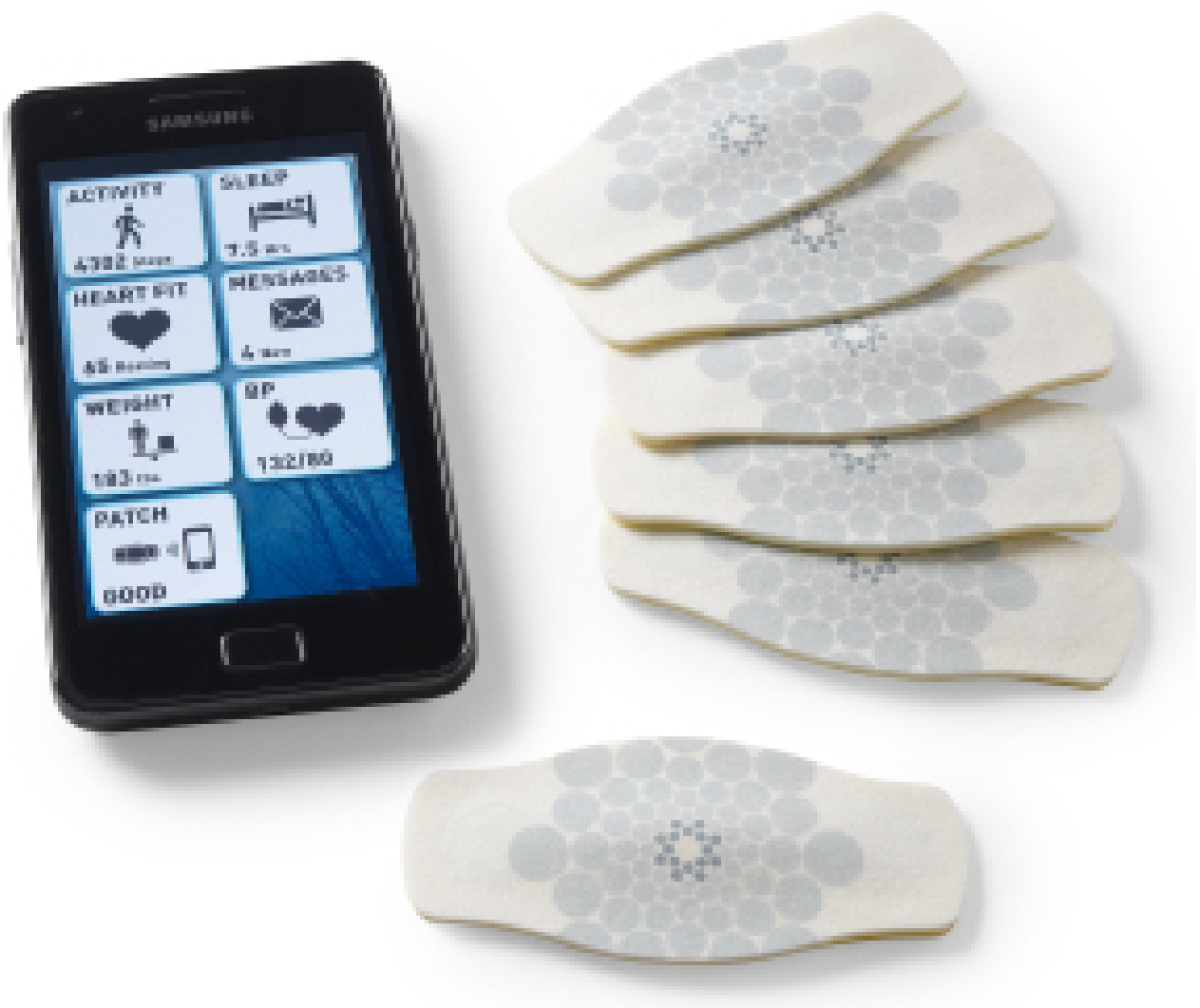

**Figure 9.1:** *Adhesive activity and heart rate monitoring patch used in the study (Proteus Digital Health, Redwood City, CA, USA). Adopted from the company marketing materials.*

## 9.3 Data analysis approach

### 9.3.1 Data pre-processing

#### 9.3.1.1 The Nuffield data set

Because actigraphic data were collected with different epoch lengths (one and two minutes) and record duration, the data were pre-processed for analysis according to the following rules:

1. Records were re-sampled for two minutes epoch according to the Actiwatch-L User Manual (CamNtech Ltd, 2008).

2. Records with less than 20 complete days (median length of unemployed controls category) were excluded from analysis.

3. Records were inspected visually and periods without significant missing data (watch-off intervals) were selected.

4. First and last days of every record were discarded as transition periods.

5. Periods of missing data were taken as zero activity. These periods usually represent periods of bathing. Although correction with a non-zero value may at first seem appropriate, it is unclear how a relevant non-zero value could be picked, since we may just be picking an activity value that is similar to normality when the patient is abnormal, or vice versa. A value of zero is unique and simply indicates that something different is happening. Removing the values and concatenating the data either side of the removed segment would not be appropriate as this would affect the meaning of the temporally-related metrics we used.

#### 9.3.1.2 The Proteus data set

Locomotor activity data (recorded in arbitrary units, ranged from zero to one) and heart rate data (recorded in beats per minute, bpm) were collected with a variable recording rate and length. Since most analysis algorithms require evenly sampled data, the data were processed according to the following rules:

1. Data points with an interval, exceeding 1.5 $\times$ data collection rate (15 minutes for HR and 7.5 minutes for activity) were labelled as low quality and removed.

2. HR values lower than 20 bpm were labelled as low quality and removed, based on normative values of the human heart rate (Jensen-Urstad et al., 2007).

3. HR data were re-sampled to exactly 10 minute intervals using zero-order hold (sample-and-hold) interpolation. Sample-and-hold interpolation approximates the next value in time series by replicating the previous value.

4. Locomotor activity data were re-sampled to exactly 5 minute intervals using zero-order hold (sample-and-hold) interpolation.

5. For analysis, where synchronous activity and heart rate data were necessary, activity data were down-sampled to 10 minute intervals using linear interpolation. Linear interpolation approximates the value in time series as a mean of previous and the next value.

6. The data from the device are influenced by motion artifacts, and high levels of activity are correlated to motion, including all data would reduce our ability to separate the independent influences of physiology and motion on the classifier performance. To avoid this bias, an automated quality assessment on the HR data was performed and subjects with more than 10% artifacts removed, consistent with Li *et al.* (Li et al., 2008). From complete recordings, the 10 days with the lowest percentage of missing data for that subject were selected. Recordings, where more than 10% of data were missing or were of low quality, were discarded.

### 9.3.2 Analysis methods

The Nuffield data set was analysed using Multiscale Entropy characteristics. For the analysis of Proteus data set four feature groups were selected, representing different dimensions of schizophrenia symptoms, and are described in the next sections (9.3.2.1 through 9.3.2.4).

#### 9.3.2.1 Statistical characteristics

Both locomotor activity and HR have been shown to be affected in schizophrenia (Walther et al., 2009; Hauge et al., 2011; Bär et al., 2008; Rachow et al., 2011), so statistical characteristics of HR and activity were calculated, including mean, median, mode, standard deviation (STD) and inter-quartile range (IQR).

#### 9.3.2.2 Rest-activity characteristics

Recent studies indicate that circadian rhythm disruption plays a significant role in schizophrenia (Wulff et al., 2012), so the rest-activity characteristics of HR and activity, including levels during the Least Active 5 hours, Most Active 10 hours, Relative Amplitude (see

Section 5.5.4), as well as Inter-day Stability and Intra-day Variability (see Section 5.5.5) were included in the analysis.

#### 9.3.2.3 Multiscale Entropy

Activity disorganisation is one of the diagnostic criteria for schizophrenia (American Psychiatric Association, 2000). Hauge *et al.* (Hauge et al., 2011) indicated that entropy can be used to estimate such disorganisation. For the analysis of the Nuffield data set, Multiscale Entropy (MSE) analysis (see Section 5.4.1) was applied to actigraphy and values of entropy at scales from two minutes up to 12 hours were used as features directly. For the analysis of the Proteus data set MSE of activity and HR was included in the analysis and first five scales were calculated and coefficients of a $3^{rd}$ degree polynomial fitted in a least squares sense into these scales, were used as features for further analysis.

#### 9.3.2.4 Transfer Entropy

To estimate the possible coupling between locomotor and cardiovascular activity, Transfer Entropy (TE) between HR and activity signals were calculated as follows

$$T_{X \to Y} = H(y_i | y_{i-t}^{(l)}) - H(y_i | y_{i-t}^{(l)}, y_{i-\tau}^{(k)}) \tag{9.1}$$

where $i$ is a given point in time, $\tau$ and $t$ are the time lags in $X$ and $Y$, and $k$ and $l$ are the block lengths of past values in $X$ and $Y$.

TE estimates directional coupling between signals and measures the reduction of uncertainty, given past knowledge of a secondary signal in comparison with knowledge of a primary signal only. The algorithm is parametrised by primary and secondary signal lags (or the amount of history to take into account) (Schreiber, 2000).

Values of $k$ and $l$ were taken as $k = 1$ and $l = 1$ as recommended by Lee *et al.* (Lee et al., 2012). The Darbellay-Vajda signal partitioning scheme (Darbellay and Vajda, 1999; Lee et al., 2012) was used for probability density function estimation. Coefficients of a $3^{rd}$ degree polynomial, fitted in a least squares sense into the TE values with $t = 1$ and $\tau = \{1, 2, 3, 4, 5\}$, were used as features for further analysis.

#### 9.3.2.5 Feature selection and classification

For the analysis of the Nuffield data set the best features to distinguish between schizophrenia and healthy controls subject categories were selected by ordering MSE values according to the absolute value of the standardised u-statistic of a two-sample unpaired Wilcoxon test (RANK). For Proteus data set, a multivariate minimum-Redundancy-Maximum-Relevance (mRMR) criterion (see Section 6.3.1.2) was used to identify the most important features for classification (Peng et al., 2005).

For classification into schizophrenia and normal controls categories, a Support Vector Machine (SVM) with a Gaussian Radial Basis Function (RBF) kernel (see Section 6.3.2.2) was used (Cortes and Vapnik, 1995), $\sigma = 4$, selected using grid search. Due to the limited number of samples in both schizophrenia and control classes, LOOCV was performed for the analysis of the Nuffield data set and 2-fold cross-validation with repeated random sub-sampling (Kohavi, 1995) was used to estimate the classification performance in the Proteus data set, where samples were randomly separated into the training and testing set and 1000 classification experiments performed to estimate the classification performance.

For the Proteus data set, to evaluate the influence of combination of physiological and locomotor activity features, three feature selection and classification experiments were performed:

- Using HR features alone.
- Using locomotor activity features alone.
- Using HR, locomotor activity and Transfer Entropy features.

To evaluate the models, Receiver Operating Characteristic (ROC) curves were created and the Area Under Curve (AUC) was calculated for each model.

## 9.4 Comparative results of activity, HR, and a combination of both

### 9.4.1 The Nuffield data set

Classification results for separation between schizophrenia and healthy control classes varied depending on the number of features. The best accuracy of 95.12% was achieved using the best 22 features, 90.24% using two or six best features and usage of only the SampEn on a scale of two minutes provided 87.8% classification accuracy as presented in Table 9.3.

**Table 9.3:** *Ordering of the best entropy features and characteristics of classifier, using the top 25 selected features (scales of MSE calculation, h is hours and m is minutes). Wilcoxon rank-sum test results for features, different with 5% significance level with Bonferroni correction for multiple testing, are denoted by †.*

| Nr. | MSE scale | p-value | Accuracy | Specificity | Sensitivity |
|---|---|---|---|---|---|
| 1 | 0h 2m | 0.000011 † | 0.878049 | 0.800000 | 0.952381 |
| 2 | 0h 4m | 0.000020 † | 0.902439 | 0.800000 | 1.000000 |
| 3 | 10h 34m | 0.024065 | 0.878049 | 0.800000 | 0.952381 |
| 4 | 0h 6m | 0.000087 † | 0.878049 | 0.800000 | 0.952381 |
| 5 | 11h 38m | 0.040606 | 0.853659 | 0.800000 | 0.904762 |
| 6 | 11h 8m | 0.055192 | 0.902439 | 0.850000 | 0.952381 |
| 7 | 11h 54m | 0.065947 | 0.902439 | 0.850000 | 0.952381 |
| 8 | 11h 52m | 0.092512 | 0.878049 | 0.850000 | 0.904762 |
| 9 | 6h 28m | 0.000665 † | 0.853659 | 0.800000 | 0.904762 |
| 10 | 11h 26m | 0.103075 | 0.853659 | 0.800000 | 0.904762 |
| 11 | 11h 30m | 0.103075 | 0.853659 | 0.800000 | 0.904762 |
| 12 | 6h 50m | 0.000882 † | 0.829268 | 0.750000 | 0.904762 |
| 13 | 10h 38m | 0.114543 | 0.804878 | 0.750000 | 0.857143 |
| 14 | 11h 56m | 0.123827 | 0.829268 | 0.750000 | 0.904762 |
| 15 | 11h 22m | 0.127043 | 0.829268 | 0.750000 | 0.904762 |
| 16 | 5h 50m | 0.001277 † | 0.804878 | 0.750000 | 0.857143 |
| 17 | 7h 16m | 0.001530 † | 0.853659 | 0.750000 | 0.952381 |
| 18 | 8h 16m | 0.001528 † | 0.853659 | 0.800000 | 0.904762 |
| 19 | 11h 6m | 0.158986 | 0.853659 | 0.800000 | 0.904762 |
| 20 | 10h 2m | 0.162880 | 0.853659 | 0.750000 | 0.952381 |
| 21 | 13h 16m | 0.162880 | 0.902439 | 0.800000 | 1.000000 |
| 22 | 13h 8m | 0.166847 | 0.951220 | 0.900000 | 1.000000 |
| 23 | 5h 22m | 0.001828 † | 0.902439 | 0.850000 | 0.952381 |
| 24 | 7h 30m | 0.001828 † | 0.951220 | 0.900000 | 1.000000 |
| 25 | 10h 10m | 0.187772 | 0.951220 | 0.900000 | 1.000000 |

### 9.4.2 The Proteus data set

After pre-processing, four records of schizophrenia subjects with an amount of missing data exceeding the 10% threshold were discarded. The missing data were probably caused by the poor contact of adhesive wearable sensor with patient's skin and motion artifacts, and not related to the diagnosis of the patient. A total of 12 records of schizophrenia patients and 19 records of normal controls were processed for further analysis, with results presented in Tables 9.4 and 9.5.

**Table 9.4:** *Accuracy of Proteus data set classification using locomotor activity, physiology and combined features subsets.*

| | Accuracy | Sensitivity | Specificity | Nr. features |
|---|---|---|---|---|
| Heart rate alone | 78.5% | 82.3% | 72.2% | 2 |
| Locomotor activity alone | 85.5% | 85.3% | 85.9% | 5 |
| Combined features | 95.3% | 98% | 91.1% | 4 |

**Table 9.5:** *Features of HR and locomotor activity of schizophrenia and control subjects (Mean ± STD, arbitrary units for activity and bpm for HR) together with mRMR ranking. The items different with 1% significance level according to two-sided Wilcoxon rank sum test marked with bold font and †.*

| | Heart Rate | | | Activity | | |
|---|---|---|---|---|---|---|
| | **Schizophrenia** | **Controls** | **Rank** | **Schizophrenia** | **Controls** | **Rank** |
| Mean | $82.01 \pm 10.46$ | $76.11 \pm 6.19$ | 4 | $0.04 \pm 0.02$ | $0.06 \pm 0.01$ | **15†** |
| Median | $79.33 \pm 10.83$ | $74.70 \pm 7.10$ | 27 | $0.01 \pm 0.00$ | $0.02 \pm 0.00$ | 17 |
| Mode | $72.27 \pm 12.71$ | $69.23 \pm 8.19$ | 20 | $0.01 \pm 0.00$ | $0.01 \pm 0.00$ | **2†** |
| STD | $13.95 \pm 2.70$ | $14.11 \pm 2.23$ | 21 | $0.08 \pm 0.03$ | $0.13 \pm 0.03$ | **1†** |
| IQR | $18.50 \pm 5.00$ | $19.71 \pm 4.25$ | 29 | $0.02 \pm 0.02$ | $0.03 \pm 0.01$ | 8 |
| L5 | $1952.19 \pm 249.99$ | $1831.15 \pm 162.50$ | 12 | $1.17 \pm 0.51$ | $1.61 \pm 0.33$ | **9†** |
| M10 | $4311.61 \pm 599.30$ | $4016.32 \pm 322.46$ | 7 | $7.15 \pm 2.71$ | $10.05 \pm 5.25$ | 13 |
| RA | $0.38 \pm 0.03$ | $0.37 \pm 0.02$ | 18 | $0.71 \pm 0.08$ | $0.70 \pm 0.06$ | 33 |
| IS | $0.30 \pm 0.14$ | $0.36 \pm 0.12$ | 23 | $0.16 \pm 0.04$ | $0.18 \pm 0.03$ | 16 |
| IV | $0.60 \pm 0.21$ | $0.70 \pm 0.23$ | 19 | $1.23 \pm 0.13$ | $1.23 \pm 0.23$ | 28 |
| MSE(1) | $< 0.01$ | $< 0.01$ | 5 | $< 0.01$ | $< 0.01$ | 31 |
| MSE(2) | $-0.02 \pm 0.12$ | $0.05 \pm 0.13$ | 24 | $-0.02 \pm 0.04$ | $-0.01 \pm 0.03$ | 35 |
| MSE(3) | $0.09 \pm 0.34$ | $-0.20 \pm 0.36$ | 14 | $0.13 \pm 0.13$ | $0.12 \pm 0.09$ | 26 |
| MSE(4) | $1.78 \pm 0.38$ | $2.35 \pm 0.34$ | **3†** | $0.30 \pm 0.11$ | $0.31 \pm 0.09$ | 30 |
| TE(1) | $< 0.01$ | $< 0.01$ | 10 | $< 0.01$ | $< 0.01$ | 32 |
| TE(2) | $0.02 \pm 0.04$ | $0.00 \pm 0.03$ | 22 | $0.00 \pm 0.06$ | $0.01 \pm 0.04$ | 36 |
| TE(3) | $-0.07 \pm 0.11$ | $-0.01 \pm 0.07$ | 6 | $0.01 \pm 0.17$ | $0.00 \pm 0.13$ | 25 |
| TE(4) | $0.19 \pm 0.15$ | $0.11 \pm 0.05$ | **11†** | $0.18 \pm 0.11$ | $0.17 \pm 0.12$ | 34 |

When using HR derived features only, ordered by the mRMR algorithm, the two best features (including mean and MSE(4)) resulted in a 78.5% classification accuracy with 82.3% sensitivity and 72.2% specificity with an AUC of 0.85.

When using locomotor activity features only, ordered by the mRMR algorithm, the five best features, (STD, mode, IQR, IS and L5) resulted in a 85.5% classification accuracy with 85.3% sensitivity and 85.9% specificity with an AUC of 0.9. Note, that despite of previous finding (Hauge et al., 2011) the MSE of activity was not identified as one of predictive features.

When HR, locomotor activity and Transfer Entropy features (ordered by the mRMR algorithm) were used to train an SVM, the four best features, (STD of locomotor activity, mode of locomotor activity, MSE(4) of HR and mean of HR) resulted in a 95.3% classification accuracy with 98% sensitivity and 91.1% specificity with an AUC of 0.99. The use of SVN classifier could potentially provide over-fitted results, however the use of cross validation with repeated random sub-sampling (as described in Section 9.3.2.5) provides a guarantee that our results are reliable estimate of classifier accuracy based on selected features.

## 9.5 Discussion and conclusions

Using a machine learning framework, it was possible to achieve classification accuracy of 95.12% using locomotor features alone (the Nuffield data set) and 85.5% (the Proteus data set). The combination of HR and locomotor activity features provided the best classification accuracy with almost a 10% increase in accuracy above using locomotor features alone (85.5% versus 95.3%) and almost 17% improvement on using heart rate related features alone (78.5% versus 95.3%).

The optimal features set in the Proteus data set included different feature classes, representing levels of HR, locomotor activity and Multiscale Entropy dynamics across the time scales of HR. Consistent with previous results and theoretical expectations (Bär et al., 2008; Rachow et al., 2011), HR was found to be elevated in schizophrenia patients and the level of locomotor activity decreased. Despite the fact that HR elevation is not significant, it was an independent characteristic of schizophrenia group, as found by mRMR feature selection. The STD of locomotor activity was both significantly different and important from the point of view of classification, as well as the $4^{th}$ coefficient of a polynomial fit of Multiscale Entropy for HR.

However, these results should be taken with caution. Recent analysis O'Regan *et al.* (O'Regan et al., 2015) performed on 4750 participants organised in four groups, including depressed and on antidepressants (N=80, 59.4±0.75), depressed and not on antidepressants (N=317, age 60.04±0.43), not depressed and on antidepressants (N=185, age 62.6±0.62) and without depression and not on antidepressants (N=4107, age 61.7±0.13), identified that depressed patients not taking antidepressants did not differ significantly from controls.

But even if the heart rate-based measures reflect purely the influence of medication, then an alternative approach to using this information can be taken. The models based upon heart rate only can be used as a measure of medication adherence and the actigraphy-based model can be used for evaluation the effectiveness of a therapy (if the former model indicates the medication is being taken). Disentangling the two would require a carefully controlled study involving both compliant and non-compliant responders and non responders to treatment.

As indicated by Teicher in his review of actigraphy and motion analysis use in psychiatry (Teicher, 1995), accurate monitoring of activity levels and circadian rhythms has a potential to aid clinicians in diagnosis, however very few studies examined sensitivity and specificity of these features. The analysis of this chapter aimed to fill this gap and added another promising dimension by fusing the HR-based measures of ANS function with locomotor activity analysis and showed that physiological and physical activity signals provide complimentary information for classification of mental health.

While the above analysis was performed on a data set including only schizophrenia patients and controls, and mood disorders symptoms related to the HR changes will be different, the results are encouraging and suggest that continuous HR acquisition may be an important signal to include in future studies and models of mood disorders.

# Chapter 10

# Conclusions and future work

## 10.1 Validity of the approach

In this work a novel approach to psychiatric diagnosis and disorder management was presented, where assessment of symptoms was performed using objectively collected physical activity and derived behaviour features.

### 10.1.1 The need for objective mental health metrics

In current clinical practice, the diagnosis of mental disorders is primarily based on patient self-reports and infrequent clinical observations (American Psychiatric Association, 2013). This often leads to erroneous conclusions and may result in unnecessary medication, incorrect treatment and decreased quality of life for patients (Regier et al., 2013). Improvements in the diagnostic process, including the introduction of more objective diagnostic criteria suitable for ambulatory assessment is a promising avenue for improving mental health care, and is recognised as a necessary further step in development of the psychiatric field (Singh and Rose, 2009; Insel et al., 2010; Phillips and Kupfer, 2013).

The approach adopted by this thesis was based on the dimensional symptoms framework described by Peter Liddle (Liddle, 2001) and summarised earlier in Table 2.1. In this framework five dimensions of symptoms are defined, including reality distortion, dis-organisation, psychomotor, mood and anxiety dimensions. The research of this thesis started with the observation that at least three of these dimensions include symptom clusters with manifestations in physical activity or behaviour, thus the usage of physiological signals may give an insight into these symptoms, as can be seen from Table 10.1.

**Table 10.1:** *Signals for objective assessment of mental health symptoms.*

| Signal | Symptom cluster | Dimension |
|---|---|---|
| Physical activity | Disorganised and bizarre behaviour,<br>Decreased voluntary motor activity,<br>Motor agitation,<br>Decreased or increased need for sleep,<br>Inappropriate affect | Disorganisation<br>Psychomotor<br>Psychomotor<br>Mood<br>Disorganisation |
| Heart rate | Feelings of unease, fear or dread,<br>Overactivity of the sympathetic nervous system | Anxiety<br>Anxiety |
| Electrodermal activity | Feelings of unease, fear or dread,<br>Overactivity of the sympathetic nervous system | Anxiety<br>Anxiety |

With the latest developments in quantified-self domain and wearable technologies it may be possible to objectively assess human behaviours (and symptoms) continuously and nearly in real-time, providing the patient and the clinician with a powerful tool for accurate diagnosis and management of mental disorders, free from subjective perception bias and accurately capturing the complete temporal picture of symptoms.

### 10.1.2 The approach of this thesis

This thesis approached the topic of objective mental health from the perspective of identifying "quantifiable symptoms" of disorders and clinical states, based on an accepted set of clinical practice definitions (American Psychiatric Association, 2013). These definitions were converted into hypothesised mathematical descriptions of objectively collected physical activity data (activity features), based on state-of-the-art methods from sleep and physical activity research, as well as the latest developments in psychophysiology.

Models were built to reproduce existing clinical scores using the above mentioned features, and a number of machine learning techniques were assessed for creating accurate models to (a) distinguish between the disorder and healthy state and (b) identify clinical state within the disorder.

Physical activity data were selected for this analysis due to its descriptive power, which were capable of characterising different groups of symptoms, and simplicity of acquisition, with different modalities available for capturing locomotor activity signals, including research grade wearable sensors and consumer mobile phones with acceleration sensors.

The resulting framework followed the dimensional model of symptomatology and has a potential benefit of not only being diagnostically accurate, but also providing intuitively clear and actionable information for both clinical specialists and patients.

### 10.1.3 Summary of main contributions

Chapter 3 describes a platform for the collection of activity and behaviour data for the purpose of mental health symptoms analysis used by the AMoSS study. This client-server platform was developed and deployed by the author, and used to collect data from mobile phones of patients with mental disorders and healthy controls (a total of 150 subjects) in an ambulatory environment and for the duration of more then one year, representing a substantial improvement compared to previous studies, where data acquisition was limited to three months (see Table 3.1). The ongoing technical support of participants was provided by study assistants from non-technical backgrounds, thus proving the possibility of using smart-phone and wearable devices as clinical tools outside of research settings.

Chapters 4 and 5 of this work proposed a set of activity data pre-processing methods and a principled framework for computational analysis of mental health symptoms by mathematical description of symptomatic behaviours. To the author's knowledge, this is the first attempt to define such a framework based on a dimensional approach to psychiatric symptomatology, as opposed to the modality-based (activity, HR, EDA) or phenomena-based (entropy, activity level) approaches of other studies. Based on solid foundations of accelerometry data analysis from population-level physical activity studies, this work built a framework for the extraction of high level behaviour features, such as the characteristics of sleep or stability of 24-hour activity patterns, using modern signal processing and machine learning techniques.

Chapter 6 assessed the difference between health and disorder, the main question of psychiatric diagnosis. Correctly identifying a disorder is important for providing adequate treatment and achieving best patient outcomes. However, it is often difficult to distinguish between different conditions, given that insight in patients is often impaired and correct diagnosis requires a good anamnesis. Objective examination of symptomatic activity patterns may help to distinguish between specific conditions that are difficult to assess using self-reports. In this study the differences between healthy subjects, bipolar and borderline personality disorder patients were analysed, and an accuracy of 70% was achieved for separation between healthy controls and borderline personality disorder, 67% for separation between healthy controls and bipolar disorder, and 80% for separation between bipolar disorder and borderline personality disorder using LOOCV. Note, that these results were achieved using only one least symptomatic week of activity data.

Chapter 7 analysed the difference in objective behaviour characteristics between clinical states, specifically mood states in bipolar disorder where mood tends to be well expressed and represents the main characteristic of the illness. Early identification of deterioration in mood is critical for management of this disorder, but may be a complicated task due to infrequent contact with clinical specialists in the course of ambulatory observations and due to impaired patient insight. This study demonstrates that behaviour correlates may allow for the identification of clinically significant emotional states with 85% accuracy for differentiation between euthymic state and depression, 80% for differ-entiation between euthymic state and mania and 90% for differentiation between mania and depression using LOOCV. Thus, it was demonstrated that using symptomatic data can provide for a potentially better separation between patients in different emotional states and clinical conditions.

By building personalised models of mood described in Chapter 8 it was possible to achieve a mean average error of prediction of ASRM scores of 2.13±1.63 points, QIDS-16SR scores of 2.65±1.47 points and GAD-7 scores of 2.51±1.31 points, *i.e.* at the granularity level of clinical state identification with psychiatric questionnaires, and overall average correlation of self-reported and estimated mood scores between 0.27 and 0.54, although below the reported correlation between different psychiatric scales (r=0.72 for ASRM and MRS and r=0.84 for QIDS-16SR and HAM-D24) (Altman et al., 1997), indicates a promising opportunity for replicating self-reports with behavioural or physiological measurements once data from wearable devices are available to evaluate the complete symptoms assessment framework.

Chapter 9 added physiology into the picture of objective mental health symptoms by analysing combined locomotor activity and heart rate in schizophrenia. The combination of HR and locomotor activity features provided classification accuracy with almost a 10% increase above using locomotor features alone (85.5% versus 95.3%) and almost 17% above using HR related features alone. The optimal features set included different feature classes, representing levels of HR, locomotor activity and Multiscale Entropy dynamics across the time scales of HR. While this analysis was performed on a data set including only schizophrenia patients and controls, and mood disorders symptoms related to the HR changes will be different, the results are encouraging and suggest that continuous HR acquisition may be an important signal to include in future studies and models of mood disorders. Even if the physiological measures reflect purely the influence of medication, then they may be used as a method for identifying medication adherence, separate to the effectiveness of the treatment (measured by actigraphy).

## 10.2 Current study limitations

The problems of defining objective physiological features for psychological phenomena and assessing mental state using objective measurements were described by Hinton in 1966 (Hinton, 1966), but are still valid today as discussed in Section 2.1.2. The present study addresses many of these problems, but due to the nature of the analysed phenomena has certain limitations, described in detail below.

### 10.2.1 Link between behaviour and psychological phenomena

Psychological constructs (such as mood) were often based solely on subjectively-reported phenomena and direct behaviour-based measures were not possible to define, therefore indirect correlates were used. For mood, this thesis adopted sleep quality as the best known predictor, based on prior research on the relationship between sleep and mood (Kahn et al., 2013), diagnostic criteria (American Psychiatric Association, 2013) and dimensional symptoms framework (Liddle, 2001). This link is not direct and may be influenced by many external factors, such as work-rest routine and physical illness. However, a person's sleep when adversely affected by external factors may in turn lead to mood changes, and substantial further research is needed on the nature of these interconnections. This is beyond the scope of the presented thesis.

Other psychological constructs with no direct behaviour correlate that may show bias during reporting, include syndromes of flat and labile affect, as well as most subjective experiences, such as fatigue, elevated or low mood, and hopelessness. In most cases there are no defined objective measures for these phenomena to allow continuous ambulatory monitoring of symptom severity, and the subjective bias can be disorder specific.

Even if a psychological construct has a direct measurable behaviour equivalent, questionnaire based assessments do not consistently provide a reliable ground truth. It was reported for example that physical activity levels, as assessed by individuals themselves or external observers, do not provide a reliable estimation of the actual physical activity (Ham et al., 2007; Rosenberg et al., 2015). For retrospective mood self-reports, there was positive and negative recall bias identified, as well as recall inaccuracies depending on diagnosis and clinical state (Ebner-Priemer et al., 2006; Ben-Zeev et al., 2009). Inaccuracies in subjective self-reports are also associated with age and mild cognitive impairments (Weston et al., 2011).

### 10.2.2 Unreliable methods of measurement

Development of micro-electro-mechanical sensor technologies, miniaturisation of electronics and improvement of battery life made continuous behaviour and physiology data acquisition feasible. However research grade instrumentation is not currently suitable for long-term use (for example the GENEActiv device needs to be connected to a computer in order to download acquired data) and consumer-grade devices still suffer from limited usability, with smart phones not collecting activity data during the night and Fitbit wearable sensors being used only 60% of the time, with only 60% of nights covered during usage periods (only 36% of nights assessed). Moreover, the low cost of consumer devices may lead to higher inter-device quality variance.

There are promising new developments in behaviour-sensing taking place in the field of smart watch technology that could provide easy-to-use instruments addressing multiple user needs, and therefore leading to better compliance with usage. Smart watch technology development is still at an early stage, with the need to re-charge every two days being the main limiting usability factor (based on the author's experience).

### 10.2.3 Ill-defined psychological constructs

Existing definitions of symptoms are mostly based on subjective impressions about patient behaviours, and are often vague and difficult to reformulate in measurable terms, especially if behaviour is assessed continuously and non-invasively in an ambulatory environment. For example, the behavioural construct of "impulsivity" is widely acknowledged as an important characteristic in many mental disorders and clinical states, including the attention deficit hyperactivity disorder, borderline personality disorder and mania, and defined as

> *"behaviour without adequate thought, the tendency to act with less forethought than do most individuals of equal ability and knowledge, or a predisposition toward rapid, unplanned reactions to internal or external stimuli without regard to the negative consequences of these reactions"*

It is clear that such a definition is ambiguous and subjective ("adequate thought" is not defined), and also not specific. The existing objective impulsivity measures, including the Balloon Analogue Risk Task (BART) (Lejuez et al., 2002), Cued Go No-Go (Marczinski and Fillmore, 2003) and Immediate and Delayed Memory Tasks (IMT/DMT) (Dougherty et al., 2002), are suitable only for laboratory settings and address mostly attentional and immediate reaction aspects of impulsivity. The most comprehensive Barratt Impulsiveness Scale (BIS-11) (Barratt and Patton, 1995) covers all factors contributing to a persons' impulsivity, but is based on self-reporting of otherwise not measured behaviours.

A similar construct defined for the assessment of schizophrenia is "behaviour disorganisation" with a dedicated cluster allocated to it in the dimensional symptoms framework of Liddle (Liddle, 2001). The link between impulsivity and disorganisation is not very well understood, and was never assessed in a principled manner, making the definition of objective behaviour measures of impulsivity and disorganisation potentially unreliable.

The test-retest reliability of a diagnosis based on DSM-5 criteria for most diagnoses has moderate-to-good inter-rater agreement with an inter-class Kappa of 0.40-0.79 (Regier et al., 2013). Diagnostic criteria for many mental disorders, although well understood by clinicians, often do not help in distinguishing diagnostic entities due to differences in possible symptom manifestations and dependence on clinical states. For example for bipolar disorder and borderline personality disorder current clinical practice is reported to be inadequate to reliably distinguish between the two (Saunders et al., 2015).

### 10.2.4 Influence of external factors

As was previously discussed in the context of sleep, external factors may influence behaviour measures. This problem may not be of the highest importance for the analysis of mental health, as mainly slow variations and semi-permanent changes in behaviour characteristics are symptomatic, and therefore the appropriate filtering techniques introduced in Chapter 8 may address the issue. However, in some real-world scenarios, such as for example a patient going on vacation, changes in behaviour may not be easily distinguishable from symptoms of manic episode. Further research is necessary on the introduction of context information into the behaviour and mood models, for example mobile phones with in-built location sensing may be able to help in this regard.

### 10.2.5 Difference in individual characteristics

Even if all of the issues above are properly addressed, there is a substantial difference in individual behaviour manifestation of mood states. The important limitation of the current study is therefore the limited number of subjects and data points for clinical states assessment that may not reflect the complete picture of possible symptomatic behaviours.

Despite being perhaps the most comprehensive project of its kind, the AMoSS study includes only 50 patients with bipolar disorder and 30 patients with borderline personality disorder (less for the presented analysis, which deals with a less complete data set). This cohort does not represent the complete spectrum of possible symptom manifestations, and may not provide a substantial number of clinically validated disorder episodes. The current study is therefore of an exploratory nature, assessing the the viability of the described approaches and providing useful information for future researchers on promising techniques and research directions. Despite these limitations, to the author's knowl-edge the study in this thesis provides the most in-depth study of automated psychiatric symptom evaluation so far published.

## 10.3 Future work

The most important problem in the field of objective mental health analysis is the growing number of studies with a small number of subjects, where data were collected in different conditions, including ambulatory and hospital, with participants often being medicated and in different phases of their illness. Furthermore, activity and physiology information has often been collected using different methods, and analysis techniques varied greatly, making the task of results generalisation almost impossible.

These limitation serve as a prompt for the collection of large scale data sets that could be shared by researchers in the field, including patients representing the complete spectrum of symptoms and allowing application of different analysis methods to compare results in well-controlled conditions. Such collection is already feasible using a range of consumer devices, such as smart phones, fitness trackers and smart watches. Once such data sets become available, the results of this work may be further validated and objective activity features related to specific symptoms of mental health.

Study cohorts should allow stratification by age, gender, medication and sociodemographic status to allow eliminating confounding factors and more precisely identifying behaviours related to disorder within certain population or treatment regimen. Compliance remains to be an issue, specifically in mental health patients, thus the data acquisition modalities shall be usable and useful outside of the research context, and not carrying stigma on study participants. The example of such devices are smart watches, where technology allows for collection of a range of potentially useful signals continuously and improving on conventional watches, thus bringing a direct benefit to participants.

These data sets should include not only information about locomotor activity (and if it is included - it must go beyond simple step counts), but also physiological data, such as heart rate and variability (which show the capability for improvement of diagnostic results), and electrodermal activity, for complete assessment of individual symptom correlations.

Given the range of possible individual symptoms of mental disorders, essentially covering all possible aspects of individual behaviours and social interaction patterns, it is important to also acquire and analyse non-conventional potential symptom correlates, such as information about social networking (including both mobile text messaging and online social networking) and geo-location data, to analyse possible deviations from individual normal behaviour patterns.

Such information may also be used as an important source of information about possible confounding factors, influencing the accuracy of actigraphic analysis. For example, knowing that an individual has changed his base location (for example while going on holiday) we may expect that his or her model of normality will change as well, and introduce this information into the diagnostic model to improve the accuracy of analysis results.

Last, but not least, is the need for improved accuracy and reliability of data acquisition instrumentation, including both technical characteristics and perceived user benefit, which transforms into compliance. Unified data formats should become available as well as reference data processing algorithms, making possible comparison of results between different studies. The use of proprietary algorithms to process actigraphy make it difficult to create analysis methods that work on more than one (version of) a particular device.

Further advances in the field will be problematic without close cooperation between clinical experts and data scientists to identify more specific and objective behaviour patterns, relevant to particular symptoms of mental health. This would allow the development of computational data analysis methods to identify activity or behaviour features, that are characteristic specifically to these symptoms. Thus, experimental psychiatry and biomedical data analysis should develop hand-in-hand, and, it is to be hoped, contribute to the development of the emerging field of "computational psychiatry".

# Bibliography

Activinsights Ltd. (2012). GENEActiv Instruction Manual v. 1.2. Available at: http://www.geneactiv.org/wp-content/uploads/2014/03/geneactiv_instruction_manual_v1.2.pdf.

Adams, R. P. and MacKay, D. J. C. (2007). Bayesian Online Changepoint Detection. Technical report, University of Cambridge, Cambridge, UK. Available at: https://hips.seas.harvard.edu/files/adams-changepoint-tr-2007.pdf.

Afonso, P., Figueira, M. L., and Paiva, T. (2013). Sleep-wake patterns in schizophrenia patients compared to healthy controls. *The World Journal of Biological Psychiatry*, 15(7):517–524.

Agelink, M., Boz, C., Ullrich, H., and Andrich, J. (2002). Relationship between major depression and heart rate variability: Clinical consequences and implications for antidepressive treatment. *Psychiatry Research*, 113(1–2):139–149.

Akiskal, H. S., Bourgeois, M. L., Angst, J., Post, R., Möller, H.-J., and Hirschfeld, R. (2000). Re-evaluating the prevalence of and diagnostic composition within the broad clinical spectrum of bipolar disorders. *Journal of Affective Disorders*, 59:S5–S30.

Altman, E. G., Hedeker, D., Peterson, J. L., and Davis, J. M. (1997). The Altman self-rating mania scale. *Biological Psychiatry*, 42(10):948–955.

Altschule, M. D. (1953). *Bodily physiology in mental and emotional disorders.* Grune & Stratton, New York.

American Psychiatric Association (2000). *Diagnostic and Statistical Manual of Mental Disorders, Fourth Edition (DSM-IV-TR)*. American Psychiatric Publishing, Arlington, VA.

American Psychiatric Association (2013). *Diagnostic and Statistical Manual of Mental Disorders, Fifth Edition (DSM-5)*. American Psychiatric Publishing, Arlington, VA.

Ancoli-Israel, S., Cole, R., Alessi, C., Chambers, M., Moorcroft, W., and Pollak, C. P. (2003). The role of actigraphy in the study of sleep and circadian rhythms. *Sleep*, 26(3):342–392.

Andreasen, N. C. (1988). Brain imaging: Applications in psychiatry. *Science*, 239(4846):1381–1388.

Angst, J., Gamma, A., Benazzi, F., Ajdacic, V., Eich, D., and Rössler, W. (2003). Toward a re-definition of subthreshold bipolarity: Epidemiology and proposed criteria for bipolar-II, minor bipolar disorders and hypomania. *Journal of Affective Disorders*, 73(1-2):133–146.

Angst, J. and Marneros, A. (2001). Bipolarity from ancient to modern times. *Journal of Affective Disorders*, 67(1-3):3–19.

Appelhans, B. and Luecken, L. (2006). Heart rate variability as an index of regulated emotional responding. *Review of General Psychology*, 10(3):229–240.

Araújo, T., Nunes, N., Quintao, C., and Gamboa, H. (2012). Localized electroencephalography sensor and detection of evoked potentials. *2nd International Workshop on Computing Paradigms for Mental Health*, pages 41–46.

Aristotle (1984). *The complete works of Aristotle*, volume 2. Princeton University Press, Princeton, NJ.

Aritake-Okada, S., Higuchi, S., Suzuki, H., Kuriyama, K., Enomoto, M., Soshi, T., Kitamura, S., Watanabe, M., Hida, A., Matsuura, M., Uchiyama, M., and Mishima,

K. (2010). Diurnal fluctuations in subjective sleep time in humans. *Neuroscience Research*, 68(3):225–231.

Armstrong, N., Price, J., and Geddes, J. (2015). Serious but not solemn: Rebalancing the assessment of risks and benefits of patient recruitment materials. *Research Ethics*, 11(2):98–107.

Baillarger, J. G. F. (1880). *De la folie à double forme*, volume 6. impr. Donnaud.

Bär, K., Wernich, K., Boettger, S., and Cordes, J. (2008). Relationship between cardiovagal modulation and psychotic state in patients with paranoid schizophrenia. *Psychiatry Research*, 157(1–3):255–257.

Baranowski, T. (1988). Validity and reliability of self-report measures of physical activity: An information-processing perspective. *Research Quarterly for Exercise and Sport*, 59(4):314–327.

Bard, P. (1928). A diencephalic mechanism for the expression of rage with special reference to the sympathetic nervous system. *American Journal of Physiology*, 84:490–515.

Bardram, J. E., Frost, M., Szántó, K., Faurholt-Jepsen, M., Vinberg, M., and Kessing, L. V. (2013). Designing mobile health technology for bipolar disorder. In *Proceedings of the SIGCHI Conference on Human Factors in Computing Systems - CHI '13*, page 2627, New York, New York, USA. ACM Press.

Bardram, J. E., Frost, M., Szántó, K., and Marcu, G. (2012). The MONARCA self-assessment system. In *Proceedings of the 2nd ACM SIGHIT Symposium on International health informatics - IHI '12*, page 21, New York, New York, USA. ACM Press.

Barratt, E. S. and Patton, J. H. (1995). Barratt Impulsiveness Scale-11. In *Handbook of psychiatric measures*. American Psychiatric Association, Washington DC.

Beck, A. T., Steer, R. A., and Carbin, M. G. (1988). Psychometric properties of the Beck Depression Inventory: Twenty-five years of evaluation. *Clinical Psychology Review*, 8(1):77–100.

Bellivier, F., Geoffroy, P.-A., Etain, B., and Scott, J. (2015). Sleep- and circadian rhythm-associated pathways as therapeutic targets in bipolar disorder. *Expert Opinion on Therapeutic Targets*, 19(6):747–763.

Ben-Zeev, D., McHugo, G. J., Xie, H., Dobbins, K., and Young, M. A. (2012). Comparing retrospective reports to real-time/real-place mobile assessments in individuals with schizophrenia and a nonclinical comparison group. *Schizophrenia Bulletin*, 38(3):396–404.

Ben-Zeev, D., Young, M. A., and Madsen, J. W. (2009). Retrospective recall of affect in clinically depressed individuals and controls. *Cognition and Emotion*, 23(5):1021–1040.

Benca, R. M. (1992). Sleep and psychiatric disorders. *Archives of General Psychiatry*, 49(8):651.

Benca, R. M., Okawa, M., Uchiyama, M., Ozaki, S., Nakajima, T., Shibui, K., and Obermeyer, W. H. (1997). Sleep and mood disorders. *Sleep Medicine Reviews*, 1(1):45–56.

Benedetti, F. and Terman, M. (2013). Much ado about... a moody clock. *Biological Psychiatry*, 74(4):236–7.

Benson, P. J., Beedie, S. A., Shephard, E., Giegling, I., Rujescu, D., and St Clair, D. (2012). Simple viewing tests can detect eye movement abnormalities that distinguish schizophrenia cases from controls with exceptional accuracy. *Biological Psychiatry*, 72(9):716–24.

Berger, A. M., Wielgus, K. K., Young-McCaughan, S., Fischer, P., Farr, L., and Lee, K. A. (2008). Methodological challenges when using actigraphy in research. *Journal of Pain and Symptom Management*, 36(2):191–199.

Berk, M., Conus, P., Lucas, N., Hallam, K., Malhi, G. S., Dodd, S., Yatham, L. N., Yung, A., and McGorry, P. (2007). Setting the stage: from prodrome to treatment resistance in bipolar disorder. *Bipolar Disorders*, 9(7):671–8.

Berle, J. O., Hauge, E. R., Oedegaard, K. J., Holsten, F., and Fasmer, O. B. (2010). Actigraphic registration of motor activity reveals a more structured behavioural pattern in schizophrenia than in major depression. *BMC Research Notes*, 3(1):149.

Bernaola-Galván, P., Ivanov, P., Nunes Amaral, L., and Stanley, H. (2001). Scale invariance in the nonstationarity of human heart rate. *Physical Review Letters*, 87(16):168105.

Bernaola-Galván, P., Oliver, J. L., Hackenberg, M., Coronado, A. V., Ivanov, P. C., and Carpena, P. (2012). Segmentation of time series with long-range fractal correlations. *The European Physical Journal. B*, 85(6).

Bertolote, J. M. and Fleischmann, A. (2002). A global perspective in the epidemiology of suicide. *Suicidology*, 7(2):6–8.

Birchler-Pedross, A., Schröder, C. M., Münch, M., Knoblauch, V., Blatter, K., Schnitzler-Sack, C., Wirz-Justice, A., and Cajochen, C. (2009). Subjective well-being is modulated by circadian phase, sleep pressure, age, and gender. *Journal of Biological Rhythms*, 24(3):232 –242.

Birchwood, M. (2000). Schizophrenia: early warning signs. *Advances in Psychiatric Treatment*, 6(2):93–101.

Birket-Smith, M., Hasle, N., and Jensen, H. (1993). Electrodermal activity in anxiety disorders. *Acta Psychiatrica Scandinavica*, 88(5):350–355.

Bishop, C. M. (2006). *Pattern Recognition and Machine Learning*, volume 4. Springer, New York, New York, USA.

Blackwell, T., Ancoli-Israel, S., Gehrman, P. R., Schneider, J. L., Pedula, K. L., and Stone, K. L. (2005). Actigraphy scoring reliability in the study of osteoporotic fractures. *Sleep*, 28(12):1599–1605.

Blum, N., St John, D., Pfohl, B., Stuart, S., McCormick, B., Allen, J., Arndt, S., and Black, D. W. (2008). Systems Training for Emotional Predictability and Problem Solving (STEPPS) for outpatients with borderline personality disorder: A randomized controlled trial and 1-year follow-up. *The American Journal of Psychiatry*, 165(4):468–78.

Botella, C., Moragrega, I., Baños, R., and García-Palacios, A. (2011). Online predictive tools for intervention in mental illness: the OPTIMI project. *Studies in Health Technology and Informatics*, 163:86–92.

Boucsein, W. (2012). *Electrodermal Activity.* Springer US, Boston, MA.

Bowling, A. (2005). Mode of questionnaire administration can have serious effects on data quality. *Journal of Public Health*, 27(3):281–91.

Bresó, A., Martínez-Miranda, J., and García-Gómez, J. M. (2014). Leveraging adaptive sessions based on therapeutic empathy through a virtual agent. *In Proceedings of the 6th International Conference on Agents and Artificial Intelligence (ICAART 2014) - Doctoral Consortium.*

Bromundt, V., Wirz-Justice, A., Kyburz, S., Opwis, K., Dammann, G., and Cajochen, C. (2013). Circadian sleep-wake cycles, well-being, and light therapy in borderline personality disorder. *Journal of Personality Disorders*, 27(5):680–96.

Brownrigg, D. R. K. (1984). The weighted median filter. *Communications of the ACM*, 27(8):807–818.

Burton, C., McKinstry, B., Szentagotai Tătar, A., Serrano-Blanco, A., Pagliari, C., and Wolters, M. (2013). Activity monitoring in patients with depression: A systematic review. *Journal of Affective Disorders*, 145(1):21–8.

Busschaert, C., De Bourdeaudhuij, I., Van Holle, V., Chastin, S. F., Cardon, G., and De Cocker, K. (2015). Reliability and validity of three questionnaires measuring context-specific sedentary behaviour and associated correlates in adolescents, adults and older adults. *The international journal of behavioral nutrition and physical activity*, 12(1):117.

Cacioppo, J. (1991). Psychophysiological approaches to the evaluation of psychotherapeutic process and outcome, 1991: Contributions from social psychophysiology. *Psychological Assessment*, 3(3):321–336.

Cacioppo, J. T., Tassinary, L. G., and Berntson, G. (2007). *Handbook of psychophysiology*. Cambridge University Press, Cambridge, UK.

CamNtech Ltd (2008). The Actiwatch User Manual V 7.2.

Cannon, M., Jones, P., Gilvarry, C., Rifkin, L., McKenzie, K., Foerster, A., and Murray, R. M. (1997). Premorbid social functioning in schizophrenia and bipolar disorder: Similarities and differences. *American Journal of Psychiatry*, 154(11):1544–1550.

Carney, R. M., Blumenthal, J. A., Stein, P. K., Watkins, L., Catellier, D., Berkman, L. F., Czajkowski, S. M., O'Connor, C., Stone, P. H., and Freedland, K. E. (2001). Depression, heart rate variability, and acute myocardial infarction. *Circulation*, 104(17):2024–2028.

Carpenter, B. E. (1990). Kao Lien's eight treatises on the nurturing of life. *Tezukayama University Review*, 67:38–51.

Catellier, D. J., Hannan, P. J., Murray, D. M., Addy, C. L., Conway, T. L., Yang, S., and

Rice, J. C. (2005). Imputation of missing data when measuring physical activity by accelerometry. *Medicine and Science in Sports and Exercise*, 37(11 Suppl):S555–62.

Chapman, J. J., Brown, W. J., and Burton, N. W. (2015). Defining a valid day of accelerometer monitoring in adults with mental illness. *Mental Health and Physical Activity*.

Chawla, N. V., Bowyer, K. W., Hall, L. O., and Kegelmeyer, W. P. (2002). SMOTE: Synthetic Minority Over-sampling Technique. *Journal of Artificial Intelligence Research*, 16:321–357.

Chen, Y. W. and Dilsaver, S. C. (1996). Lifetime rates of suicide attempts among subjects with bipolar and unipolar disorders relative to subjects with other Axis I disorders. *Biological Psychiatry*, 39(10):896–9.

Clifford, G. D., Tsanas, A., Guy, G., and Saunders, K. (2015). Automated Monitoring of Symptoms Severity (AMoSS) Study Protocol. Technical report. Available by request.

Cohen, H., Kaplan, Z., Kotler, M., Mittelman, I., Osher, Y., and Bersudsky, Y. (2003). Impaired heart rate variability in euthymic bipolar patients. *Bipolar Disorders*, 5(2):138–143.

Cohen, S., Tyrrell, D. A., and Smith, A. P. (1991). Psychological stress and susceptibility to the common cold. *New England Journal of Medicine*, 325(9):606–612.

Cole, R. J., Kripke, D. F., Gruen, W., Mullaney, D. J., and Gillin, J. C. (1992). Automatic sleep/wake identification from wrist activity. *Sleep*, 15(5):461–469.

Cortes, C. and Vapnik, V. (1995). Support-vector networks. *Machine Learning*, 20(3):273–297.

Costa, M., Goldberger, A., and Peng, C.-K. (2005). Multiscale entropy analysis of biological signals. *Physical Review E*, 71(2):021906.

Curone, D., Bertolotti, G. M., Cristiani, A., Secco, E. L., and Magenes, G. (2010). A real-time and self-calibrating algorithm based on triaxial accelerometer signals for the detection of human posture and activity. *IEEE Transactions on Information Technology in Biomedicine*, 14(4):1098–105.

da Silva, I. C., van Hees, V. T., Ramires, V. V., Knuth, A. G., Bielemann, R. M., Ekelund, U., Brage, S., and Hallal, P. C. (2014). Physical activity levels in three Brazilian birth cohorts as assessed with raw triaxial wrist accelerometry. *International Journal of Epidemiology*, 43(6):1959–68.

Dahl, A. A. (1985). Borderline disorders–the validity of the diagnostic concept. *Psychiatric Developments*, 3(2):109–52.

Darbellay, G. A. and Vajda, I. (1999). Estimation of the information by an adaptive partitioning of the observation space. *IEEE Transactions on Information Theory*, 45(4):1315–1321.

Darwin, C. (1872). The expression of the emotions in man and animals. *The American Journal of the Medical Sciences*, 232(4):477.

Davies, P., Fried, R., and Gather, U. (2004). Robust signal extraction for on-line monitoring data. *Journal of Statistical Planning and Inference*, 122(1-2):65–78.

Dawson, M. E., Schell, A. M., Rissling, A., Ventura, J., Subotnik, K. L., and Nuechterlein, K. H. (2010). Psychophysiological prodromal signs of schizophrenic relapse: A pilot study. *Schizophrenia Research*, 123(1):64–67.

Depue, R. A., Krauss, S., Spoont, M. R., and Arbisi, P. (1989). General behavior inventory identification of unipolar and bipolar affective conditions in a nonclinical university population. *Journal of Abnormal Psychology*, 98(2):117–26.

Dougherty, D. M., Marsh, D. M., and Mathias, C. W. (2002). Immediate and delayed

memory tasks: A computerized behavioral measure of memory, attention, and impulsivity. *Behavior Research Methods, Instruments and Computers*, 34(3):391–398.

Ebner-Priemer, U. W., Kuo, J., Welch, S. S., Thielgen, T., Witte, S., Bohus, M., and Linehan, M. M. (2006). A valence-dependent group-specific recall bias of retrospective self-reports: A study of borderline personality disorder in everyday life. *The Journal of Nervous and Mental Disease*, 194(10):774–9.

El-Amrawy, F. and Nounou, M. I. (2015). Are Currently Available Wearable Devices for Activity Tracking and Heart Rate Monitoring Accurate, Precise, and Medically Beneficial? *Healthcare Informatics Research*, 21(4):315–320.

Enders, C. K. (2006). A primer on the use of modern missing-data methods in psychosomatic medicine research. *Psychosomatic Medicine*, 68(3):427–36.

Falloon, I. R., Kydd, R. R., Coverdale, J. H., and Laidlaw, T. M. (1996). Early detection and intervention for initial episodes of schizophrenia. *Schizophrenia Bulletin*, 22(2):271–82.

Falret, J.-P. (1854). Mémoire sur la folie circulaire. *Bulletin de I'Academie de Medicine*, 19:382–415.

Faulkner, G. and Biddle, S. (2013). Standing on top of the world: Is sedentary behaviour associated with mental health? *Mental Health and Physical Activity*, 1(6):1–2.

First, M. B., Gibbon, M., Spitzer, R. L., Williams, J. B. W., and S., B. L. (1997). *Structured Clinical Interview for DSM-IV Axis II Personality Disorders. (SCID-II)*. American Psychiatric Press, Washington, D. C.

First, M. B., Spitzer, R. L., Gibbon, M., and Williams, J. B. W. (1996). *Structured Clinical Interview for DSM-IV Axis I Disorders, Clinician Version (SCID-CV)*. American Psychiatric Press, Washington, D. C.

First, M. B., Spitzer, R. L., Gibbon, M., and Williams, J. B. W. (2002). *Structured Clinical Interview for DSM-IV-TR Axis I Disorders, Research Version, Patient Edition With Psychotic Screen (SCID-I/P W/ PSY SCREEN)*. Biometrics Research, New York.

First, M. B., Williams, J. B. W., Spitzer, R. L., and Gibbon, M. (2007). *Structured Clinical Interview for DSM-IV-TR Axis I Disorders, Clinical Trials Version (SCID-CT)*. Biometrics Research, New York.

Firth, J. and Torous, J. (2015). Smartphone apps for schizophrenia: A systematic review. *JMIR mHealth and uHealth*, 3(4):e102.

Fleischhacker, W., Meise, U., Günther, V., and Kurz, M. (1994). Compliance with antipsychotic drug treatment: influence of side effects. *Acta Psychiatrica Scandinavica*, 89(s382):11–15.

Foster, R. and Wulff, K. (2005). The rhythm of rest and excess. *Nature Reviews Neuroscience*, 6(5):407–414.

Fowles, D. (2007). The three arousal model: Implications of Gray's two-factor learning theory for heart rate, electrodermal activity, and psychopathy. *Psychophysiology*, 17(2):87–104.

Franks, J., Hallam-Baker, P., Hostetler, J., Lawrence, S., Leach, P., Luotonen, A., and Stewart, L. (1999). RFC 2617: HTTP authentication: Basic and digest access authentication.

Franzen, P. L., Siegle, G. J., and Buysse, D. J. (2008). Relationships between affect, vigilance, and sleepiness following sleep deprivation. *Journal of Sleep Research*, 17(1):34–41.

Freedson, P. S. and Miller, K. (2000). Objective monitoring of physical activity using motion sensors and heart rate. *Research Quarterly for Exercise and Sport*, 71(2 Suppl):S21–9.

Friedman, B. and Thayer, J. (1998). Autonomic balance revisited: Panic anxiety and heart rate variability. *Journal of Psychosomatic Research*, 44(1):133–151.

Fung, M. T., Raine, A., Loeber, R., Lynam, D. R., Steinhauer, S. R., Venables, P. H., and Stouthamer-Loeber, M. (2005). Reduced electrodermal activity in psychopathy-prone adolescents. *Journal of Abnormal Psychology*, 114(2):187–196.

Fuster-Garcia, E., Bresó, A., Martínez-Miranda, J., Rosell-Ferrer, J., Matheson, C., and García-Gómez, J. M. (2015). Fusing actigraphy signals for outpatient monitoring. *Information Fusion*, 23:69–80.

Gauss, C. F. (1963). *Theoria motus corporum celestium, No. 1809; Theory of motion of the heavenly bodies.* Dover, New York.

Geddes, J. R. and Miklowitz, D. J. (2013). Treatment of bipolar disorder. *Lancet*, 381(9878):1672–82.

Geoffroy, P. A., Boudebesse, C., Bellivier, F., Lajnef, M., Henry, C., Leboyer, M., Scott, J., and Etain, B. (2014). Sleep in remitted bipolar disorder: A naturalistic case-control study using actigraphy. *Journal of Affective Disorders*, 158:1–7.

Gerard, R. W. (1949). Physiology and psychiatry. *The American Journal of Psychiatry*, 106:161–173.

Gietzelt, M., Wolf, K.-H., Marschollek, M., and Haux, R. (2008). Automatic self-calibration of body worn triaxial-accelerometers for application in healthcare. In *2008 Second International Conference on Pervasive Computing Technologies for Healthcare*, pages 177–180. IEEE.

Goossens, P. J. J., Kupka, R. W., Beentjes, T. A. A., and van Achterberg, T. (2010). Recognising prodromes of manic or depressive recurrence in outpatients with bipo-lar disorder: A cross-sectional study. *International Journal of Nursing Studies*, 47(10):1201–7.

Grant, B. F., Chou, S. P., Goldstein, R. B., Huang, B., Stinson, F. S., Saha, T. D., Smith, S. M., Dawson, D. A., Pulay, A. J., Pickering, R. P., and Ruan, W. J. (2008). Prevalence, correlates, disability, and comorbidity of DSM-IV borderline personality disorder: Results from the Wave 2 National Epidemiologic Survey on Alcohol and Related Conditions. *The Journal of Clinical Psychiatry*, 69(4):533–45.

Gravenhorst, F., Muaremi, A., Gruenerbl, A., Arnrich, B., and Troester, G. (2013). Towards a mobile galvanic skin response measurement system for mentally disordered patients. In *Proceedings of the 8th International Conference on Body Area Networks*, pages 432–435. ACM.

Greco, A., Lanatà, A., Valenza, G., Rota, G., Vanello, N., and Scilingo, E. P. (2012). On the deconvolution analysis of electrodermal activity in bipolar patients. In *Annual International Conference of the IEEE Engineering in Medicine and Biology Society. IEEE Engineering in Medicine and Biology Society.*, volume 2012, pages 6691–4. IEEE.

Greco, A., Valenza, G., Lanata', A., Rota, G., and Scilingo, E. (2014). Electrodermal activity in bipolar patients during affective elicitation. *IEEE Journal of Biomedical and Health Informatics*, 18(6):1865–1873.

Green, M. F. (2006). Cognitive impairment and functional outcome in schizophrenia and bipolar disorder. *The Journal of Clinical Psychiatry*, 67(10):e12.

Greene, T. (2007). The Kraepelinian dichotomy: The twin pillars crumbling? *History of Psychiatry*, 18(3):361–379.

Gruenerbl, A., Osmani, V., Bahle, G., Carrasco, J. C., Oehler, S., Mayora, O., Haring, C., and Lukowicz, P. (2014). Using smart phone mobility traces for the diagnosis of depressive and manic episodes in bipolar patients. In *Proceedings of the 5th Augmented Human International Conference*, pages 1–8, New York, New York, USA. ACM Press.

Gruner, O. C. (1970). *A treatise on the canon of medicine of Avicenna, incorporating a translation of the first book.* Kelley, New York.

Grunerbl, A., Muaremi, A., Osmani, V., Bahle, G., Ohler, S., Troester, G., Mayora, O., Haring, C., and Lukowicz, P. (2014). Smart-phone based recognition of states and state changes in bipolar disorder patients. *IEEE Journal of Biomedical and Health Informatics*, 19(1):140–148.

Guidi, A., Vanello, N., Bertschy, G., Gentili, C., Landini, L., and Scilingo, E. (2014). Automatic analysis of speech F0 contour for the characterization of mood changes in bipolar patients. *Biomedical Signal Processing and Control*, 17:29–37.

Guyon, I. and Elisseeff, A. (2003). An introduction to variable and feature selection. *The Journal of Machine Learning Research*, 3:1157–1182.

Haiminen, N., Mannila, H., and Terzi, E. (2007). Comparing segmentations by applying randomization techniques. *BMC Bioinformatics*, 8(1):171.

Ham, S. A., Reis, J. P., Strath, S. J., Dubose, K. D., and Ainsworth, B. E. (2007). Discrepancies between methods of identifying objectively determined physical activity. *Medicine and Science in Sports and Exercise*, 39(1):52–8.

Hamilton, M. (1960). A rating scale for depression. *Journal of Neurology, Neurosurgery, and Psychiatry*, 23:56–62.

Hare, R. and Quinn, M. (1971). Psychopathy and autonomic conditioning. *Journal of Abnormal Psychology*, 77(3):223–235.

Hauge, E., Berle, J., and Oedegaard, K. (2011). Nonlinear analysis of motor activity shows differences between schizophrenia and depression: A study using Fourier analysis and sample entropy. *PloS One*, 6(1):e16291.

Helmes, E. and Landmark, J. (2003). Subtypes of schizophrenia: A cluster analytic approach. *Canadian Journal of Psychiatry*, 48(10):702–708.

Henry, B. L., Minassian, A., Paulus, M. P., Geyer, M. A., and Perry, W. (2010). Heart rate variability in bipolar mania and schizophrenia. *Journal of Psychiatric Research*, 44(3):168–76.

Henry, C., Mitropoulou, V., New, A. S., Koenigsberg, H. W., Silverman, J., and Siever, L. J. (2001). Affective instability and impulsivity in borderline personality and bipolar II disorders: Similarities and differences. *Journal of Psychiatric Research*, 35(6):307–312.

Hersen, M. (2006). *Clinician's handbook of child behavioral assessment.* Elsvier Academic Press, Burlington, MA.

Hildebrand, M., Van Hees, V. T., Hansen, B. H., and Ekelund, U. (2014). Age group comparability of raw accelerometer output from wrist- and hip-worn monitors. *Medicine and Science in Sports and Exercise*, 46(9):1816–24.

Hinton, J. M. (1966). Physiological measurements in psychiatry. *Proceedings of the Royal Society of Medicine*, 59(10):1028–30.

Hornero, R., Abásolo, D., Jimeno, N., Sánchez, C. I., Poza, J., and Aboy, M. (2006). Variability, regularity, and complexity of time series generated by schizophrenic patients and control subjects. *IEEE Transactions on Biomedical Engineering*, 53(2):210–218.

Hu, K., Van Someren, E. J. W., Shea, S. A., and Scheer, F. A. J. L. (2009). Reduction of scale invariance of activity fluctuations with aging and Alzheimer's disease: Involve-ment of the circadian pacemaker. *Proceedings of the National Academy of Sciences of the United States of America*, 106(8):2490–4.

Hunter, J. S. (1986). The exponentially weighted moving average. *Journal of Quality Technology*, 18(4):203–210.

Huynh, C., Guilé, J.-M., Breton, J.-J., and Godbout, R. (2015). Sleep-wake patterns of adolescents with borderline personality disorder and bipolar disorder. *Child Psychiatry and Human Development*, [Epub].

ICT4Depression Consortium (2013a). ICT4Depression Pilot in Netherlands. Technical report. Available at: http://www.ict4depression.eu/wp/wp-content/uploads/2013/10/Results-Dutch-pilot.pdf.

ICT4Depression Consortium (2013b). ICT4Depression Pilot in Sweden. Technical report. Available at: http://www.ict4depression.eu/wp/wp-content/uploads/2013/10/Results-Swedish-pilot.pdf.

Indic, P., Salvatore, P., Maggini, C., and Ghidini, S. (2011). Scaling behavior of human locomotor activity amplitude: Association with bipolar disorder. *PLoS One*, 6(5):e20650.

Insel, T., Cuthbert, B., Garvey, M., Heinssen, R., Pine, D. S., Quinn, K., Sanislow, C., and Wang, P. (2010). Research domain criteria (RDoC): Toward a new classification framework for research on mental disorders. *The American Journal of Psychiatry*, 167(7):748–51.

Iverson, G. L., Gaetz, M. B., Rzempoluck, E. J., McLean, P., Linden, W., and Remick, R. (2005). A new potential marker for abnormal cardiac physiology in depression. *Journal of Behavioral Medicine*, 28(6):507–11.

Jackson, A., Cavanagh, J., and Scott, J. (2003). A systematic review of manic and depressive prodromes. *Journal of Affective Disorders*, 74(3):209–217.

Jackson, S. W. (1969). Galen — on mental disorders. *Journal of the History of the Behavioral Sciences*, 5(4):365–384.

James, W. (1884). What is an emotion? *Mind*, 9(34):188–205.

Javelot, H., Spadazzi, A., Weiner, L., Garcia, S., Gentili, C., Kosel, M., and Bertschy, G. (2014). Telemonitoring with respect to mood disorders and information and communication technologies: Overview and presentation of the PSYCHE project. *BioMed Research International*, 2014:104658.

Jensen-Urstad, K., Saltin, B., Ericson, M., Storck, N., and Jensen-Urstad, M. (2007). Pronounced resting bradycardia in male elite runners is associated with high heart rate variability. *Scandinavian Journal of Medicine & Science in Sports*, 7(5):274–278.

Jones, S. H., Hare, D. J., and Evershed, K. (2005). Actigraphic assessment of circadian activity and sleep patterns in bipolar disorder. *Bipolar Disorders*, 7(2):176–86.

Jones, W. H. S. (1923). *Hippocrates, Volume II.* Harvard University Press (Loeb Classical Library), London, Cambridge.

Jung, C. G. (1921). *Psychologische Typen.* Rascher Verlag, Zurich.

Kahn, M., Sheppes, G., and Sadeh, A. (2013). Sleep and emotions: Bidirectional links and underlying mechanisms. *International Journal of Psychophysiology*, 89(2):218–228.

Kappeler-Setz, C., Gravenhorst, F., Schumm, J., Arnrich, B., and Tröster, G. (2011). Towards long term monitoring of electrodermal activity in daily life. *Personal and Ubiquitous Computing*, 17(2):261–271.

Keitner, G. I., Solomon, D. A., Ryan, C. E., Miller, I. W., Mallinger, A., Kupfer, D. J., and Frank, E. (1996). Prodromal and residual symptoms in bipolar I disorder. *Comprehensive Psychiatry*, 37(5):362–367.

Kessler, R. C., Andrews, G., Colpe, L. J., Hiripi, E., Mroczek, D. K., Normand, S. L. T., Walters, E. E., and Zaslavsky, A. M. (2002). Short screening scales to monitor population prevalences and trends in non-specific psychological distress. *Psychological Medicine*, 32(6):959–76.

Kiecolt-Glaser, J. K., Page, G. G., Marucha, P. T., MacCallum, R. C., and Glaser, R. (1998). Psychological influences on surgical recovery. Perspectives from psychoneuroimmunology. *The American Psychologist*, 53(11):1209–18.

Kiehl, K. A., Smith, A. M., Hare, R. D., and Liddle, P. F. (2000). An event-related potential investigation of response inhibition in schizophrenia and psychopathy. *Biological Psychiatry*, 48(3):210–221.

Klosterkötter, J., Hellmich, M., Steinmeyer, E. M., and Schultze-Lutter, F. (2001). Diagnosing schizophrenia in the initial prodromal phase. *Archives of General Psychiatry*, 58(2):158–64.

Kohavi, R. (1995). A study of cross-validation and bootstrap for accuracy estimation and model selection. In *IJCAI'95 Proceedings of the 14th International Joint Conference on Artificial Intelligence*, pages 1137–1143. Morgan Kaufmann Publishers Inc.

Kraemer, H. C., Kupfer, D. J., Clarke, D. E., Narrow, W. E., and Regier, D. A. (2012). DSM-5: How reliable is reliable enough? *American Journal of Psychiatry*, 169(1):13–15.

Kraepelin, E. (1921). Manic depressive insanity and paranoia. *The Journal of Nervous and Mental Disease*, 53(4):350.

Kraepelin, E. (1987). Dementia praecox.

Kroenke, K., Spitzer, R. L., and Williams, J. B. W. (2001). The PHQ-9. *Journal of General Internal Medicine*, 16(9):606–613.

Lacey, B. and Lacey, J. (1974). Studies of heart rate and other bodily processes in sensorimotor behavior. In P. A. Obrist, A. H. Black, J. Brener, and L. V. DiCara, editors, *Cardiovascular Psychophysiology*, pages 538–564. Aldine Transaction, New Brunswick, NJ, US.

Lacey, J. (1959). Psychophysiological approaches to the evaluation of psychotherapeutic process and outcome. In E. A. Rubinstein M. B. Parloff, editor, *Research in Psychotherapy*, pages 160–208. American Psychological Association, Washington, DC, US.

Lanata, A., Valenza, G., Nardelli, M., Gentili, C., and Scilingo, E. P. (2015). Complexity index from a personalized wearable monitoring system for assessing remission in mental health. *IEEE Journal of Biomedical and Health Informatics*, 19(1):132–9.

Lang, P. J., Bradley, M. M., and Cuthbert, B. N. (1998). Emotion, motivation, and anxiety: Brain mechanisms and psychophysiology. *Biological Psychiatry*, 44(12):1248–1263.

Larson, J. A. (1923). The cardio-pneumo-psychogram in deception. *Journal of Experimental Psychology*, 6(6):420–454.

Lathia, N., Pejovic, V., Rachuri, K. K., Mascolo, C., Musolesi, M., and Rentfrow, P. J. (2013). Smartphones for large-scale behavior change interventions. *IEEE Pervasive Computing*, 12(3):66–73.

Lauderdale, D. S., Knutson, K. L., Yan, L. L., Liu, K., and Rathouz, P. J. (2008). Self-reported and measured sleep duration: How similar are they? *Epidemiology*, 19(6):838–45.

Layard, R. (2012). How mental illness loses out in the NHS: A report by The Centre for Economic Performance's Mental Health Policy Group. CEP Special Reports CEPSP26, CEP.

Lee, J., Nemati, S., Silva, I., Edwards, B. A., Butler, J. P., and Malhotra, A. (2012). Transfer entropy estimation and directional coupling change detection in biomedical time series. *Biomedical Engineering Online*, 11(1):19.

Legendre, A. M. (1805). *Nouvelles méthodes pour la détermination des orbites des comètes.* Courcier, Paris.

Lehman, A. F., Lieberman, J. A., Dixon, L. B., McGlashan, T. H., Miller, A. L., Perkins, D. O., and Kreyenbuhl, J. (2004). Practice guideline for the treatment of patients with schizophrenia, second edition. *The American Journal of Psychiatry*, 161(2 Suppl):1–56.

Lejuez, C. W., Read, J. P., Kahler, C. W., Richards, J. B., Ramsey, S. E., Stuart, G. L., Strong, D. R., and Brown, R. A. (2002). Evaluation of a behavioral measure of risk taking: the Balloon Analogue Risk Task (BART). *Journal of Experimental Psychology.*, 8(2):75–84.

Li, Q., Mark, R. G., and Clifford, G. D. (2008). Robust heart rate estimation from multiple asynchronous noisy sources using signal quality indices and a Kalman filter. *Physiological measurement*, 29(1):15–32.

Liddle, P. (2001). *Disordered mind and brain: The neural basis of mental symptoms.* Cromwell Press, Trowbridge, UK.

Liddle, P. F. (1987). The symptoms of chronic schizophrenia. A re-examination of the positive- negative dichotomy. *The British Journal of Psychiatry*, 151(2):145–151.

Lieb, K., Zanarini, M. C., Schmahl, C., Linehan, M. M., and Bohus, M. (2004). Borderline personality disorder. *Lancet*, 364(9432):453–61.

Littner, M., Kushida, C. A., Anderson, W. M., Bailey, D., Berry, R. B., Davila, D. G., Hirshkowitz, M., Kapen, S., Kramer, M., Loube, D., Wise, M., and Johnson, S. F.

(2003). Practice parameters for the role of actigraphy in the study of sleep and circadian rhythms: An update for 2002. *Sleep*, 26(3):337–41.

Lobbestael, J., Leurgans, M., and Arntz, A. (2010). Inter-rater reliability of the Structured Clinical Interview for DSM-IV Axis I Disorders (SCID I) and Axis II Disorders (SCID II). *Clinical Psychology and Psychotherapy*, 18(1):75–9.

Lorber, M. (2004). Psychophysiology of aggression, psychopathy, and conduct problems: A meta-analysis. *Psychological Bulletin*, 130(4):531–552.

Lötters, J., Schipper, J., Veltink, P., Olthuis, W., and Bergveld, P. (1998). Procedure for in-use calibration of triaxial accelerometers in medical applications. *Sensors and Actuators A: Physical*, 68(1-3):221–228.

Løvendahl, P. and Chagunda, M. G. G. (2010). On the use of physical activity monitoring for estrus detection in dairy cows. *Journal of Dairy Science*, 93(1):249–59.

Lu, S., Zhao, H., Ju, K., Shin, K., and Lee, M. (2008). Can photoplethysmography variability serve as an alternative approach to obtain heart rate variability information? *Journal of Clinical Monitoring and Computing*, 22(1):23–29.

Maj, M. (1998). Critique of the DSM-IV operational diagnostic criteria for schizophrenia. *The British Journal of Psychiatry*, 172(6):458–458.

Maj, M. (2005). "Psychiatric comorbidity": An artefact of current diagnostic systems? *The British Journal of Psychiatry*, 186(3):182–4.

Marczinski, C. A. and Fillmore, M. T. (2003). Preresponse cues reduce the impairing effects of alcohol on the execution and suppression of responses. *Experimental and Clinical Psychopharmacology*, 11(1):110–117.

Mathers, C. D., Fat, D. M., and Boerma, J. T. (2008). *The global burden of disease: 2004 update*. World Health Organization.

Mayers, A. G. and Baldwin, D. S. (2015). The relationship between sleep disturbance and depression. *International Journal of Psychiatry in Clinical Practice*, 10(1):2–16.

McCall, W. V. (2015). A rest-activity biomarker to predict response to SSRIs in major depressive disorder. *Journal of Psychiatric Research*, 64:19–22.

McClung, C. A. (2013). How might circadian rhythms control mood? Let me count the ways... *Biological Psychiatry*, 74(4):242–9.

McCraty, R., Atkinson, M., Tomasino, D., and Stuppy, W. (2001). Analysis of twenty-four hour heart rate variability in patients with panic disorder. *Biological Psychology*, 56(2):131–150.

McCrone, P. (2008). Paying the price: The cost of mental health care in England to 2026.

McElroy, S. L., Kotwal, R., Malhotra, S., Nelson, E. B., Keck, P. E., and Nemeroff, C. B. (2004). Are mood disorders and obesity related? A review for the mental health professional. *The Journal of Clinical Psychiatry*, 65(5):634–51.

McSharry, P., Clifford, G., Tarassenko, L., and Smith, L. (2002). Method for generating an artificial RR tachogram of a typical healthy human over 24-hours. In *Computers in Cardiology*, pages 225–228. IEEE.

Migo, E. M., Haynes, B. I., Harris, L., Friedner, K., Humphreys, K., and Kopelman, M. D. (2014). mHealth and memory aids: Levels of smartphone ownership in patients. *Journal of Mental Health*, 24(5):266–270.

Miklowitz, D. J., Price, J., Holmes, E. A., Rendell, J., Bell, S., Budge, K., Christensen, J., Wallace, J., Simon, J., Armstrong, N. M., McPeake, L., Goodwin, G. M., and Geddes, J. R. (2012). Facilitated integrated mood management for adults with bipolar disorder. *Bipolar Disorders*, 14(2):185–97.

Minassian, A., Henry, B. L., Geyer, M. A., Paulus, M. P., Young, J. W., and Perry, W. (2010). The quantitative assessment of motor activity in mania and schizophrenia. *Journal of Affective Disorders*, 120(1):200–206.

Mohr, D. C., Schueller, S. M., Araya, R., Gureje, O., and Montague, E. (2014). Mental health technologies and the needs of cultural groups. *The Lancet Psychiatry*, 1(5):326–327.

Molnar, G., Feeney, M. G., and Fava, G. A. (1988). Duration and symptoms of bipolar prodromes. *The American Journal of Psychiatry*, 145(12):1576–8.

Moore, P. J., Little, M. A., McSharry, P. E., Geddes, J. R., and Goodwin, G. M. (2012). Forecasting depression in bipolar disorder. *IEEE Transactions on Biomedical Engineering*, 59(10):2801–7.

Moore, P. J., Little, M. A., McSharry, P. E., Goodwin, G. M., and Geddes, J. R. (2014). Correlates of depression in bipolar disorder. *Proceedings. Biological Sciences / The Royal Society*, 281(1776):20132320.

Morriss, R. (2004). The early warning symptom intervention for patients with bipolar affective disorder. *Advances in Psychiatric Treatment*, 10(1):18–26.

Mueser, K. T., Corrigan, P. W., Hilton, D. W., Tanzman, B., Schaub, A., Gingerich, S., Essock, S. M., Tarrier, N., Morey, B., Vogel-Scibilia, S., and Herz, M. I. (2002). Illness management and recovery: A review of the research. *Psychiatric Services*, 53(10):1272–84.

Mueser, K. T., Torrey, W. C., Lynde, D., Singer, P., and Drake, R. E. (2003). Implementing evidence-based practices for people with severe mental illness. *Behavior Modification*, 27(3):387–411.

Murphy, K. P. (2007). Conjugate bayesian analysis of the gaussian distribution. Techni-cal report. Available at: https://www.cs.ubc.ca/~murphyk/Papers/bayesGauss.pdf.

Narrow, W. E., Clarke, D. E., Kuramoto, S. J., Kraemer, H. C., Kupfer, D. J., Greiner, L., and Regier, D. A. (2013). DSM-5 field trials in the United States and Canada, Part III: Development and reliability testing of a cross-cutting symptom assessment for DSM-5. *The American Journal of Psychiatry*, 170(1):71–82.

Naslund, J. A., Aschbrenner, K. A., Barre, L. K., and Bartels, S. J. (2015). Feasibility of popular m-health technologies for activity tracking among individuals with serious mental illness. *Telemedicine Journal and E-health*, 21(3):213–6.

Oakley, N. R. (1997). Validation with polysomnography of the Sleepwatch sleep/wake scoring algorithm used by the Actiwatch activity monitoring system. *Bend: Mini Mitter, Cambridge Neurotechnology*. Available by request.

Öhman, A. (1981). Electrodermal activity and vulnerability to schizophrenia: A review. *Biological Psychology*, 12(2–3):87–145.

O'Regan, C., Kenny, R. A., Cronin, H., Finucane, C., and Kearney, P. M. (2015). Antidepressants strongly influence the relationship between depression and heart rate variability: Findings from The Irish Longitudinal Study on Ageing (TILDA). *Psychological Medicine*, 45(3):623–36.

Ortiz-Tudela, E., Martinez-Nicolas, A., Campos, M., Rol, M. Á., and Madrid, J. A. (2010). A new integrated variable based on thermometry, actimetry and body position (TAP) to evaluate circadian system status in humans. *PLoS Computational Biology*, 6(11):e1000996.

Osipov, M., Behzadi, Y., Kane, J. M., Petrides, G., and Clifford, G. D. (2015). Objective identification and analysis of physiological and behavioral signs of schizophrenia. *Journal of Mental Health*, 24(5):276–282.

Osipov, M., Wulff, K., Foster, R. G., and Clifford, G. D. (2013). Multiscale entropy of physical activity as an objective measure of activity disorganization in a context of schizophrenia. In *IX Congreso Internacional de Informática en Salud 2013*, pages 1–9.

Osmani, V., Maxhuni, A., Grünerbl, A., Lukowicz, P., Haring, C., and Mayora, O. (2013). Monitoring activity of patients with bipolar disorder using smart phones. In *Proceedings of International Conference on Advances in Mobile Computing & Multimedia - MoMM '13*, pages 85–92, New York, New York, USA. ACM Press.

Ouzir, M. (2013). Impulsivity in schizophrenia: A comprehensive update. *Aggression and Violent Behavior*, 18(2):247–254.

Palmius, N., Osipov, M., Goodwin, G. M., Saunders, K., Bilderbeck, A. C., Tsanas, A., and Clifford, G. D. (2014). A multi-sensor monitoring system for objective mental health management in resource constrained environments. In *Appropriate Healthcare Technologies for Low Resource Settings (AHT 2014)*, pages 4–4. Institution of Engineering and Technology.

Paradiso, R., Bianchi, A. M., Lau, K., and Scilingo, E. P. (2010). PSYCHE: Personalised monitoring systems for care in mental health. In *Annual International Conference of the IEEE Engineering in Medicine and Biology Society.*, volume 2010, pages 3602–5.

Peng, C.-K., Buldyrev, S., Havlin, S., Simons, M., Stanley, H., and Goldberger, A. (1994). Mosaic organization of DNA nucleotides. *Physical Review E*, 49(2):1685–1689.

Peng, H., Long, F., and Ding, C. (2005). Feature selection based on mutual information criteria of max-dependency, max-relevance, and min-redundancy. *IEEE Transactions on Pattern Analysis and Machine Intelligence*, 27(8):1226–1238.

Perez-Diaz de Cerio, D., Ruiz Boque, S., Rosell-Ferrer, J., Ramos-Castro, J., and Castro, J. (2012). A wireless sensor network design for the Help4Mood european project. In *European Wireless, 2012. EW. 18th European Wireless Conference*, pages 1–6.

Perry, A., Tarrier, N., Morriss, R., McCarthy, E., and Limb, K. (1999). Randomised controlled trial of efficacy of teaching patients with bipolar disorder to identify early symptoms of relapse and obtain treatment. *BMJ*, 318(7177):149–153.

Phillips, M. L. and Kupfer, D. J. (2013). Bipolar disorder diagnosis: Challenges and future directions. *Lancet*, 381(9878):1663–71.

Pompili, M., Gonda, X., Serafini, G., Innamorati, M., Sher, L., Amore, M., Rihmer, Z., and Girardi, P. (2013). Epidemiology of suicide in bipolar disorders: A systematic review of the literature. *Bipolar Disorders*, 15(5):457–490.

Pomrock, Y. and Ginath, Y. (1982). Differences between schizophrenics and normal controls using chirological (hand) testing. *The Israel Journal of Psychiatry and Related Sciences*, 19(1):5–22.

Prociow, P., Wac, K., and Crowe, J. (2012). Mobile psychiatry: Towards improving the care for bipolar disorder. *International Journal of Mental Health Systems*, 6(1):5.

Rabiner, L. (1989). A tutorial on hidden Markov models and selected applications in speech recognition. *Proceedings of the IEEE*, 77(2):257–286.

Rachow, T., Berger, S., and Boettger, M. (2011). Nonlinear relationship between electrodermal activity and heart rate variability in patients with acute schizophrenia. *Psychophysiology*, 48(10):1323–1332.

Rachuri, K. K., Musolesi, M., Mascolo, C., Rentfrow, P. J., Longworth, C., and Aucinas, A. (2010). EmotionSense: A mobile phones based adaptive platform for experimental social psychology research. In *Proceedings of the 12th ACM international conference on Ubiquitous computing*, Ubicomp '10, pages 281–290, New York, NY, USA. ACM.

Rasmussen, C. E. and Williams, C. K. I. (2006). *Gaussian processes in machine learning*, volume 14 of *Lecture Notes in Computer Science*. The MIT Press.

Rechlin, T., Weis, M., Spitzer, A., and Kaschka, W. (1994). Are affective disorders associated with alterations of heart rate variability? *Journal of Affective Disorders*, 32(4):271–275.

Rechtschaffen, A. and Kales, A. (1968). *A manual of standardized terminology, techniques and scoring system for sleep stages of human subjects.* UCLA Brain Information Service, Brain Research Institute, Los Angeles.

Refinetti, R., Cornélissen, G., and Halberg, F. (2007). Procedures for numerical analysis of circadian rhythms. *Biological Rhythm Research*, 38(4):275–325.

Regier, D. A. (2007). Time for a fresh start? Rethinking psychosis in DSM-V. *Schizophrenia Bulletin*, 33(4):843–5.

Regier, D. A., Narrow, W. E., Clarke, D. E., Kraemer, H. C., Kuramoto, S. J., Kuhl, E. A., and Kupfer, D. J. (2013). DSM-5 field trials in the United States and Canada, Part II: Test-retest reliability of selected categorical diagnoses. *The American Journal of Psychiatry*, 170(1):59–70.

Rescorla, E. (2000). RFC 2818: HTTP over TLS.

Richman, J. and Moorman, J. (2000). Physiological time-series analysis using approximate entropy and sample entropy. *American Journal of Physiology-Heart and Circulatory Physiology*, 278(6):H2039–2049.

Ritschel, L. A. and Kilpela, L. S. (2014). Borderline personality disorder. In *The Encyclopedia of Clinical Psychology*. John Wiley & Sons, Inc., New Jersey, USA.

Roberts, S., Osborne, M., Ebden, M., Reece, S., Gibson, N., and Aigrain, S. (2013). Gaussian processes for time-series modelling. *Philosophical Transactions. Series A, Mathematical, Physical, and Engineering sciences*, 371(1984):20110550.

Rock, P., Goodwin, G., Harmer, C., and Wulff, K. (2014). Daily rest-activity patterns in the bipolar phenotype: A controlled actigraphy study. *Chronobiology International*, 31(2):290–296.

Roebuck, A., Monasterio, V., Gederi, E., Osipov, M., Behar, J., Malhotra, A., Penzel, T., and Clifford, G. D. (2013). A review of signals used in sleep analysis. *Physiological Measurement*, 35(1):R1–R57.

Rosa, A. R., Singh, N., Whitaker, E., de Brito, M., Lewis, A. M., Vieta, E., Churchill, G. C., Geddes, J. R., and Goodwin, G. M. (2014). Altered plasma glutathione levels in bipolar disorder indicates higher oxidative stress; a possible risk factor for illness onset despite normal brain-derived neurotrophic factor (BDNF) levels. *Psychological Medicine*, 44(11):1–10.

Rosenberg, D. E., Bellettiere, J., Gardiner, P. A., Villarreal, V. N., Crist, K., and Kerr, J. (2015). Independent associations between sedentary behaviors and mental, cognitive, physical, and functional health among older adults in retirement communities. *The Journals of Gerontology. Series A*, 71(1):78–83.

Rothemund, Y., Ziegler, S., and Hermann, C. (2012). Fear conditioning in psychopaths: Event-related potentials and peripheral measures. *Biological Psychology*, 90(1):50–59.

Rowlands, A. V., Fraysse, F., Catt, M., Stiles, V. H., Stanley, R. M., Eston, R. G., and Olds, T. S. (2014a). Comparability of measured acceleration from accelerometry-based activity monitors.

Rowlands, A. V., Olds, T. S., Hillsdon, M., Pulsford, R., Hurst, T. L., Eston, R. G., Gomersall, S. R., Johnston, K., and Langford, J. (2014b). Assessing sedentary behavior with the GENEActiv: introducing the sedentary sphere. *Medicine and science in sports and exercise*, 46(6):1235–47.

Rubin, D. B. (1976). Inference and missing data. *Biometrika*, 63(3):581–592.

Rush, A., Trivedi, M. H., Ibrahim, H. M., Carmody, T. J., Arnow, B., Klein, D. N., Markowitz, J. C., Ninan, P. T., Kornstein, S., Manber, R., Thase, M. E., Kocsis, J. H., and Keller, M. B. (2003). The 16-Item quick inventory of depressive symptomatology (QIDS), clinician rating (QIDS-C), and self-report (QIDS-SR): A psychometric evaluation in patients with chronic major depression. *Biological Psychiatry*, 54(5):573–583.

Rush, A. J. J., First, M. B., and Blacker, D., editors (2008). *Handbook of psychiatric measures*. American Psychiatric Publishing, Arlington, VA, 2 edition.

Sabia, S., van Hees, V. T., Shipley, M. J., Trenell, M. I., Hagger-Johnson, G., Elbaz, A., Kivimaki, M., and Singh-Manoux, A. (2014). Association between questionnaire- and accelerometer-assessed physical activity: The role of sociodemographic factors. *American Journal of Epidemiology*, 179(6):781–90.

Sadeh, A. (2011). The role and validity of actigraphy in sleep medicine: An update. *Sleep Medicine Reviews*, 15(4):259–267.

Sadeh, A., Keinan, G., and Daon, K. (2004). Effects of stress on sleep: The moderating role of coping style. *Health Psychology*, 23(5):542–5.

Sadeh, A., Sharkey, K., and Carskadon, M. (1994). Activity-based sleep-wake identification: An empirical test of methodological issues. *Sleep*, 17(3):201–207.

Sansone, R. A. and Sansone, L. A. (2011). Gender patterns in borderline personality disorder. *Innovations in Clinical Neuroscience*, 8(5):16–20.

Santos, R., Sousa, J., Marques, C. J., Gamboa, H., and Silva, H. (2012). Towards emotion related feature extraction based on generalized source-independent event detection. *Proceedings of the 2nd International Workshop on Computing Paradigms for Mental Health (MindCare 2012)*, pages 71–78.

Saunders, K., Bilderbeck, A., Price, J., and Goodwin, G. (2015). Distinguishing bipolar disorder from borderline personality disorder: A study of current clinical practice. *European Psychiatry*, 30(8):965–974.

Sayers, J. (2001). The world health report 2001 - mental health: New understanding, new hope. *Bulletin of the World Health Organization*, 79(11):1085.

Scargle, J. D., Norris, J. P., Jackson, B., and Chiang, J. (2013). Studies in astronomical time series analysis. VI. Bayesian block representations. *The Astrophysical Journal*, 764(2):167.

Scarpa, A. and Raine, A. (1997). Psychophysiology of anger and violent behavior. *Psychiatric Clinics of North America*, 20(2):375–394.

Schaefer, C. A., Nigg, C. R., Hill, J. O., Brink, L. A., and Browning, R. C. (2014). Establishing and evaluating wrist cutpoints for the GENEActiv accelerometer in youth. *Medicine and Science in Sports and Exercise*, 46(4):826–833.

Schafer, J. L. and Olsen, M. K. (1998). Multiple imputation for multivariate missing-data problems: A data analyst's perspective. *Multivariate Behavioral Research*, 33(4):545–571.

Schou, M., Baastrup, P. C., Grof, P., Weis, P., and Angst, J. (1970). Pharmacological and clinical problems of lithium prophylaxis. *The British Journal of Psychiatry*, 116(535):615–619.

Schreiber, T. (2000). Measuring information transfer. *Physical Review Letters*, 85(2):461–464.

Scott, J. (2011). Clinical parameters of circadian rhythms in affective disorders. *European Neuropsychopharmacology*, 21 Suppl 4:S671–5.

Senin, P. (2008). Dynamic time warping algorithm review. *Information and Computer Science Department University of Hawaii at Manoa Honolulu, USA*, pages 1–23.

Shedler, J., Mayman, M., and Manis, M. (1993). The illusion of mental health. *American Psychologist*, 48(11):1117–1131.

Sierra, P., Livianos, L., Arques, S., Castelló, J., and Rojo, L. (2009). Prodromal symptoms to relapse in bipolar disorder. *Australian and New Zealand Journal of Psychiatry*, 41(5):385–91.

Silva, H., Sousa, J., and Gamboa, H. (2012). Study and evaluation of palmar blood volume pulse for heart rate monitoring in a multimodal framework. *Proceedings of the 2nd International Workshop on Computing Paradigms for Mental Health (MindCare 2012)*, pages 35–40.

Singh, I. and Rose, N. (2009). Biomarkers in psychiatry. *Nature*, 460(7252):202–7.

Smeraldi, E., Benedetti, F., Barbini, B., Campori, E., and Colombo, C. (1999). Sustained antidepressant effect of sleep deprivation combined with pindolol in bipolar depression. A placebo-controlled trial. *Neuropsychopharmacology*, 20(4):380–5.

Sokolove, P. and Bushell, W. (1978). The chi square periodogram: Its utility for analysis of circadian rhythms. *Journal of Theoretical Biology*, 72(1):131–160.

Spitzer, R. L., Kroenke, K., Williams, J. B. W., and Löwe, B. (2006). A brief measure for assessing generalized anxiety disorder: The GAD-7. *Archives of Internal Medicine*, 166(10):1092–7.

Stanley, B. and Wilson, S. T. (2006). Heightened subjective experience of depression in borderline personality disorder. *Journal of Personality Disorders*, 20(4):307–18.

Steenari, B. M.-R. and Aronen, E. (2002). Actigraph placement and sleep estimation in children. *Sleep*, 25(2):235–237.

Steiner, H. (2011). *Handbook of developmental psychiatry.* World Scientific Publishing Company, Singapore.

Stone, A. A. and Shiffman, S. (2002). Capturing momentary, self-report data: A proposal for reporting guidelines. *Annals of Behavioral Medicine*, 24(3):236–243.

Sung, M., Marci, C., and Pentland, A. (2005). Objective physiological and behavioral measures for identifying and tracking depression state in clinically depressed patients. Technical report, Massachusetts Institute of Technology. Available at: http://hd.media.mit.edu/tech-reports/TR-595.pdf.

Swann, A. C. (2009). Impulsivity in mania. *Current Psychiatry Reports*, 11(6):481–7.

Swartz, A. M., Strath, S. J., Bassett, D. R., O'Brien, W. L., King, G. A., and Ainsworth, B. E. (2000). Estimation of energy expenditure using CSA accelerometers at hip and wrist sites. *Medicine and Science in Sports and Exercise*, 32(9 Suppl):S450–456.

Taillard, J., Lemoine, P., Boule, P., Drogue, M., and Mouret, J. (1993). Sleep and heart rate circadian rhythm in depression: The necessity to separate. *Chronobiology International*, 10(1):63–72.

Tandon, R., Keshavan, M. S., and Nasrallah, H. A. (2008). Schizophrenia, "just the facts" what we know in 2008. 2. Epidemiology and etiology. *Schizophrenia Research*, 102(1-3):1–18.

Tandon, R., Nasrallah, H. A., and Keshavan, M. S. (2009). Schizophrenia, "just the facts" 4. Clinical features and conceptualization. *Schizophrenia Research*, 110(1-3):1–23.

Task Force of the European Society of Cardiology the North American Society of Pacing Electrophysiology (1996). Heart rate variability: Standards of measurement, physiological interpretation, and clinical use. *Circulation*, 93(5):1043–1065.

te Lindert, B. H. W. and Van Someren, E. J. W. (2013). Sleep estimates using microelectromechanical systems (MEMS). *Sleep*, 36(5):781–9.

Teicher, M. H. (1995). Actigraphy and motion analysis: New tools for psychiatry. *Harvard Rev Psychiatry*, 3:18–35.

Tibshirani, R. (1996). Regression shrinkage and selection via the lasso. *Journal of the Royal Statistical Society. Series B (Methodological)*, pages 267–288.

Torous, J., Staples, P., and Onnela, J.-P. (2015). Realizing the potential of mobile mental health: New methods for new data in psychiatry. *Current Psychiatry Reports*, 17(8):602.

Tracy, D. J., Xu, Z., Choi, L., Acra, S., Chen, K. Y., and Buchowski, M. S. (2014). Separating bedtime rest from activity using waist or wrist-worn accelerometers in youth. *PloS One*, 9(4):e92512.

Troiano, R. P., Berrigan, D., Dodd, K. W., Mâsse, L. C., Tilert, T., and McDowell, M. (2008). Physical activity in the United States measured by accelerometer. *Medicine and Science in Sports and Exercise*, 40(1):181–8.

Troiano, R. P., McClain, J. J., Brychta, R. J., and Chen, K. Y. (2014). Evolution of accelerometer methods for physical activity research. *British journal of sports medicine*, 48(13):1019–23.

Trost, S. G. (2001). Objective measurement of physical activity in youth: Current issues, future directions. *Exercise and Sport Sciences Reviews*, 29(1):32–6.

Tsanas, A., Little, M. A., McSharry, P. E., and Ramig, L. O. (2010). Accurate telemonitoring of Parkinson's disease progression by noninvasive speech tests. *IEEE Transactions on Biomedical Engineering*, 57(4):884–893.

Tsoi, D. T.-Y., Hunter, M. D., and Woodruff, P. W. (2008). History, aetiology, and symptomatology of schizophrenia. *Psychiatry*, 7(10):404–409.

Tucker, J. M., Welk, G. J., and Beyler, N. K. (2011). Physical activity in U.S.: Adults compliance with the Physical Activity Guidelines for Americans. *American Journal of Preventive Medicine*, 40(4):454–61.

Tudor-Locke, C., Brashear, M. M., Johnson, W. D., and Katzmarzyk, P. T. (2010). Accelerometer profiles of physical activity and inactivity in normal weight, overweight, and obese U.S. men and women. *The International Journal of Behavioral Nutrition and Physical Activity*, 7:60.

Tukey, J. W. (1977). Exploratory data analysis. *Analysis*, 2(1999):688.

Turner, R., Saatci, Y., and Rasmussen, C. E. (2009). Adaptive sequential Bayesian change point detection. Technical report. Available at: http://mlg.eng.cam.ac.uk/pub/pdf/TurSaaRas09.pdf.

Valenza, G., Citi, L., Gentili, C., Lanata, A., Scilingo, E. P., and Barbieri, R. (2015). Characterization of depressive states in bipolar patients using wearable textile technology and instantaneous heart rate variability assessment. *IEEE Journal of Biomedical and Health Informatics*, 19(1):263–74.

Valenza, G., Gentili, C., Lanatà, A., and Scilingo, E. P. (2013). Mood recognition in bipolar patients through the PSYCHE platform: Preliminary evaluations and perspectives. *Artificial Intelligence in Medicine*, 57(1):49–58.

Valenza, G., Nardelli, M., Bertschy, G., Lanatà, A., Barbieri, R., and Scilingo, E. P. (2014a). Maximal-radius multiscale entropy of cardiovascular variability: A promising biomarker of pathological mood states in bipolar disorders. In *Annual International Conference of the IEEE Engineering in Medicine and Biology Society.*, volume 2014, pages 6663–6.

Valenza, G., Nardelli, M., Bertschy, G., Lanata, A., and Scilingo, E. P. (2014b). Complexity modulation in heart rate variability during pathological mental states of bipolar disorders. In *2014 8th Conference of the European Study Group on Cardiovascular Oscillations (ESGCO)*, pages 99–100. IEEE.

Valkonen-Korhonen, M., Tarvainen, M. P., Ranta-Aho, P., Karjalainen, P. A., Partanen,

J., Karhu, J., and Lehtonen, J. (2003). Heart rate variability in acute psychosis. *Psychophysiology*, 40(5):716–726.

van de Ven, P., Henriques, M. R., Hoogendoorn, M., Klein, M., McGovern, E., Nelson, J., Silva, H., and Tousset, E. (2012). A mobile system for treatment of depression. *Proceedings of the 2nd International Workshop on Computing Paradigms for Mental Health (MindCare 2012)*, pages 47–58.

Van de Water, A. T. M., Holmes, A., and Hurley, D. A. (2011). Objective measurements of sleep for non-laboratory settings as alternatives to polysomnography - a systematic review. *Journal of Sleep Research*, 20(1.1):183–200.

van Hees, V. T., Fang, Z., Langford, J., Assah, F., Mohammad, A., da Silva, I. C. M., Trenell, M. I., White, T., Wareham, N. J., and Brage, S. (2014). Autocalibration of accelerometer data for free-living physical activity assessment using local gravity and temperature: An evaluation on four continents. *Journal of Applied Physiology*, 117(7):738–44.

van Hees, V. T., Gorzelniak, L., Dean León, E. C., Eder, M., Pias, M., Taherian, S., Ekelund, U., Renström, F., Franks, P. W., Horsch, A., and Brage, S. (2013). Separating movement and gravity components in an acceleration signal and implications for the assessment of human daily physical activity. *PloS One*, 8(4):e61691.

van Hees, V. T., Renström, F., Wright, A., Gradmark, A., Catt, M., Chen, K. Y., Löf, M., Bluck, L., Pomeroy, J., Wareham, N. J., Ekelund, U., Brage, S., and Franks, P. W. (2011). Estimation of daily energy expenditure in pregnant and non-pregnant women using a wrist-worn tri-axial accelerometer. *PloS One*, 6(7):e22922.

Van Someren, E. (1999). Bright light therapy: Improved sensitivity to its effects on rest-activity rhythms in Alzheimer patients by application of nonparametric methods. *Chronobiology International*, 16(4):505–518.

Vancampfort, D., Knapen, J., Probst, M., Scheewe, T., Remans, S., and De Hert, M. (2012). A systematic review of correlates of physical activity in patients with schizophrenia. *Acta Psychiatrica Scandinavica*, 125(5):352–62.

Ventura, J., Nuechterlein, K. H., Subotnik, K. L., Gutkind, D., and Gilbert, E. A. (2000). Symptom dimensions in recent-onset schizophrenia and mania: A principal components analysis of the 24-item Brief Psychiatric Rating Scale. *Psychiatry Research*, 97(2-3):129–135.

Walther, S., Horn, H., and Razavi, N. (2009). Quantitative motor activity differentiates schizophrenia subtypes. *Neuropsychobiology*, 60(2):80–86.

Walther, S., Ramseyer, F., Horn, H., Strik, W., and Tschacher, W. (2014). Less structured movement patterns predict severity of positive syndrome, excitement, and disorganization. *Schizophrenia Bulletin*, 40(3):585–91.

Warmerdam, L., Riper, H., Klein, M., van den Ven, P., Rocha, A., Ricardo Henriques, M., Tousset, E., Silva, H., Andersson, G., and Cuijpers, P. (2012). Innovative ICT solutions to improve treatment outcomes for depression: The ICT4Depression project. *Studies in Health Technology and Informatics*, 181:339–43.

Webster, J. B., Kripke, D. F., Messin, S., Mullaney, D. J., and Wyborney, G. (1982). An activity-based sleep monitor system for ambulatory use. *Sleep*, 5(4):389–99.

Weissman, M. M. (1996). Cross-national epidemiology of major depression and bipolar disorder. *The Journal of the American Medical Association*, 276(4):293.

Westen, D., Moses, M. J., Silk, K. R., Lohr, N. E., Cohen, R., and Segal, H. (1992). Quality of depressive experience in borderline personality disorder and major depression: When depression is not just depression. *Journal of Personality Disorders*, 6(4):382–393.

Weston, A., Barton, C., Lesselyong, J., and Yaffe, K. (2011). Functional deficits among patients with mild cognitive impairment. *Alzheimer's and Dementia*, 7(6):611–4.

Widiger, T. A. and Weissman, M. M. (1991). Epidemiology of borderline personality disorder. *Psychiatric Services*, 42(10):1015–1021.

Winters, P. R. (1960). Forecasting sales by exponentially weighted moving averages. *Management Science*, 6(3):324–342.

Wirz-Justice, A. (2006). Biological rhythm disturbances in mood disorders. *International Clinical Psychopharmacology*, 21 Suppl 1:S11–5.

Wirz-Justice, A. (2007a). Chronobiology and psychiatry. *Sleep Medicine Reviews*, 11(6):423–427.

Wirz-Justice, A. (2007b). How to measure circadian rhythms in humans. *Medicographia*, 29(1):84–90.

Wirz-Justice, A., Haug, H., and Cajochen, C. (2001). Disturbed circadian rest-activity cycles in schizophrenia patients: An effect of drugs? *Schizophrenia Bulletin*, 27(3):497 –502.

Wirz-Justice, A., Schröder, C. M., Gasio, P. F., Cajochen, C., and Savaskan, E. (2010). The circadian rest-activity cycle in Korsakoff psychosis. *The American Journal of Geriatric Psychiatry*, 18(1):33–41.

Witting, W., Kwa, I., and Eikelenboom, P. (1990). Alterations in the circadian rest-activity rhythm in aging and Alzheimer's disease. *Biological Psychiatry*, 27(6):563–572.

Wolters, M., Burton, C., Matheson, C., Bresó, A., Szentagotai, A., Martinez-Miranda, J., Fuster, E., Rosell, J., Pagliari, C., and McKinstry, B. (2013). Help4Mood - supporting joint sense making in the treatmet of major depressive disorder. *Workshop on Interactive Systems in Healthcare (WISH), Washington DC.*

Wulff, K., Dijk, D.-J., and Middleton, B. (2012). Sleep and circadian rhythm disruption in schizophrenia. *The British Journal of Psychiatry*, 200(4):308–316.

Wulff, K., Gatti, S., Wettstein, J. G., and Foster, R. G. (2010). Sleep and circadian rhythm disruption in psychiatric and neurodegenerative disease. *Nature Reviews. Neuroscience*, 11(8):589–99.

Wulff, K., Joyce, E., Middleton, B., Dijk, D.-J., and Foster, R. G. (2006). The suitability of actigraphy, diary data, and urinary melatonin profiles for quantitative assessment of sleep disturbances in schizophrenia: A case report. *Chronobiology International*, 23(1-2):485–495.

Young, R. C., Biggs, J. T., Ziegler, V. E., and Meyer, D. A. (1978). A rating scale for mania: reliability, validity and sensitivity. *The British Journal of Psychiatry*, 133:429–35.

Zaragozá, I., Rey, B., Botella, C., Baños, R., Moragrega, I., Castilla, D., and Alcañiz, M. (2011). A user-friendly tool for detecting the stress level in a person's daily life. In *Human-Computer Interaction. Design and Development Approaches. Lecture Notes in Computer Science*, volume 6761, pages 423–431. Springer Berlin Heidelberg.

Zhang, S., Rowlands, A. V., Murray, P., and Hurst, T. L. (2012). Physical activity classification using the GENEA wrist-worn accelerometer. *Medicine and Science in Sports and Exercise*, 44(4):742–8.

Zhang, X., Hu, B., Zhou, L., Moore, P., and Chen, J. (2013). An EEG based pervasive depression detection for females. In *Pervasive Computing and the Networked World. Lecture Notes in Computer Science*, volume 7719, pages 848–861. Springer Berlin Heidelberg.

# Appendices

# Appendix A

# The Altman Self-Rating scale for Mania (ASRM)

## A.1 Instructions

1. On this questionnaire are groups of five statements; read each group of statements carefully.
2. Choose the one statement in each group that best describes the way you have been feeling for the past week.
3. Circle the number next to the statement you picked.

*Please note:* The word "occasionally" when used here means once or twice; "often" means several times or more; "frequently" means most of the time.

## A.2 Questions

1. Positive mood:
   (a) I do not feel happier or more cheerful than usual.
   (b) I occasionally feel happier or more cheerful than usual.

(c) I often feel happier or more cheerful than usual.

(d) I feel happier or more cheerful than usual most of the time.

(e) I feel happier or more cheerful than usual all of the time.

2. Self-confidence:

    (a) I do not feel more self-confident than usual.

    (b) I occasionally feel more self-confident than usual.

    (c) I often feel more self-confident than usual.

    (d) I feel more self-confident than usual.

    (e) I feel extremely self-confident all of the time.

3. Sleep patterns:

    (a) I do not need less sleep than usual.

    (b) I occasionally need less sleep than usual.

    (c) I often need less sleep than usual.

    (d) I frequently need less sleep than usual.

    (e) I can go all day and night without any sleep and still not feel tired.

4. Speech:

    (a) I do not talk more than usual.

    (b) I occasionally talk more than usual.

    (c) I often talk more than usual.

    (d) I frequently talk more than usual.

    (e) I talk constantly and cannot be interrupted.

5. Activity level:

(a) I have not been more active (either socially, sexually, at work, home or school) than usual.

(b) I have occasionally been more active than usual.

(c) I have often been more active than usual.

(d) I have frequently been more active than usual.

(e) I am constantly active or on the go all the time.

# Appendix B

# The Quick Inventory of Depressive Symptomatology, Self-Report (QIDS-SR16)

## B.1 Instructions

Please circle the one response to each item that best describes you for the past seven days.

## B.2 Questions

1. Falling Asleep:

    (a) I never take longer than 30 minutes to fall asleep.

    (b) I take at least 30 minutes to fall asleep, less than half the time.

    (c) I take at least 30 minutes to fall asleep, more than half the time.

    (d) I take more than 60 minutes to fall alseep, more than half the time.

2. Sleep During the Night:

(a) I do not wake up at night.

(b) I have a restless, light sleep with a few brief awakenings each night.

(c) I wake up at least once a night, but I go back to sleep easily.

(d) I awaken more than once a night and stay awake for 20 minutes or more, more than half the time.

3. Waking Up Too Early:

   (a) Most of the time, I awaken no more than 30 minutes before I need to get up.

   (b) More than half the time, I awaken more than 30 minutes before I need to get up.

   (c) I almost always awaken at least one hour or so before I need to, but I go back to sleep eventually.

   (d) I awaken at least one hour before I need to, and can't go back to sleep.

4. Sleeping Too Much:

   (a) I sleep no longer than 7–8 hours/night, without napping during the day.

   (b) I sleep no longer than 10 hours in a 24-hour period including naps.

   (c) I sleep no longer than 12 hours in a 24-hour period including naps.

   (d) I sleep longer than 12 hours in a 24-hour period including naps.

5. Feeling Sad:

   (a) I do not feel sad.

   (b) I feel sad less than half the time.

   (c) I feel sad more than half the time.

   (d) I feel sad nearly all of the time.

6. Decreased Appetite:

(a) There is no change in my usual appetite.
(b) I eat somewhat less often or lesser amounts of food than usual.
(c) I eat much less than usual and only with personal effort.
(d) I rarely eat within a 24-hour period, and only with extreme personal effort or when others persuade me to eat.

7. Increased Appetite:
   (a) There is no change from my usual appetite.
   (b) I feel a need to eat more frequently than usual.
   (c) I regularly eat more often and/or greater amounts of food than usual.
   (d) I feel driven to overeat both at mealtime and between meals.

8. Decreased Weight (Within the Last Two Weeks):
   (a) I have not had a change in my weight.
   (b) I feel as if I've had a slight weight loss.
   (c) I have lost 2 pounds or more.
   (d) I have lost 5 pounds or more.

9. Increased Weight (Within the Last Two Weeks):
   (a) I have not had a change in my weight.
   (b) I feel as if I've had a slight weight gain.
   (c) I have gained 2 pounds or more.
   (d) I have gained 5 pounds or more.

10. Concentration/Decision Making:
    (a) There is no change in my usual capacity to concentrate or make decisions.

(b) I occasionally feel indecisive or find that my attention wanders.

(c) Most of the time, I struggle to focus my attention or to make decisions.

(d) I cannot concentrate well enough to read or cannot make even minor decisions.

11. View of Myself:

    (a) I see myself as equally worthwhile and deserving as other people.

    (b) I am more self-blaming than usual.

    (c) I largely believe that I cause problems for others.

    (d) I think almost constantly about major and minor defects in myself.

12. Thoughts of Death or Suicide:

    (a) I do not think of suicide or death.

    (b) I feel that life is empty or wonder if it's worth living.

    (c) I think of suicide or death several times a week for several minutes.

    (d) I think of suicide or death several times a day in some detail, or I have made specific plans for suicide or have actually tried to take my life.

13. General Interest:

    (a) There is no change from usual in how interested I am in other people or activities.

    (b) I notice that I am less interested in people or activities.

    (c) I find I have interest in only one or two of my formerly pursued activities.

    (d) I have virtually no interest in formerly pursued activities.

14. Energy Level:

    (a) There is no change in my usual level of energy.

    (b) I get tired more easily than usual.

(c) I have to make a big effort to start or finish my usual daily activities (for example, shopping, homework, cooking or going to work).

(d) I really cannot carry out most of my usual daily activities because I just don't have the energy.

15. Feeling Slowed Down:

    (a) I think, speak, and move at my usual rate of speed.

    (b) I find that my thinking is slowed down or my voice sounds dull or flat.

    (c) It takes me several seconds to respond to most questions and I'm sure my thinking is slowed.

    (d) I am often unable to respond to questions without extreme effort.

16. Feeling Restless:

    (a) I do not feel restless.

    (b) I'm often fidgety, wringing my hands, or need to shift how I am sitting.

    (c) I have impulses to move about and am quite restless.

    (d) At times, I am unable to stay seated and need to pace around.

# Appendix C

# The Generalised Anxiety Disorder 7-item scale (GAD-7)

## C.1 Instructions

Over the last 2 weeks, how often have you been bothered by the following problems?

## C.2 Questions

1. Feeling nervous, anxious, or on edge:
   (a) Not at all.
   (b) Several days.
   (c) More than half the days.
   (d) Nearly every day.
2. Not being able to stop or control worrying:
   (a) Not at all.
   (b) Several days.

(c) More than half the days.

(d) Nearly every day.

3. Worrying too much about different things:

   (a) Not at all.

   (b) Several days.

   (c) More than half the days.

   (d) Nearly every day.

4. Trouble relaxing:

   (a) Not at all.

   (b) Several days.

   (c) More than half the days.

   (d) Nearly every day.

5. Being so restless that it is hard to sit still:

   (a) Not at all.

   (b) Several days.

   (c) More than half the days.

   (d) Nearly every day.

6. Becoming easily annoyed or irritable:

   (a) Not at all.

   (b) Several days.

   (c) More than half the days.

   (d) Nearly every day.

7. Feeling afraid as if something awful might happen:

   (a) Not at all.

   (b) Several days.

   (c) More than half the days.

   (d) Nearly every day.